# Ultracold Atomic Gases
# in Artificial Magnetic Fields

Von der Fakultät für Mathematik und Physik der

Gottfried Wilhelm Leibniz Universität Hannover

zur Erlangung des Grades

**Doktor der Naturwissenschaften**

— Doctor rerum naturalium —

(Dr. rer. nat.)

genehmigte

**Dissertation**

von

**Dipl. Phys. Klaus Osterloh**

geboren am 15. April 1977 in Herford in Nordrhein-Westfalen

2006

*To Anna-Katharina*

*"A thing of beauty is a joy forever, its loveliness increases, it will never pass into nothingness."*

John Keats

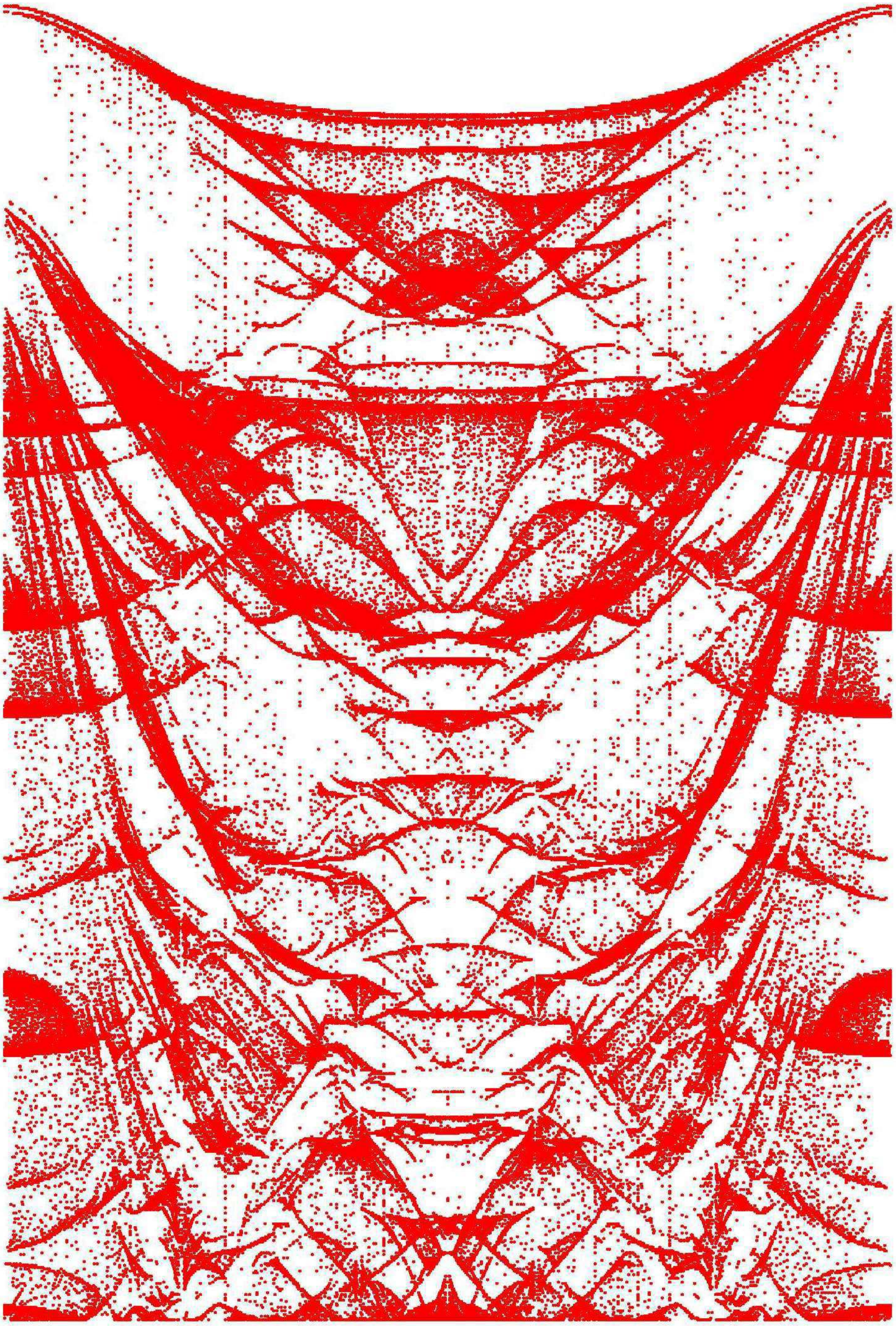

# Abstract

A phenomenon can hardly be found that accompanied physical paradigms and theoretical concepts in a more reflecting way than magnetism. From the beginnings of metaphysics and the first classical approaches to magnetic poles and streamlines of the field, it has inspired modern physics on its way to the classical field description of electrodynamics, and further to the quantum mechanical description of internal degrees of freedom of elementary particles. Meanwhile, magnetic manifestations have posed and still do pose complex and often controversially debated questions. This regards so various and utterly distinct topics as quantum spin systems and the grand unification theory. This may be foremost caused by the fact that all of these effects are based on correlated structures, which are induced by the interplay of dynamics and elementary interactions. It is strongly correlated systems that certainly represent one of the most fascinating and universal fields of research. In particular, low dimensional systems are in the focus of interest, as they reveal strongly pronounced correlations of counterintuitive nature. As regards this framework, the quantum Hall effect must be seen as one of the most intriguing and complex problems of modern solid state physics. Even after two decades and the same number of Nobel prizes, it still keeps researchers of nearly all fields of physics occupied. In spite of seminal progress, its inherent correlated order still lacks understanding on a microscopic level. Despite this, it is obvious that the phenomenon is thoroughly fundamental of nature. To resolve some puzzles of this nature is a key topic of this thesis. The chosen approach might be, at first sight, confusing, as ultracold atomic gases are electrically neutral objects, and their quantum mechanical characteristics only pave the way for successful cooling in optical and magnetic traps. Yet, it is because of the rapid development of quantum optical experimental techniques during the last decade that there are methods at hand to grant access to various interesting models and problems of the theory of condensed matter. Exactly the amalgamation of different physical disciplines, here atomic and molecular physics and quantum optics, have contributed to this decisively. Not only is it possible to trap atoms in nearly perfect periodic crystals of light, in which their interactions can be determined in strength and type quasi at will, but also to subject them to artificial magnetic fields by phase imprinting and rotation-induced centrifugal potentials. In such systems, the trapped particles react as if they were charged objects and are subject to effective vector potentials. The dimensionality of the trap systems can also be controlled. Paradigms of strongly correlated systems, such as the Mott insulator and the one-dimensional Tonks Girardeau gas, have recently been experimentally achieved and detected. Experiments with rotating Bose gases have continuously drawn nearer to the quantum Hall regime. Great effort is being made to achieve a stable rotation of fermions soon, and to venture into the lowest strongly correlated Landau level in microscopic atomic probes. A main focus of this thesis is devoted to the ground state structures appearing in this regime and their elementary excitations on the crossover from weakly interacting states to strongly correlated Laughlin liquids. The possibilities of quantum-mechanical manipulation of ultracold atomic gases have, as regards the above phenomena, yet by far not been exhausted. Completely novel physical systems, which cannot be found in nature in this form, can be achieved by mixing atomic species or by simultaneously controlling the internal atomic degrees of freedom. The latter is facilitated by the realization of effective isospins, which can be exposed to state-selective vector potentials in optical lattices. These gauge potentials act on the space of atomic isospins and can be accordingly modeled as operators of symmetry groups. Thereby, non-Abelian vector potentials can be generated, inducing complex correlated atom dynamics, which can be solely observed in this form in elementary field theories. The resulting opportunities for generalized quantum Hall systems and simulation of lattice gauge theories conclude the investigations of this thesis.

# Zusammenfassung

Kaum ein Phänomen hat die Entwicklung physikalischer Gedankenmodelle und theoretischer Konzepte reflektierender begleitet als der Magnetismus. Von den Anfängen der Metaphysik und ersten klassischen Beschreibungen magnetischer Pole und Feldlinien inspirierte er die moderne Physik auf ihrem Weg zur klassischen Feldbeschreibung der Elektrodynamik bis hin zur quantenmechanischen Beschreibung innerer Freiheitsgrade der Elementarteilchen. Magnetische Erscheinungsformen werfen hierbei bis zum heutigen Tage komplexe und oftmals kontrovers diskutierte Fragestellungen auf, und dies in so vielfältigen und scheinbar grundverschiedenen Gebieten wie Quantenspinsystemen und der großen Vereinheitlichungstheorie. Dies mag vor allem darin begründet sein, dass diesen Effekten korrelierte Strukturen zugrunde liegen, die durch das Wechselspiel von Dynamik und elementaren Wechselwirkungen induziert werden. Stark korrelierte Systeme stellen wohl eines der spannendsten und universellsten Forschungsgebiete der modernen Physik. Insbesondere niederdimensionale Systeme sind hierbei von besonderem Interesse, da in diesen Korrelationen besonders ausgeprägt und von kontraintuitiver Natur sind. Der Quanten-Hall Effekt ist in diesem Rahmen sicher eines der fesselndsten und komplexesten Probleme moderner Festkörperphysik und beschäftigt selbst nach zwei Jahrzehnten und der gleichen Zahl ausgelobter Nobelpreise Forscher aus nahezu allen Teilbereichen der Physik. Die ihm inhärente magnetisch korrelierte Ordnung ist trotz wegweisender Fortschritte mikroskopisch weiterhin nur unvollständig verstanden. Offenkundig ist hingegen, dass dieses Phänomen von ausgesprochen fundamentaler Natur ist. Dieser ein wenig näher zu kommen, ist ein Kernthema dieser Arbeit. Der Zugang scheint hierbei zunächst verwirrend, da ultrakalte atomare Gase elektrisch neutrale Objekte sind und ihre quantenmagnetischen Charakteristika lediglich den Weg zu ihrer erfolgreichen Kühlung in optischen und magnetischen Fallen bahnen. Aufgrund der rapiden Entwicklung quantenoptischer Experimentiertechniken im letzten Jahrzehnt stehen jedoch mittlerweile Methoden zur Verfügung, die einen Zugang zu vielfältigen interessanten Modellen und Problemen der Theorie kondensierter Materie gewährleisten. Gerade die Verschmelzung verschiedener physikalischer Disziplinen wie die der Atom- und Mo-lekülphysik mit der Quantenoptik hat hierzu entscheidend beigetragen. Nicht nur lassen sich heutzutage Atome in nahezu perfekten periodischen Potentialstrukturen fangen, in denen ihre Wechselwirkung quasi nach Belieben in Stärke und Art durchstimmbar ist, sondern darüber hinaus durch Phasenaufprägung und rotationsinduzierte Zentrifugalpotentiale dem Effekt artifizieller Magnetfelder unterwerfen. In solchen Systemen verhalten sich die gefangenen Teilchen als wären sie geladene Objekte und stehen in Wechselwirkung mit effektiven Vektorpotentialen. Die Dimensionalität der Fallensysteme kann hierbei ebenfalls kontrolliert werden. Paradigmen stark wechselwirkender Systeme wie der Mott Isolator und das eindimensionale Tonks-Girardeau Gas sind kürzlich experimentell realisiert und detektiert worden. Experimente mit rotierenden Bose Gasen nähern sich kontinuierlich dem Quanten-Hall Regime. Große Anstrengungen werden unternommen, bald auch Fermionen stabil rotieren zu können und gerade in mikroskopischen atomaren Proben ins stark korrelierte niedrigste Landau Niveau vorzustoßen. Den in diesem Regime auftretenden Grundzustandsstrukturen und ihren elementaren Anregungen im Übergang von schwach wechselwirkenden Zuständen zu stark korrelierten Laughlin Flüssigkeiten ist ein Schwerpunkt dieser Arbeit gewidmet. Die Analyse behandelt hierbei kurzreichweitig wechselwirkende Bose und dipolare Fermi Gase. Im Fokus der Untersuchungen stehen hierbei bosonische Vortexstrukturen und eine kritische Gegenüberstellung der verschiedenen Wechselwirkungen. Die Möglichkeiten quantenmechanischer Manipulation ultrakalter atomarer Gase sind im Rahmen der obigen Phänomene jedoch bei weitem nicht erschöpft. Vollkommen neuartige Quantensysteme, die in dieser Form in der Natur nicht vorkommen, lassen sich durch Mischung atomarer Spezies oder die simultane Kontrolle innerer atomarer Freiheitsgrade verwirklichen. Letzteres ermöglicht insbesondere die Realisierung effektiver Isospins, die zustandsselektiven Vektorpotentialen in optischen Gittern ausgesetzt werden können. Diese wirken nun auf dem Raum der atomaren Isospins und sind zum Beispiel als Operatoren von Symmetriegruppen modellierbar. Dadurch lassen sich nicht-Abelsche Vektorpotentiale generieren, die eine komplex korrelierte atomare Dynamik bewirken wie sie in solcher Form nur in elementaren Feldtheorien vorzufinden ist. Die sich hierdurch bietenden Möglickeiten im Hinblick auf verallgemeinerte Quanten-Hall Systeme und die Simulation von Gittereichtheorien bilden den Abschluß der dieser Arbeit zugrunde liegenden Untersuchungen.

The scientific results underlying this thesis are based on the following articles

*“Fractional Quantum Hall States*
*in Ultracold Rapidly Rotating Dipolar Fermi Gases”*,
M. A. Baranov, K. Osterloh, and M. Lewenstein,
Phys. Rev. Lett. **94**, 070404 (2005).

*“Cold Atoms in Non-Abelian Gauge Potentials:*
*From the Hofstadter ”Moth” to Lattice Gauge Theory”*,
K. Osterloh, M. Baig, L. Santos, P. Zoller, and M. Lewenstein,
Phys. Rev. Lett. **95**, 010403 (2005).

*“Ordered Structures in Rotating Ultracold Bose Gases”*,
N. Barberán, M. Lewenstein, K. Osterloh, and D. Dagnino,
Phys. Rev. A **73**, 063623 (2006).

*“Strongly Correlated States*
*of Ultracold Rotating Dipolar Fermi Gases”*,
K. Osterloh, N. Barberán, and M. Lewenstein,
Phys. Rev. Lett. **99**, 160403 (2007).

Contributions to further publications

*“Travelling to Exotic Places with Ultracold Atoms”*,
M. Lewenstein, A. Kubasiak, J. Larson, C. Menotti,
G. Morigi, K. Osterloh, and A. Sanpera,
AIP Proceedings, Melville (2006), vol. **869**, pp. 201-211.

*“Symmetry Breaking in Small Rotating Clouds of Trapped Ultracold Bose Atoms”*,
D. Dagnino, N. Barberán, M. Lewenstein, K. Osterloh, and A. Riera,
Phys. Rev. A **76**, 013625 (2007).

# Contents

*"Die Theorie ist hübsch, aber ob auch etwas Wahres dran ist?"*

"The theory is beautiful, but is there any truth in it?"

Albert Einstein in a letter on Bose-Einstein condensation to Paul Ehrenfest in 1924.

# Introduction

In the ongoing search of truth and perception towards a fundamental understanding of nature, scientific curiosity has always been inspired by the observation of puzzling phenomena. Theoretical concepts, which aim at their explanation, often tend to produce even more mysterious offsprings. One of the most famous examples is certainly Planck's solution of the black-body radiation problem [1]. The formulation of his quantum hypothesis gave rise to the concept of the photon, which earned Albert Einstein the Nobel prize [2]. Subsequent studies of Bose on a proper statistical description of this novel particle [3] culminated in Einstein's seminal quantum theory of the ideal Bose gas [4, 5]. A direct prediction of this theory is the existence of a new collective state of quantum matter, nowadays known as Bose-Einstein condensate. Einstein was struck by the beauty of his result, yet, at the same time questioned, if it was but an artificial construct of his model. He should not live up to the first experimental observation of condensed atomic samples in rubidium [6], sodium [7], and lithium [8, 9].

Yet another paradigmatic phenomenon started being a constant companion to scientists, long before the first mathematical and conceptually logical theoretical approaches had been developed. Magnetism has astonished, confused, and excited physicists in every scientific era. On the path towards the demystification of the lodestone [10], the closed formulation of electromagnetism in terms of Maxwell's classical field equations [11] already marks a very modern step. These equations unified certain former approaches and ruled out others at the same time. They provided for the first time a conceptual separation between cause and effect of electromagnetic fields. This was more thoroughly conceived on the level of gauge symmetric vector potentials defining the fields. From this modern point of view, Maxwell's equations even contain the interpretation of the fields as a direct consequence of moving charges and embody special relativity [12]. Still, the elementary sources of the fields were unknown. After the discovery of the electron at the dawn of the 20th century [13], the nature of deeply rooted magnetic struc-tures in the constituents of matter finally became concrete. Consecutive detailed investigations of the notorious inconsistencies, which plagued the classical theory of magnetism, initiated the demise of classical physics, and the flourishing era of the quantum world began. Research on magnetic quantum phenomena is assuredly one of the most active and complex fields of modern condensed matter theory [14, 15]. Various manifestations of magnetism in material sciences nowadays raise complex problems, which are controversially discussed in a wide range of physical disciplines. This diversity is based on the fact that correlated structures are particularly pronounced in magnetic systems.

The present work deals with physical phenomena, where both of the above quantum worlds, cold quantum gases and magnetism, merge. To motivate this approach, this introductory chapter is separated into two parts consisting of four sections. Sec. (a) is devoted to strongly correlated phenomena in general, which are arguably the most exciting and universal, yet, at the same time least understood field of physics. In this context, particularly lower-dimensional systems are in the focus of interest. There, correlations are ubiquitous and of counterintuitive nature. Besides the fascinating properties of low-dimensional spin-systems, the quantum Hall effect is certainly one of the most intriguing phenomena of condensed matter theory [16, 17]. Despite seminal discoveries in the last two decades, several of them even awarded with the Nobel prize [18, 19, 20], the inherent, magnetically correlated structures still lack a rigorous microscopic understanding. But it is well-known is instead that the underlying puzzles arise from fundamental physical principles. The understanding of some specific characteristics and the investigation of experimentally feasible manifestations of this effect in ultracold gaseous systems are core issues of this thesis. The necessary basic knowledge on quantum Hall states is provided by a self-contained short introduction given in Sec. (b).

At first glance, the chosen approach to the quantum Hall effect might seem confusing. Quantum gases are neutral systems, and in most cases, the magnetic spin degrees of freedom of the atomic compound play a role solely in the sophisticated trapping and cooling techniques [21, 22]. Yet, the rapid progress in quantum optical experiments made powerful manipulation and control schemes available, which could only be dreamt of a decade ago. These schemes considerably extended the scope of the pioneering experiments, which concentrated on and finally succeeded in the achievement of quantum degeneracy in bosonic gases. It has to be stressed that this vast development mostly became possible since various disciplines of physics joined their strengths and efforts. Two of the most important ingredients which are required to subject neutral atoms to the effects of artificial magnetic fields are presented in the second part of the introduction. First, Sec. (c) illuminates the phenomenological world of rotating gases, before Sec. (d) concludes with a brief presentation of optical lattice potentials.

The main part of the thesis is devoted to the analysis of particle dynamics in the presence of magnetic vector potentials. In Chapter 1, the experimental feasibility of incompressible fractional quantum Hall-like states in ultra-cold two-dimensional and rapidly rotating dipolar Fermi gases is demonstrated. In particular, a substantial energy gap is identified in the quasi-particle excitation spectrum of the most prominent Laughlin state at filling fraction $\nu = 1/3$. In this context, dipolar atomic gases are discussed as natural candidates of systems that allow for the realization these highly interesting strongly correlated states in future experiments. Particularly, the importance and effects of a truly long-range interaction are addressed in terms of experimental observability and detection.

Presently, promising experimental approaches to reach the lowest Landau level regime aim at multiple copies of samples with but a few particles. In such arrangements, effects of mesoscopic or even microscopic system size strongly influence the particle dynamics. The unique possibility to observe such small systems allows for a close view at the fundamental crossover phenomena on the path to a strongly correlated quantum Hall system. The analysis of Chapter 2 is concentrated on the evolution of ordered ground state structures in these microscopic rotating samples.

The first part of Chapter 2 is devoted to short-range interacting bosons. It is demonstrated how incipient vortex structures are nucleated due to effects of broken symmetry. In contrast to experiments with a large number of atoms, vorticity is only expected for specific values of the rotational frequency, where ground states of different angular momentum become degenerate. Furthermore, the nature of the bosonic Laughlin state at filling $\nu = 1/2$ is investigated which shows “crystalline” order for microscopic numbers of particles. This distinctness of ”crystallization” ceases to be manifest when the bulk structure of the system becomes important. This behavior is accompanied by an increase of inherent correlations. Thus, for bigger particle numbers, the ground state at Laughlin filling factors evolves towards a true quantum liquid.

The second part of this chapter focuses on dipolar interacting Fermi gases and ties up with the results of Chapter 1. The investigation follows the conceptual approach applied to the bosons. Within this context, comparisons with previously obtained results are made and extensions are discussed. As a starting point, general spectral features are illuminated. Then, the evolution of ground state structures towards the strongly correlated regime is investigated. In this crossover regime, frustration of competing series of states leads to ground states with an incipient quasi-hole in the center. These may be linked with proposals of an effective theory of composite fermions. The analysis concludes with the detailed investigation of dipolar Laughlin states. The effect of long-range interparticle correlations is contrasted with their short-range counter-part, and the non-trivial identification of elementary excitations is addressed. In contrast to the precedent section, the analysis is concentrated on the manifestation of bulk phenomena. For this reason, much larger systems have been exactly diagonalized.

In the framework of Chapter. 3, a completely novel method is proposed to subject atomic samples to the effects of artificial non-Abelian magnetic vector potentials. The method employs atoms with multiple internal states and laser assisted state-sensitive tunneling, described by unitary operators, which act on the atomic "isospins". The single particle dynamics in the case of intense $U(2)$ vector potentials leads to a generalized Hofstadter butterfly spectrum which shows a complex "moth"-like structure of robust holes. It is argued conclusively that non-Abelian Aharonov-Bohm interferometry can be realized in these samples and that many-body dynamics of ultracold matter in external lattice gauge fields can be studied.

The last chapter summarizes the results of this thesis and puts them into further context. Unresolved issues are stressed and directions of possible future research are proposed. Technical details are deferred to appendices A-C.

# SECTION (a)

# Strongly Correlated Systems

In the historical development of physics, the class of systems which is assuredly best understood shares one fundamental property. The physics in these systems can be somehow mapped to a quasi free problem or, more precisely speaking, to a weakly interacting theory. The many-body description is hereby obtained in the same simple manner used in a quantum mechanical product ansatz. In other words, the partition function of the system factorizes to single particle partitions, or equivalently stated

$$1 + 1 = 2\,.$$

This free ansatz is weakly perturbed by external fields and interactions between the particles. Relevant quantities of such theories are calculated via the identification of a tractable power series in the corresponding small parameter. Well-established theoretical frameworks are mean-field methods and Landau's theory of quantum Fermi liquids [23]. The wide range of phenomena covered by perturbative approaches in condensed matter, atomic, molecular, and high energy physics includes band theory of semiconductors, Bose-Einstein condensation and quantum electrodynamics.

Even conventional superconductivity belongs to the above class of problems. Though the dominant property in these systems is the interaction between electrons and ions of the crystal lattice, the effect can be successfully described in the limit of weak coupling. This leads to far-ranged correlations that favor the formation of weakly interacting quasi-particles, the well-known Cooper pairs. Their dynamics is governed by the famous Bardeen-Cooper-Schrieffer Hamiltonian [24].

The enthusiasm about the achievements of these methods is significantly weakened by the simple fact that they are incapable of explaining, how a little iron button fixes a piece of paper, telling “Please do not disturb – examination!”, to the doorframe. Quantum ferromagnetism is certainly one of the well-known examples where correlations between particles are dominating the physical properties. The particular problem of band ferromagnetism in conducting materials is still unsolved.

Other prominent candidates include quark confinement, high temperature superconductivity, Mott insulators, and various lower dimensional problems. Among the latter class, especially 1+1- and 2+0-dimensional systems are highly interesting. There, quantum fluctuation and correlations are most pronounced. The nature of these specific strongly correlated systems is simultaneously far better understood than higher-dimensional problems. This is simply caused by the fact that the majority of applicable methods is restricted to this case.

## a.1 Theoretical Approaches

In the above species of systems, the dynamics of two electrons is no longer determined by the pair of them, but by all electrons in the "vicinity". If this argumentation is continued, the conclusion is simply that the behavior of one particle is influenced by all others and vice versa. The perturbative approach is inherently unsuitable to describe physics of this kind in a proper way [25] . Despite this apparent lack, enormous effort has been made to gain results within this theoretical framework, e. g., by implementing concepts of universality as renormalization group flow [26, 27] or symmetry constraints as Ward identities [28]. For a quite general set of problems, however, it was proven 'that, how, and most importantly, why' this approach fails [29].

This problem is circumvented, when a mapping of the complex interacting system to a weakly interacting effective theory exists. In low-dimensional systems, the most prominent and successful discovery of such a transformation is bosonization [30, 31, 32, 33] which was inspired by the pioneering work of Jordan and Wigner [34]. In this approach, fermions are reformulated in terms of vertex operators, i. e., exponentials in bosonic fields. Correlations are absorbed by the nonlinearity of these objects. The application of bosonization made a large class of problems tractable and led, among other achievements, to the identification of a universal effective theory of one-dimensional metals in the spectral sector of low energy, named Tomonaga-Luttinger liquids [35, 36, 37, 38].

Internal degrees of freedom are difficult in the formalism of bosonization, because distinct spin states have to be mapped to bosonic fields which have to preserve the original symmetries. The resolution to this difficulty was achieved by a non-Abelian generalization of the bosonization method, where the internal symmetry is represented by the corresponding group algebra on the level of currents [39, 40, 41, 42].

Yet, there exists a variety of systems in the space of Hamiltonians, which is governed by highly non-trivial scattering processes. Correlations in this class of systems can be sufficiently pronounced that they exceed the predictive power of the above methods. Despite this, a subset of these Hamiltonians may be treated by an inherently different approach. For a comprehensive introduction to the consecutively discussed field of integrable systems and, in particular, to the Hubbard model, the reader is referred to [43].

The pioneer in this field of studies was Hans Bethe, who managed to reformulate the problem of the half-spin Heisenberg chain in terms of a finite set of algebraic equations [44]. This mathematical treatment miraculously allowed for the exact derivation of ground state properties and excitations. The underlying deeply rooted mathematical features of this ansatz gave birth to the theoretical discipline of integrable models. Successively, a variety of problems, most of them 1+1-dimensional, was identified to be solvable by this approach, e. g., the two-dimensional Ising model [45], the contact interacting 1D Bose gas [46], and six-vertex models [47, 48].

The generalization to internal degrees of freedom proved again to be very difficult, but was finally accomplished in terms of a more profound ansatz governed by the Yang-Baxter equations [49, 50, 51, 52]. The constraint for its successful application is the factorization of multi-particle scattering processes to the level of particle pairs. Soon after this breakthrough, the algebraic structure of these equations and the related transfer matrices were discovered to be deeply linked with the existence of an infinite set of conserved quantities [53]. In this context, "integrability" of a Hamiltonian system could be understood from first principles. Consecutively, a systematic approach termed 'Quantum Inverse Scattering Method' was developed in order to construct and solve integrable Hamilton operators [54].

Integrals of motion are fundamentally connected to symmetries of the Lagrangian [55]. Starting from an algebraic level, a system with an infinite number of commuting operators is supposed to be integrable. Conformal invariance in 1+1-dimensions guarantees the existence of such an infinite algebra. In this way, conformal field theories are a very special class of integrable systems and led to a revolution in the physics of low-dimensional systems in various fields of research [56].

Though integrable systems are quite unique objects in the space of Hamiltonians, their qualitative characteristics should hold for Hamiltonians in the "vicinity" of finite-size-scaled couplings, which follow the flow of renormalization group transformations. Furthermore, they provide a fundamental and non-perturbative understanding of paradigmatic physical phenomena. In this way, they allow for an intuitive picture of strong correlations, which certainly to some extent can be applied to more general classes of systems.

The somehow orthogonal treatment of strongly correlated systems is numerics. In this approach, the analysis aims at a robust description of physical properties for a wide range of couplings in the space of Hamilton operators. Unfortunately, strong correlations are mostly accompanied by large correlation lengths. This demands for a sufficiently large size of the modeled system. Very rapidly, the limits of available computational ressources are met. In this way, most of nowadays available efficient algorithms are restricted to lower dimensions. The discussion on these routines, their underlying concepts and successive applications by far exceed the scope of this work. The reader is thus referred to he following excellent reviews on Quantum Monte Carlo [57, 58], Density Matrix Renormalization Group [59, 60] and the very recently developed Vidal algorithm [61, 62]. In fact, integrable models provide an excellent benchmark for the stability and accuracy of these algorithms. In this thesis, the calculations on microscopic atomic systems in Chapter. 2 rely on the very basic, severely hardware-limited, but, for the considered problem very suitable, approach of exact (block) diagonalization. This is introduced to some extent in App. B.

Yet another possibility to handle strong correlations is the variational ansatz. It is in principle similar to bosonization, but sets in at a different level. In the Fermion-Boson mapping, correlations are absorbed in terms of the nonlinear vertex operator formalism. Instead, the variational ansatz starts from suitable wave functions, which are assumed to span the relevant Hilbert space and represent properties of expected ground states. They are by construction correlated objects and in general difficult to be conceived. The most prominent examples are the proposals by Jastrow [63], Gutzwiller [64], and Laughlin [20]. The latter ansatz wave functions were proposed as ground state candidates for the fractional quantum Hall effect and made their inventor a Nobel laureate. The underlying exciting phenomenon is introduced in Sec. (b).

## a.2 Strongly Correlated Systems in Cold Atomic Gases

Different fields of physics are sometimes not as distinct as they may seem at first glance. As a consequence, methodical concepts apply in more or less equivalent form to inherently distinct phenomena. The spectrum of physical fields linked in this manner ranges from condensed matter physics to quantum field theory, cosmology and even string theory. It may be conjectured that there is some deep principle which reveals itself in different surfaces. The above parallels are remarkably pronounced in the theory of strongly correlated systems. In this way, an improved understanding of correlated phenomena may establish a kind of unified understanding of the mysteries of nature.

Unfortunately, as it has been discussed in the previous section, the majority of strongly correlated systems even lacks a properly established theoretical basis. The few exceptions, e. g., integrable models, are very specific and unique systems, and, generally, do not exist in their pure form in nature. Moreover, experimental boundary conditions very often impose strong constraints and restrict the set of observable quantum phases. To overcome these obstacles, the best means to have is a perfect quantum simulator as it has been conceived by Feynman [65]. With such a machine, it would be possible to realize even the most exotic quantum systems, choose the favorable species or any desired mixture of particles and tune the interactions at will. Perfect quantum engineering is of course but a dream, yet it is one that has come closer to reality in the recent years.

The advent of optical lattices in experiments of ultracold atomic quantum gases certainly marks a breakthrough in this direction. Optical lattices, which are discussed in Sec. (c), allow for the realization of perfect crystals of light. Their geometrical topology, well depth, and even the level of perfectness are experimentally adjustable. Finally, due to the rapid advance in laser cooling schemes, ultracold atomic species could be transferred to these lattices and kinetically confined. This paved the way for an exciting new class of experiments which allows for a pure realization of models investigated in condensed matter theory.

The exciting phase diagram of the Bose Hubbard model [66] stimulated the idea how a weakly interacting Bose gas could be converted to a strongly interacting Mott insulator in a suitable optical lattice setup [67]. When this transition was indeed observed a few years later [68], the stone was set rolling.

As the optical lattice scheme allows for the creation of quasi lower dimensional quantum gases, the problem of particle equivalence in one-dimension could be challenged. There, in contrast to a 3D system, the effective strength of interaction $V_{\mathrm{int}}$ is inverse proportional to the density. In the limit of strong interactions $V_{\mathrm{int}} \to \infty$, the bosons become hardcore, i. e. impenetrable, and behave as if they were fermions. A highly elongated ultracold Bose gas was recently observed deep in the Tonks-Girardeau regime, very close to the limit of fermionization [69]. Sufficient aspect ratios are achieved in a weak harmonically confined gas which is subject to a deep 2D optical lattice. Thus, the first step towards strongly correlated one-dimensional quantum systems has been made. The list of further challenges is numerous, and certainly the frontiers of quantum optics and condensed matter theory slowly but continuously converge. In some cases, limitations of solid matter are even exceeded. The following list of examples is solely meant to provide a rough impression on the variety of achievements which have been made in the last decade. A very recent and detailed review, which addresses this topic, is given by [70].

Nowadays, bosons, fermions, and even mixtures have been confined to lattice systems. Novel quantum degenerate species as molecular [71, 72, 73] and dipolar interacting gases [74] have just become available. Artificial or even truly random potentials may be superimposed with the lattice [75, 76, 77]. These allow for the exciting investigation of the competition between correlations and disorder. Furthermore, the strength of (contact) interactions can be tuned by the implementation of Feshbach resonances [78, 79]. These resonances can be seized to even change the character of interaction. Thus, they can induce specific BCS-BEC transitions which may shed light on the puzzle of high-$T_c$ superconductivity. Among other spin systems, frustrated anti-ferromagnetic systems are feasible [80].

Yet, one important ingredient is apparently missing — charge. By this, sensitivity to magnetic fields is absent in all of the above approaches. This is very unfortunate since ordered structures, which are imposed by quantum magnetism, are certainly one of the most exciting topics in the context of strongly correlated systems. How even this barrier can be overcome, is the topic of this thesis and is discussed thoroughly throughout this work. To a large extent, the focus of interest is concentrated on fractional quantum Hall phenomena of various types.

# SECTION (b)

# The Quantum Hall Effect

The quantum Hall effect is an intriguing and amazing phenomenon in the field of condensed matter physics which initiated a strong interest in two-dimensional electron systems. In the last two decades, a wide range of sophisticated concepts have been employed in nearly all domains of modern theoretical research to uncover its nature. The scope ranges from microscopic Hamiltonian theories to topological and conformal field theory approaches.

This section introduces both, the integer and the fractional quantum Hall effect. Starting from basic quantum mechanics, the integer phenomenon is discussed as a one-electron system involving disorder. The intrinsically different nature of the fractional effect as a strongly correlated electron system is illuminated in the second part. In the last part, both phenomena are linked in terms of Jain's effective composite fermion model. Suitable introductions to the theory are provided by [16, 17, 81, 82].

## b.1 The Integer Quantum Hall Effect

The integer quantum Hall effect was discovered by Klaus v. Klitzing in 1980 [18]. He studied the charge-transport behavior of high mobility two-dimensional electron gases at very low temperatures and strong magnetic fields. Von Klitzing found that — for certain values of the magnetic field $B$ — the longitudinal resistance of the semiconductor sample becomes negligibly small while the plot of the transverse conductance $\sigma_{\mathrm{H}}$ exhibits plateaus as a function of the magnetic field $B$. These plateaus turned out to be centered around integer multiples of the natural unit $\mathrm{e}^2/h$. This quantization is observed with amazing precision (up to $10^{-8}$). Due to experimental circumstances, e.g., macroscopic sizes and shapes of the probes, disorder, and finite temperature effects, this is even more surprising and leads to the conclusion that fundamental quantum physical properties are revealed. In 1985, von Klitzing was honored with the Nobel prize for his discovery, and the accuracy of the quantum Hall effect made it the etalon of electric resistance.

In order to understand the effect in a proper way, it is convenient to start from Landau's analysis of the quantum dynamics of an electron moving in a perpendicular and uniform magnetic field. The Hamiltonian reads

$$\mathcal{H} = \frac{1}{2M}\Big(\vec{p} - \frac{\mathrm{e}}{c}\vec{A}(\vec{r})\Big)^2 \;, \tag{b.1}$$

where $M$ and $\mathrm{e}$ are mass and charge of the electron, $\mathrm{c}$ is the speed of light, and $\vec{A}(\vec{r})$ is the vector potential chosen in the Landau gauge

$$\vec{A}(\vec{r}) = \big(-By,\, 0,\, 0\big) \;. \tag{b.2}$$

The electron is restricted to move in the $xy$-plane. Since $[H,\, p_x] = 0$, the natural ansatz $\Psi(\vec{r}\,) = \exp(\frac{\mathrm{i}}{\hbar}p_x x)\psi(y)$ inserted in the Schrödinger equation yields the differential equation of a one-dimensional harmonic oscillator

$$\psi''(y) + \frac{2m}{\hbar^2}\Big[\varepsilon - \frac{m}{2}\omega_c^2(y-y_0)^2\Big]\psi(y) = 0 \;, \tag{b.3}$$

where $\omega_c = \frac{\mathrm{e}B}{mc}$ is the cyclotron frequency and $y_0 = \frac{c\hbar k_x}{\mathrm{e}B}$ is the centre of the cyclotron orbit in classical terms. The solution is given by

$$\Psi_n(y-y_0) = \exp\Big[-\frac{m\omega_c}{2\hbar}(y-y_0)^2\Big]\mathrm{H}_n\Big(\sqrt{\tfrac{m\omega_c}{\hbar}}(y-y_0)\Big) \;, \varepsilon_n = \hbar\omega_c(n+\frac{1}{2}) \;. \tag{b.4}$$

Here, $\mathrm{H}_n$ are the Hermite polynomials. The energy levels $\varepsilon_n$ are called Landau levels and are highly degenerate due to $y_0$. Their degeneracy $N(n)$ per given area $A$ is related to the total magnetic flux $\Phi = BA$ perpendicularly piercing the electron gas and equals

$$N(n) = \frac{\Phi}{A\Phi_0} = \frac{B}{hc/e} \;. \tag{b.5}$$

Thus, the degeneracy is a constant for all Landau levels. It depends linearly on $B$, and is measured in units of the magnetic flux quantum $\Phi_0$.

If it is assumed that electron-electron interactions can be neglected, the above results can be extended to a system of $n$ electrons. To classify this system properly, it is reasonable to define the filling fraction $\nu$

$$\nu = \frac{\text{number of electrons}}{\text{number of Landau states}} = \frac{hc}{\mathrm{e}B}n_{\mathrm{e}} \;, \tag{b.6}$$

FIGURE 1: Hall conductance and electron density over Fermi energy

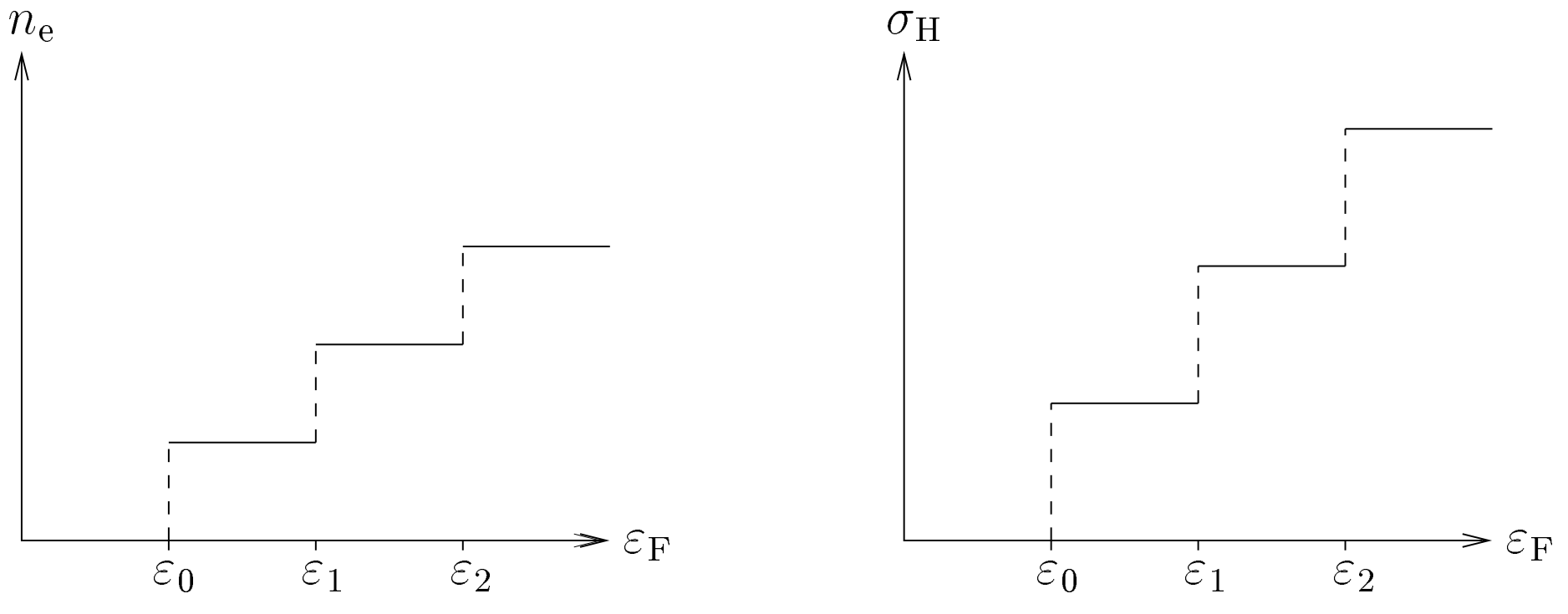

where $n_e$ is the surface density of the electrons. To calculate the Hall conductance $\sigma_H$, an electric field $\vec{E}(\vec{r}) = (0,\, E,\, 0)$ has to be included in (b.1). It solely leads to a shift of $y_0$ and of the energy $\varepsilon_n$:

$$y_0 \longrightarrow y_0' = y_0 + \frac{eE}{m\omega_c^2} \;, \quad \varepsilon_n \longrightarrow \varepsilon_n' = \varepsilon_n + eEy_0 + \frac{m}{2}\left(\frac{cE}{B}\right)^2 \;. \tag{b.7}$$

The expectation value $\langle v_x \rangle$ is derived from (b.4) under consideration of (b.7)

$$\sigma_H = -\frac{n_e e \langle v_x \rangle}{E} = -\frac{n_e ce}{B} = -\nu \frac{e^2}{h} \,. \tag{b.8}$$

This dependence of $\sigma_H$ on the electron density $n_e$ and the filling fraction $\nu$, respectively, has to be analyzed in detail. In a gedankenexperiment, the Fermi energy $\varepsilon_F$ of the system is continuously varied. Then, $n_e$ remains unchanged until the next Landau level is reached. Exceeding $\varepsilon_n$ fills the whole level. Since each electron state of the system equally contributes to the Hall current, $\sigma_H$ shows the same behavior. The graphs are illustrated in figure 1. If combined, they yield the linear dependence (b.8) and no quantum Hall effect is expected. This is resolved if disorder of the system is taken into account. Impurities lift the degeneracy of the Landau levels which broaden into bands. These bands consist of localized states bound by defects of the probe and extended states carrying the Hall current. It is assumed and can be shown for several types of potentials that these latter states exist in the quantum Hall regime and that they are located around the centre of the Landau band [83, 84]. Thus, by varying the Fermi energy, the electron density is continuously increased while in the region of localized states (separating the extended ones) $\sigma_H$ remains constant. In addition, as it was first shown by Prange [85] for the case of impurities represented by $\delta$-functional potentials, the total current carried by a Landau level is unchanged, since an extended state exactly compensates for the loss due to localization effects. As a consequence, (b.8) remains valid in the domain of a plateau. The behavior of $\sigma_{\mathrm{H}}$ is indicated in figure 2.

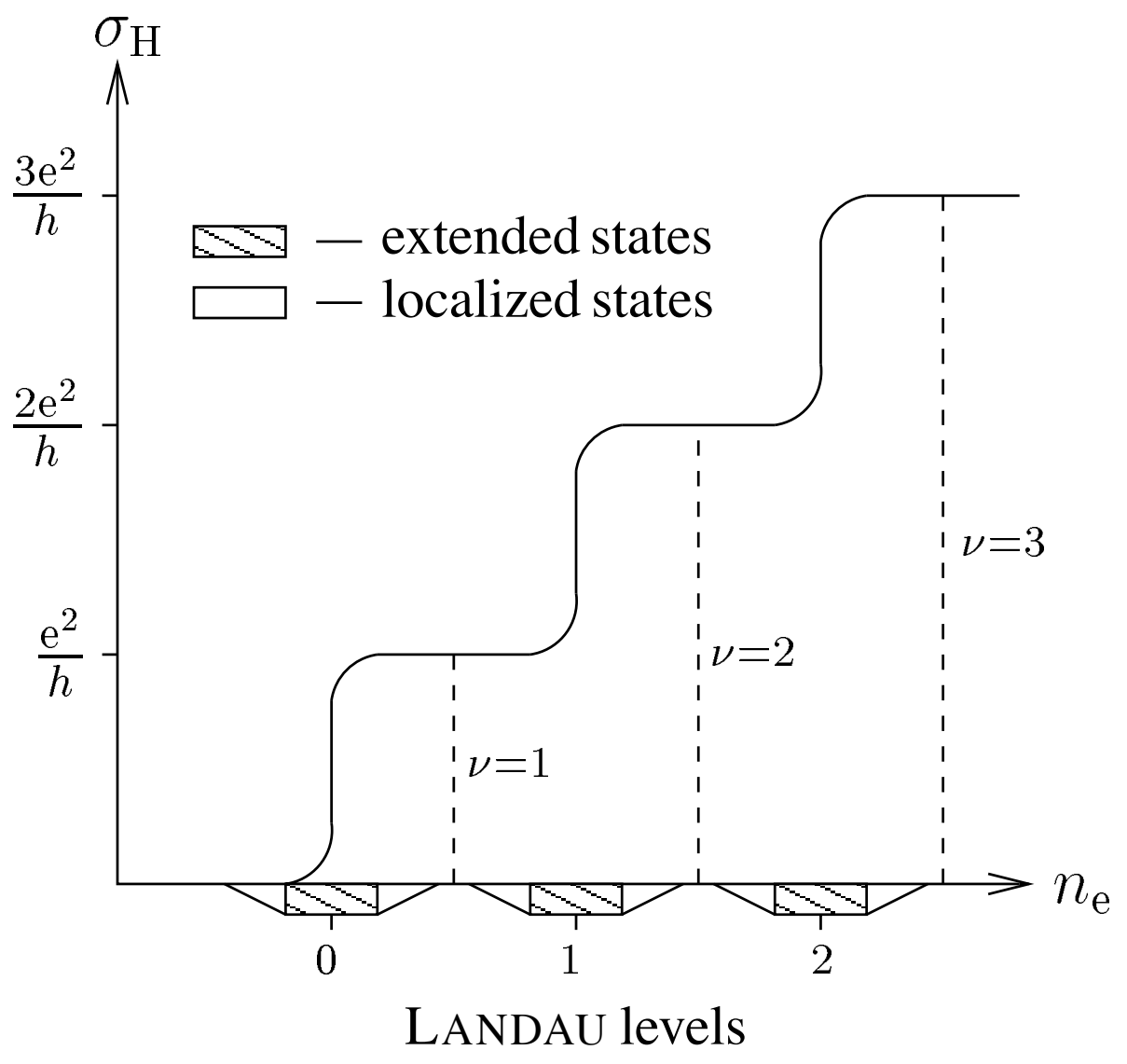

FIGURE 2: Hall conductance over electron density

The Hall conductance shows plateaus with centers located at integral filling fractions $\nu$ and is quantized in units of $e^2/h$. In principle, this explains the results of the experiments, but is not capable of describing the amazing accuracy of the effect. As indicated above, the quantization of $\sigma_{\mathrm{H}}$ remains exact. This even holds up to macroscopic length scales and more complicated types of disorder. The issue is resolved by relating the conductance to gauge covariance, first proposed by Laughlin [86] and later on extended by Halperin [83]. It is shown that $\sigma_{\mathrm{H}}$ is a topological invariant if the Fermi level lies in a (mobility) gap, i.e., the domain of localized states. By this, it is assumed that quantum Hall states have to be incompressible. This is well supported by perturbative methods and numerical research. Therefore, the quantization of $\sigma_{\mathrm{H}}$ (b.8) is based on fundamental physical principles independent of experimental circumstances and devices. The discussion on topological invariance of the Hall conductance exceeds the scope of this work, a detailed access is provided in the first chapter of [82], which reviews diverse approaches [87, 88] in a closed context.

There is a certainty even if a microscopic theory has not been discovered so far: the inte-ger quantum Hall effect is to be understood as a one-electron effect involving disorder where electron-electron interactions can be neglected. The corresponding ground states have to be incompressible quantum liquids involving a non-trivial geometrical setting. This yields integer quantization if and only if the ground state is non-degenerate.

## b.2 The Fractional Quantum Hall Effect

In 1982, three years before Klaus v. Klitzing was awarded the Nobel prize, theoretical physicists believed they understood the quantization of the Hall conductance in natural units. Therefore, it was rather surprising when Tsui, Störmer and Gossard discovered a plateau of the Hall conductance $\sigma_{\mathrm{H}}$ at $\nu = 1/3$ and indications for another one at $\nu = 2/3$ [19]. This 'anomalous' behavior of quantization was inconsistent with respect to the theory of the integer effect. It soon became obvious that fractional Hall states cannot be described by single-electron quantum mechanics. Since the Fermi energy resides within a Landau level, the energy gap necessary to establish a plateau is due to a strongly correlated electron movement reducing the Coulomb interaction. Therefore, the corresponding states are expected to be of completely different geometrical and topological nature. The complexity of fractions nowadays observed in high-mobility semiconductors is illustrated in Fig. 3

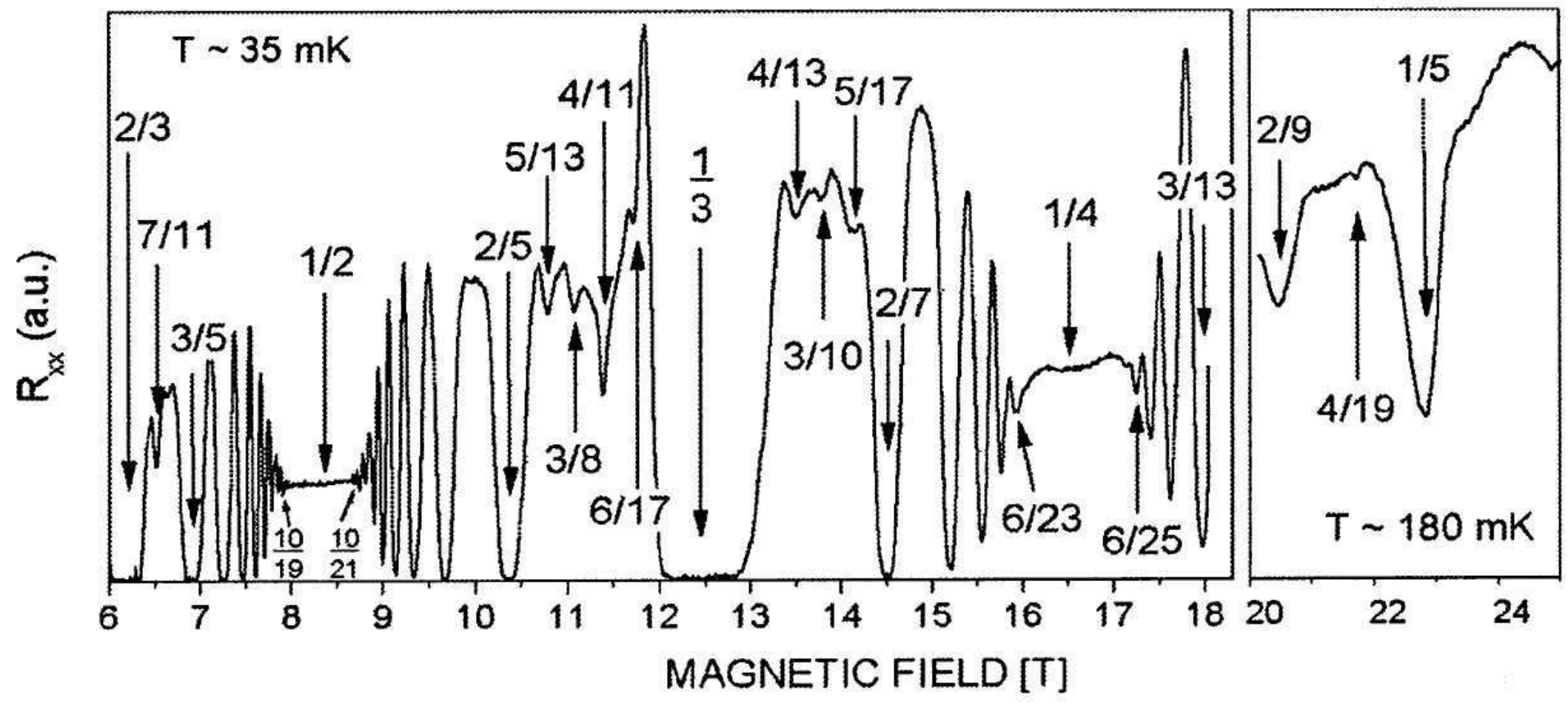

FIGURE 3: Fractional quantum Hall states observed as local minima of the longitudinal resistance over the magnetic field (*by courtesy of H. Störmer* [89]).

To begin the analysis of the fractional quantum Hall effect and study its topological features it is advisable to complexify the theory introduced in the first chapter by $z = x + \mathrm{i}y$ and its complex conjugate $\bar{z}$, yielding

$$x = \frac{1}{2}(z + \bar{z}) , \qquad y = \frac{1}{2\mathrm{i}}(z - \bar{z}) \ ,$$
$$\partial_z = \frac{1}{2}(\partial_x - \mathrm{i}\partial_y) , \qquad \partial_{\bar{z}} = \frac{1}{2}(\partial_x + \mathrm{i}\partial_y) \ . \tag{b.9}$$

Using the symmetric gauge, $\vec{A}(\vec{r}) = \big( -\frac{1}{2}By, \, \frac{1}{2}Bx, \, 0\big)$, the Hamiltonian (b.1) becomes

$$\mathcal{H} = -\frac{2\hbar^2}{m}\big(\partial_z - \frac{1}{4l^2}\bar{z}\big)\big(\partial_{\bar{z}} + \frac{1}{4l^2}z\big) + \frac{\hbar}{2}\omega_c \ . \tag{b.10}$$

where $l = \sqrt{\hbar c / \mathrm{e}B}$ is the magnetic length unit. In the following, $l \equiv 1$ for reasons of simplicity. Since $\hbar\omega_c/2$ is the ground state energy of (b.1), it follows from (b.10) that any wave functions $\Psi(z, \, \bar{z})$ satisfying

$$\Big(\partial_{\bar{z}} + \frac{1}{4}z\Big)\Psi(z, \, \bar{z}) = 0 \ , \tag{b.11}$$

describe a lowest Landau level state. They are derived as

$$\Psi(z, \, \bar{z}) = \mathrm{f}(z) \exp\Big( - \frac{1}{4} \left|z\right|^2 \Big) \ . \tag{b.12}$$

In fact, the space of lowest Landau level wave functions is equivalent to the Bargmann space of analytic functions [90] with inner product

$$\langle \mathrm{f}(z, \, \bar{z}) | \mathrm{g}(z, \, \bar{z}) \rangle = \int \mathrm{d}^2 z \, \bar{\mathrm{f}}(z, \, \bar{z}) \mathrm{g}(z, \, \bar{z}) \exp\Big( - \frac{1}{2} \left|z\right|^2 \Big) \ , \tag{b.13}$$

## b.3 Laughlin States

The first big step forward in order to solve the puzzle of the geometric structure of fractional quantum Hall states was conducted by Laughlin by presenting his trial wave functions [20]:

$$\Psi_{\text{Laughlin}}(z_1, \ldots, z_n) = \mathcal{N} \prod_{k<l}^{n} (z_k - z_l)^{2p+1} \exp\Big( - \frac{1}{4} \sum_{i}^{n} \left|z_i\right|^2 \Big) \ , \tag{b.14}$$

where $p \in \mathbb{N}$, $z_i$ is the position of the $i$-th electron in unified complex coordinates (b.9), and $\mathcal{N}$ is a normalization factor. They were conceived as the variational ground state wave functions for the model Hamiltonian

$$\mathcal{H} = \sum_{k}^{n} \left[ \frac{1}{2m} \left( \frac{\hbar}{\mathrm{i}} \nabla_k - \frac{\mathrm{e}}{c} \vec{A}(\vec{r}_k) \right)^2 + V_{\mathrm{bg}}(\vec{r}_k) \right] + \sum_{k<l}^{n} \frac{\mathrm{e}^2}{|\vec{r}_k - \vec{r}_l|} \ , \tag{b.15}$$

with the vector potential taken in the symmetric gauge. Here, $V_{\mathrm{bg}}$ is a potential of a background charge distribution that neutralizes the electrons' Coulomb repulsion. This guarantees the stability of the system. Despite their simple structure, Laughlin's wave functions include amazing features. Firstly, referring to (b.12), they are an element of the Bargmann space and thus describe a state in the lowest Landau level. Secondly, since $p \in \mathbb{N}$, they are completely anti-symmetric satisfying the Pauli principle, and thirdly, due to the zeroes in the polynomial factor, the electrons are widely separated from each other. This is a crucial condition for the stability of the state with respect to electron-electron interactions in strongly correlated systems. Additionally, they are exact ground states of various short-ranged $\delta$-potential Hamiltonians.

For further investigation, it is important to realize that the modulus squared of the wave function is equivalent to the Boltzmann distribution of a two-dimensional one-component plasma.

$$|\Psi|^2 = \exp(-\beta\Phi) \ , \quad \beta = \frac{1}{2p+1} \ ,$$

$$\Phi = -2(2p+1)^2 \sum_{k<l}^{n} \ln|z_k - z_l| + \frac{2p+1}{2} \sum_{k}^{n} |z_k|^2 \ . \tag{b.16}$$

Even if the analysis is far from being easy, the main advantage of this identification is to investigate the thermodynamic limit. It turns out that with respect to charge neutrality, the electron density corresponds to filling fractions $\nu = 1/(2p+1)$. Since $p \in \mathbb{N}$, the Pauli principle is directly related to odd-denominator fractions. Furthermore, the thermodynamic behavior for small $p$ reveals that the system is an incompressible liquid rather than a Wigner crystal. This property yields the existence of plateaus in the Hall conductance $\sigma_{\mathrm{H}}$. Finally, numerical calculations for systems of finite size show an excellent overlap with (b.14) of more than 99.5%.

The Laughlin ground state can be extended with respect to quasi-hole excitations by introducing a simple polynomial factor

$$\Psi_{\mathrm{exc.}} = \mathcal{N}(\zeta_i) \prod_{k,l} (z_k - \zeta_l) \prod_{r<s} (z_r - z_s)^{2p+1} \exp\Big( -\frac{1}{4} \sum_i |z_i|^2 \Big) \ . \tag{b.17}$$

Here, the $\zeta_i$ denote the positions of the quasi-hole excitations. With respect to (b.16) the excited states, in contrast to the ground states, have a non-uniform charge distribution. In comparison with the two-dimensional plasma picture, a charge deficit of $\mathrm{e}/(2p+1)$ is found at the point $\zeta_i$, which shows that the quasi-holes are fractionally charged.

In order to analyze the quasi-hole statistics more carefully, the Berry connection has to be derived from the normalization factor. This was first stated by Arovas et al. [91] (a detailed comment on the derivation is provided in [82], chapter 2):

$$\Psi_{\mathrm{exc}} = \mathcal{N} \prod_{k,l} (z_k - \zeta_l) \prod_{r<s} (z_r - z_s)^{2p+1} (\zeta_r - \zeta_s)^{\frac{1}{2p+1}} \exp\left(-\mathrm{F}\left(z_i, \zeta_i\right)\right) \ , \qquad \text{(b.18)}$$

$$\mathrm{F}\left(z_i, \zeta_i\right) = \frac{1}{4} \sum_i \left( |z_i|^2 + \frac{1}{2p+1} |\zeta_i|^2 \right) \ .$$

If a quasi-particle at $\zeta_i$ encircles another one at $\zeta_j$

$$(\zeta_i - \zeta_j) \longrightarrow (\zeta_i - \zeta_j) \exp(2\pi \mathrm{i}) \ ,$$

a phase of $2\pi/(2p+1)$ is picked up. This mapping is equivalent to exchanging them twice. Thus, they obey fractional statistics

$$\theta = \frac{\pi}{2p+1} \ . \qquad \text{(b.19)}$$

To stress another important feature: the non-holomorphic factors in (b.18) describing quasi-particle interactions lead to multi-valued wave functions and give rise to the complex geometry the Laughlin states are built on. Despite its fundamental importance, this one-to-one correspondence between statistics and analyticity is often omitted in the discussion of the fractional quantum Hall effect. However, in a suitable field theoretical description it has to be considered precisely.

## b.4 Beyond Laughlin

A strongly correlated electron system underlies the fractional quantum Hall effect. In such systems interactions dominate the physics and long range effects take place. Well known examples are superconductivity and the Hubbard model which can be described in terms of effective theories. A common feature of these theories is the demand for the existence of effective particles in the system, e.g., Cooper pairs (superconductivity) or spinons and holons (Hubbard model).

FIGURE 4: Four flux composite fermion

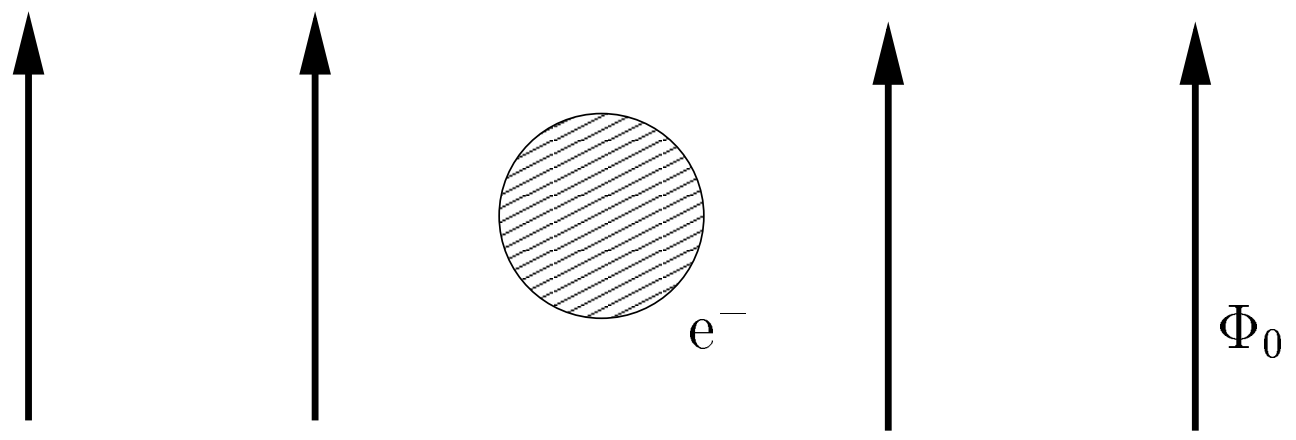

Concerning the fractional quantum Hall effect, one widely accepted effective theory with direct correspondence to experimental facts was developed by Jain [92, 93, 94]. He explained the fractional effect by proposing the composite fermion model. A composite fermion consists of one electron with a number of pairs of flux quanta of the magnetic field attached to it, e.g., as in figure 4. Jain showed that the fractional quantum Hall effect can be expressed in terms of an effective integer quantum Hall effect for the composite fermions.

In order to explain this in a proper way, the results of the integer effect have to be reconsidered. In the symmetric gauge, the single-electron ground state wave functions (b.4) are first expressed in unified complex coordinates $z_i$. They are classified by two suitable quantum numbers $n,\, m \in \mathbb{N}_0$, which label the Landau level and quanta of angular momentum, respectively. This leads to

$$\begin{aligned} \Psi_{n,m} &= \mathcal{N} \exp\big(+\tfrac{1}{4}\,|z|^2\big)\partial_z^n z^m \exp\big(-\tfrac{1}{2}\,|z|^2\big)\ , \\ \text{e.g.,}\quad \Psi_{0,m} &= \mathcal{N}_0\, z^m \exp\big(-\tfrac{1}{4}\,|z|^2\big)\ , \\ \Psi_{1,m} &= \mathcal{N}_1\, z^{m-1}(2m - |z|^2) \exp\big(-\tfrac{1}{4}\,|z|^2\big)\ . \end{aligned} \tag{b.20}$$

The integer effect wave function $\Psi_{\mathrm{I}}$ (filling fraction $\nu_{\mathrm{I}} = \mathrm{I} \in \mathbb{N}$) is obtained by taking the Slater determinant of $(\Psi_{I,0},\, \ldots,\, \Psi_{I,N-1})$ where $N$ is the degeneracy of the Landau level, e.g.,

$$\begin{aligned} \Psi_1 &= \begin{vmatrix} 1 & 1 & \cdots & 1 \\ z_1 & z_2 & \cdots & z_N \\ \vdots & \vdots & \ddots & \vdots \\ z_1^{N-1} & z_2^{N-1} & \cdots & z_N^{N-1} \end{vmatrix} \exp\Big(-\tfrac{1}{4}\sum_{i=1}^{N}|z_i|^2\Big) \\ &= \prod_{k<l}^{N}(z_k - z_l) \exp\Big(-\tfrac{1}{4}\sum_{i=1}^{N}|z_i|^2\Big)\,. \end{aligned} \tag{b.21}$$

The composite fermion trial wave functions $\Psi_{\rm CF}$ are obtained by multiplying the integer quantum Hall effect wave function, e.g., (b.21) with a polynomial Jastrow factor. This analytically represents the attachment of $p$ pairs of flux quanta to the electron

$$\Psi_{\rm CF} = \prod_{i<j}^{N} (z_i - z_j)^{2p} \Psi_{\rm I} \ . \tag{b.22}$$

Starting from the the wave function $\Psi_{\rm I}$, which describes $I$ filled Landau levels, the corresponding composite fermion filling is derived as

$$\nu_{\rm CF} = \frac{I}{2pI+1} \ . \tag{b.23}$$

It can be shown that this procedure neither destroys the correlations of the system nor the incompressibility of the state. Laughlin's wave functions are the simplest examples of this scheme. Starting from a $\nu = 1$ integer quantum Hall state (b.21), $p$ pairs of flux quanta are attached. This yields:

$$\Psi_{\rm Laughlin} = \mathcal{N} \prod_{i<j}^{N} (z_i - z_j)^{2p} \underbrace{\prod_{i<j}^{N} (z_i - z_j) \exp\Big( -\tfrac{1}{4} \sum_i |z_i|^2 \Big)}_{\Psi_1}, \ \ \nu = \frac{1}{2p+1} \ . \tag{b.24}$$

With respect to states beyond the main Laughlin series, the crucial point in Jain's approach is that higher Landau levels contribute to states with $\nu \leq 1$. This might seem confusing and has to be investigated in more detail. It is obvious from (b.20) that higher Landau level wave functions depend explicitly on $\bar{z}$. This makes them more complicated to deal with since a lot of numerical results and field theoretical approaches reveal to be valid solely in the lowest Landau level approximation. Apart from this, it is naturally expected that a $\nu \leq 1$ state is dominated by its overlap with the lowest level. Following these considerations, the wave functions (b.22) have to be mapped for further analysis using the lowest Landau level projector:

$$\widehat{\rm P}_{\rm LLL} = \sum_{k=0}^{\infty} \frac{z^k}{2\pi k! 2^k} \exp\big( -\tfrac{1}{4} |z|^2 \big) \int {\rm d}^2 z' \, (\bar{z}')^k \exp\big( -\tfrac{1}{4} |z'|^2 \big) \ldots \ , \ \widehat{\rm P}^2_{\rm LLL} = \widehat{\rm P}_{\rm LLL} \ . \tag{b.25}$$

Wave functions with I $> 1$ consist of monomials of the form

$$\rho(z) = z^n (\bar{z})^m \exp\big( -\tfrac{1}{4} |z|^2 \big) \ . \tag{b.26}$$

These are projected to

$$\widehat{\mathrm{P}}_{\mathrm{LLL}}\big[\rho(z)\big] = \binom{n}{m} m!\, z^{n-m} \exp\big(-\tfrac{1}{4}\,|z|^2\big) \ . \tag{b.27}$$

It is obvious from (b.27) that monomials with $m > n$ are identically mapped to zero.

Jain showed that the wave functions (b.22) have a large overlap with the lowest Landau level for a small number of electrons, but in comparison with the Laughlin series there exists no analogue that carries this argument to the thermodynamic limit. The question according to the injectivity of the projection (b.25) is even more difficult to answer. However, a pure analytic wave function is arguably the more natural and accurate ansatz to describe lowest Landau level states in a suitable scheme of hierarchy.

# SECTION (c)

# Rotating Gases

Rotations in the quantum world are accompanied by a variety of unexpected and counterintuitive phenomena. The source of these is a manifestation of a deeply rooted and pure quantum behavior of the underlying particle dynamics. The question of the dynamical response of a fluid to a rotating bucket has already attained considerable attention on the level of classical mechanics, though rather from a more or less philosophical perspective in the context of absolute reference systems [95]. Certainly, a bucket filled with sufficiently cold helium would have caused a lot of confusion in those times.

Superfluidity, which has first been observed in liquid He II [96, 97], leads to an intrinsic robustness to transversal excitations and moving perturbations of any kind. Thus, below a certain critical velocity of rotation, no response of the superfluid medium is experimentally observed. If this threshold is surpassed, singular objects manifest themselves in the velocity field. These vortices are topological defects and carry quanta of angular momentum. In this way, they contribute to the rotation of the superfluid, more precisely, solely their existence allows for it [98]. Vorticity has thus been an object of extensive studies in condensed matter physics. If the rotation is further increased, more and more of these objects are nucleated [99], which after sufficient equilibration arrange themselves on a regular triangular Abrikosov lattice [100].

The exploration of these phenomena in the context of ultracold gaseous systems has been motivated by several issues. First, the existence of a critical threshold velocity is a direct proof of the superfluid nature of a Bose-Einstein condensate. On the other hand, dilute atomic gases provide an excellent theoretically tractable playground, and the underlying physics can be seized to check the validity of the theory of interacting inhomogeneous Bose systems on small length scales. Another advantage is the direct observability of vortices. Due to a larger healing length in cold atomic systems, their core size is considerably larger. This is even further enhanced by ballistic expansion used for imaging.

Motivated by these perspectives, intensive studies on vortex phenomena in these systems have been initiated, soon after the first quantum degenerate Bose gases had been experimentally realized. An excellent overview is provided by [101] and references therein. Since then, there has been a tremendous theoretical and experimental progress in this field. The strength of the approach via cold gaseous systems is the pure and controllable preparation of the desired system. This allows for very clean and detailed experimental observations, which cannot be achieved to this extent in an experiment of condensed matter. In this manner, a variety of effects has been investigated in atomic systems, e. g., the realization of Abrikosov vortex lattices [102] and the nucleation and evolution of Tkachenko oscillations [103].

Over the last few years, considerable interest has been attained by experiments, which arrive at very high rotational rates. There, the frequency of rotation approaches the radial trap frequency. As a general feature, rotation subjects the cold atomic sample to an effective "artificial" magnetic field while shallowing the effective trap strength. In the above centrifugal limit, the trapping potential is finally absorbed and the gas becomes unstable [104]. On the crossover from slow to fast rotational frequencies, the Thomas-Fermi mean-field description finally breaks down, the vortex lattice melts and strongly correlated states are predicted to appear [105, 106, 107].

The first part of this section introduces the theoretical concepts which illustrate how quantized vortex structures manifest themselves in ultracold atomic gases. Consecutively, experimental methods are explained how these topological objects are successfully nucleated. The last part of this section addresses rapid rotations. This regime, which is certainly at the frontier of experimental research in ultracold atomic systems, is in the focus of this thesis.

For a detailed theoretical treatment of ultracold rotating gases, the reader is referred to the excellent book [108]. A very nice and compact experimental review is given by [109].

## c.1 Vortices

The low temperature dynamics of a nonuniform dilute Bose gas constituted of a large number $N$ of particles is governed by the Gross-Pitaevskii equation [110, 111]

$$\mathrm{i}\hbar\partial_t\psi_0(\vec{r},\, t) = \left(-\frac{\hbar^2}{2M}\Delta + \mathcal{V}_{\text{ext}} + g\left|\psi_0(\vec{r},\, t)\right|^2\right)\psi_0(\vec{r},\, t)\,, \tag{c.1}$$

where $M$ is the mass of an atom, $\mathcal{V}_{\rm ext}$ denotes the external potential, and the effective interaction is mediated by the coupling $g$. This may be expressed solely in terms of the $s$-wave scattering length in the above limits

$$g = \frac{4\pi\hbar^2}{M} a_s \,. \tag{c.2}$$

To derive this equation, the particle field operator of the Schrödinger equation has been replaced by the classical field $\psi_0(\vec{r},\, t)$ which is often referred to as the order parameter or the wave function of the condensate. If the gas is assumed to be sufficiently dilute

$$n(\vec{r},\, t)\, |a_s|^3 \ll 1 \,, \tag{c.3}$$

effects of thermal and quantum depletion can be neglected. Then, $\psi_0(\vec{r},\, t)$ is normalized to the number of particles. This implies that its modulus squared equals the density of the gas in this approximation

$$n(\vec{r},\, t) = |\psi_0(\vec{r},\, t)|^2 \;. \tag{c.4}$$

Effectively, the above approach reduces the dynamics of the full quantum system to a single particle problem. From this point of view, amplitude and phase of the condensate wave function can be defined as

$$\psi_0(\vec{r},\, t) \;=\; \sqrt{n(\vec{r},\, t)}\, \exp\Big(\mathrm{i}\Phi(\vec{r},\, t)\Big) \,. \tag{c.5}$$

The phase $\Phi$ is related to the current density by

$$\vec{j}(\vec{r},\, t) \;=\; -\mathrm{i}\frac{\hbar}{2m}\Big(\psi_0^*\nabla\psi_0 - \psi_0\nabla\psi_0^*\Big) \;=\; \frac{\hbar}{M} n(\vec{r},\, t)\nabla\Phi(\vec{r},\, t) \,. \tag{c.6}$$

From this equation, the velocity field of the condensate can be read off directly

$$\vec{v}(\vec{r},\, t) = \frac{\hbar}{M}\nabla\Phi(\vec{r},\, t) \,. \tag{c.7}$$

Apparently, the curl of the velocity field vanishes

$$\nabla \times \vec{v}(\vec{r},\, t) = 0 \,. \tag{c.8}$$

This imposes the so-called condition of irrotationality on the condensate which is a distinguishing characteristic of superfluidity.

A stationary vortical solution of the Gross-Pitaevskii equation cannot be strictly derived by the previously pursued approach, as it will violate the condition (c.8). Nonetheless, any stationary solution can be written in terms of (c.5). If the gas is assumed to be confined in a container of cylindrical symmetry, the ansatz for a condensate wave function which represents a rotation of the gas around the longitudinal symmetry axis of the cylinder reads

$$\psi_0(\vec{r}) = \sqrt{n(\rho,\, z)}\,\exp(\mathrm{i}m\varphi)\,, \tag{c.9}$$

where $\rho$, $\varphi$ and $z$ is the set of cylindrical coordinates, and $m \in \mathbb{N}$ to guarantee the single-valuedness of the order parameter. The wave function (c.9) is an eigenstate of the angular momentum operator $\hat{L}^z = -\mathrm{i}\hbar\partial_\varphi$ with eigenvalue $\hbar m$. This yields a total angular momentum of $L^z = N\hbar m$, which is shown to be carried by the vortex. According to (c.7), the velocity field, which represents the flow of the condensate, is derived as

$$\vec{v}(\vec{r},\, t) = m\frac{\hbar}{M}\frac{\vec{\mathrm{e}}_\varphi}{r}\,. \tag{c.10}$$

It is oriented in direction of the polar unit vector $\vec{\mathrm{e}}_\Phi = (-\sin(\phi),\, \cos(\phi),\, 0)$ and purely tangential in the $xy$-plane. Apparently, $\vec{v}$ is singular at the origin. This leads to a non-vanishing curl on the axis of rotation

$$\nabla \times \vec{v}(\vec{r},\, t) = 2\pi m\frac{\hbar}{M}\delta^{(2)}(\vec{r}_{xy})\vec{\mathrm{e}}_z\,. \tag{c.11}$$

Here, $\vec{r}_{xy}$ is the projection of $\vec{r}$ to the $xy$-plane. The circulation of the velocity field is thus concentrated on this vortex line and quantized in units of $\hbar/M$. Outside this line, the condition of irrotationality (c.8) is strictly fulfilled. As a consequence, (c.9) describes a pure condensate with a topological defect at the origin, i. e., a vortex which carries $Nm$ quanta of angular momentum. From the Gross-Pitaevskii equation, the explicit form of $n(\rho,\, z)$ is determined. The extension of the vortex core is derived to equal few multiples of the healing length $\xi = \hbar/\sqrt{2mgn_0}$, where $n_0$ equals the density of the unperturbed uniform Bose gas.

The above calculation can be naturally generalized to a trapped real Bose gas, whose geometry, atom number, *et cetera* influence the quantitative results while preserving qualitative properties. It has to be stressed that the dynamical formation of vortices and their influence on the condensate constitute a field of study on their own. Suitable mean-field treatments have been developed to properly describe multi-vortex evolutions and analyze arising quantum phases.

## c.2 Phase Control, Stirring and Spinning

Two classes of methods have been proposed to nucleate quantized vortices in experiments with ultracold atomic gases. The first is motivated by the state configuration itself. A vortex is created if the phase pattern of (c.9) is properly imprinted on the condensate and the density is suppressed to zero at the center. This may be achieved in various ways. The straightforward ansatz implies the use of optical beams which possess an appropriate topology to directly imprint the necessary phase. The disadvantage of such an approach is that it relies on dissipative processes. This is due to the fact that driving the density of the condensate to zero at the origin while keeping the proper phase leads to some uncertain excited state configuration, which somehow has to relax to the vortex state. To what extent such processes preserve the necessary phase configuration is not predictable. This drawback is circumvened, if the formation of the vortex configuration is achieved in terms of a coherent process. Based on the theoretical proposal of [112], vortices have been coherently created in a spatially and temporally controlled transfer process between two internal spin states of $^{87}$Rb [113]. These are simultaneously trapped in a magnetic potential, and a two-photon microwave field induces transitions between them. This conversion is controlled by a focussed and spatially moving laser beam which shifts the internal energy levels of the atoms by optical forces, on which some details are given in Sec. (d) of this introduction. As indicated in Fig. 5, the microwave is detuned from the transition $|1\rangle \rightarrow |2\rangle$, and the laser is rotated at the appropriate frequency to guarantee resonance. Due to this condition, atoms which change their internal state, are driven to do so and have to adapt the spatial symmetry of the driving laser. In contrast to atoms in the core region, which do not experience a temporal change of phase, atoms near the boundary are affected by a periodic phase delay linked to the polar angle of the circumference. The imprinted circulation is hereby determined by the sign of the microwave detuning and not by the orientation of rotating of the laser beam. The remaining fraction of unconverted atoms still fills the central region and supports a "pinning" of the vortex. This method allows for a very efficient single vortex creation and manipulation. In particular, stable higher quantized vortices can be created because of strong pinning, and precession phenomena for a non-central vortex lines may be studied. A similar scheme achieves phase imprinting by adiabatic inversion of the magnetic field in the core region of the condensate [114].

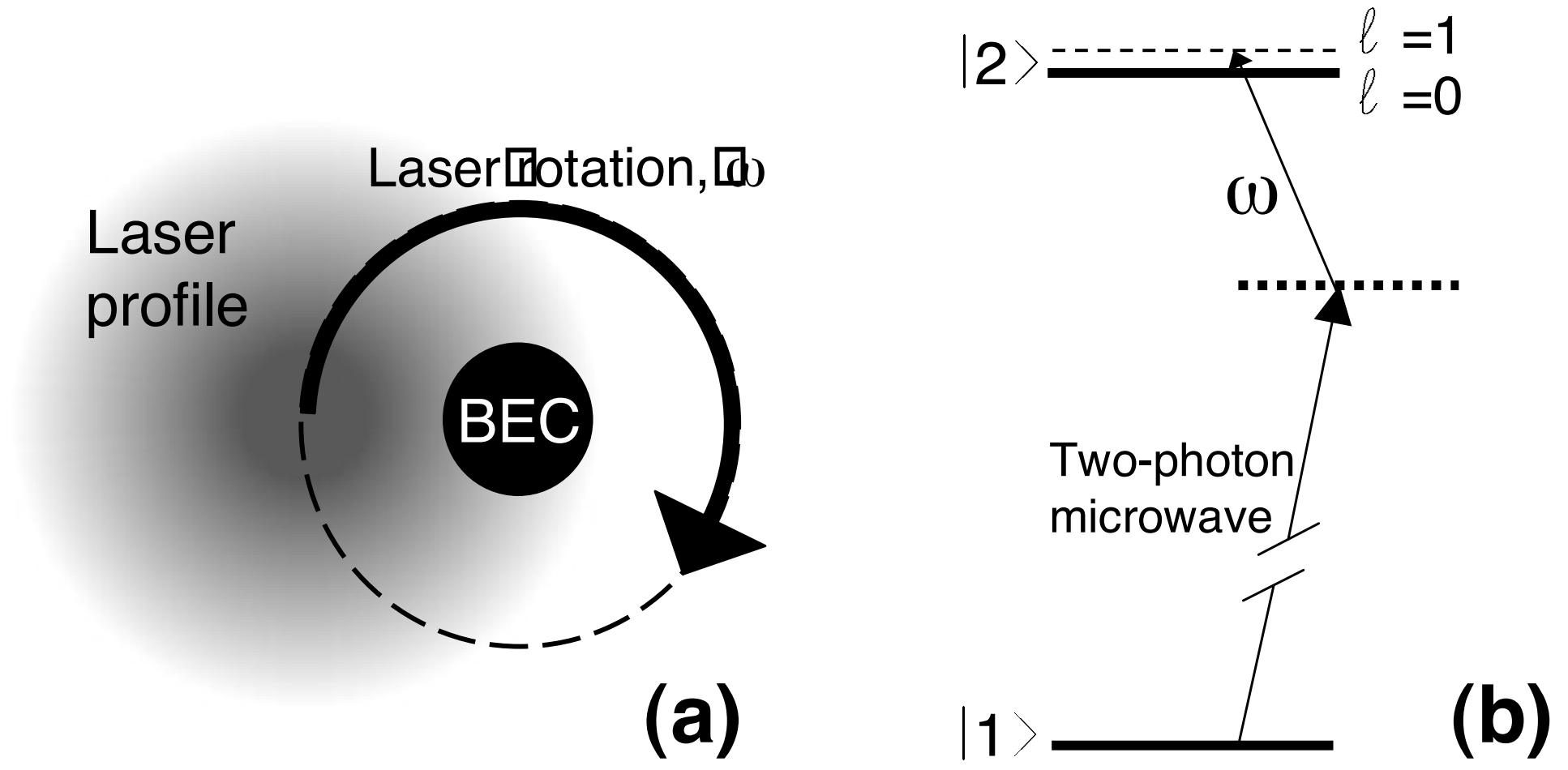

FIGURE 5: (a) Scheme to imprint phase on a condensate to create a vortex. An off-resonant laser provides a rotating gradient in the AC Stark shift. The transition is driven by a microwave drive of detuning $d$. (b) Level diagram with slightly detuned microwave from the state $|2\rangle$. The modulation due to the rotation of the laser couples only to the $l = 1$ state when $\omega = \delta$. The energy splitting between $l = 1$ and $l = 0$ is exaggerated (*by courtesy of E. Cornell* [113]).

The second method closely follows the idea of the rotating bucket and induces a stirring process at the boundary of the condensate. In the seminal experiment [115], this was achieved by adding a rotating anisotropy to an axially symmetric harmonic trap with radial and axial frequencies $\omega_\perp$ and $\omega_z$, respectively. This is implemented by a stirring "spoon" laser aligned along the symmetry axis of the trap. The motion of this beam consists of two components. The fast part of motion refers to a rapid oscillation of the spoon's axis between two intersection points with the plane of the gas. This movement is superimposed by a slow overall rotation, which circulates these intersection points around the condensate. In this way, the time-dependent potential induced by the laser can be approximated by

$$\mathcal{V}_{\text{stir}} = \frac{\varepsilon_{\text{aniso}}}{2} M\omega_\perp \Big[(x^2 - y^2)\cos(2\Omega t) + 2xy\sin(2\Omega t)\Big] . \qquad \text{(c.12)}$$

Here, $\varepsilon_{\text{aniso}}$ measures the relative strength of the anisotropy with respect to the trap frequency and ranges from $2 - 10\%$ in typical experimental setups. In the above scheme, the stirring laser is turned on in the last phase of evaporative cooling and continuously applied, while the atoms thermally equilibrate after the ramp. Then, its intensity is slowly decreased to avoid excitations in the atomic sample. The magnetic field is turned off and after gravitationally induced ballistic expansion, the density profile is measured by absorption imaging techniques. This approach leads to the formation of stationary vortex states which become true ground state configurations in the rotating frame of reference. There, the stirring potential becomes time-independent

$$\mathcal{V}_{\text{stir}} = \frac{\varepsilon_{\text{aniso}}}{2} M\omega_\perp(\hat{X}^2 - \hat{Y}^2)\,, \tag{c.13}$$

where $\left(\hat{X},\, \hat{Y},\, \hat{Z}\right)$ is the set of coordinates which rotate as $\Omega t$ with respect to the laboratory frame

$$\left(\hat{X},\, \hat{Y},\, \hat{Z}\right) = \left(x\cos(\Omega t) + y\sin(\Omega t),\; y\cos(\Omega t) - x\sin(\Omega t),\; z\right). \tag{c.14}$$

After the spoon laser has been ramped down, this state survives as an excited state of the radially symmetric Hamiltonian due to angular momentum conservation and is consecutively imaged after expansion. As apparent in Fig. 6, configurations with many vortex lines can be generated.

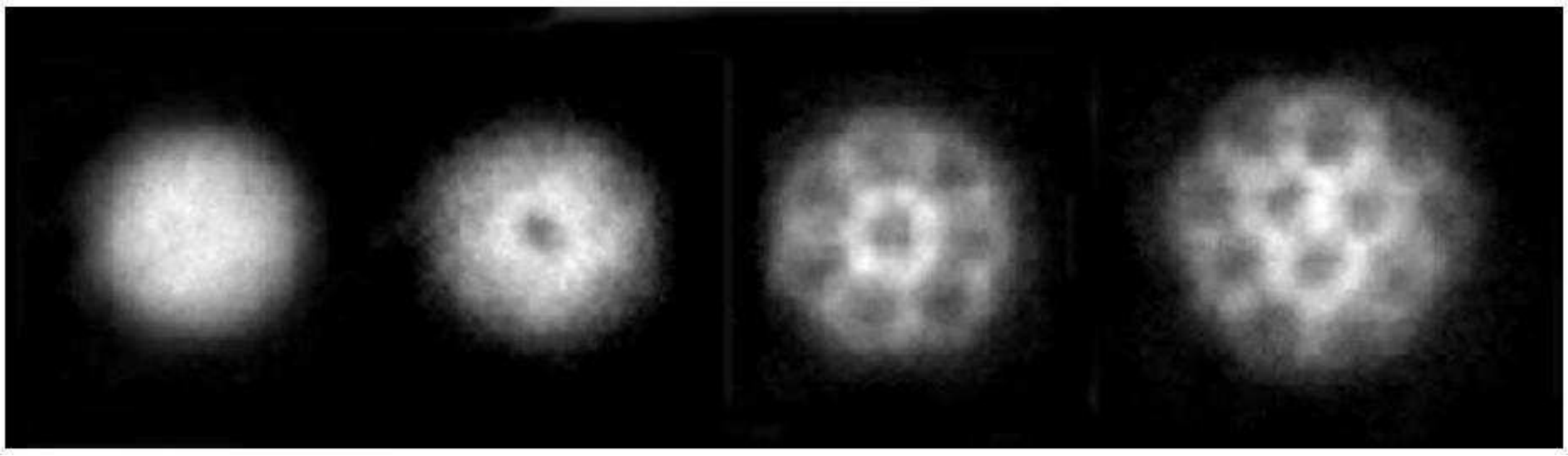

FIGURE 6: Quantized vortices. Absorption images of a stirred $^{87}$Rb Bose-Einstein condensate. The rotational frequency $\Omega$ increases from the left to the right (*by courtesy of J. Dalibard* [115]).

Such mechanical stirring processes have also been successfully implemented by several other groups, either by means of optical techniques [102] or by manipulation of magnetic fields [116, 117]. The latter of these papers motivated yet another method to arrive at sufficiently high rotational rates of the condensate. After a normal gas cloud slightly above $T_c$ has been successfully set into rotation, the frequency of rotation is accelerated by spatially selective evaporation.

It is implemented by the removal of axially displaced atoms in a cigar-shaped trap. These rotate more slowly than the statistic average. This method has been further developed to a true "evaporative spinup" technique which allows for rotation rates larger than $99\%$ of the centrifugal limit [118]. In this immediate vicinity of critical rotation, the gas is only marginally confined in the plane of rotation which is discussed in detail in the consecutive section. To experimentally control and equilibrate such a labile atomic system is a delicate business. Furthermore, over-critical rotation is physically impossible though interesting phases may be expected in this regime. With this in mind, an additional higher harmonic potential can be superimposed with the rotating gaseous sample which guarantees the stability [119]. Such a potential has to be chosen with great care in order not to significantly perturb the original system.

## c.3 Rapid Rotation and Landau Levels

The pioneering experiments on rotating atomic gases, which have been presented in the previous section, recently attracted considerable interest, not only in the community of quantum optics. Due to the possibility to create and control a rotating gas close to the centrifugal limit, there are promising indications that a novel class of systems may be encountered soon at this frontier of ultracold atom experiments.

Theoretically, the range of rotational frequencies $\Omega$ may be divided into three regimes. They can be characterized by the quantity of the filling factor which is defined as the ratio of the number of particles to vortices

$$\nu = N/N_V \,. \tag{c.15}$$

The synonymous notation is not chosen by accident as shown below. High fillings $\nu > 100$ correspond to the previously discussed regime of vortex lattices which was identified as a mean-field quantum Hall regime [120, 121, 122]. Due to this analogy, the definition of a filling factor is qualitatively reasonable, though it should not be mistaken for the quantity known from condensed matter physics, especially for bosons. If more and more vortices are nucleated in the system, transversal excitations of the vortex lattice decrease in frequency. This long-wavelength behavior of Tkachenko modes is a precursor to the predicted lattice melting below $\nu \approx 10$, which is expected to be driven by quantum fluctuations [123, 124, 125]. In this crossover regime, exotic vortex phases are expected to replace the Abrikosov lattice [126]. If $N_V$ eventually exceeds the number of particles in the vicinity of critical rotation, strongly correlated quantum phases are expected to appear, in particular, fractional quantum Hall liquids [105, 106, 107, 123, 124, 127, 128].

To investigate the origin of these phenomena and understand the problems encountered within the theoretical framework, the stationary state structure of the rotating system has to be studied. This is solely possible in the rotating frame of reference, where the anisotropic "stirring" potentials were shown to become time-independent (c.13). The analysis starts from the Hamiltonian of a non-interacting particle of mass $M$ in an axially symmetric and harmonic trap

$$\mathcal{H}_0 = \frac{p_x^2 + p_y^2}{2M} + \frac{1}{2}M\omega_\perp^2(x^2 + y^2) + \frac{p_z^2}{2M} + \frac{1}{2}M\omega_z^2 z^2 \,. \tag{c.16}$$

If this system is considered to axially rotated at frequency $\Omega\vec{e}_z$, the transformation to the co-rotating frame only provides a single centrifugal term

$$\mathcal{H}_{\text{rot}}^{xy} = \frac{\vec{p}_\perp^{\,2}}{2M} + \frac{1}{2}M\omega_\perp^2\vec{r}_\perp^{\,2} - \Omega L^z + \mathcal{H}_0^z \,, \tag{c.17}$$

where $\vec{p}_\perp = (p_x,\, p_y,\, 0)$ and $\vec{r}_\perp = (x,\, y,\, 0)$ are the two-dimensional momentum and position vectors, and $L^z$ is the projected angular momentum of the particle with respect to the $z$-axis. Due to the orientation of $\vec{\Omega}$, the dynamics in the axial coordinate $\mathcal{H}_z$ decouple from the rest of the problem and yield the standard harmonic oscillator energies $\varepsilon_z = \hbar\omega_z(n_z + 1/2)$. The above Hamiltonian may be algebraically transformed to

$$\mathcal{H}_{\text{rot}}^{xy} = \frac{1}{2M}\Big(\vec{p}_\perp - M\vec{\Omega}\times\vec{r}_\perp\Big)^2 + \frac{1}{2}M\Big(\omega_\perp^2 - \Omega^2\Big)\vec{r}_\perp^{\,2} \,. \tag{c.18}$$

Its first term is formally equivalent to the Landau Hamiltonian of charged particles moving in a perpendicular magnetic field (b.1), which is discussed in the context of the fractional quantum Hall effect in Sec. (b). Thus, the rotation subjects the particles to an effective magnetic field represented by the "artificial" vector potential

$$\vec{A} = (M\Omega c/e)\vec{e}_z \times \vec{r}_\perp \,, \tag{c.19}$$

where the elementary charge and the speed of light are solely introduced for reasons of algebraic equivalence. As it can be inferred from (c.18), the trap strength is simultaneously reduced by the centrifugal term $-M\Omega^2\vec{r}_\perp^2$. In the limit $\Omega \to \omega_\perp$, the gas is no more confined and becomes centrifugally unstable.

For spectral analysis, (c.18) may be rewritten as

$$\mathcal{H}_{\mathrm{rot}}^{xy} = \frac{1}{2M}\Big(\vec{p}_\perp - M\omega_\perp \vec{e}_z \times \vec{r}_\perp\Big)^2 + \Big(\omega_\perp - \Omega\Big) L^z \,. \tag{c.20}$$

This form shows that $L^z$ is a suitable quantum number of the system, which still holds when radially symmetric interactions of the particles are introduced. The spectrum of $\mathcal{H}_{\mathrm{rot}}^{xy}$ can be read off directly

$$\varepsilon_{xy} = 2\hbar\omega_\perp\Big(n_{xy} + \frac{1}{2}\Big) + \hbar\Delta\Omega(m_{xy} - n_{xy})\,, \tag{c.21}$$

where the frequency difference from critical rotation $\Delta\Omega \equiv \omega_\perp - \Omega$ was introduced.

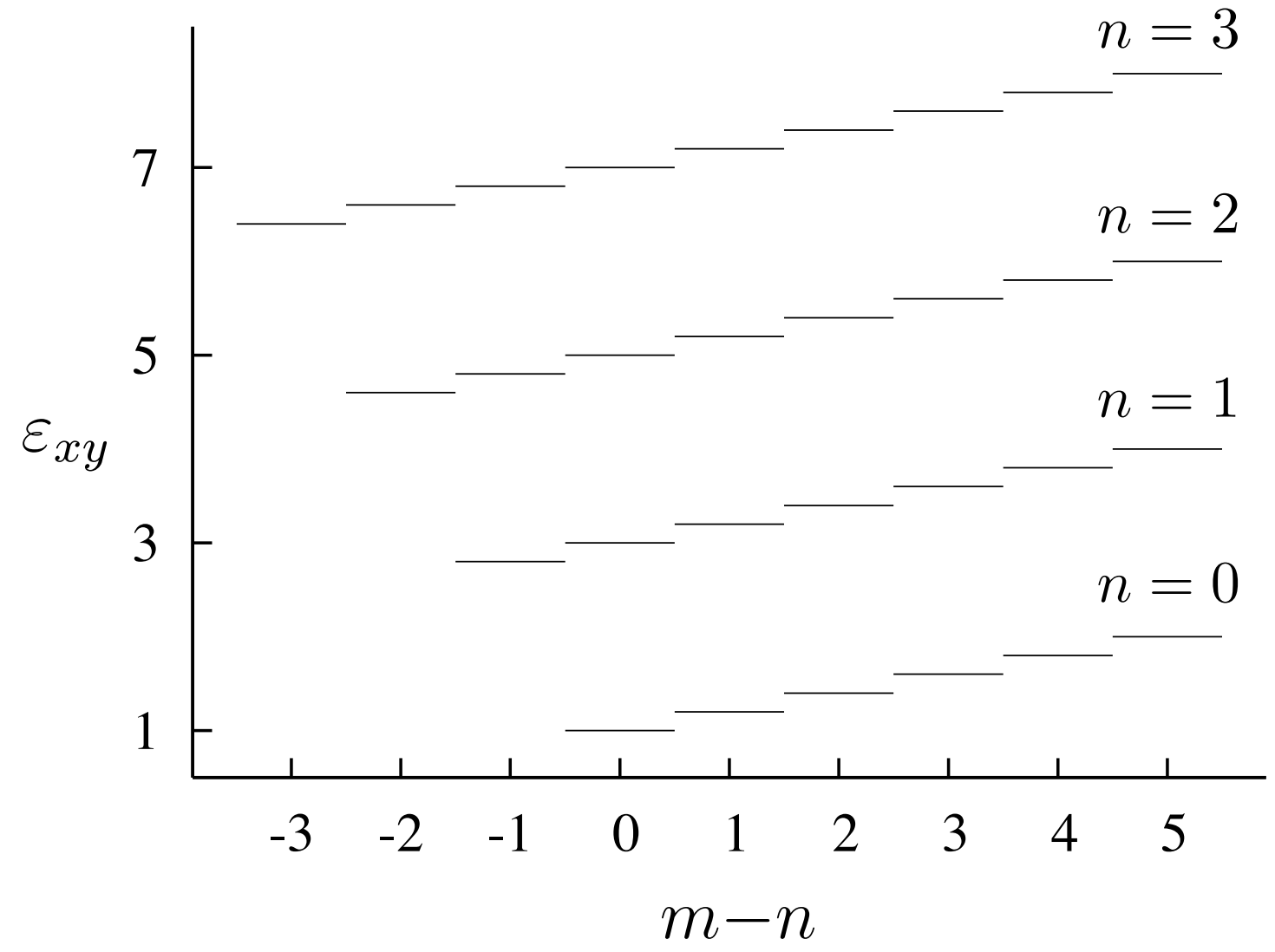

FIGURE 7: Single particle Landau energy spectrum for motion in the $xy$-plane expressed in units of $2\hbar\omega_\perp$. $(n,\, m)$ are the Landau quantum numbers, $\hbar(m - n)$ denotes the $z$-projected angular momentum, and the rotational frequency is set to $\Omega = 0.8\omega_\perp$.

The corresponding eigenstates are the previously discussed Landau states

$$\Psi_{n,m} = \frac{1}{\sqrt{2^{n+m+1}\pi n! m!}} \exp\left[\frac{(x^2+y^2)}{4}\right] \Big(\partial_x + \mathrm{i}\partial_y\Big)^m \Big(\partial_x - \mathrm{i}\partial_y\Big)^n \exp\left[-\frac{(x^2+y^2)}{2}\right] \tag{c.22}$$

where $n,\, m\, \in \mathbb{N}_0$, distances are measured in units of the magnetic length $l_0 \equiv \sqrt{\hbar/(2M\omega_\perp)}$, and the index $xy$ has been dropped for clarity. As depicted in Fig. 7, non-critical rotation leads to a tilting of the Landau levels with slope $\Delta\Omega$. Thus, mixing processes are expected to occur for sufficiently large angular momenta $L^z$.

If the rotation is sufficiently slow, a large number of levels is occupied, and the quantized level structure significantly changes because of interaction processes. Nonetheless, the residual magnetic field imposes ordered structures on the quantum system. These are observed, e. g., in terms of the Abrikosov lattice. If instead $\Omega$ approaches $\omega_\perp$, the planar confinement will be continuously relaxed. Due to centrifugal expansion, the average density of the atomic gas decreases. In this limit, the interaction energy per particle becomes negligible compared to the level spacing $2\hbar\omega_\perp$. Yet, in principle, it can be still sufficiently large to couple the whole lowest Landau level. Then, the Landau states $|0, m\rangle$ constitute the relevant basis, if the temperature of the system guarantees $T \ll 2\hbar\omega_\perp$. If, furthermore, the $z$-degree of freedom is frozen out to the Gaussian ground state of $\mathcal{H}_z$, the particle dynamics will be quasi two-dimensional, and they are subjected to an ultra-strong effective magnetic field perpendicular to their plane of motion. In this regime, strongly correlated fractional quantum Hall liquids are known to exist as stable ground states. The characteristics and stability of these exciting species of states are key topics of this thesis.

# SECTION (d)

# Optical Lattices

The rapid advance in laser techniques and in the manipulation of atomic matter by coherent light fields nowadays allows for quantum engineering with an impressive level of precision. Seizing the wide range of cooling mechanisms, quantum degenerate gases of bosonic [6, 7], fermionic [129, 130, 131] and even molecular constituents [71, 72, 73] became accessible at ultracold temperatures of the order of nano Kelvin. In this regime, an old suggestion [132] to confine atoms in wavelength-sized regions of a standing wave could be pursued. Historically, this idea stimulated yet another experimental approach to laser cooling [133, 134]. Following this concept, experimental success in cooling atoms in such configurations led to spectacular phenomena, e. g., quantized atomic center of mass motion and samples revealing long range order imposed by the light fields. These effects led to the interpretation that the physical behavior of these systems is determined and dominated by atomic movement in periodic crystalline potentials. This concept of optical lattices opened the door to simulate and explore phenomena of condensed matter systems in a pure and controllable way. Perfectly periodic and rigid crystals are realized by modern techniques, hosting more than a million particles. Lattice depth, geometry, dimensionality, and, by employing Feshbach resonances [78, 79], even the strength and character of interactions are under full control of the experimentalist. As a concesquence, these systems offer a unique possibility to investigate various models of condensed matter theory, verify conjectures and resolve open questions. They may even be seized to create completely novel quantum systems which have no analogue in nature, e. g., Bose-Fermi mixtures [135] or matter dynamics in non-Abelian vector potentials, the latter discussed in Chapter 3.

## d.1 Optical Forces

Two fundamental forces arise when atomic samples are subject to coherent light waves [21, 22]. First, the dissipative Doppler force used for efficient laser cooling and secondly, the conservative dipole force which allows for a purely optical trapping, given that the atomic molasses is sufficiently cold [136, 137]. To understand the nature of these forces, a very simple model can be consulted, which motivates many features of real atomic systems, i. e., a two-level atom in a monochromatic coherent light field.

The analysis starts from the quantum mechanical two-level Rabi problem [138]. Let $\mathcal{H}_0$ be the time-independent atomic Hamiltonian with spectrum $\varepsilon_\mathrm{J}$ and eigenfunctions $\Phi_\mathrm{J}(\vec{r})$, where J denotes a complete index set of quantum numbers. The time evolution is determined by the full Hamiltonian

$$\mathcal{H}(t) = \mathcal{H}_0 + \mathcal{V}_\mathrm{rad}(t)\,, \tag{d.1}$$

where $\mathcal{V}_\mathrm{rad}(t)$ is the radiative interaction potential. A general state $|\psi\rangle = \sum_\mathrm{J} c_\mathrm{J}|\Phi_\mathrm{J}\rangle$ evolves as

$$\mathrm{i}\hbar\frac{\mathrm{d}}{\mathrm{d}t}c_\mathrm{J}(t) = \sum_K c_\mathrm{K}(t)\mathcal{H}_\mathrm{J\,K}(t)\mathrm{e}^{\mathrm{i}(\omega_\mathrm{J}-\omega_\mathrm{K})}\,, \tag{d.2}$$

where $\mathcal{H}_\mathrm{J\,K}(t) \equiv \langle\Phi_\mathrm{J}|\tilde{\mathcal{V}}_\mathrm{rad}|\Phi_\mathrm{K}\rangle$. Here, $\tilde{\mathcal{V}}_\mathrm{rad}$ is the off-diagonal coherence potential derived from the full radiative operator by absorption of the diagonal elements in $\mathcal{H}_0$. In the following, the Hilbert space is truncated to two dimensions spanned by the non-degenerate ground state $|\psi_\mathrm{g}\rangle$ and the excited state $|\psi_\mathrm{e}\rangle$, which are gapped by $\hbar\omega_\Delta$ and coupled via the relevant part of $\tilde{\mathcal{V}}_\mathrm{rad}$

$$\mathcal{V}_\mathrm{rel} = -\mathrm{e}\vec{r}\cdot\vec{E}(\vec{r},\,t)\,. \tag{d.3}$$

To solve the coupled differential equations (d.2), two further approximations have to be applied. The first is the rotating wave approximation which demands the laser detuning $\Delta = \omega_\mathrm{L} - \omega_\Delta$ to be negligible compared to the laser frequency $\Delta \ll \omega_\mathrm{L}$. The second assumes the electric field to be constant over the spatial extent of the wave function. It implies a certain orientation of the electric dipole depending on the polarizability of the atom. In case of a two-level system, it is parallel to the applied field. For a plane wave propagating in the $\vec{e}_z$ direction and polarized according to the unit vector $\hat{u}$, the electric field satisfies

$$\vec{E}(\vec{r},\,t) = E_0\hat{u}\cos(kz - \omega_\mathrm{L}t)\,, \tag{d.4}$$

and the solution for the occupations (d.2) with initial condition $c_\mathrm{g} = 1$ reads

$$c_\mathrm{g}(t) = \left(\cos\frac{\tilde{\Omega}t}{2} - \mathrm{i}\frac{\Delta}{\tilde{\Omega}}\sin\frac{\tilde{\Omega}t}{2}\right)\mathrm{e}^{\mathrm{i}\Delta t/2} \tag{d.5a}$$

$$c_\mathrm{e}(t) = -\mathrm{i}\frac{\Omega}{\tilde{\Omega}}\sin\frac{\tilde{\Omega}t}{2}\mathrm{e}^{-\mathrm{i}\Delta t/2}\,, \tag{d.5b}$$

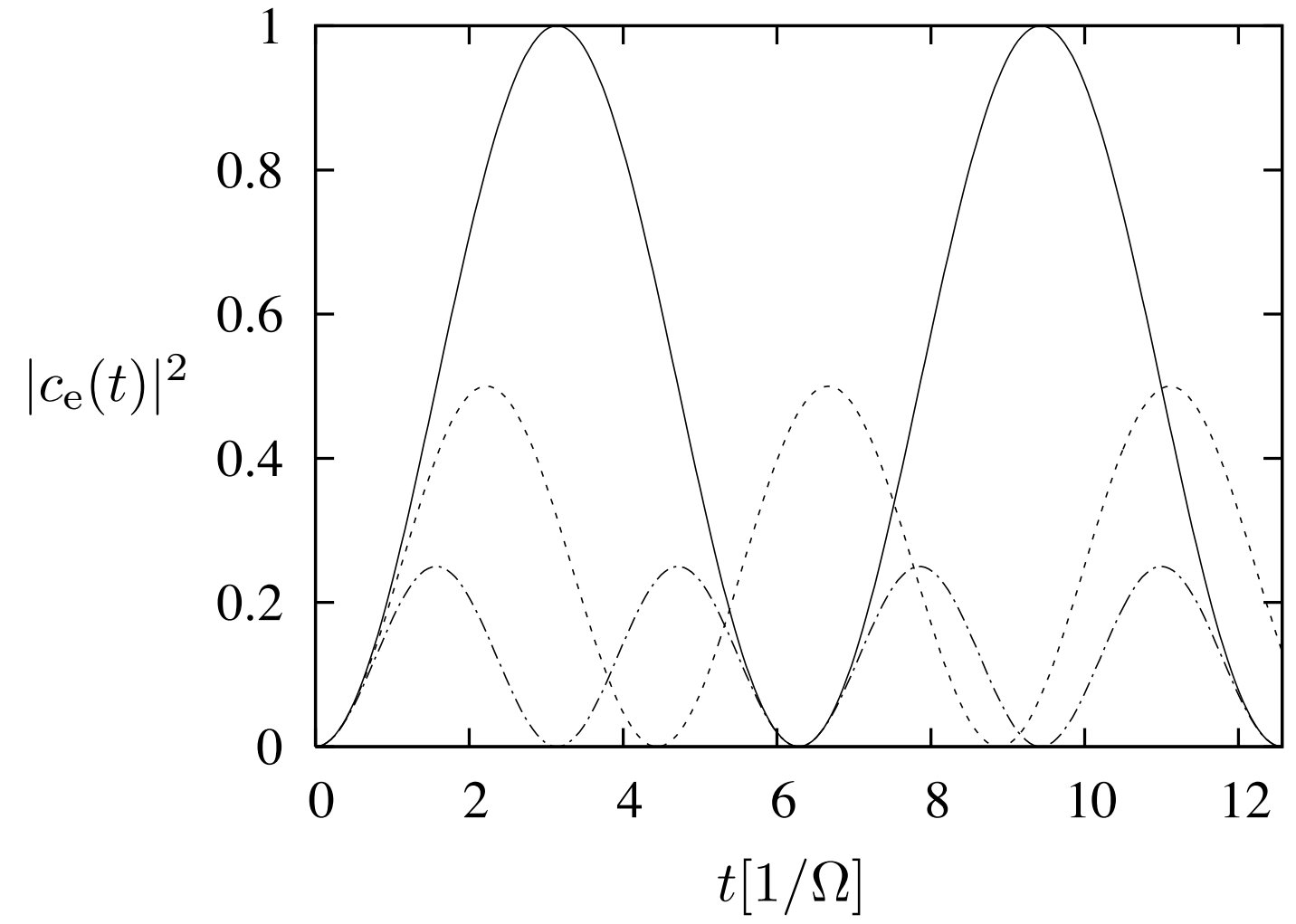

FIGURE 8: Probability $|c_{\mathrm{e}}(t)|^2$ to detect an atom in the excited state $|\mathrm{e}\rangle$ for different detunings: $\Delta = 0\,(\mathrm{solid})$, $\Omega\,(\mathrm{dotted})$, $\sqrt{3}\Omega\,(\mathrm{dash-dotted})$. Time is measured in units of $1/\Omega$.

where the Rabi frequency $\Omega$ is defined by the transition matrix element $\hbar\Omega \equiv -\mathrm{e}E_0\langle\Phi_{\mathrm{e}}|r|\Phi_{\mathrm{g}}\rangle$ and the generalized Rabi frequency equals $\tilde{\Omega} \equiv \sqrt{\Omega^2 + \Delta^2}$.

As depicted in Fig. 8, the probabilities $|c_{\mathrm{e,\,g}}|^2$ oscillate at $\tilde{\Omega}$ which is an increasing function of the absolute value of the detuning $\Delta$. Moreover, the influence of the photonic field shifts the eigenvalues of the unperturbed atomic states, which is the AC equivalent process to the quadratic Stark effect. For the electric field (d.4) and red detuning $\delta < 0$, the levels are shifted apart by

$$\Delta E_{\mathrm{e,\,g}} = \hbar(\tilde{\Omega} - |\delta|)/2\,. \tag{d.6}$$

In the limit $\Omega \ll |\delta|$, $\Delta E$ is proportional to the intensity of the field $\Omega^2$. For this reason, the above phenomenon is referred to as light shift. More generally, the light shift is derived as

$$\Delta E_{\mathrm{e,\,g}} = \sum_{i,\,j=x,\,y,\,z} \alpha_{ij}(\omega_{\mathrm{L}})\langle E_i(\vec{r},\,t)E_j(\vec{r},\,t)\rangle \sim I(\vec{r})/\Delta\,, \tag{d.7}$$

where $\alpha_{ij}(\omega_{\mathrm{L}})$ is the polarizability tensor of the atom, which has to be experimentally determined, and $\langle\ldots\rangle$ averages over the rapid temporal oscillations of the electric field. In case of an optical potential with an intensity gradient, e. g., a standing wave, this shift is spatially varying and leads to a conservative force. This dipole force is proportional to the intensity gradient of the coherent light field. For red detuning $\Delta < 0$, the atoms are attracted to the maxima of the intensity, the opposite holds for blue detuning. The Doppler force instead arises from phase variations of the light field and is dissipative in nature. In standing wave configurations, it depends on the velocity of the atoms and, for red detuning, it induces a viscous damping. This damping process strongly relies on the rate of spontaneous emission. In an intuitive picture, it can be understood as follows. For $\delta < 0$, atoms moving in the direction of one of the beams "see" the light Doppler-shifted towards resonance, whereas the light of the other beam is shifted away by the same magnitude. As a consequence, they absorb photons from the counterpropagating beam with higher probability. Spontaneous emission couples to all possible modes of the photonic field $\hbar\vec{k}$. This undirectioned scattering averages to an effective Doppler energy loss, which narrows the width of the density distribution function of the atoms. This mechanism of optical molasses is one of the most important tools in laser cooling.

## d.2 Optical Lattice Potentials

The simple model discussed in the previous section motivates the appearance of the two fundamental forces induced on atomic matter by coherent light waves. Depending on the structure of the electromagnetic field and the kinetics of the atom, the response to these forces is of different nature. In real atomic systems, where a variety of states is involved in the quantum dynamical process, the basic concepts are still applicable. Nonetheless, effects as Sysiphus, sub-Doppler, and sub-Recoil cooling are solely understood from the fully quantized treatment of the atomic system and the photonic field. This equivalently applies to the appearance of a manifest dipolar trapping force with non-vanishing spatial expectation value. The necessary theoretical description exceeds analytical treatment and as well the scope of this thesis. The reader is referred to the following excellent reviews and references therein [139, 140, 141].

The simplest realistic periodic dipolar potential is obtained when two counterpropagating red-detuned focussed Gaussian beams are overlapped. The intensity profile for $\vec{e}_z$ propagation is given by

$$I(\rho,\, z) = \frac{2P}{\pi\sigma^2(z)}\mathrm{e}^{-2\rho^2/\sigma^2(z)}\,, \tag{d.8}$$

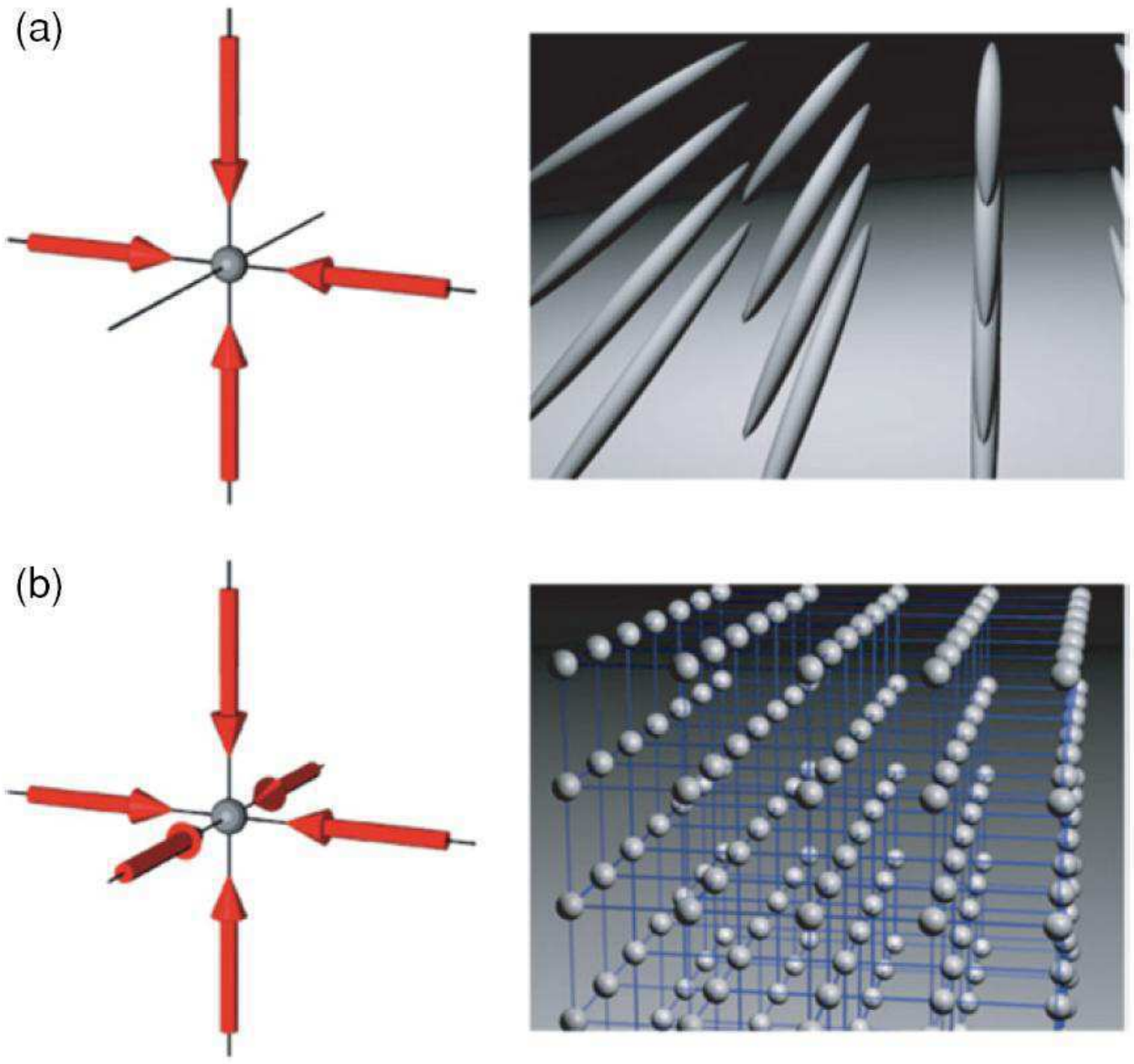

FIGURE 9: Optical lattice setups in a) 2D and b) 3D (*by courtesy of I. Bloch* [142]).

where $P$ denotes the laser power and $\sigma(z)$ is the standard radial width. The resulting 1D optical potential reads

$$V_{\rm dip}(\rho,\, z) = -V_0 \mathrm{e}^{-2\rho^2/\sigma^2(z)} \sin^2(kz)\,. \tag{d.9}$$

Higher dimensional configurations, as illustrated in Fig. 9, are realized when two standing waves are overlapped with an enclosed angle $\phi$. In the simplest case, $\phi = \pi/2$ and neglecting the radial curvature of the Gaussian profiles, the lattice potential equals

$$-V_0 \left[\cos^2(kx) + \cos^2(ky) + 2\hat{u}_1\hat{u}_2 \cos\Theta \cos(kx)\cos(ky)\right] \,. \tag{d.10}$$

Here, $\hat{u}_j$ is the polarization of the $j$-th standing wave and $\Theta$ denotes the temporal phase between them. Unless $\hat{u}_1 \perp \hat{u}_2$, it is essential to guarantee the phase stability of the beam pairs at high precision. Various methods may be employed, most of them include additional lasers. The generalization to three dimensions is straightforward. Nowadays, a vast variety of lattice topographies has been experimentally realized, e. g., the Kagomè geometry [80]. Even so-called superlattices have been implemented [143], where a standing wave is superimposed with an aligned wave of different frequency $\omega_{\rm S}$. This leads to a beating of the intensity. If $\omega_{\rm S}$ is sufficiently incommensurate with the original lattice frequency and given that the intensity of the superlattice is sufficiently small, a quasi-disordered system is realized. Another possibility is to tune the lattice frequency in between the fine or even hyperfine structure of two atomic states. Under specific polarization conditions, two state selective lattices with adjustable translational offset are obtained. These can be used for state-dependent transport or trapping of different hyperfine states.

Apart from the realization of rigid crystal potentials for neutral atoms, various further manipulations are feasible. The lattice may be tilted by a homogeneous acceleration of the lattice or by an additonal DC electric field. Thus, kinetic tunneling may be suppressed in any desired direction and replaced by tunneling-assisted Raman transitions, where the phase acquired by the atom is under full control of the experimentalist. The possibilities are numerous and seem to be solely restricted by the creativity of the experimentalist, a good enough accessibility of the trapped gas by a suitable setup of the optical table and the technical status quo.

# CHAPTER 1

# Fractional Quantum Hall States in Rotating Dipolar Fermi Gases

In this chapter, rotating quasi two-dimensional gaseous Fermi systems with dipole-dipole interactions are presented as a suitable candidate for the observation of fractional quantum Hall-like states. The chance to create these strongly correlated states in a controllable quantum optics experiment with various experimental degrees of freedom opens up the remarkable possibility to shed light on this intriguing phenomenon of condensed matter physics, which, even after more than two decades of intensive research, lacks a fundamental understanding.

## 1.1 The Quantum Hall Regime in Cold Gaseous Systems

As it was in Sec. (c) of the introduction, the Hamiltonian of a system of $N$ interacting particles confined to two-dimensional motion in the $xy$-plane in a rotating frame of reference with frequency $\vec{\Omega} = \Omega\vec{\mathbf{e}}_z$ reads

$$\mathcal{H} = \sum_{j=1}^{N} \frac{1}{2M} \left(\vec{p}_j - M\omega_\perp \vec{\mathbf{e}}_z \times \vec{r}_j\right)^2 + \left(\omega_\perp - \Omega\right) L_j^z + \mathcal{V}_{\text{int}} \tag{1.1}$$

where $\omega_\perp \ll \omega_z$ are the radial and $z$-axial frequency of the confining trap potential, respectively, $M$ is the mass of the particles, $\vec{r}_j$ and $\vec{p}_j$, are the 2D position and momentum vectors of the $j$-th particle, $L_j^z$ is the projected angular momentum of the $j$-th particle with respect to the $z$-axis, and $\mathcal{V}_{\text{int}}$ is the interaction potential. In the limit of critical rotation $\vec{\Omega} \to \vec{\omega_\perp}$, the second part of the above Hamiltonian $\mathcal{H}_\Delta \equiv \sum_j \left(\omega_\perp - \Omega\right) L_j^z \to 0$, and (1.1) is algebraically equivalent to the Hamiltonian of charged particles subject to a constant magnetic field of strength $B = 2\omega_\perp \frac{e}{cm}$, perpendicular to their plane of motion. Given this limit and strong enough $B$ that particles solely occupy the lowest kinetic energy manifold, i.e., the lowest Landau level, the system is dominated by interparticle interactions. According to Sec. (b) of the introduction, it is this strongly correlated regime where the fractional quantum Hall effect is experimentally observed in MOS-type field effect transistors and semiconductor heterojunctions. The microscopic details of this intriguing effect are yet to be theoretically understood. Furthermore, there is so far no accepted experimental proof of anyonic quasi-particles with fractional charge and statistics. These are predicted to appear as excitations in such a system. The fascinating features arising from their lower dimensional nature, for which the relativistic link of spin and statistics [147] does not apply, are not only interesting from a fundamental point of view. They are eagerly discussed in terms of a feasible implementation of quantum computing [145, 146]. Motivated by the possibility to map a fermionic fractional quantum Hall system to a bosonic version at adapted filling factors $\nu_B = \nu_F/(1-\nu_F)$, the appearance of highly interesting correlated phases in ultracold bosonic experiments has been thoroughly analyzed [105, 106, 123, 148]. In such an approach, the important features as off-diagonal long range order have to be preserved [149].

Consecutively, a path to experimentally create, manipulate and detect $\pi/2$-anyonic excitations of the bosonic $\nu = 1/2$ Laughlin state was proposed. However, because of the short-ranged character of interparticle interactions determined by the s-wave scattering length, it was found that atomic fractional quantum Hall states are detectable only for a small number of particles. This is due to the fact that Laughlin-like states do not play any specific role in a macroscopic system, when the interaction is short-ranged, because they are energetically degenerate with other states. In that case, the Jastrow prefactor in the corresponding wave functions, $\prod_{i<j}(z_i - z_j)^p$, where $z_j = x_j + \mathrm{i}y_j$ is the coordinate of the $j$-th particle, and $p$ is an integer (even for bosons and odd for fermions), significantly reduces the effects of a short-range interaction and, as a consequence, gapless excitations appear. This contrasts to the case of electrons where the Coulomb interaction favors fractional quantum Hall phases by lifting the degeneracy of the ground state and provides a gap for single-particle excitations [16]. Though, if more than one Landau level is occupied in the composite particle description of the fractional quantum Hall effect [92]), the situation may be improved [150] if the strength of interaction is drastically enhanced by a Feshbach resonance in the $p$-wave channel [151]. However, experiments performed in the vicinity of such a resonance are up to now inevitably accompanied by dramatic particle losses and their feasibility requires a more careful analysis.

## 1.2 Rotating Dipolar Fermi Gases

In this section, it is demonstrated that the above difficulties can be overcome when a rotating quasi two-dimensional gaseous system with dipole-dipole interactions is considered. The dipole moments are assumed to be aligned perpendicular to the plane of motion by external fields, and the interaction reads

$$V_d = \sum_{j<k}^{N} \frac{d^2}{|\vec{r}_j - \vec{r}_k|^3}, \tag{1.2}$$

with dipolar moment $d$. In particular, it is shown that dipole-dipole interactions favor fractional quantum Hall phases by creating a substantial gap in the single-particle excitation spectrum leading to macroscopically incompressible states. The analysis is concentrated on a quasi-hole excitation in the most famous fermionic Laughlin state at filling $\nu = 1/3$ in a homogeneous quasi 2D dipolar rotating Fermi gas with dipolar moments polarized perpendicular to the plane of motion.

In this case, the trial wave functions for the ground and quasi-hole excited states can be taken in the variational form proposed by Laughlin [20]

$$\psi_{\mathrm{L}}(\{z_j\}) = \mathcal{N} \prod_{k<l}^{N} (z_k - z_l)^{2m+1} \exp\Big(- \sum_{i=1}^{N} |z_i|^2 / 4l_0^2\Big), \tag{1.3}$$

$$\psi_{\mathrm{qh}}(\{z_j\}, \zeta_0) = \mathcal{N}_0 \prod_{j=1}^{N} (z_j - \zeta_0) \Psi_{\mathrm{L}}, \tag{1.4}$$

where $\mathcal{N}$ is a normalization factor, $z_j = x_j + \mathrm{i}y_j$ denotes the position of the $j$-th particle in complexified coordinates, $\zeta_0$ is the position of the quasi-hole, $l_0 = \sqrt{\hbar/(2\hbar\omega_\perp)}$ is the unit of magnetic length and $m$ is an integer which determines the filling fraction $\nu = 1/(2m+1)$. The choice of these wave functions in case of the considered system with dipole-dipole interactions can be justified as follows.

They are exact eigenfunctions for specific short-range interactions and are proven to be very good trial wave functions for the Coulomb interaction problem. Actually, as it was shown by Haldane (see the corresponding contribution in [16]), the Laughlin states are essentially unique and rigid at the corresponding filling factors, especially for $\nu = 1/3$. They are favored by strong short-range pseudopotential components, which are particularly pronounced in case of a dipolar potential, even more than in the Coulomb case.

Other possible candidates for the ground state, historically speaking the first proposed ones for the Coulomb problem, are charge-density wave(CDW) states. Though experimental and theoretical issues [152, 153, 154] excluded these states as suitable candidates for higher lowest Landau level fillings $\nu > 1/9$, it is not clear *ab initione* if this argumentation holds in case of a dipolar system. Especially the most probable candidate among these CDW states for electrons, i. e., the 2D Wigner crystal [156], has to be considered as a competing state. It is known that for a non-rotating dipolar Fermi gas in a 2D trap, this state has lower energy than the gaseous state for sufficiently high densities. The estimate of the stability region can be obtained from the Lindemann criterion demanding that the ratio $\gamma$ of the mean square difference of displacements in neighbouring lattice sites to the square of the interparticle distance $a$

$$\gamma = \left\langle (\mathbf{u}_i - \mathbf{u}_{i-1})^2 \right\rangle / a^2 \,, \tag{1.5}$$

should be less than some critical value $\gamma_c$ [157]. The results of Ref. [158] indicate that $\gamma_c \approx 0.07$. For zero temperature, $\gamma$ could be estimated as $\gamma \sim \hbar/ma^2\omega_D$, where $\omega_D$ is the characteristic Debye frequency of the lattice phonons, $\omega_D^2 \sim 36d^2/ma^5$. As a result, the dipolar crystal in a non-rotating gas is stable, if the interparticle distance $a = (\pi n_f)^{-1/2}$ satisfies the condition $a < a_d(6\gamma_c)^2 \ll a_d$, where $a_d = md^2/\hbar^2$ can be considered as the characteristic range of dipolar interactions. The above condition assumes the gas to be in the strongly correlated regime, $V_d \sim d^2/a^3 \sim (a_d/a)(\hbar^2/ma^2) \gtrsim \varepsilon_F/(6\gamma_c)^2 \gg \varepsilon_F$. A strong magnetic field with the cyclotron frequency larger than the Debye frequency, $\omega_c > \omega_D$, favors the crystalline state by modifying the vibrational spectrum of the crystal. In this case, $\gamma \sim \hbar/ma^2\omega_c$ [156], and the corresponding critical value is $\gamma_c \approx 0.08$ [158]. Therefore, the crystalline state should be stable if $\gamma \sim \nu/2 < \gamma_c$. This limits the filling factor $\nu$ to small values $\nu < 1/6$.

The above estimates are strongly supported by a recent detailed investigation on dipolar Wigner crystals [159]. Considering the influence of the experimentally present extension of the gas in the rotational direction $\vec{e}_z$, it is proven that a dipolar Wigner crystals has a lower energy than the Laughlin state for $\nu \leq 1/7$. Calculations included Monte Carlo simulations for up to $N = 512$ particles. In order to truly find a crystalline state for smaller fillings, it is crucial to analyze its stability with respect to phonon modes. In the limit of a strong magnetic field, a self-consistent iterative integration of the phonon-phonon vertex Dyson equation predicts a melting point at $\nu \approx 4/23 > 1/7$.

As a result, the Wigner crystal is a promising candidate for lower fillings $\nu \leq 1/7$, whereas the Laughlin wave function (1.3) is strongly supported to properly describe the ground state at $\nu = 1/3$. Especially, the more pronounced short-range character of dipolar interactions suggests $\psi_\mathrm{L}$ to be an even better approximation than in the Coulomb case.

In order to prove the robust incompressibility of the state $\psi_\mathrm{L}$, in a quasi 2D dipolar Fermi gas, the energy gap $\Delta\varepsilon_\mathrm{qh}$ for the quasi-hole excitation is calculated. This gap can be expressed in terms of the pair correlation functions of the ground state $g_0(z_1,\, z_2)$ and the quasi-hole excited state $g_\mathrm{qh}(z_1,\, z_2)$, respectively, as

$$\Delta\varepsilon_\mathrm{qh} = \int \mathrm{d}^2 z_1 \mathrm{d}^2 z_2 \frac{\mathrm{d}^2}{|z_1 - z_2|^3} \Big[ g_\mathrm{qh}(z_1,\, z_2) - g_0(z_1,\, z_2) \Big] \, . \tag{1.6}$$

For the states (1.3) and (1.4) with the quasi-hole located at $\zeta_0 = 0$, the functions $g_0$ and $g_\mathrm{qh}$ were calculated by Monte Carlo methods [90]. In the thermodynamic limit, an excellent analytical approximation has been deduced by Girvin [160, 161]

$$g_0(z_1,\, z_2) = \frac{\nu^2}{(2\pi)^2} \left( 1 - \mathrm{e}^{-\frac{|z_1 - z_2|^2}{2}} - 2 \sum_j^{\mathrm{odd}} \frac{C_j(\nu)}{4^j j!} \, |z_1 - z_2|^{2j} \, \mathrm{e}^{-\frac{|z_1 - z_2|^2}{4}} \right) , \tag{1.7a}$$

$$g_\mathrm{qh}(z_1,\, z_2) = \frac{\nu^2}{(2\pi)^2} \left[ \prod_{j=1}^{2} \left( 1 - \mathrm{e}^{-\frac{|z_j|^2}{2}} \right) - \mathrm{e}^{-\frac{|z_1|^2 + |z_2|^2}{2}} G(z_1,\, z_2) \right] , \tag{1.7b}$$

$$G(z_1,\, z_2) = \left| \mathrm{e}^{\frac{z_1 \bar{z}_2}{2}} - 1 \right|^2 + 2 \sum_j^{\mathrm{odd}} \frac{C_j(\nu)}{4^j j!} \sum_{k=0}^{\infty} \frac{|F_{j,k}(z_1,\, z_2)|^2}{4^k k!} \, , \tag{1.7c}$$

$$F_{j,k}(z_1,\, z_2) = \frac{z_1 z_2}{2} \sum_{r=0}^{j} \sum_{s=0}^{k} \binom{j}{r} \binom{k}{s} \frac{(-1)^r z_1^{r+s} z_2^{j+k-(r+s)}}{\sqrt{(r+s+1)(j+k+1-(r+s))}} \, , \tag{1.7d}$$

where the values of the expansion coefficients $C_j(\nu)$ can be found in Ref. [161]. In Fig. 1.1, the difference $g_\mathrm{qh} - g_0$ is plotted for the particular choice $z_1 = 3$ and $\zeta_0 = 0$.

With a close look at (1.7), the correlation functions can be decomposed into two parts. The first is — up to the normalization coefficient in front — independent from the filling factor, i.e., from the coefficients $C_j(\nu)$, and the second consists of the modulation terms of the correlation functions, which are state-specifically determined by exactly these coefficients.

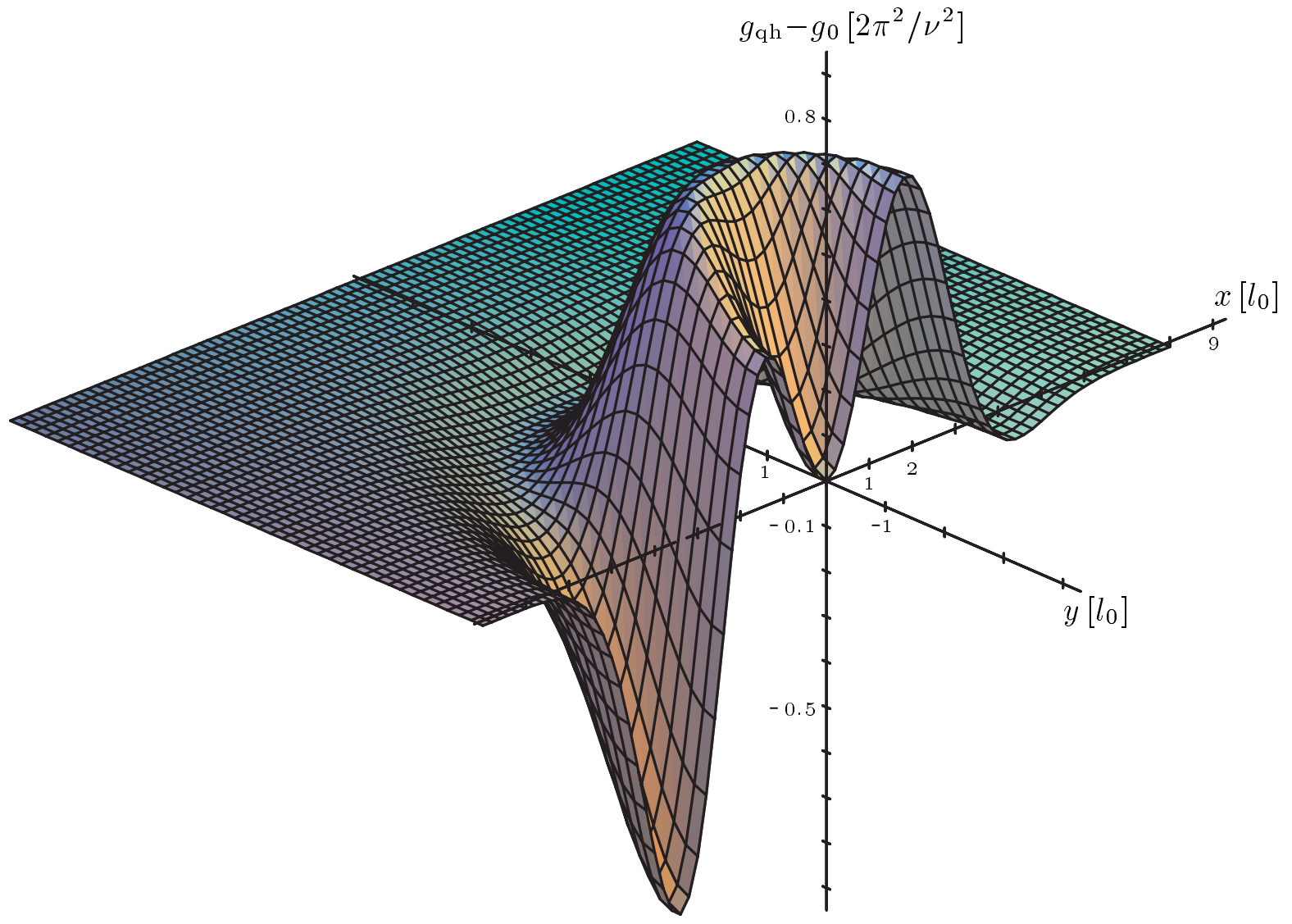

FIGURE 1.1: The difference $g_{\rm qh} - g_0$ as a function of $z \equiv z_2 - z_1$ for $\zeta_0 = 0$ and $z_1 = 3$. Both particles strongly avoid each other, and rotational invariance is broken by the quasi-hole.

By virtue of the translationally invariant interaction, the integration is performed in center of mass coordinates $z_1 = z_{\rm cm} + h$ and $z_2 = z_{\rm cm} - h$ in both cases. This demands a delicate proceeding, as translational invariance is explicitly broken in $g_{\rm qh}$ due to the presence of the quasi-hole, whereas $g_0$ is only a function of the relative coordinate $h$. Naively integrating the latter over $z_{\rm cm}$ leads to a pole divergence which has to be countered by terms of $g_{\rm qh}$.

Regardless of this fact, the first part of the correlation functions

$$P_1(z_1,\, z_2) = \frac{\nu^2}{(2\pi)^2}\left[\prod_{j=1}^{2}\left(1 - \mathrm{e}^{-\frac{|z_j|^2}{2}}\right) - \mathrm{e}^{-\frac{|z_1|^2+|z_2|^2}{2}}\left|\mathrm{e}^{\frac{z_1 z_2^\star}{2}} - 1\right|^2 - \left(1 - \mathrm{e}^{-\frac{|z_1-z_2|^2}{2}}\right)\right] \tag{1.8}$$

can be integrated analytically

$$Q_1 = \int \mathrm{d}^2 z_1 \mathrm{d}^2 z_2 \frac{\mathrm{d}^2}{|z_1 - z_2|^3} P_1(z_1,\, z_2) = \sqrt{2\pi}\nu^2 \mathrm{d}^2\,. \tag{1.9}$$

As indicated by the above comments, the second part

$$P_2(z_1,\, z_2) = \frac{2\nu^2}{(2\pi)^2}\sum_j^{\rm odd}\frac{C_j(\nu)}{4^j j!}\left[|z_1 - z_2|^{2j}\,\mathrm{e}^{-\frac{|z_1-z_2|^2}{4}} - \mathrm{e}^{-\frac{|z_1|^2+|z_2|^2}{2}}\sum_{k=0}^{\infty}\frac{|F_{j,k}(z_1,\, z_2)|^2}{4^k k!}\right] \tag{1.10}$$

involves major difficulties. In order to identify and regularize all singularities, the coordinate transformation demands the analytic treatment of $\sum_k$ in (1.7c) as follows

$$F_{j,k}(z_1,\, z_2) = \frac{z_1 z_2}{2\pi}\int \mathrm{d}\alpha_1 \mathrm{d}\alpha_2 \left(z_2 \mathrm{e}^{-\alpha_2^2} - z_1 \mathrm{e}^{-\alpha_1^2}\right)^{j+k}\mathrm{e}^{-(\alpha_1^2+\alpha_2^2)} \tag{1.11}$$

By this, $\sum_k$ can be evaluated explicitly via

$$\sum_{k=0}^{\infty} \frac{\left(z_2 \mathrm{e}^{-\alpha_2^2} - z_1 \mathrm{e}^{-\alpha_1^2}\right)^k}{4^k k!} = \exp\left[\frac{1}{4}\left(z_2 \mathrm{e}^{-\alpha_2^2} - z_1 \mathrm{e}^{-\alpha_1^2}\right)\right] . \tag{1.12}$$

Analogously, the complex conjugated part involved in the absolute value of (1.10) is converted, for which two further integrations $\int \mathrm{d}\beta_1 \mathrm{d}\beta_2$ are introduced, which lead to another term equal to the complex conjugate of (1.12) with $\alpha_j$ replaced by $\beta_j$. Now, (1.10) is transformed to center of mass coordinates and multiplied by a regularization parameter $\exp(-\epsilon\,|z_{\mathrm{cm}}|)$. Then, all polynomials in $z_{\mathrm{cm}}, \bar{z}_{\mathrm{cm}}$ are expressed as derivatives with respect to auxiliary parameters $\eta, \bar{\eta}$. Finally, $z_{\mathrm{cm}}$ is integrated out, the derivatives are applied, and $\eta, \bar{\eta}$ are set to zero

$$\begin{aligned}
Q_2 &= \int \mathrm{d}^2 z_1 \mathrm{d}^2 z_2 \frac{\mathrm{d}^2}{|z_1 - z_2|^3} P_2(z_1,\, z_2) \\
&= \frac{\mathrm{d}^2 \nu^2}{32\pi^3} \sum_j^{\mathrm{odd}} \frac{C_j(\nu)}{4^j j!} \int \mathrm{d}^2 h \frac{1}{|h|^3} \Bigg( \int \mathrm{d}\alpha_i \mathrm{d}\beta_i \Bigg\{ \frac{2^{2j+1}\,|h|^{2j}\,\mathrm{e}^{-|h|^2}}{(\epsilon + \frac{\vec{\alpha}^2 + \vec{\beta}^2}{2})^3} - \frac{\mathrm{e}^{-(\vec{\alpha}^2 + \vec{\beta}^2)}}{A^{3+j}(\epsilon)} \mathrm{e}^{-B(\epsilon)|h|^2} (a_- b_-)^j \\
&\times \sum_{k=0}^{j+2} (j+2-k)!\,|h|^{2k}\,A^k(\epsilon)(K_1 K_2)^k \Bigg[ \binom{j}{k-2}^2 (K_1 K_2)^{-2} \gamma_1 \delta_1 + \binom{j}{k-1}^2 (K_1 K_2)^{-1} \gamma_2 \delta_2 \\
&+ \binom{j}{k}^2 + \binom{j}{k-1}\binom{j}{k-2} (K_1^{-2} K_2^{-1} \gamma_1 \delta_2 + K_1^{-1} K_2^{-2} \gamma_2 \delta_1) + \binom{j}{k}\binom{j}{k-1} (K_1^{-1} \gamma_2 + K_2^{-1} \delta_2) \\
&+ \binom{j}{k}\binom{j}{k-2} (K_1^{-2} \gamma_1 + K_2^{-2} \delta_1) \Bigg] \Bigg\} \Bigg) \Bigg|_{\epsilon = 0} .
\end{aligned} \tag{1.13}$$

Here,

$$\begin{aligned}
A(\epsilon) &= 1 + \epsilon - \frac{a_+ b_+}{4} & B(\epsilon) &= 1 - \frac{a_- b_-}{4}\left[1 + a_+ b_+ / 4A(\epsilon)\right] \\
K_1 &= a_+ b_- / 4A(\epsilon) + b_0 & K_2 &= a_- b_+ / 4A(\epsilon) + a_0 \\
\gamma_1 &= a_+^2 b_-^2 / 16 A^2(\epsilon) - 1 & \delta_1 &= a_-^2 b_+^2 / 16 A^2(\epsilon) - 1 \\
\gamma_2 &= a_+ b_- / 2A(\epsilon) & \delta_2 &= a_- b_+ / 2A(\epsilon)
\end{aligned} \tag{1.14}$$

with $a_i = \exp(-\alpha_i^2)$, $a_\pm = a_2 \pm a_1$, $a_0 = a_+/a_-$ and analogously for $b_i$. Even though (1.13) may seem a bit complex, its underlying symmetry allows to identify the regularization counterterms. Interpreting $\int \mathrm{d}\alpha_i \mathrm{d}\beta_i$ as a four-dimensional spherical integral, it is obvious that divergencies in $\epsilon$ solely occur when $(\alpha_1,\, \alpha_2,\, \beta_1,\, \beta_2) \to \vec{0}$. From the set of parameters (1.14), solely the $K_i$ diverge because of $a_-$ and $b_-$ in the above limit, respectively. The leading term of the quasi-hole correlation function is identified as

$$\frac{\mathrm{e}^{-(\vec{\alpha}^2+\vec{\beta}^2)}}{A^{3+j}(\epsilon)}\mathrm{e}^{-B(\epsilon)}(a_-b_-)^j 2!\,|h|^{2j}\,A^j(\epsilon)(K_1K_2)^j\binom{j}{j}^2 \longrightarrow \frac{2^{2j+1}\,|h|^{2j}\,\mathrm{e}^{-|h|^2}}{(\epsilon+\frac{\vec{\alpha}^2+\vec{\beta}^2}{2})^3} \tag{1.15}$$

which exactly cancels the pole in $\epsilon$ of the homogeneous term derived as

$$\frac{1}{2\pi}\int \mathrm{d}\alpha_i\mathrm{d}\beta_i\frac{1}{(\epsilon+\frac{\vec{\alpha}^2+\vec{\beta}^2}{2})^3}=\frac{\pi}{\epsilon} \tag{1.16}$$

There are further radial divergencies proportional to $\ln(\epsilon)$, but these are always cancelled by integration over the four-dimensional solid angle. Nonetheless, for a numerical treatment of (1.13) in terms of Monte Carlo methods, it is crucial to identify all logarithmic terms analytically.

Calculations have been performed for the primary Laughlin state at filling $\nu = 1/3$, which possesses the most pronounced gap. In that case, it is sufficient to consider solely the first two coefficients $C_1(1/3)=1$ and $C_3(1/3)=-1/2$ in order to achieve an accuracy better than $98\%$. Having identified all pole and logarithmic counterterms, Eq. (1.6) is integrated numerically, and the result reads

$$\Delta\varepsilon_{\mathrm{qh}} = (0.9271 \pm 0.019)\, d^2/l_0^3 \tag{1.17}$$

for the energy gap in the spectrum of quasi-holes. Naturally, a gap on the same order of magnitude is to be expected in the spectrum of quasi-particles (quasi-electrons, in the language of the fractional quantum Hall effect) although calculations in this case are much more difficult. This is based on the fact that no proper closed or even approximate expression for the corresponding pair correlation function exists. In order to make an experimental prediction based on this result, (1.17) is rewritten to

$$\Delta\varepsilon_{\mathrm{qh}} = (0.9271 \pm 0.019)\hbar\omega_c(a_d/l_0),$$

For a dipole moment on the order of $0.5\mathrm{Debye}$ and a mass $m \sim 30$ atomic mass units, the value of $a_d$ is on the order of $10^3\mathring{A}$. Given a trap frequency of $\omega_\perp \sim 2\pi 10^3\mathrm{Hz}$, a gap of the order $\Delta\varepsilon_{\mathrm{qh}} \sim 30\mathrm{nK}$ is obtained while $\Delta\varepsilon_{\mathrm{qh}}/\hbar\omega_c < 1$.

This result shows (see Fig. 1.2) that on the one hand, the interparticle interaction does not mix different Landau levels, and thus guarantees the reliability of the lowest Landau level approximation used to apply the Laughlin wave function ansatz. On the other hand, it guarantees that the neglected term $\mathcal{H}_\Delta$ in the Hamiltonian (1.1) is indeed small and does not influence the trial wave functions. This is an important consequence, as $\mathcal{H}_\Delta$ is inevitably present in experimental realizations for stability reasons.

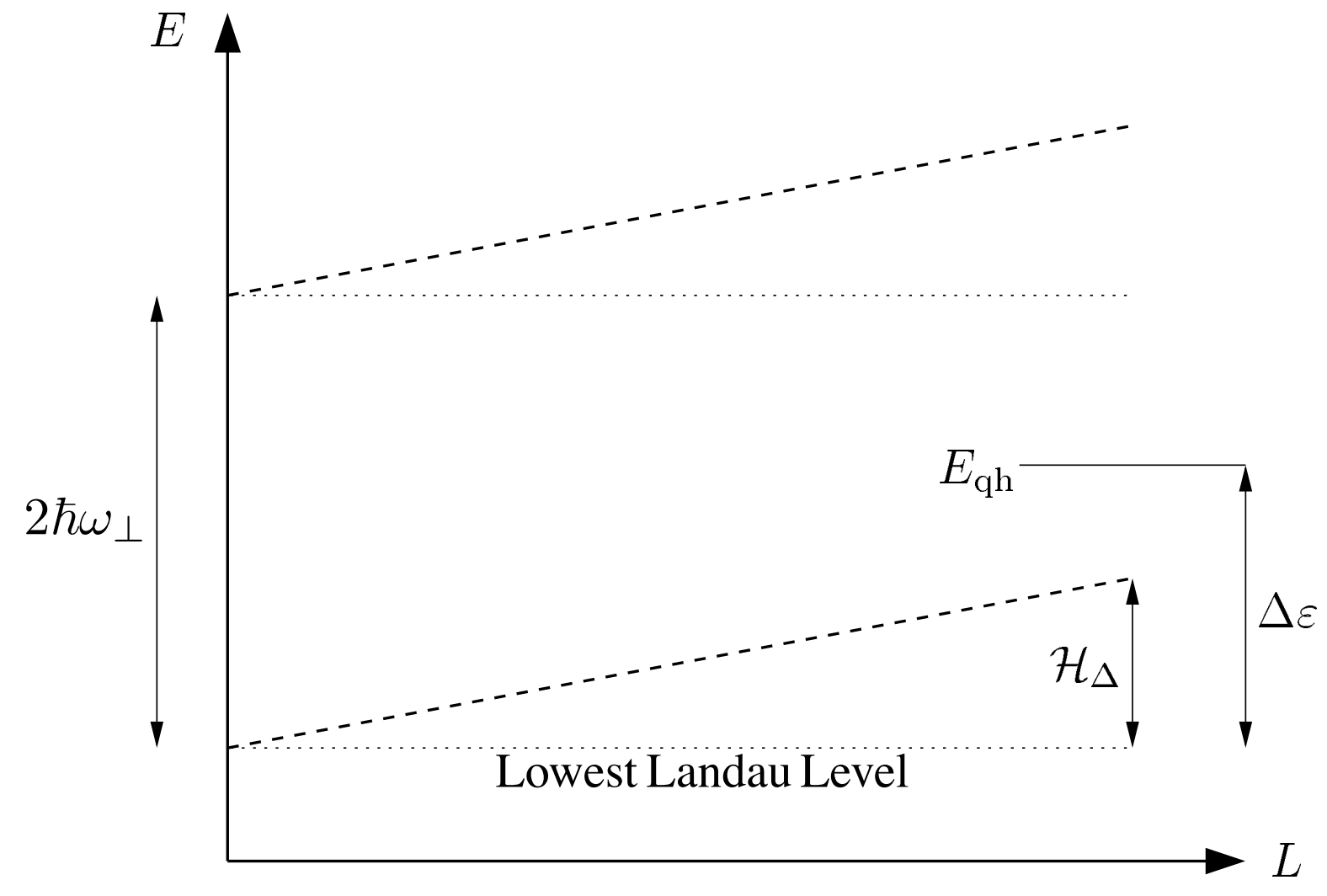

FIGURE 1.2: Single-particle energy levels of the Hamiltonian $\mathcal{H} = \mathcal{H}_{\rm Landau} + \mathcal{H}_\Delta$ versus their angular momentum $L$.

This result has to be compared with the gap in the quasi-particle spectrum in the corresponding bose-system with contact interactions [128]. In that case, in order to obtain a sustantial gap $\Delta\varepsilon \approx 0.5\hbar\omega_c$, the $s$ -wave scattering length needs to be comparable with the magnetic length. Typically, such large values could be achieved by using Feshbach resonance techniques. However, performing an experiment with the system tuned close to a Feshbach resonance, requires a thorough analysis of inelastic losses.

## 1.3 Experimental Realization and Detection

Given this substantial gap in a dipolar system, possible ways to experimentally realize, detect and manipulate a cold atom gas in the fractional quantum Hall regime have to be discussed. As discussed in the introductory Sec. (c), there exist two experimental methods to create rapidly rotating gas clouds. In the first approach, a rotating Bose-Einstein condensate is evaporatively spinned up in a harmonic trap [117, 144]. In this case, the term $\mathcal{H}_\Delta$ is inevitably present in the Hamiltonian and limits the total number of particles $N$. The condition $\mathcal{H}_\Delta < \Delta\varepsilon_{\rm qh} \lesssim \hbar\omega_c$ and the fact that single-particle states with angular momenta up to $L_z = 3N(N-1)/2$ contribute to the states (1.3) and (1.4), impose the constraint $3N(N-1)/2 < \Delta\varepsilon_{\rm qh}/\Delta\omega$. For the experimentally realized $\Delta\omega/\omega_\perp = 10^{-3}$, it gives $N < 30$. It has to be stressed that this bound is obtained by completely neglecting interactions. These drastically change the simplified picture of the spectrum as shown in 1.2. The above constraint is thus to be taken as a very conservative lower bound of possible particle numbers. Apart from this, it is expected to be large enough to guarantee the validity of the above calculations, which assumed a homogeneous gas in the thermodynamic limit.

In the second class of experiments the bosonic gas sample is stirred by a moving optical or magnetic anisotropy [102, 115, 116, 117] One of the above groups has managed to superimposed the harmonic potential of the optical trap with a higher harmonic, i. e., quartic confining potential [119]. This allows to reach and even exceed the critical value $\omega_c$. Thus, the term $\mathcal{H}_\Delta$ can be neglected and the number of particles is only limited by the radial size of the gas cloud confined in an effective harmonic potential.

With an experiment of this type, it is possible to impose an additional quenched disorder potential in the rotating frame, generated by speckle radiation from a rotating diffractive mask [162]. The rotation of this mask is to be synchronized with the stirring laser. The generated disorder localizes single-particle excitations that appear in the system when the filling factor $\nu$ deviates from the value $1/3$. This provides fractional quantum Hall states with the necessary robustness for experimental observation.

The question remains how to experimentally detect fractional quantum Hall states in a feasible manner. One possibility is certainly the introductorily discussed measurement of the statistics of quasi-holes using a Ramsey-type interferometric method similar to that proposed in Ref. [107]. Further specific characteristics are properties of low energy surface, i.e., edge modes, in analogy to chiral edge states of electrons in the semiconductor quantum Hall effect. The corresponding analysis for a rotating bosonic cloud was performed in Ref. [163]. Yet another way is the detection of collective modes similar to magnetorotons and magnetoplasmons in electronic quantum Hall systems (see, e.g., Girvin's contribution to Ref. [16]).

So far, no rotating Fermi gas has been realized experimentally. Both of the above experiments are focussed on bosonic species. This is mainly due to the fact that cooling methods are momentarily far more efficient and technically established in these systems. Furthermore, the pioneering experiments in cold gaseous matter aimed at the condensation of atoms, and from this starting point, a wide range of phenomena has been studied in the bosonic mean-field regime of weak interactions. The consecutive theoretical investigation on the fascinating branch of vortex dynamics naturally led to ENS and JILA experiments. Yet, in recent years, fermionic gases and mixtures have been discovered as promising systems opening up a rich field of completely novel phenomena never to be found in condensed matter systems. With this rewarding perspective, strong efforts are made to improve engineering and control of these systems, and the realization of the first rotating Fermi gas is just a matter of time.

At least equally challenging was the historic path towards the creation of degenerate dipolar quantum gases. Initially, the large magnetic moments of chromium($6\mu_{\mathrm{B}}$) and europium$7\mu_{\mathrm{B}}$ made them interesting and suitable elements for magnetic trapping [164, 165], before they were considered fascinating candidates for a novel type of Bose-Einstein condensate promising an enriched variety of phenomena to be observed due to the competition between long- and short-ranged interactions [166]. Different techniques as buffer gas [165] and laser cooling [167] were implemented. The latter turned out to be the more successful one, but it took six more years before a degenerate quantum gas of chromium could become reality [74]. Shortly after, the manifestation of dipolar effects on ultracold atomic gases was detected for the first time [168].

Despite this breakthrough, it has to be stated that if the fractional quantum Hall effect is to be realized in an experiment of rotating chromium atoms, the quasi-hole excitation gap $\Delta\varepsilon$ is smaller by two orders of magnitude, and temperatures below $1\mathrm{nK}$ have to be achieved to make it visible. Yet, a variety of alternative methods for the experimental realization of dipolar gases has been proposed (for a review, see [155]), and exciting perspectives are opened by recent experiments on polar molecules [169, 170, 171, 172, 173, 174]. There, dipolar moments of several Debye are found shifting the gap temperature to the $\mu\mathrm{K}$ scale.

Strong efforts are being made to meet the above requirements, nonetheless the different disciplines have to be unified in a single pioneering experiment. This will open the door to a new generation of cold atomic experiments engineering a full-feature condensed matter scenario, i.e., long-range interacting particles subject to magnetic fields even in periodic potentials.

For the time being, as dipolar gases are not disposable for rotating experiments, maybe the most promising approach to reach the lowest Landau level regime aims at samples with but a few particles. To access and analyze such a system, a sufficient strength of signal is needed. For this, multiple coherent copies are needed which may be nucleated in, e. g., arrays of two-dimensional rotating microtraps in an optical lattice [175] or equivalently stacks of rotating microlenses [176]. Yet another option is the manipulation of microscopically sized traps created by tightly focused lasers [177]. In such arrangements, it will be natural to study mesoscopic, or even microscopic systems of few atoms whose physical properties are strongly affected by finite size effects. The latter have to be thoroughly studied to understand the fundamental crossover phenomena on the path to a strongly correlated system.

# CHAPTER 2

# Ordered Structures in Rotating Atomic Microsamples

In this chapter, the properties of quasi two-dimensional rotating atomic systems constituted of but a few number of atoms are studied in the lowest Landau level regime. The focus is set on the evolution of ground state structures on the crossover from the weakly to the strongly interacting regime. These structures are revealed when, e. g., the symmetry of the Hamiltonian is explicitly broken so that the one-particle density already exhibits order due to coherent mixing of degenerate states. Equivalently, ordered patterns hidden in a pure state with well-defined angular momentum become evident on the level of the pair correlation function. Calculations were performed by means of exact diagonalization. This formalism turns out to be the appropriate method to deal with small systems, for which the assumptions made in mean field theories are not applicable. In contrast to other branches of physics, such systems are not only tractable models, but are experimentally accessible. Both density and pair correlation functions, are measurable by various experimental techniques.

After the introduction and motivation of general concepts, bosonic systems with short-ranged interactions are analyzed before an investigation of fermionic dipolar gases subsequently ties up to the results and discussion of the previous chapter.

## 2.1 Ordered Structures

Ordered structures, and in particular hidden ordered structures, have been a subject of intensive studies in the physics of Bose-Einstein condensates, and more generally, in the physics of ultracold atoms.

The paradigm of such structures is realized in the interference of two Bose-Einstein condensates observed in seminal experiments of Ref. [178]. In a gedankenexperiment, where despite the superselection rule, two condensates could be prepared in coherent atomic quantum states, these states would be characterized by fluctuating atom numbers $N_1$, $N_2$, but sharply defined phases $\phi_1$, $\phi_2$. Thus, they would minimize the Heisenberg uncertainty relation for number and phase operators and the phase difference $\Delta\phi = \phi_1 - \phi_2$ would determine the position of the interference fringes. Similarly, if two condensates were prepared by splitting a parent condensate with a fixed number of atoms $N = N_1 + N_2$, it would be possible to arrive at sharp values of both $\Delta\phi$ and $N_1 - N_2$. Amazingly, the interference pattern will also appear if the two Bose-Einstein condensates in Fock states with fixed $N_1$ and $N_2$ overlap. As pointed out in Ref. [179, 180, 181], this is deeply rooted in the fact that as soon as atoms are detected without knowledge which condensate they originate from, the measurement will introduce the necessary uncertainty of the atom numbers and narrow the relative phase distribution. As a consequence, an interference pattern with a sharply defined $\Delta\phi$ is obtained in each realization of the measurement. By this, the measurement process itself uncovers the otherwise hidden interference pattern in the two-point first order correlation function of atomic creation and annihilation field operators $\langle\hat{\Psi}^\dagger(\vec{r})\hat{\Psi}(\vec{r}\,')\rangle$. If experimentally averaged over many realizations, this interference pattern vanishes, since each realization leads to a different and completely random $\Delta\phi$. Similar measurement induced structures, and the interplay between single shot and averaged results were also discussed in the context of dark solitons in quantum degenerate Bose gases [182].

Other types of ordered structures occur in rotating Bose-Einstein condensates. In the standard scenario, as the rotational frequency increases, more and more vortices appear in form of regular structures [102, 108, 115]. As their number grows, they organize themselves in a triangular Abrikosov lattice [100]. This manifestation may seem confusing, since the ground state of a rotating system in a harmonic trap is expected to be rotationally invariant and to have a fixed total angular momentum. *Ergo*, it should not at all exhibit any structures that break rotational symmetry as does the Abrikosov lattice. In reality though, as discussed in Sec. (c) of the introduction, the preparation of vortices is performed by a "laser stirring" process which explicitly breaks rotational symmetry and introduces significant couplings between states with different total angular momenta. Here, the preparation process (which may also be regarded as a sort of measurement) reveals otherwise hidden structures in the density of the condensate, i.e., in the one-point first order correlation function $\langle\hat{\Psi}^\dagger(\vec{r})\hat{\Psi}(\vec{r})\rangle$.

As it is very well known from quantum optics, measurements of first order correlation functions, i. e., first order "coherences", do not always reveal the underlying structures. If not, higher order coherences have to be measured, e. g. $\langle\hat{\Psi}^\dagger(\vec{r}_1)\hat{\Psi}^\dagger(\vec{r}_2)\hat{\Psi}(\vec{r}_3)\hat{\Psi}(\vec{r}_4)\rangle$. As discussed in [183], the archetype for this necessity goes back to Michelson interferometry which is restricted to first order coherences and is sensitive to atmospheric fluctuations. This deficiency stimulated Brown and Twiss to measure the intensity-intensity correlations of the radiation coming from Sirius [184], which in turn allowed them to precisely determine the coherence length and the angular size of this star.

Measurements of second order correlations play an important role in the physics of ultracold gases (for earlier works on atomic beams, see [185]). The most directly accessible quantity is the density-density correlation, in the following named pair correlation function,

$$\rho^{(2)}\left(\vec{r}_1,\, \vec{r}_2\right) = \langle\hat{\Psi}^\dagger(\vec{r}_1)\hat{\Psi}^\dagger(\vec{r}_2)\hat{\Psi}(\vec{r}_2)\hat{\Psi}(\vec{r}_1)\rangle\,, \tag{2.1}$$

which formally is the two-point second order correlation function of the atomic field operators. It was directly measured in a recent atom counting experiment [186], for the first time demonstrating an atomic Hanbury Brown-Twiss effect for thermal atoms and the second order coherence of a Bose-Einstein condensate. Earlier, a 4-point second order correlation function was detected [187, 188] where density-density correlations of interfering condensates have been monitored in order to precisely determine the phase coherence length of quasi 1D condensates, in full analogy to the Hanbury Brown and Twiss method.

Recently, yet another tool, i.e., noise interferometry was proposed to analyze visible and hidden structures appearing in various quantum phases of ultracold gases [189, 190, 191, 192]. This method allows to determine density-density correlations, and was employed by several groups to study interference phenomena of independent condensates [193], residual coherence and lattice order in Mott insulators [194], as well as pair correlations of fermionic atoms in a Fermi superfluid [195].

At this point it is necessary to mention that the double (spatial and temporal) Fourier transform of the pair correlation function is known to be a dynamical structure factor [108], and is accessible, e. g., in Bragg scattering experiments [196].

## 2.2 Rotating Bose Gases

This section concentrates on phenomena to be observed in systems of $N$ bosonic atoms trapped in a rotating parabolic potential and confined to two-dimensional motion. Along the lines of the introductory Sec. (c), the corresponding Hamiltonian in a rotating frame of reference with frequency $\vec{\Omega} = \Omega\vec{\mathrm{e}}_z$ reads

$$\mathcal{H} = \sum_{j=1}^{N} \frac{1}{2M} \left(\vec{p}_j - M\omega_\perp \vec{\mathrm{e}}_z \times \vec{r}_j\right)^2 + \left(\omega_\perp - \Omega\right) L_j^z + g \sum_{k<l} \delta\left(\vec{r}_k - \vec{r}_l\right) , \tag{2.2}$$

where $\omega_\perp$ is the radial trap frequency, $M$ is the mass of the particles, $\vec{r}_j$ and $\vec{p}_j$, are the 2D position and momentum vectors of the $j$-th particle, $L_j^z$ is the projected angular momentum of the $j$-th particle with respect to the $z$-axis, and $g$ is the interaction coefficient that approximates the potential of the Van der Waals forces between the atoms. In the very dilute limit, these are properly described by the $s$-wave scattering length $a_s$ via

$$g = \frac{4\pi\hbar^2 a_s}{M} \tag{2.3}$$

The rotation subjects the particles to an effective magnetic field described by the vector potential $\vec{A}^\star = (M\Omega\mathrm{c}/\mathrm{e})\vec{\mathrm{e}}_z \times \vec{r}$, where the elementary charge and the speed of light are solely introduced for reasons of algebraic equivalence.

In the following, it is assumed that particles only occupy the lowest Landau level. The validity of this approach bases on the interplay between effective strength of the magnetic field, interaction strength and rotational frequency and relies on sufficiently low experimental temperatures [197]. It is applicable for the system considered.

Restricted to the lowest Landau level and rewritten in second quantized form, (2.2) reads

$$\hat{\mathcal{H}} = \hbar\omega_\perp \hat{N} + \hbar\left(\omega_\perp - \Omega\right) \hat{L}^z + \frac{1}{2} \sum_{m_1,\ldots,m_4} V_{1234}\, \mathrm{a}^\dagger_{m_1} \mathrm{a}^\dagger_{m_2} \mathrm{a}_{m_4} \mathrm{a}_{m_3} \tag{2.4}$$

where $\hat{N}$ and $\hat{L}^z$ are the total number and z-component angular momentum operators, $a^\dagger_{m_i}$ and $a_{m_i}$ create and annihilate a bosonic particle with angular momentum $m_i$, respectively, and $V_{1234} = \langle m_1\, m_2 |\, g \sum_{k<l} \delta\left(\vec{r}_k - \vec{r}_l\right)\, | m_3\, m_4 \rangle$ is the matrix element of interaction expressed in the Fock–Darwin single particle angular momentum basis[198, 199]

$$\langle z \mid m \rangle = \frac{1}{\lambda\sqrt{\pi m!}} \left(\frac{z}{\lambda}\right)^m e^{-|z|^2/2\lambda^2} , \tag{2.5}$$

with $\lambda = \sqrt{\hbar/(M\omega_\perp)}$, and generalized complex coordinates $z = x + iy$. It has to be stressed that $\lambda$ is linked to the magnetic length via $\lambda = \sqrt{2}l_0$. This is adapted to the convention used in the majority of publications on this topic. For the contact interaction, the above matrix elements are calculated as

$$V_{1234} = \frac{g}{\lambda^2\pi} \frac{\delta_{m_1+m_2,m_3+m_4}}{\sqrt{m_1!m_2!m_3!m_4!}} \frac{(m_1+m_2)!}{2^{m_1+m_2+1}} . \tag{2.6}$$

The radial symmetry of (2.4) can be used to diagonalize subspaces of well-defined total $z$-component of angular momentum $L^z = \sum_{j=1}^{N} m_j$. This significantly decreases the dimension of the individual Hilbert spaces to be conquered by the numerical routines. Calculations were performed for $N = 3$ to $10$ particles with complete and Davidson block diagonalization techniques, for details see App. B.

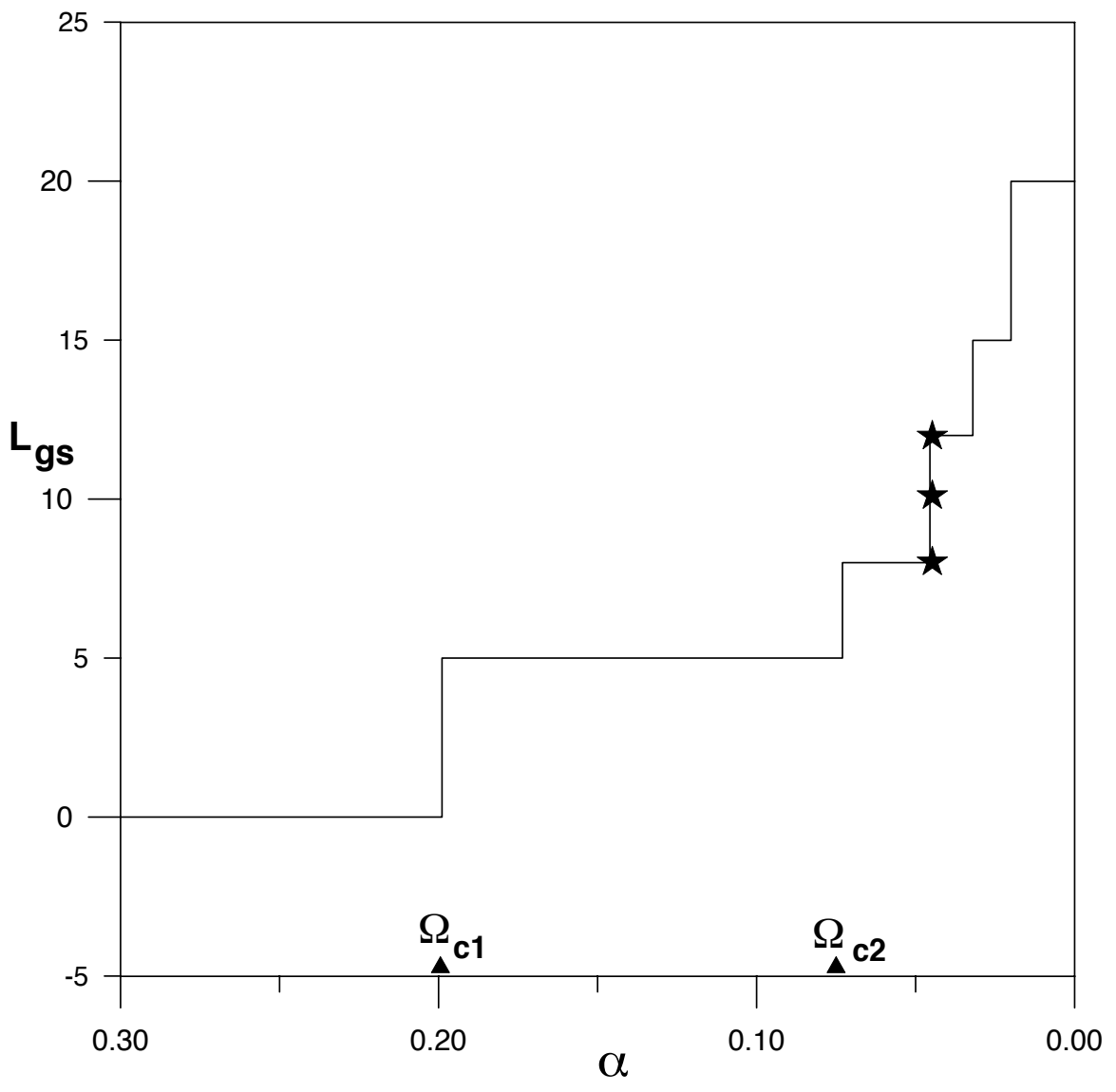

FIGURE 2.1: Change of the ground state angular momentum $L_{\text{gs}}$ for $N = 5$ as the rotation frequency increases; transitions take place at critical values $\Omega_{\text{cn}}$ of the rotational frequency. $\alpha \equiv \hbar(\omega_\perp - \Omega)$ in units of $\hbar\omega_\perp$; it lowers as the rotation frequency approaches $\omega_\perp$.

Apart from the constant kinetic contribution from $\hbar\omega_\perp\hat{N}$ and the diagonal linear "tilting" term $(\omega_\perp - \Omega)\, L^z$, only the interaction spectrum has to be calculated, which is a monotonously decreasing function of the chosen quantum number $L^z$. With tilting included, the first important quantity is obtained, i. e., the total angular momentum of the ground state, while $\Omega$ grows from zero to $\omega_\perp$, illustrated in Fig. 2.1 for a system of $N = 5$ particles. Without higher order confining potential on top, the system becomes centrifugally unstable at critical rotation $\Omega = \omega_\perp$. As an obvious feature, the ground state angular momentum remains constant for a finite range of $\Omega$. At critical values $\Omega_{\rm cn}$, transitions to new angular momenta take place. Even though only specific so-called "magic" $L^z$ become unique ground states, different $L^z$ may be energetically degenerate or, more precisely, quasi-degenerate on discrete steps, e. g., for $L^z = 8, 10, \text{and } 12$ as indicated by stars in Fig. 2.1. The last plateau of this series is the bosonic $\nu = 1/2$ Laughlin state at $L^z = N(N-1)$, for which the interaction energy equals zero, as the wave function strictly avoids relative angular momenta associated with the lowest Haldane pseudo-potential coefficient (all higher orders vanish in case of the contact interaction). This can be easily deduced from the analytical expression of the many-body wave function

$$\Psi_{\rm Laughlin} = \mathcal{N} \prod_{i<j} (z_i - z_j)^2 e^{-\sum |z_i|^2/2\lambda^2}\,. \tag{2.7}$$

In the following, the ground state of the system in the $L^z$-subspace is denoted by $\Psi_L$. In order to study its nature, it is useful to analyze the expectation values of relevant operators. As stated above, the density operator defined in first quantization as

$$\hat{\rho}^{(1)}(\vec{r}) = \sum_{i=1}^{N} \delta(\vec{r} - \vec{r}_i) \tag{2.8}$$

does not exhibit any interference pattern when calculated for a definite $L^z$. This can be inferred from its analytical expression in second quantized form

$$\hat{\rho}^{(1)}(\vec{r}) = \sum_{ij} \langle \phi_i(\vec{r}\,') \mid \delta(\vec{r} - \vec{r}\,') \mid \phi_j(\vec{r}\,') \rangle a_i^\dagger a_j, \tag{2.9}$$

where $|\phi_i(\vec{r})\rangle = |m_i\rangle$ as in Eq. (2.5). Due to conservation of angular momentum, $a_i^\dagger a_j$ uniquely selects a specific single particle state. All information is contained in products of different amplitudes, and, consequently, the interference pattern is lost. It has to be stressed that a symmetric manifestation of the ground state may not be concluded by this. Rather, $\hat{\rho}^{(1)}(\vec{r})$ is by definition restricted to represent cylindrically symmetric distributions and merely the information of individual densities are preserved

$$\rho^{(1)}(\vec{r}) = \langle \Psi_L \mid \hat{\rho}(\vec{r}) \mid \Psi_L \rangle = \sum_{i}^{N} O_i \mid \phi_i(\vec{r}) \mid^2, \tag{2.10}$$

where $O_i$ is the total occupation of the single particle state $|m_i\rangle$. These occupations are the eigenvalues of the single particle density matrix.

### 2.2.1 Ordered structures in pair correlation functions

In order to analyze the internal structure of relevant states, the following operator is considered

$$\hat{\rho}^{(2)}(\vec{r},\vec{r}_0) = \sum_{i<j}^{N} \delta(\vec{r}_i - \vec{r}_0)\delta(\vec{r}_j - \vec{r}). \tag{2.11}$$

It yields the conditional probability to find an atom at $\vec{r}$, when another is simultaneously situated at $\vec{r}_0$. $\hat{\rho}^{(2)}(\vec{r},\vec{r}_0)$ provides information which originates from the amplitudes of single particle wave functions instead of their density, in contrast to the previously discussed $\hat{\rho}^{(1)}(\vec{r})$. In second quantized formalism, the expected value of (2.11) with respect to $\Psi_L$ reads

$$\rho^{(2)}(\vec{r},\vec{r}_0) = \sum_{ijkl}\sum_{pp'} \alpha_p^* \alpha_{p'}\, \phi_i^*(\vec{r})\phi_j^*(\vec{r}_0)\phi_k(\vec{r})\phi_l(\vec{r}_0)\,\langle \Phi_p \mid a_i^\dagger a_j^\dagger a_l a_k \mid \Phi_{p'}\rangle, \tag{2.12}$$

where

$$\mid \Psi_L\rangle = \sum_{p=1}^{n_d} \alpha_p \mid \Phi_p\rangle\,, \tag{2.13}$$

and $\mid \Phi_p\rangle$ are the bosonic Fock $N$-body states of the basis in the $L^z$-subspace of dimension $n_d$. Angular momentum conservation implies $i + j = k + l$. It should be stressed that $\rho^{(2)}(\vec{r},\vec{r}_0)$ deviates from the single particle density matrix

$$n^{(1)}(\vec{r},\vec{r}\,') = \langle \hat{\Psi}^+(\vec{r})\hat{\Psi}(\vec{r}\,')\rangle, \tag{2.14}$$

which defines the off-diagonal long-range order that characterizes Bose condensation [108]. The operator $\hat{\rho}^{(2)}$ is a two-particle operator, in contrast to $\hat{n}^{(1)}(\vec{r},\vec{r}\,') = \hat{\Psi}^+(\vec{r})\hat{\Psi}(\vec{r}\,')$, which solely affects a single particle. In particular, the single particle density is obtained by

$$\rho^{(1)}(\vec{r}) = n^{(1)}(\vec{r},\vec{r}) = \frac{1}{N-1}\int \mathrm{d}\vec{r}_0\, \rho(\vec{r},\vec{r}_0)\,. \tag{2.15}$$

Strictly speaking, more profound information on ordered structures is revealed by $\hat{\rho}^{(2)}(\vec{r},\vec{r}_0)$. Eq. (2.12) can be interpreted as the sum of products of amplitudes at $\vec{r}$ weighted by a function that depends on $\vec{r}_0$ and on the ground state coefficients $\alpha_p$.

The role of the parameter $\vec{r}_0$ in $\langle \Psi_L \mid \hat{\rho}^{(2)}(\vec{r}, \vec{r}_0) \mid \Psi_L \rangle$ as a function of $\vec{r}$ is understood from

$$\phi_n^*(\vec{r})\phi_j^*(\vec{r}_0)\phi_k(\vec{r})\phi_l(\vec{r}_0) = \frac{1}{\pi^2} \frac{r^{m_n} r_0^{m_j} r^{m_k} r_0^{m_l}}{\sqrt{m_n! m_j! m_k! m_l!}} e^{i(m_k - m_n)\theta} e^{i(m_l - m_j)\theta_0} e^{-(r^2 + r_0^2)}, \tag{2.16}$$

expressed in units of $\lambda$. Angular momentum conservation demands $l - j = n - k$ and the angular dependence is derived as

$$e^{i(m_k - m_n)\theta} e^{i(m_l - m_j)\theta_0} = e^{i(m_k - m_n)(\theta - \theta_0)} . \tag{2.17}$$

Evidently, if $r_0$ is fixed, the change of $\theta_0$ is nothing but a rigid rotation of the function. Any arbitrary choice of $\theta_0$ fixes the origin of angles and breaks cylindrical symmetry. This happens in analogy to experiments which perform a single shot measurement. The measurement and the choice of $\theta_0$ are equivalent processes (see for instance [189, 194]).

In the particular case $\vec{r}_0 = \vec{0}$, cylindrical symmetry is recovered, since $l = j = 0$ is the unique non-zero contribution. This demands $i = k$ and yields

$$\rho^{(2)}(\vec{r}, \vec{0}) = |\phi_0(\vec{0})|^2 \sum_i |\phi_i(\vec{r})|^2 \sum_{pp'} \alpha_p^* \alpha_{p'} \langle \Phi_p \mid a_i^\dagger a_0^\dagger a_0 a_k \mid \Phi_{p'} \rangle, \tag{2.18}$$

which is independent from $\theta$.

### 2.2.2 Ordered structures in the density

In this subsection, ordered structures in ground states with no well-defined angular momentum are discussed within the framework of two approaches. On the one hand, motivated by arguments of perturbation theory, a thermal or stirring induced anisotropic potential is assumed to couple (quasi-)degenerate $L^z$ manifolds, while being small enough not to excite states other than the $\Psi_L$ involved. In a simplified model, an effective Hamiltonian of dimension $d_L$ is obtained, where $d_L$ equals the number of mixed subspaces. The above potential leads to off-diagonal terms that within first order are considered to be of same magnitude. In case of but two coupled angular momentum spaces with ground states $\Psi_{L_1}$ and $\Psi_{L_2}$, the eigenstates are

$$|\Psi_\pm\rangle = \frac{1}{\sqrt{2}} \left( \Psi_{L_1} \pm \Psi_{L_2} \right) . \tag{2.19}$$

The ground state is determined by the sign of the mixing potential. Generalized to the case of larger $d_L$ with weighted off-diagonal coupling strength, physical *ad hoc* linear combinations of the $\Psi_{L_j}$ may be deduced. Such combinations have been previously used in exact diagonalization calculations to obtain two vortices [105, 106].

In the second approach, an anisotropic term is added to the Hamiltonian, which mimics the deformation introduced by the stirring laser. Numerical diagonalization is performed without the restriction of angular momentum conservation. In this way, the explicit structures of the true ground states are revealed on the level of density due to broken symmetry. This procedure was suggested by experiments in order to explain the nucleation of multiple vortices [115, 200]. These states are generated as equilibrium states of a Bose condensate rotating under the action of a stirring laser that produces an anisotropic potential in the $xy$-plane. Subject to this anisotropy, the state with vortices becomes the true ground state and survives during the time of flight expansion and detection as an excited state of the restored symmetric Hamiltonian after the trap has been switched off.

As an ansatz for the stirring process, the additional term $\hat{V}_p = A\sum_{i=1}^{N}(x^2 - y^2)_i$ is included in (2.2). It adequately describes the experimental procedure of Ref. [115], and has several distinct advantages. Its simple analytical form allows for efficient calculations. Furthermore, it describes other kinds of "stirring", e. g., the process which uses an elliptical, rotating mask to generate the above optical potential. In second quantized form, $\hat{V}_p$ reads

$$\hat{V}_p = \frac{A}{2}\lambda^2 \sum_m \left[\sqrt{m(m-1)}a_m^{+}a_{m-2} + \sqrt{(m+1)(m+2)}a_m^{\dagger}a_{m+2}\right] . \tag{2.20}$$

To be a small perturbation of the system, $A\lambda^2/2 \ll \hbar(\omega_\perp - \Omega)$ has to be guaranteed.

It has to be stressed that the structures caused by the above symmetry breaking, e. g., vortex formations, are only obtained in the very specific situation when the ground state is a combination of (quasi-)degenerate $\Psi_L$-states coupled by a perturbing term slightly larger than their energy difference, but much smaller than low-lying excitations.

More precisely, the unique situation where vortices are nucleated in the system corresponds to the plateau steps in the $L_{\mathrm{gs}}$ dependence on $\Omega$, where a degeneracy of states with different $L^z$ takes place at $\Omega_{\mathrm{cn}}$. This result seemingly does not support the experimental results of Ref. [200]. However, it can be attributed to an essentially different behavior of macroscopic systems in contrast to microscopic ones. As $N$ grows, the size of some of the plateaus drastically shrinks in such a way that finite intervals of $\Omega$ with energetically degenerate states become possible.

Fig. 2.2 shows the appearance of such micro-plateaus obtained for $N = 6$, $7$, $8$, and $9$.

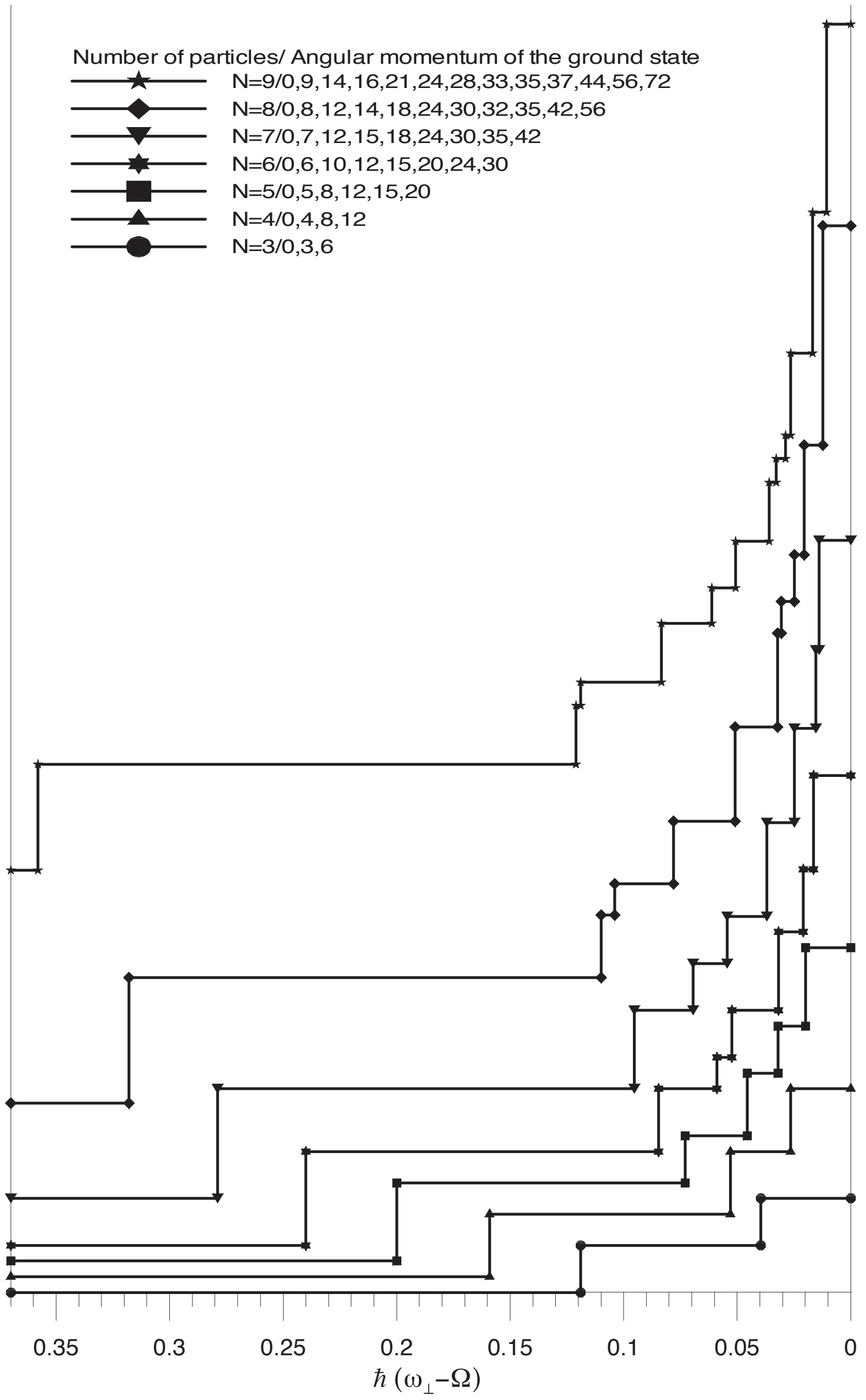

FIGURE 2.2: Change of ground state angular momentum $L_{\rm gs}$, as the rotation frequency increases for $N = 3$ to $9$ from bottom to top. Graphs are vertically shifted for clarity.

### 2.2.3 Numerical results

In this part, the results obtained from exact diagonalization are discussed. Whenever states $\Psi_L$ with well-defined angular momentum are analyzed, the tilting strength $\alpha \equiv \hbar(\omega_\perp - \Omega)$ is assumed to be adjusted appropriately to guarantee $\Psi_L$ as a ground state of the system. If a mixture of different $L^z$ manifolds is investigated, the specific value of $\alpha$ is chosen, which ensures the energetic degeneracy of the contributing subspaces. Correlation functions are displayed in pairs, a 3D and a contour plot, unless otherwise specified. $\lambda$ and $\hbar\omega_\perp$ are defined as units of length and energy, respectively. The relative interaction strength is set to $g\lambda^{-2} = 1$ in these units.

The analysis starts from the most transparent and intuitively accessible case of $N = 3$. In Fig. 2.3-2.6, the main results including the ground state evolution are illustrated. Fig. 2.3a shows the lowest eigenenergies for each $L^z$ in case of no tilting, the so-called Yrast line. The initial points of the plateaus at $L^z = 0, 3$, and $6$ are the unique possible ground states apart from degeneracies at the steps as indicated by Fig. 2.3b. The plateau previous to the Laughlin state (here, $L^z = 3,\ 4,\ 5$) consists of $N$ points, which is a general feature.

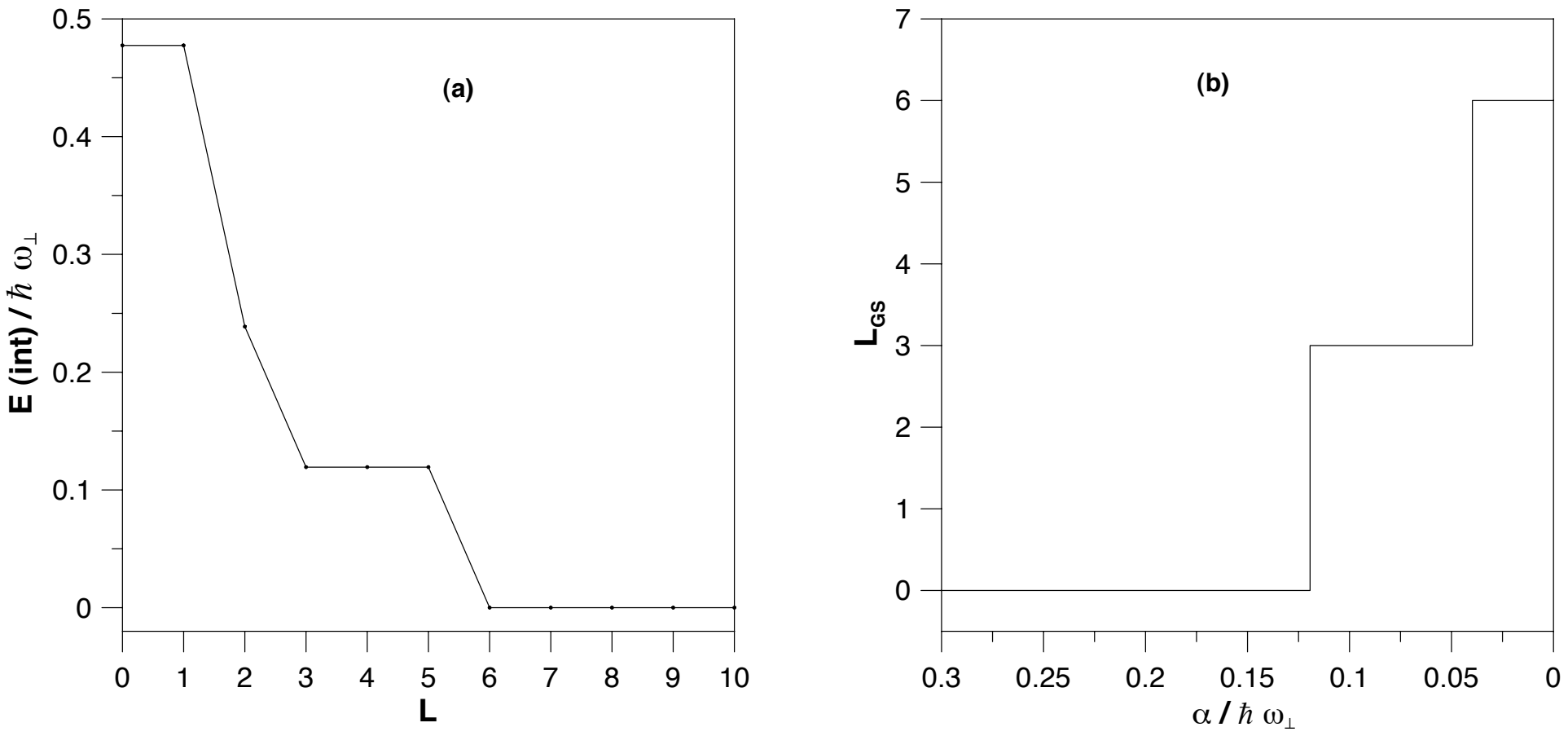

FIGURE 2.3: For $N = 3$, a) lowest interaction contribution to the energy as a function of total angular momentum (Yrast line). b) Angular momentum of the ground state over $\alpha$. The critical values for $\alpha$ at the steps are: 0.1194 and 0.0398.

Fig. 2.4 shows the densities from $L^z = 0$ to $9$. Finally, the pair correlation function for $L^z = 0, 3, 4, 5, 6$, and $9$ are displayed in Fig. 2.5, with the position $r_0$ of the second particle set to the maximum of the density unless otherwise stated. The system evolves from a completely

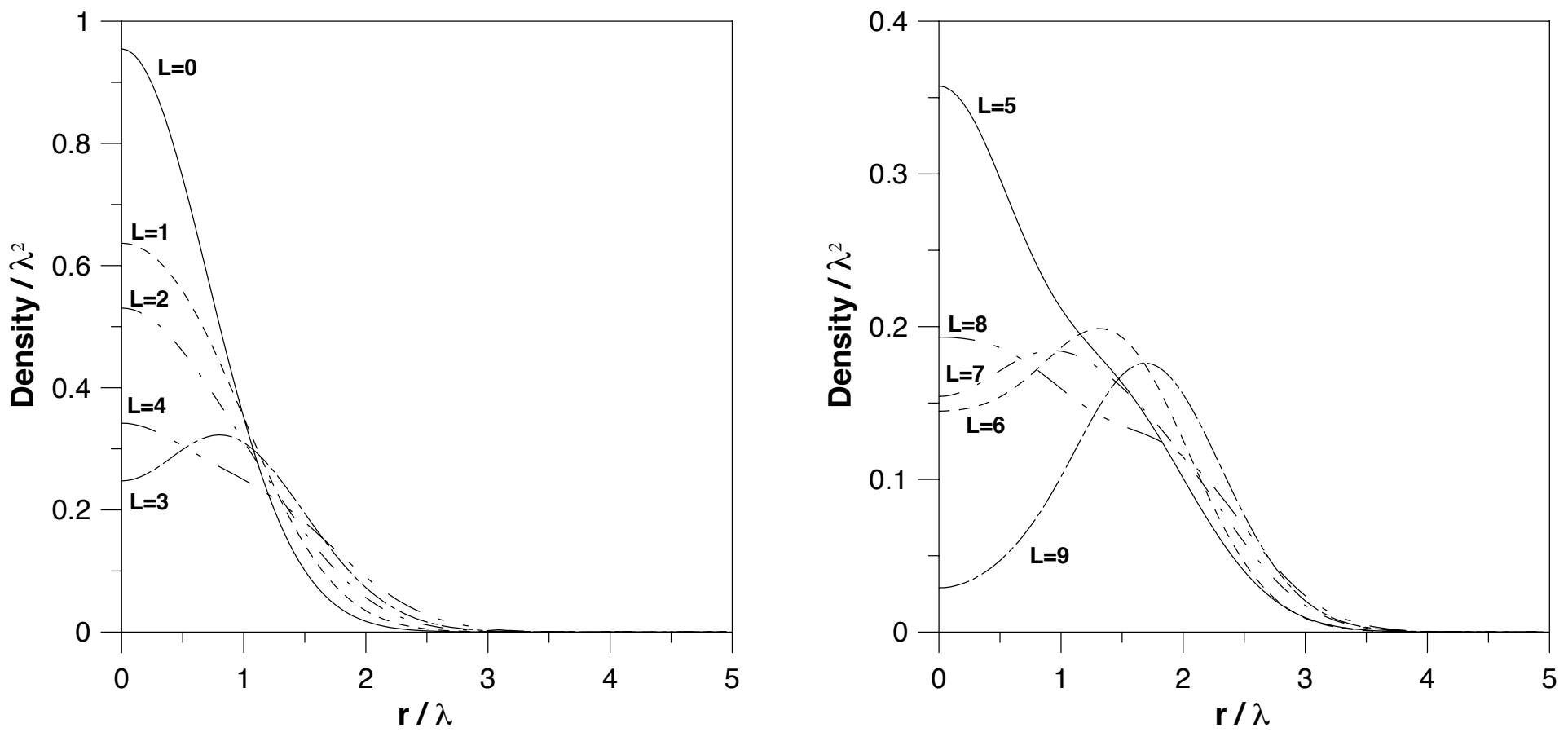

FIGURE 2.4: Density of the $\Psi_L$ states for $N = 3$.

"condensed" system at $L^z = 0$ to the Laughlin state at $L^z = 6$, where a clear triangular structure appears with the third particle included at $r_0$. This loss of condensation is directly related to the increase of space correlations. The evolution of single particle occupations is provided by Fig. 2.6. It demonstrates that "macroscopic" occupation of a specific single particle wave function vanishes as $L^z$ increases.

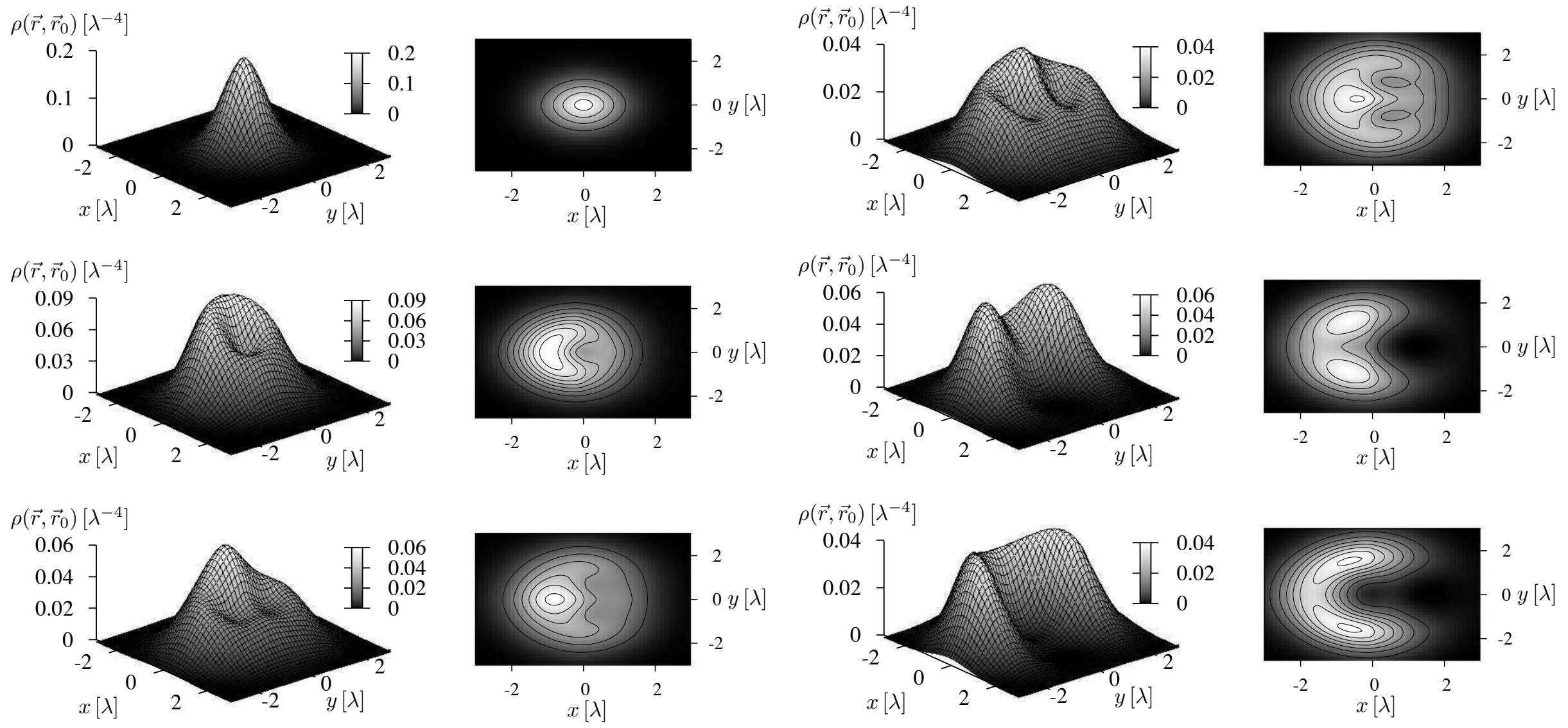

FIGURE 2.5: Pair correlation function for $N = 3$; From top to bottom: $L^z = 0$, 3, and 4 on the l. h. s., and $L^z = 5, 6$, and 9 on the r. h. s. In the same order: $r_0 = 1.0, 0.8, 1.0, 1.0, 1.3$ and 1.7.

In order to see how the previous general tendency evolves as $N$ increases, the above results are contrasted to the case $N = 5$. Fig. 2.7 shows the densities of the ground states from $L^z = 0$ to the Laughlin state at $L^z = 20$, and Fig. 2.8 displays their pair correlation functions. The same tendency towards space ordering at the Laughlin state is obvious. In addition, it can be inferred that correlations are stronger for nearest neighbors as a manifestation of partial long range order in finite systems. In Fig. 2.9, the occupations of single particle states are identified.

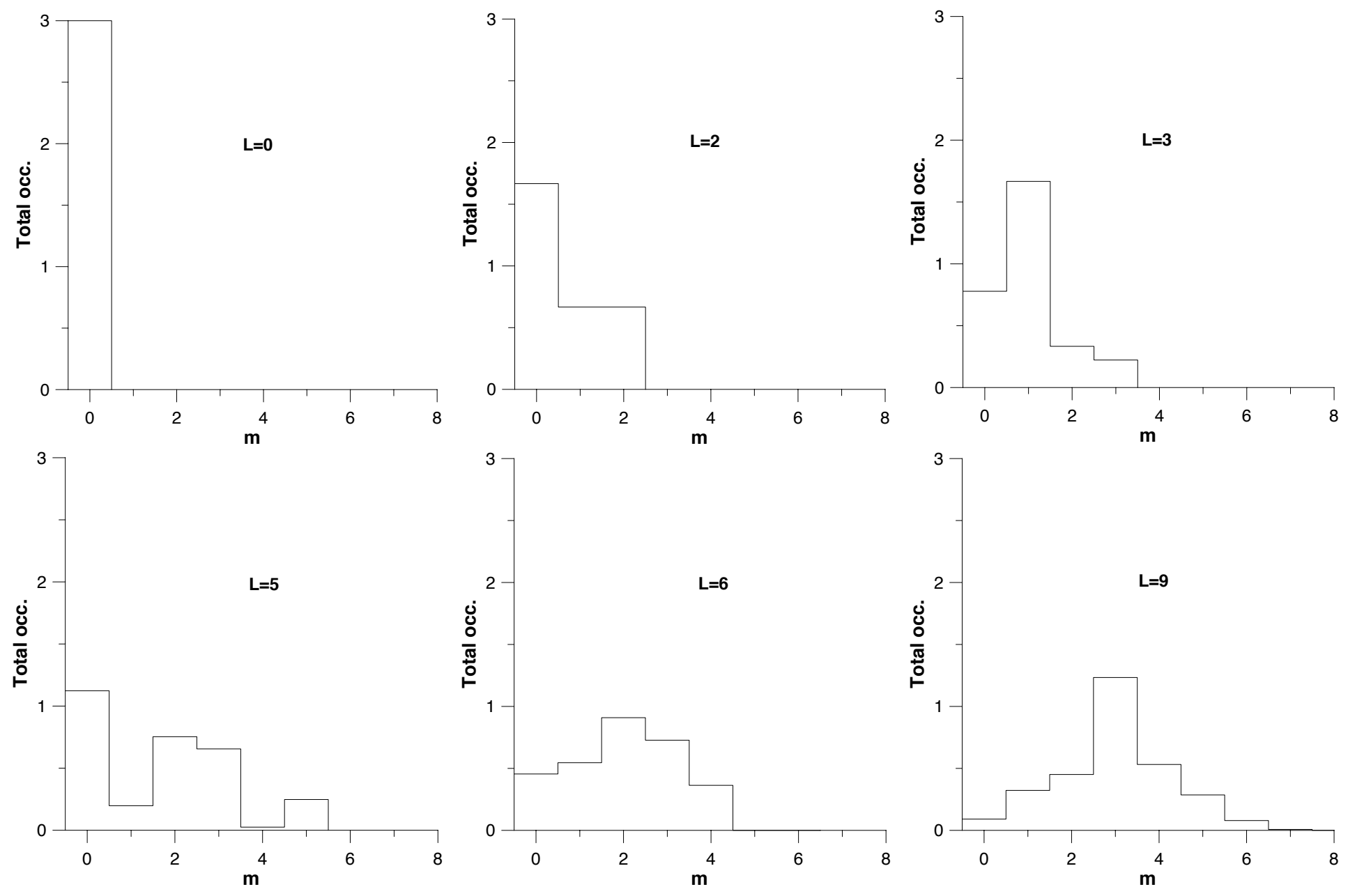

FIGURE 2.6: Single particle state occupations of specific $\Psi_L$ for $N = 3$.

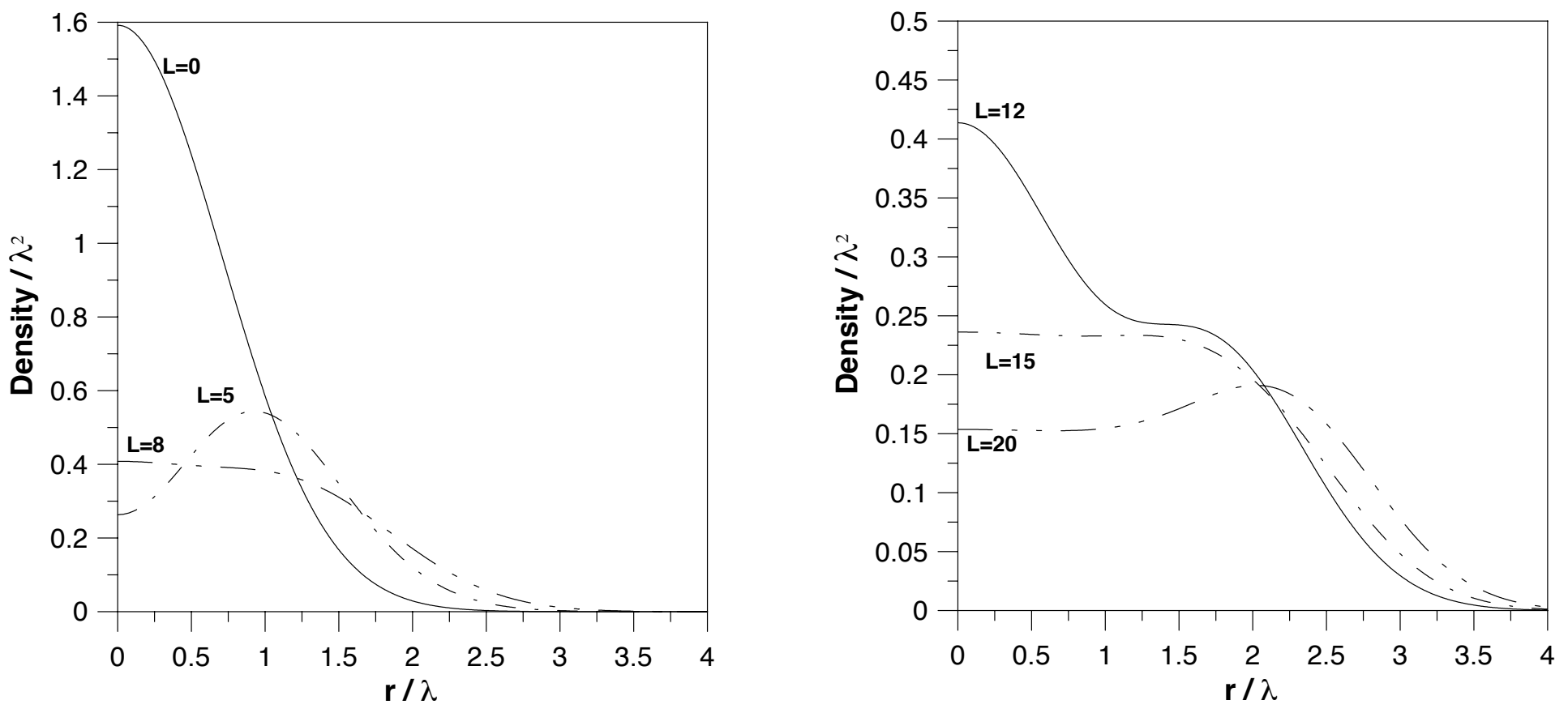

FIGURE 2.7: Density of the $L_{\rm gs}$ ground states for $N = 5$.

Remarkably, some indications of the Laughlin state, which typical for large systems, e. g., a flat density at the central part and a hump at the edge, are already manifested in such small systems. This can be seen in the last graph of Fig. 2.7. Moreover, the density at the origin is very close to $1/(2\pi) = 0.16$, as necessary for a homogeneous system at filling factor $1/2$.

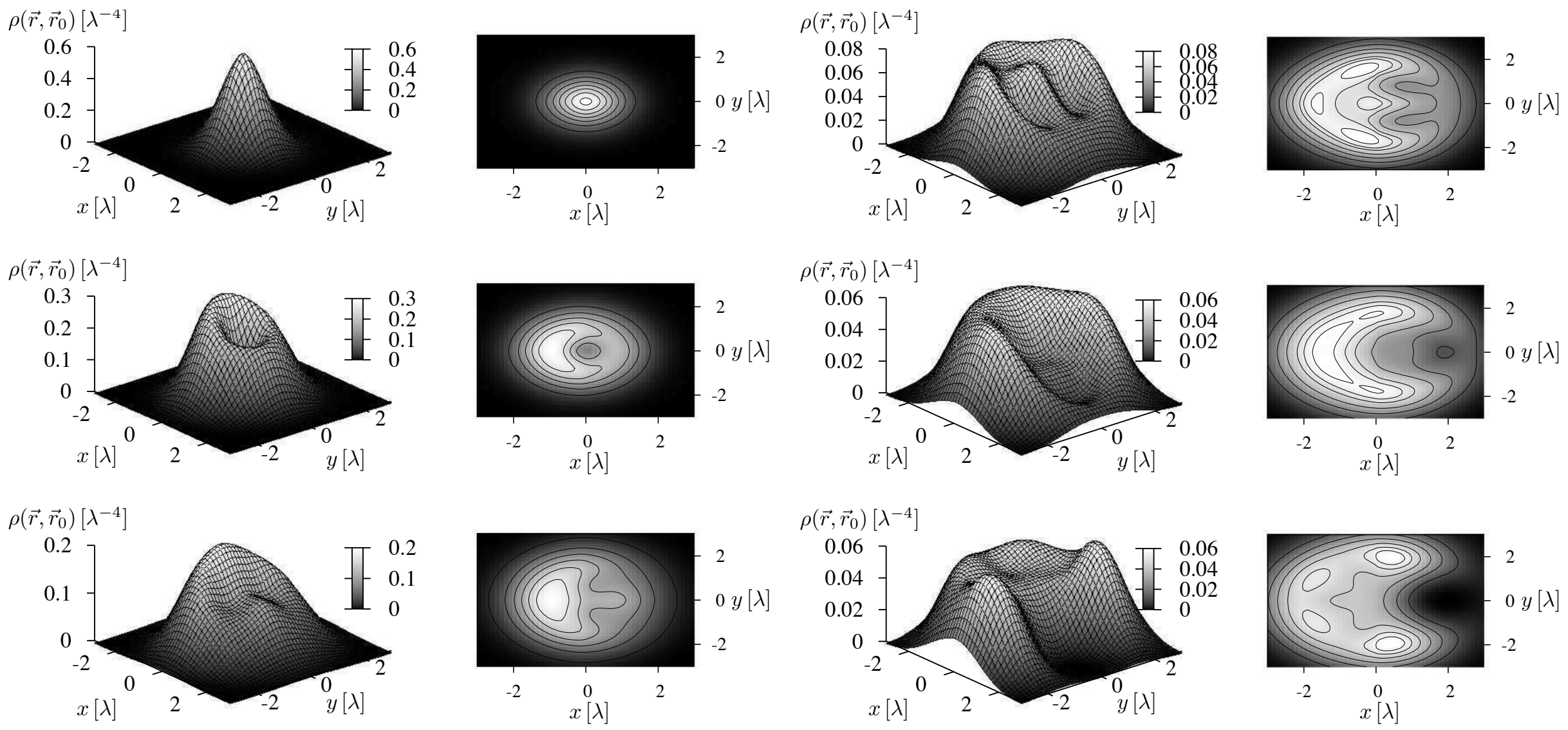

FIGURE 2.8: Pair correlation function for $N = 5$. From top to bottom: (l. h. s.) $L^z = 0, 5$, and 8; (r. h. s.) $L^z = 12, 15$, and 20. In the same order: $r_0 = 1.0, 0.9, 1.0, 1.0, 1.0$, and 2.0.

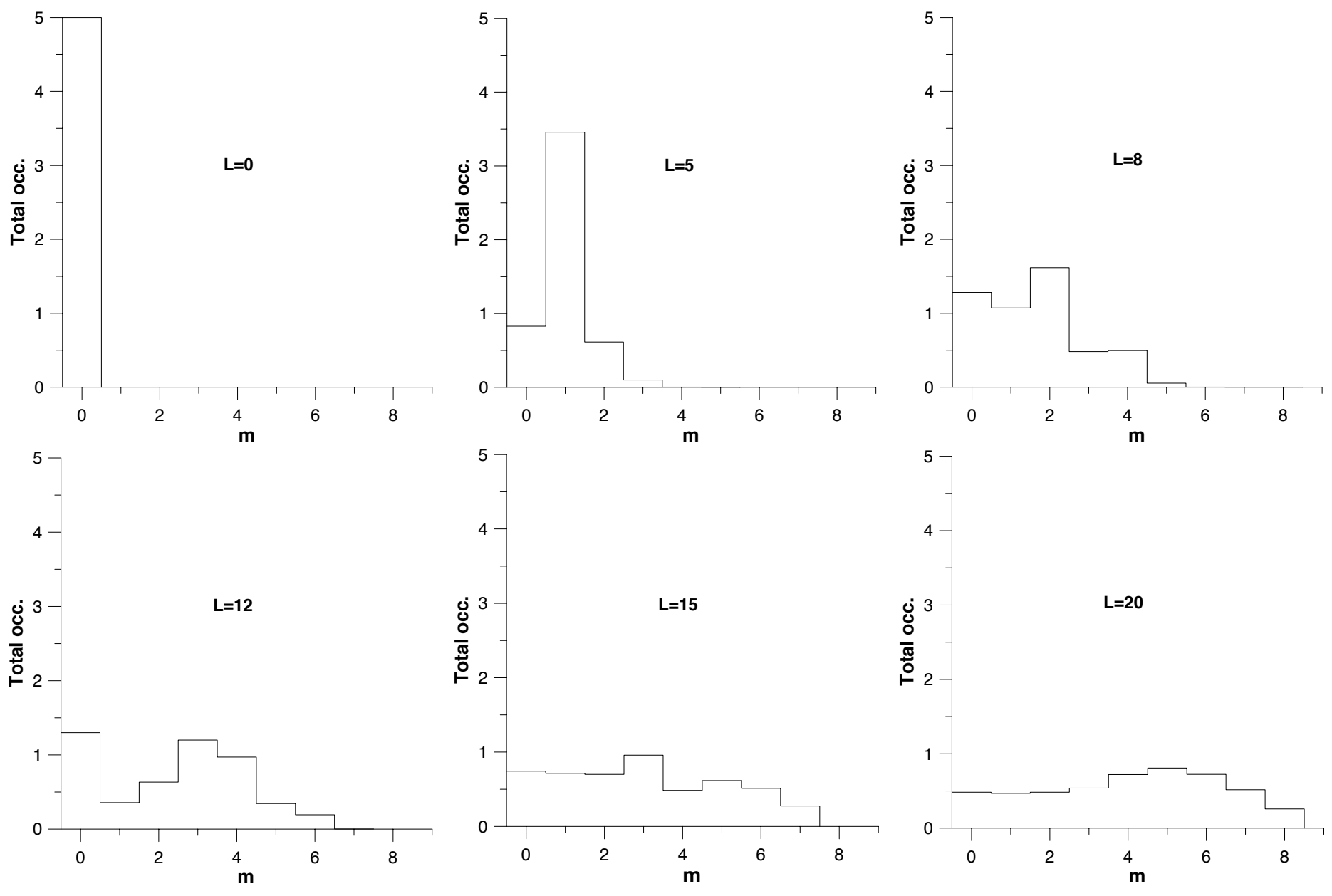

FIGURE 2.9: Single particle state occupations of distinct $\Psi_L$ for $N = 5$.

To understand the evolution of ground state structures, the series $\Psi_{L_{\rm gs}}$ has to be investigated in more detail while the rotational frequency $\Omega$ increases. Starting from the fully condensed state at $L^z = 0$, the system is soon driven through a first transition which occurs at the initial step of $L^z_{gs}(\Omega)$ from 0 to $N$.

It is exactly the corresponding $\Omega_{c1} = \omega_\perp - gN/(8\pi)$, where in a mean field analysis for $N \gg 1$, the first vortex is dynamically nucleated at the edge. Further acceleration stabilizes this vortex at the centre of the atomic sample. The associated state is circular symmetric without any spatial correlations and is characterized by a high occupancy of the $m = 1$ single particle state. In the given spectral and thus static analysis for a finite number of particles, $\Psi_N$ is indeed a robust ground state in the interval $\Omega_{c1} < \Omega < \Omega_{c2}$ (see Fig. 2.1). But in contrast to the thermodynamic limit, a vortex is not clearly manifested unless a considerably large number of atoms is considered. This is shown in Fig. 2.10, where the pair correlation of $\Psi_N$ is displayed for $N = 3$, 10 and 20. In all the cases, the density has a minimum at the origin and $r_0$ is set to its maximum. The slow tendency to recover the behavior of large condensed systems is obvious from the occupation of the $m = 1$ single particle state. Due to the finite size of the microscopic sample, the angular momentum of the state is not completely contributed by the vortex. This result shares similarities with Ref. [201], where the formation of the first vortex, from its initial nucleation at the cloud boundary towards its final stabilization at the centre, was studied for a large number of atoms. In this dynamical situation, the vortex is not yet fully inside the trap for small $N$, but approaches the trap centre from the boundary of the cloud as $N$ increases.

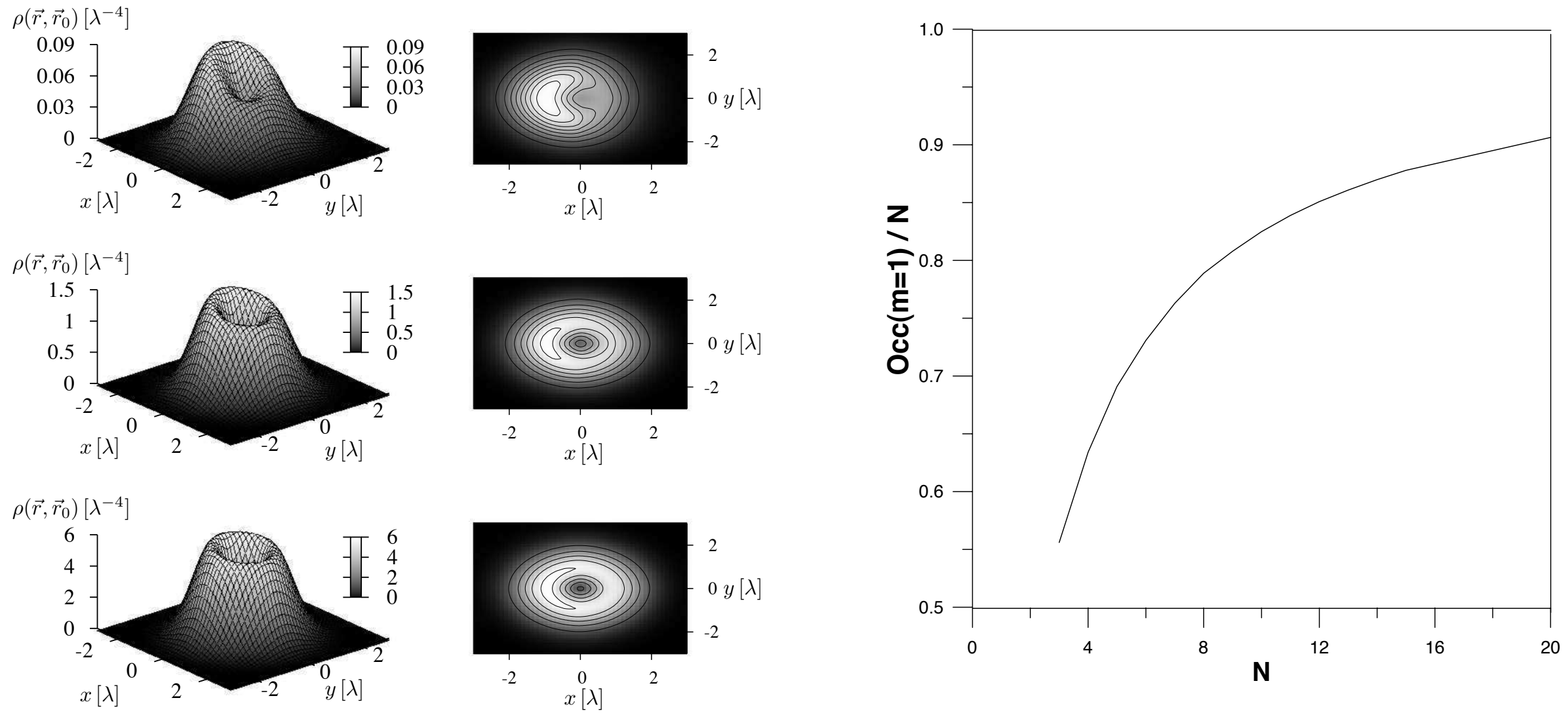

FIGURE 2.10: Pair correlation function for $N = 3, 10,$ and $20$ of $\Psi_N$ with $r_0$ set to $0.8$, $0.95$, and $1.0$, respectively; (r. h. s.) per particle occupation of the m=1 single particle state over N.

As it is well known from experimental results, more vortices are expected to be nucleated when the rotational frequency $\Omega$ increases. Their arrangement in the energetically favorable triangular Abrikosov lattice breaks the symmetry of the Hamiltonian. In the microscopic systems discussed, this is only possible at the steps in $L_{\text{gs}}$, where different $\Psi_{L_j}$ become energetically degenerate due to thermal or stirring-induced coupling between different $L^z$-subspaces. With the anisotropy potential $V_p$ of Eq. (2.20), this possibility is even further constrained, as solely $L^z$ and $L^z \pm 2$ are connected. Nonetheless, as shown in Fig. 2.11 patterns of two incipient vortices are obtained from the full diagonalization of $H + V_p$ for $N = 5, 6$ at $\alpha = 0.0458$ and $\alpha = 0.05904$, respectively. This choice corresponds to the third step of $L_{\text{gs}}$ in both cases. Instead, for e. g. $N = 7$, no vortex structures appear with the above restriction. This more or less artificial result implies no physical consequences, unless an experiment could be controlled with a precision that guarantees only quadrupolar couplings.

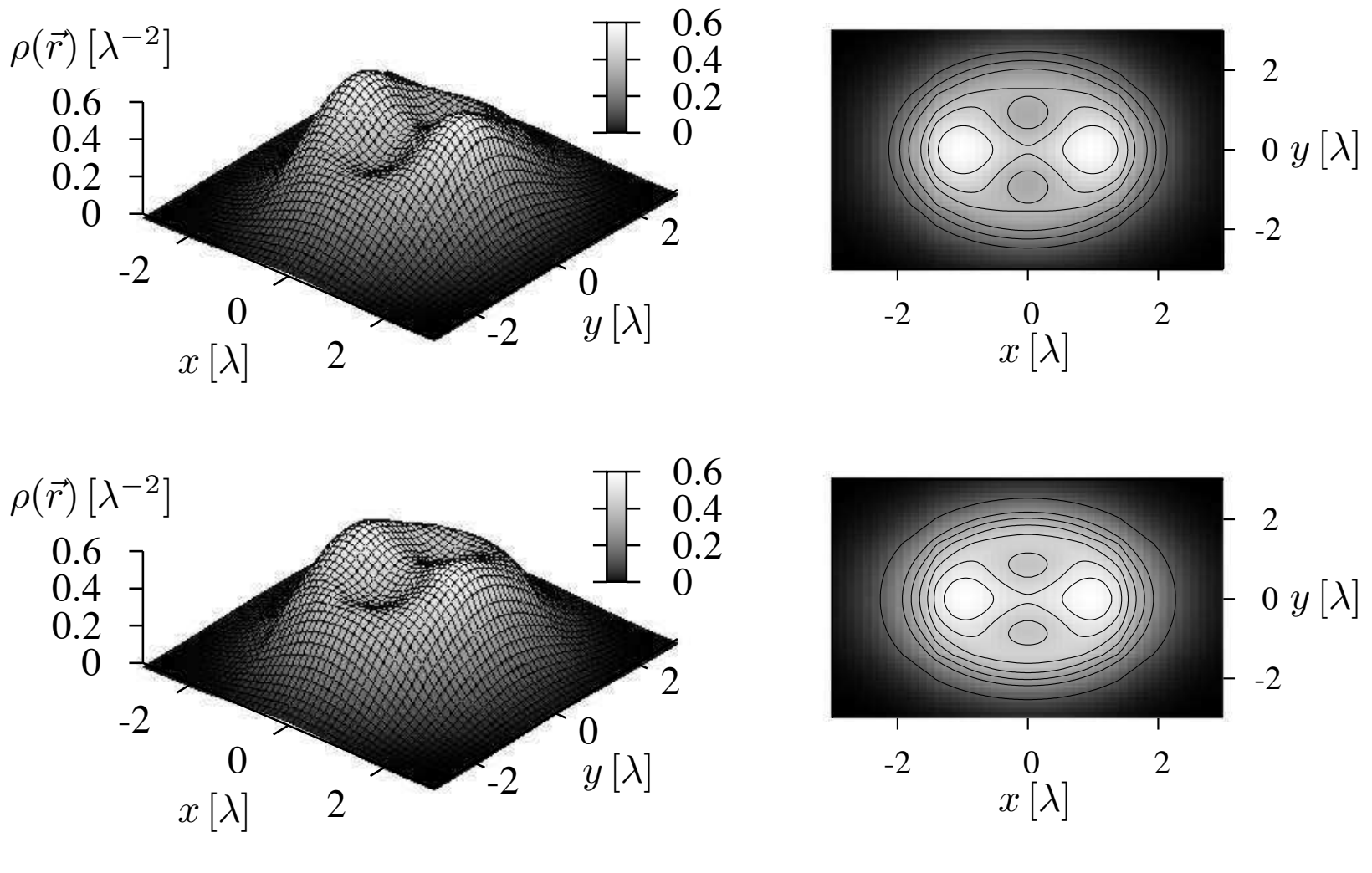

FIGURE 2.11: Densities of two-vortex structures for $N = 5$ at $L^z = 8+10+12$, and for $N = 6$ at $L^z = 10 + 12 + 14$.

For $N = 5$ where $L = 8, 10$ and 12 are involved, the weights of $\Psi_8$ and $\Psi_{10}$ within the expansion of the ground state are significantly larger than the weight of $\Psi_{12}$, the same holds for $N = 6$. Thus, the expected vortex angular momentum is lower than $2N$, in agreement with the results demonstrated in Ref. [202], where it is concluded that the contribution of a vortex to the total angular momentum depends on its distance from the origin. It runs from $N$ at the centre to zero at the trap boundary. Even more vortices appear for higher rotational rates. Fig. 2.12 contrasts the two- and four-vortex patterns for $N = 6$.

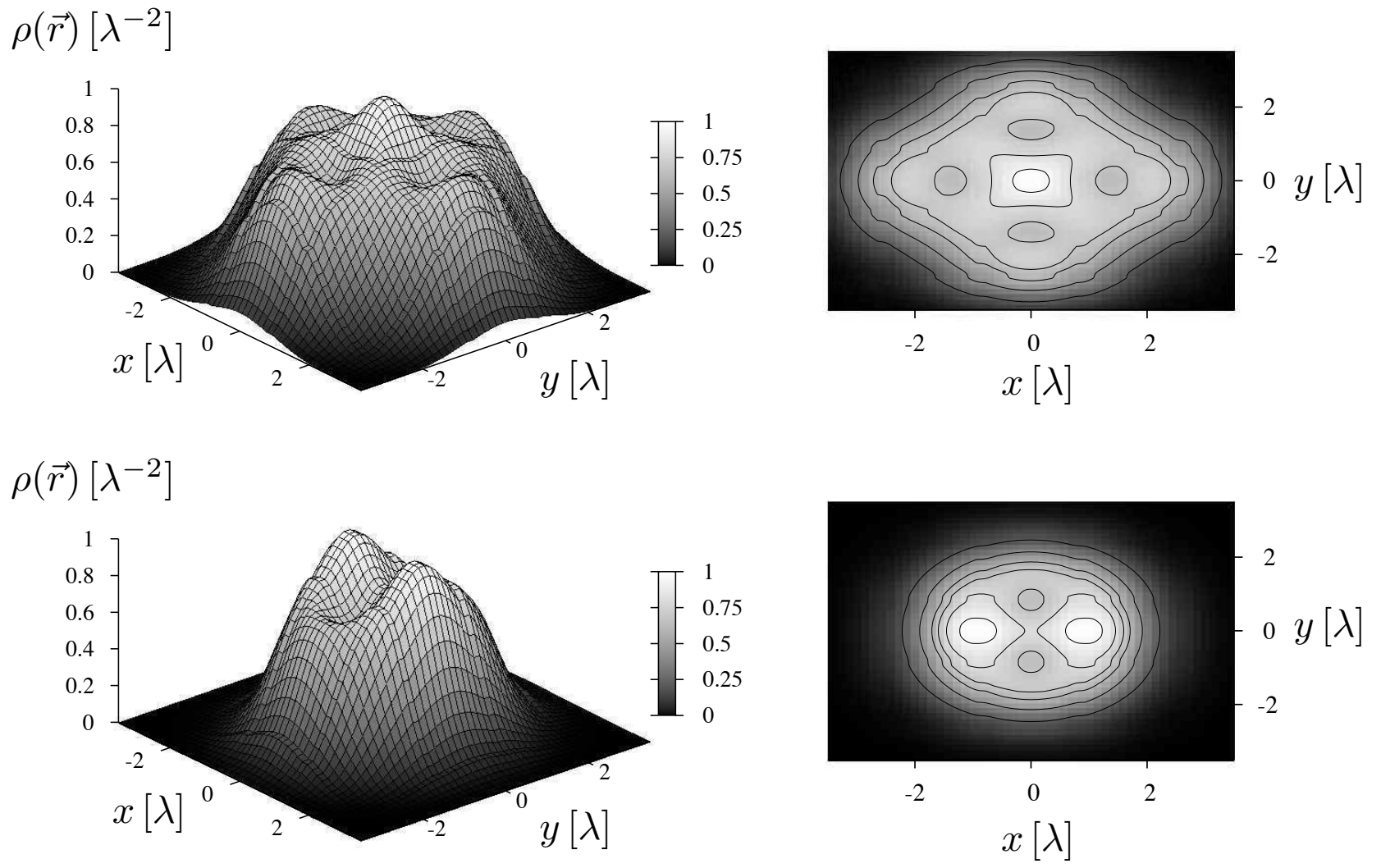

FIGURE 2.12: Density of two- and four-vortex patterns for $N = 6$, $L^z = 10 + 12 + 14$ and $L^z = 20 + 22 + 24$, respectively.

In accordance with the results commented in the previous paragraph and independent from the specific model of coupling potentials, the frequency $\Omega$ must be carefully tuned around $\Omega_{\mathrm{cn}}$ in order to generate incipient vortices in the density distribution. This has important experimental consequences on the dependence of vortex contributions to the angular momentum as a function of $\Omega$. While this function is monotonously increasing for large systems, it becomes a function with peaks around $\Omega_{\mathrm{cn}}$ and wide minima in between for microscopic samples. As the particle number increases, the peaks are expected to broaden due to the appearance of micro-plateaus which in turn lead to finite ranges of $\Omega$-values where quasi-degenerate states coexist. In this manner, the experimental result for large systems is recovered. It has to be noted that in all calculations, $g\lambda^{-2}$ was set to unity, which is large compared with typical experimental values used for large systems. Compared with the parameters for $^{87}$Rb as in Ref. [203, 204], the strength of interaction equals $g\lambda^{-2}/\hbar\omega_\perp \approx 0.02$. However, the effect of a reduced $g$ is solely a change of scale with respect to the tilting axis and does not affect any previously obtained conclusions.

The following analysis is focussed on the bosonic Laughlin states at $L^z_\mathrm{L} = N(N-1)$. As visible in their pair correlation functions(r. h. s. of Fig.2.13a), the atoms are organized in a regular polygon on the edge of the sample. To better understand this structure, the terms that contribute to broken cylindrical symmetry

$$\langle \Psi_L \mid a_i^\dagger a_j \mid \Psi_{L+M} \rangle \,, \tag{2.21}$$

with $M \geq 1$, have to be taken into account. Without tilting, all states $\Psi_L$ with $L > N(N-1)$ are energetically degenerate with the bosonic Laughlin state and have to be included on equal footing in any systematic analysis. However, a finite amount of tilting is necessary to stabilize the system, but it would always favor the lowest lying $\Psi_L$. Though none of the linear combinations from $L^z = N(N-1)$ to $L^z_\mathrm{L}-1$ reproduces the polygonal structure, including $L^z_\mathrm{L}+N = N^2$ solves the issue. Even more surprisingly at first glance, the regular structure is obtained if the Hilbert space is restricted to the two-state system $L^z_\mathrm{L}$ and $L^z = N^2$. This is caused by the necessity to include all single particle angular momenta contributing to the Laughlin wave function. It should be pointed out that this feature is restricted to low enough $N$, because bulk correlations manifest themselves for $N > 6$ that add to and compete with the pure crystalline polygonal edge structure. The value of the parameter $r_0$ in the pair correlation functions in Fig. 2.13 was obtained in a different than the previously used way. With the knowledge that the Laughlin wave function (2.7) is the exact solution of the full Hamiltonian, and given its pair correlation function to be a ring shaped structure of unknown radius $r_0$, the square modulus of the wave function may be interpreted as a probability distribution

$$P(T) = \mid \Psi_{\text{Laughlin}}(\vec{r}_1, \vec{r}_2, \ldots, \vec{r}_N) \mid^2 = \mathrm{e}^{-T} \tag{2.22}$$

with effective Boltzmann weight

$$T = \sum_i \frac{r_i^2}{2\lambda^2} - 2q \sum_{i<j} \ln \mid z_i - z_j \mid \,, \tag{2.23}$$

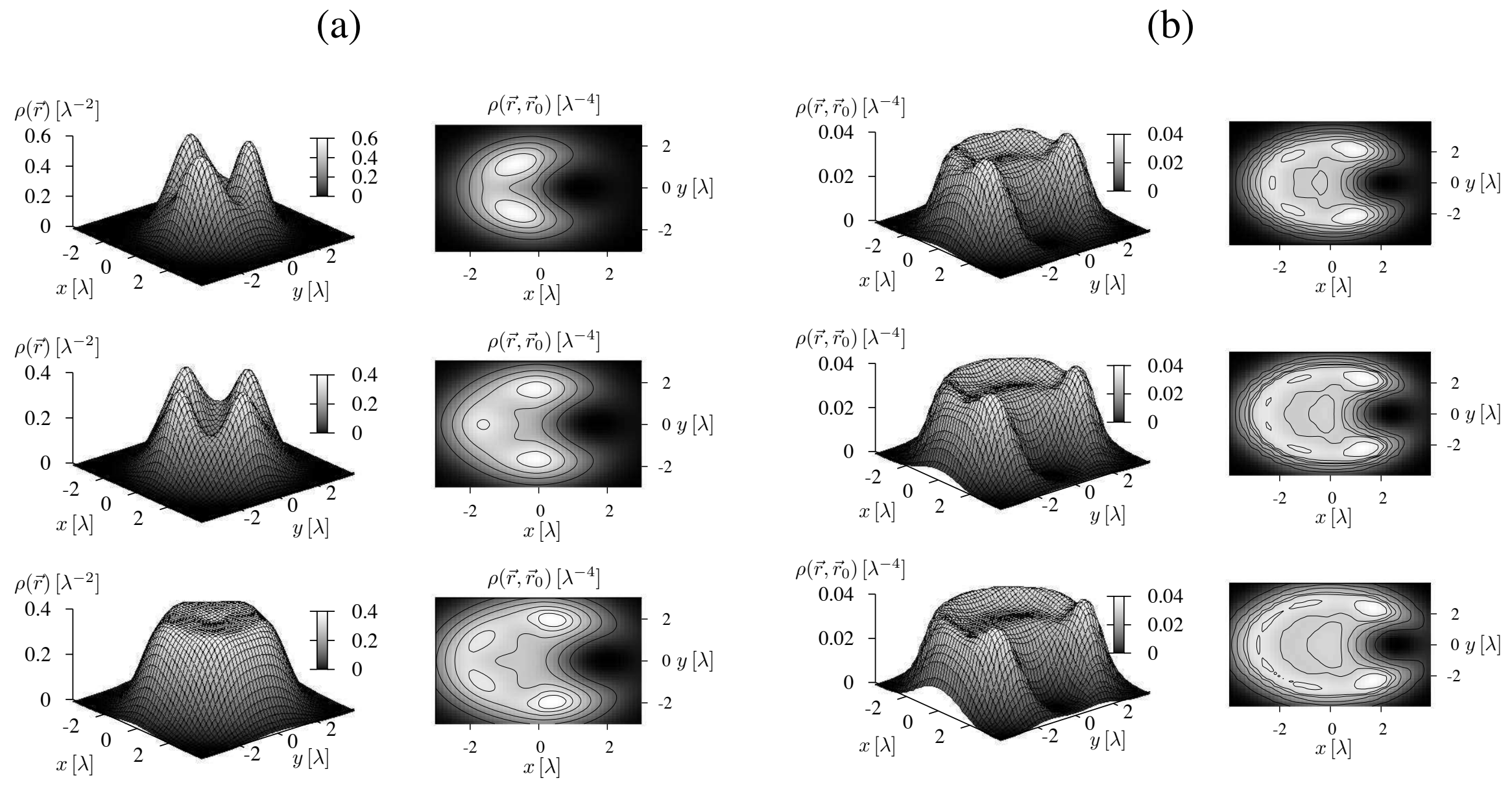

FIGURE 2.13: (a) For $N = 3$, $4$ and $5$, (l. h. s.) the surface plots show the density of equally weighted mixtures $\Psi_{L_j}$ with angular momenta $L^z_j = N(N-1)$ and $N^2$; (r. .h .s) pair correlation functions for the corresponding Laughlin states with $r_0 = \sqrt{N-1}/\lambda$
(b) Pair correlation functions of the Laughlin states for $N = 6$, $7$, and $8$, $r_0 = \sqrt{N}/\lambda$.

where $q = \nu^{-1} = 2$ for the bosonic Laughlin state. Minimization of $P(T)$ or equivalently $T$ with respect to $r_0$ yields $r_0 = \sqrt{N-1}$ in case of a pure ring and $r_0 = \sqrt{N}$ if one atom is situated at the origin as, e. g., for $N = 6$, $7$, and $8$. This radius is always smaller than the size of the system given by $R = \sqrt{4N-2}$.

To understand the evolution of the above discussed hidden ordered patterns as $N$ increases, the "degree of correlation" $C$ is introduced as the height of the maximum peak in the pair correlation function. As indicated in Fig. 2.14, it monotonously decreases in $N$. This feature is naturally expected in order to recover the quantum liquid character of the Laughlin state for large systems.

Getting back to the above discussion on polygonal structures, in the Laughlin states of but a few bosonic atoms, an explanation for this sort of Wigner molecular patterns is the following. The nature of the ground state is independent from the kinetic part as long as the lowest Landau level approximation holds. Consecutively, its structure does not result from the competition between different kinds of energy. In addition, as the repulsive interaction energy is zero in the

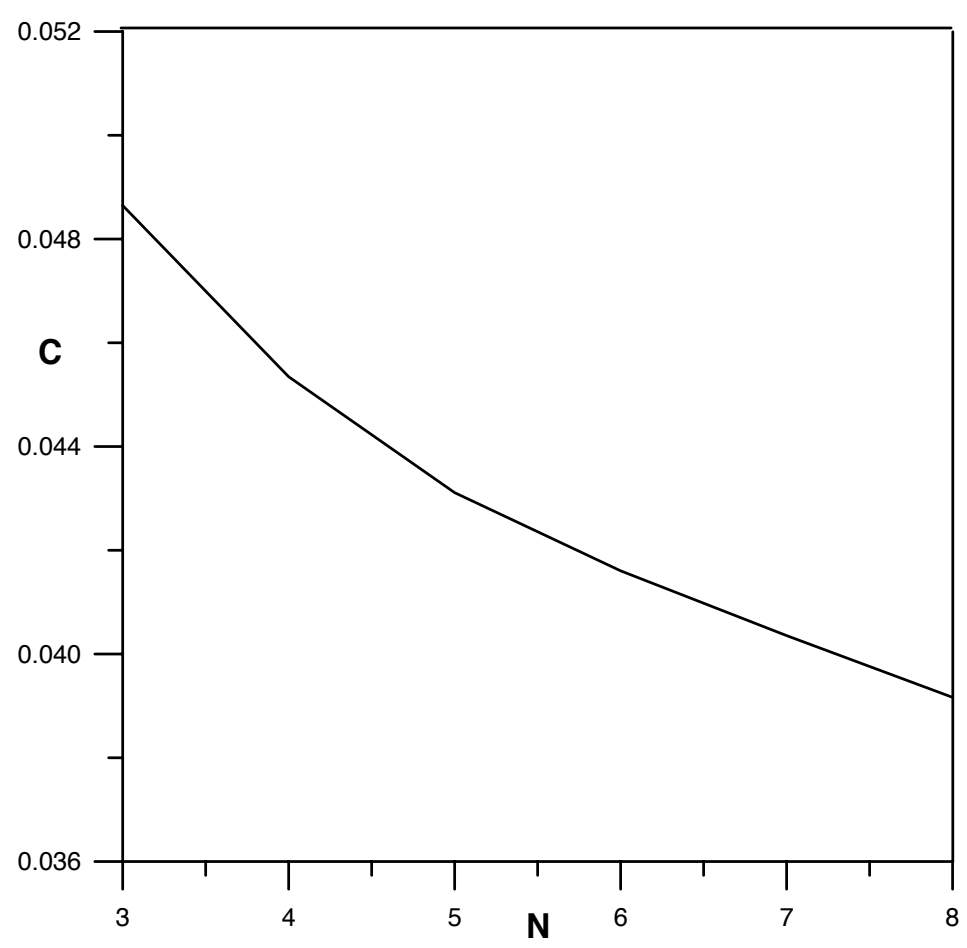

FIGURE 2.14: Degree of correlation of the Laughlin state as a function of N.

Laughlin state, the atoms seemingly choose symmetric and well-separated positions due to two conditions. Firstly, the system must have a large angular momentum given by $L^z = N(N-1)$ (which means large distances from the origin) and secondly, each atom has to be surrounded by a quasi-hole (which leads to effective mutual repulsion). This last statement is supported by the following observation. For $N = 3$, the contour plots at $L^z = 4$ and $L^z = 5$ in Fig. 2.5 suggest that in those precursory states of the Laughlin state($L^z = 6$), quasi-holes not attached to atoms are created without cost of internal energy. Their contribution to the total angular momentum evolves as $1/3$ (in $L = 4$), $2/3$ (in $L = 5$) and $3/3$, until the Laughlin structure becomes possible with one quasi-hole attached to each atom, lowering the interaction energy. The evolution of this behavior as $N$ increases provides a further conclusion. Due to the fact that the dependence of $L^z$ on $N$ is quadratic, the increase of total angular momentum is more easily accomplished for a large number of particles, and atoms do not have to be widely separated. Thus, the symmetric distribution should tend to disappear. Recently, it was proposed that the phenomenology of strongly correlated bosonic and fermionic rotating systems converges to the case of classical particles. They are supposed to finally crystallize at high rotational frequencies [205]. In the above context, traces of such crystallization are present, but it has to be pointed out that this ”crystallization” ceases to be manifest when the bulk structure starts to dominate the system. This happens for larger particle numbers, where the ground state at Laughlin angular momentum behaves more and more like a true quantum liquid.

## 2.3 Rotating Dipolar Fermions

This section deals with the analysis of rotating microscopic fermionic samples which interact by dipolar forces. The investigation follows the conceptual approach applied to bosonic atoms with short-ranged interactions. Within this context, the previously obtained results are compared with and extended. As a starting point, general spectral features are illuminated, before the evolution of ground state phases towards the strongly correlated regime is investigated. In this crossover regime, frustration of competing series of states lead to ground states with an incipient quasi-hole in the center. These may be linked with proposals of an effective theory of composite fermions. The final part concentrates on dipolar Laughlin states and their elementary excitations. In contrast to the precedent section, the focus is set on the manifestation of bulk phenomena. Within this final discussion, the necessary fundament is set to tie up to the results and conclusions of the first chapter.

The system considered is constituted of $N$ spin-polarized dipolar fermions rotating in an axially symmetric harmonic trapping potential strongly confined along the axis of rotation. Along this $\vec{e}_z$-axis, dipoles are assumed to be aligned. Low temperature $T$ and weak chemical potential $\mu$ with respect to axial confinement $\omega_z$ guarantee the gas to be effectively two-dimensional. As derived in Sec. (c), the second quantized Hamiltonian in the rotating reference frame reads

$$\hat{\mathcal{H}} = \hbar\omega_\perp \hat{N} + \hbar\left(\omega_\perp - \Omega\right)\hat{L}^z + \sum_{m_1,\dots,m_4} V_{1234}\, c^\dagger_{m_1} c^\dagger_{m_2} c_{m_4} c_{m_3} \tag{2.24}$$

Here, $\omega_\perp \ll \omega_z$ is the radial trap frequency, $\Omega$ is the frequency of rotation, $\hat{N}$ and $\hat{L}^z$ are the total number and z-component angular momentum operators, $c^\dagger_{m_j}$ and $c_{m_j}$ create and annihilate a fermion with angular momentum $m_j$ and

$$V_{1234} = \frac{1}{2}\langle m_1\, m_2|\sum_{j<k}^{N}\frac{d^2}{|\vec{r}_j - \vec{r}_k|^3}\,|m_3\, m_4\rangle \tag{2.25}$$

is the matrix element of uniaxial dipolar interaction expressed in the Fock–Darwin single particle angular momentum basis[198, 199]. In this basis, the above expectation value takes the analytical form

$$V_{1234} = \sum_{l}^{\text{odd}} \frac{\Gamma(l-\frac{1}{2})}{8l!}\, C^{n\,l}_{m_1\,m_2} C^{n\,l}_{m_3\,m_4} \delta_{m_1+m_2,\,m_3+m_4}\ , \tag{2.26}$$

where $n = m_1 + m_2 - l$ and

$$C^{\,n\;l}_{m_1\,m_2} = \sqrt{\frac{m_1!m_2!}{2^{m_1+m_2}n!l!}} \sum_{k=\max\{m_2-l,0\}}^{\min\{m_2,n\}} \binom{n}{k}\binom{l}{m_2-k}\left(-1\right)^{m_2-k}. \tag{2.27}$$

Strictly speaking, the above identity is valid if and only if the internal fermionic spin degree of freedom is frozen out such that the antisymmetric nature of creation and annihilation operators guarantees the cancellation of even Haldane pseudopotential coefficients. Without this constraint, the dipolar interaction leads to a pole divergence due to $1/r^3$ scaling. This seemingly unphysical behavior is resolved if a hard core short distance repulsive potential is included in the model. The specific characteristics of such a potential have to be determined from the atomic properties of the participating particles. In the given system, the Pauli principle ensures the validity of (2.26). As before, radial symmetry allows for a blockwise diagonalization with respect to $L^z$. Calculations have been performed for $N = 3$ to 12 particles using complete and Davidson block diagonalization techniques discussed in App. B. The strength of interaction scales as $d^2/l_0^3 = \hbar\omega_\perp(a_d/l_0)$ with $a_d = Md^2/\hbar^2$ and is set by $a_d/l_0 = 1$ in the natural units of energy $\hbar\omega_\perp$ and length $l_0$, respectively.

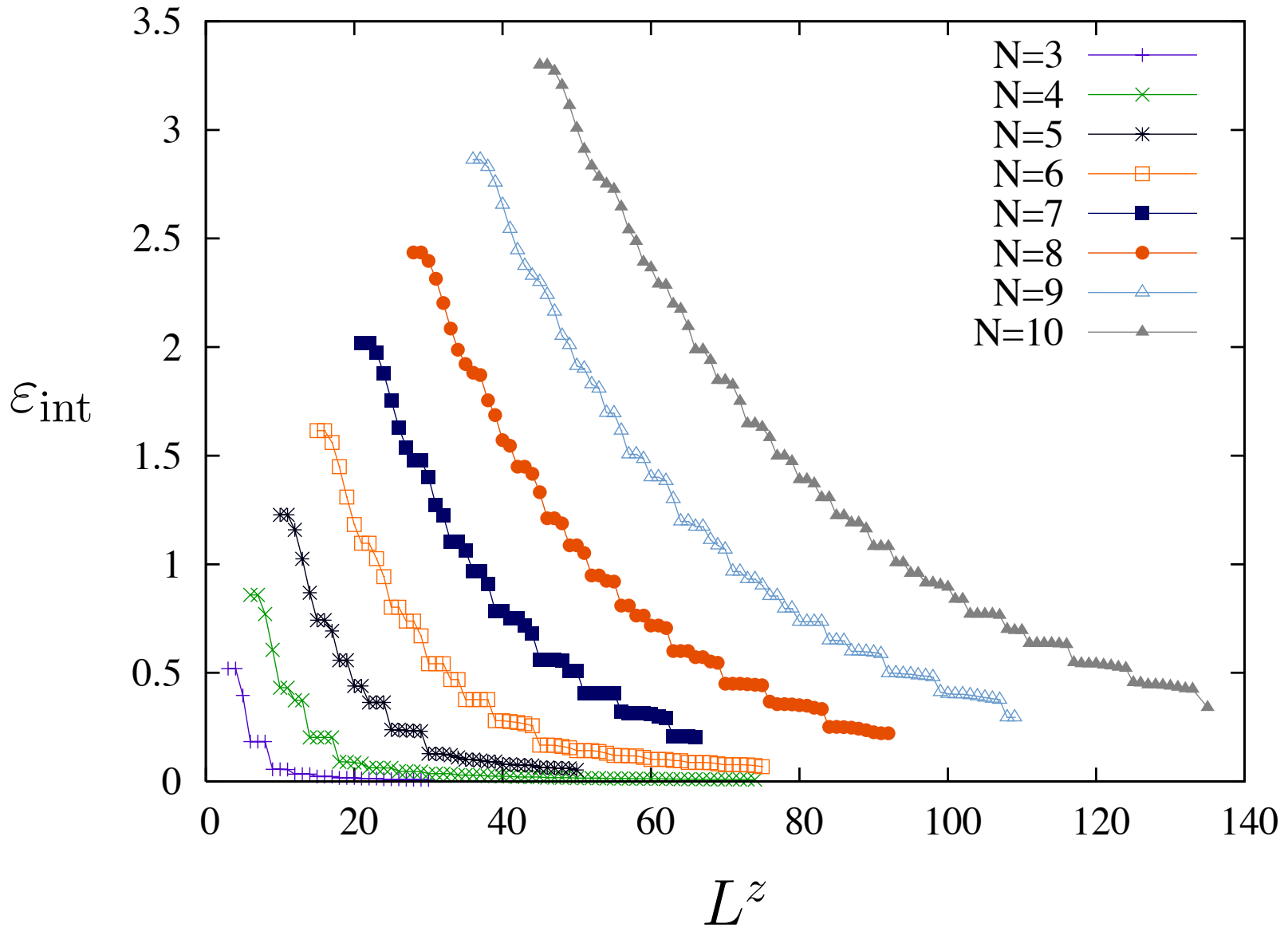

FIGURE 2.15: Yrast line: Ground state energy over $L^z$

The analysis starts from the ground state energy for a given $L^z$ which reveals the expected plateaus depicted in Fig. 2.15. Ground state candidates are the first states of these plateaus where a downward cusp occurs in the spectrum. By experimentally tuning the rotational frequency, some of these states are selected as true ground states at specific "magic" angular momenta shown in Fig. 2.16.

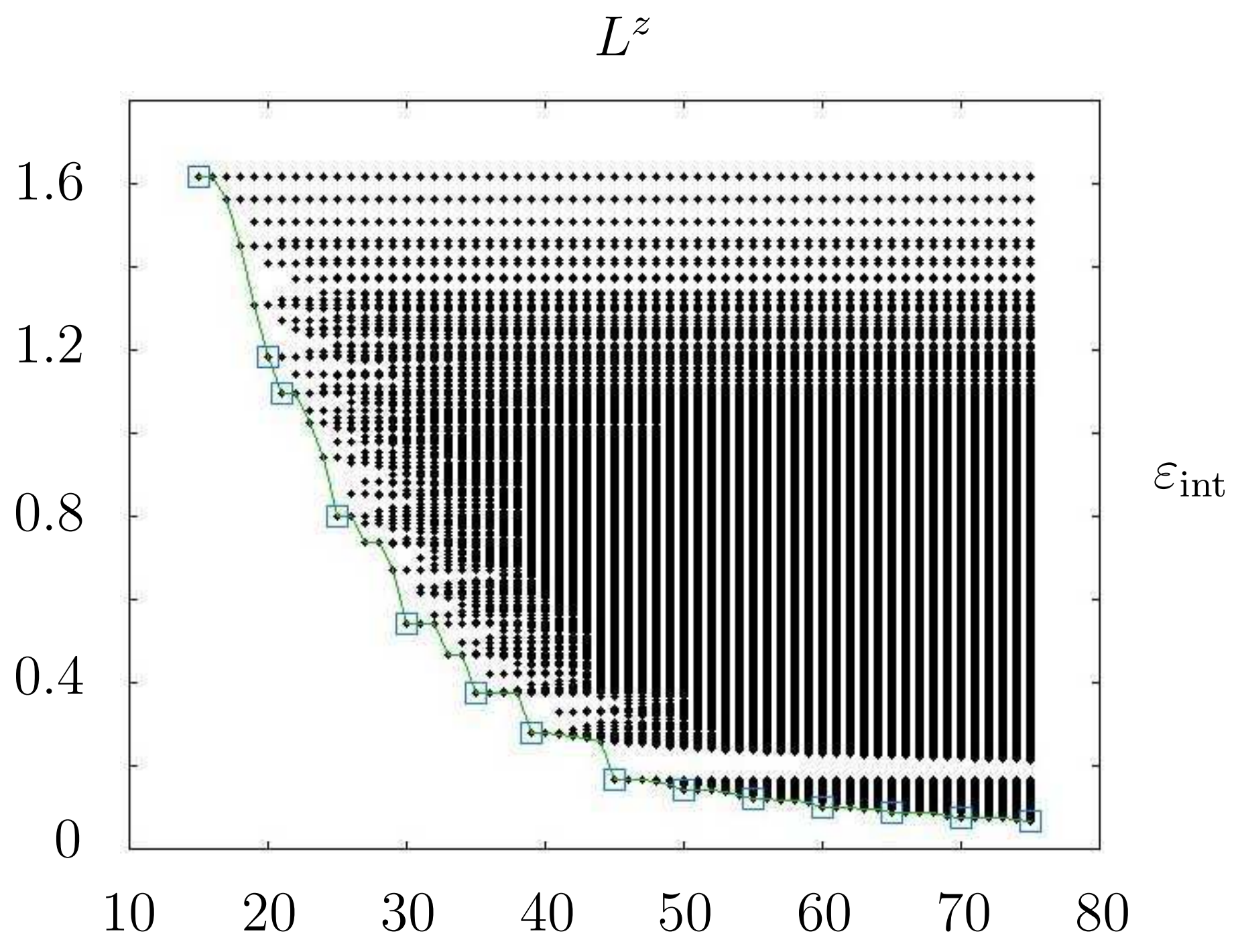

FIGURE 2.16: Interaction spectrum for $N = 6$ including green Yrast line and "magic" angular momenta indicated by blue squares.

The spectrum reveals each of the stable ground states as characterized by a substantial energy gap within its $L^z$ manifold. This type of excitation is referred to as "neutral" based on the nomenclature of condensed matter physics. As a second feature, the energy levels at a given $L_0^z$ are a subset of the spectrum of any larger angular momentum. This is a clear manifestation of the symmetry of interaction which makes the center of mass excitations gapless. It is important to note that this degeneracy is lifted by the residual trap present in case of any undercritical rotation. Under consideration of the above statements, there is one exceptional ground state situated at $L^z = 45$, where the neutral gap shows a significant robustness as it survives within the spectra of larger angular momenta. This unique ground state is manifest at all particle numbers, corresponds to a system with finite-size filling factor $\nu = 1/3$ and is deeply linked to the principal fermionic Laughlin state as subsequently discussed.

The complete set of "magic" angular momenta is illustrated in Fig. 2.17 expressed in terms of the function $L_{\mathrm{gs}}(\alpha)$ for several $N$. For $\alpha \equiv \hbar(\omega_\perp - \Omega) = 1$, i.e., no rotation, the ground state is the filled Landau level state at $\nu = 1$ which is completely insensitive to type and strength of interaction as long as the lowest Landau level approximation holds. While $\Omega$ is continuously increased, the system evolves from the weakly interacting regime to strongly correlated states. In the previously discussed case of bosonic short-range interacting gases, this process terminates at $L^z_{\mathrm{L}} = N(N-1)$, where the bosonic Laughlin state at filling $\nu = 1/2$ becomes the true ground state. The existence of a final $L^z$ is due to the fact that the ground state contact interaction energy vanishes for $L^z \geq L^z_{\mathrm{L}}$. The long range nature of dipolar interactions lifts this degeneracy. Thus, the whole principal series of fillings, i.e., $L^z_\nu = \nu^{-1}N(N-1)/2$ with $\nu = 1/(2m+1)$ for fermionic gases, is accessible. As before, internal structures of relevant states are analyzed via density-density correlations

$$\hat{\rho}^{(2)}(\vec{r}, \vec{r}_0) = \sum_{j<k}^{N} \delta(\vec{r_j} - \vec{r})\delta(\vec{r_k} - \vec{r}_0)\,. \qquad (2.28)$$

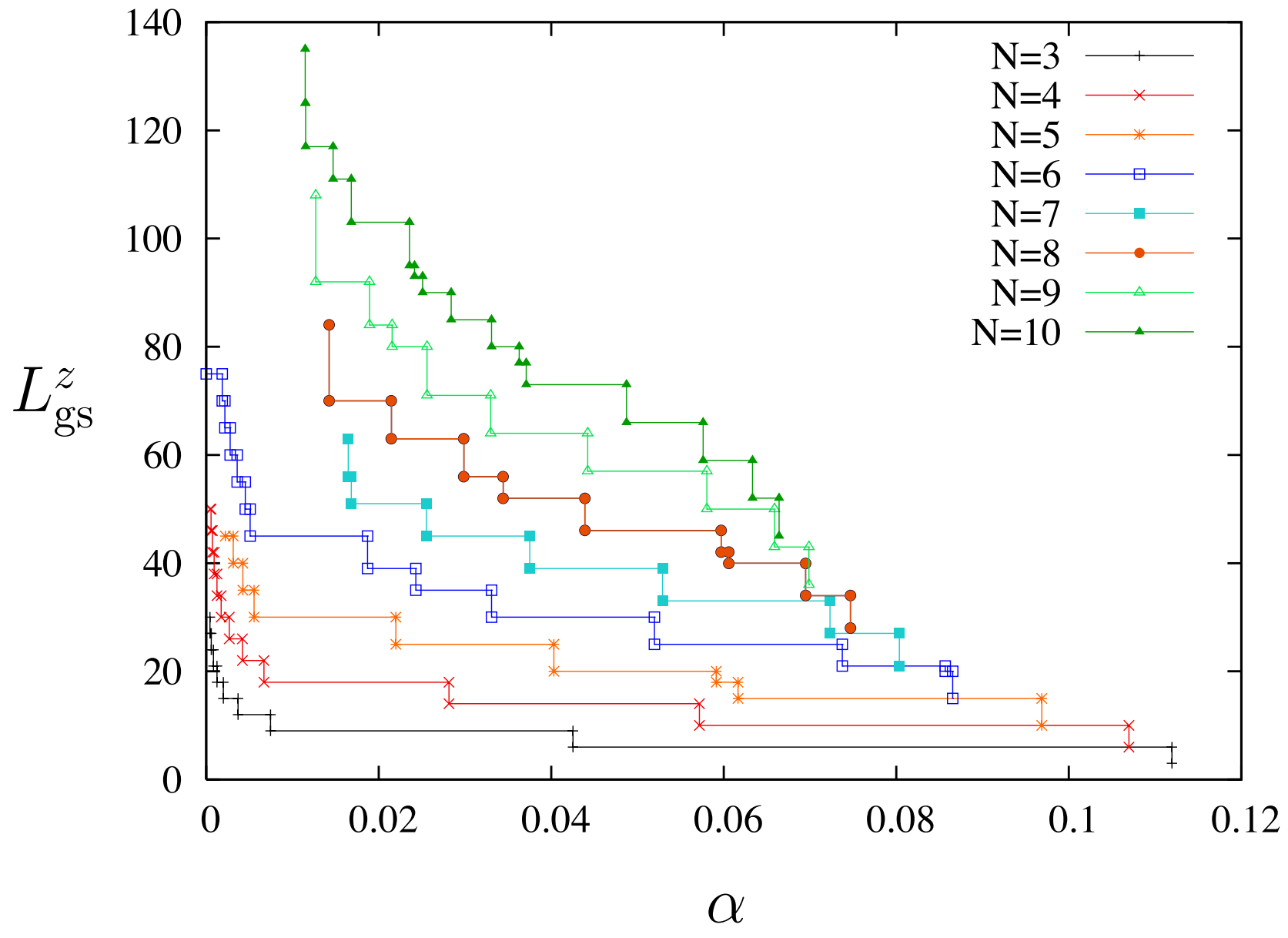

FIGURE 2.17: Ground state series $L^z_{\mathrm{gs}}$ as a function of $\alpha \equiv \hbar(\omega_\perp - \Omega)$

Fig. 2.18 shows $\hat{\rho}^{(2)}$ for the complete series of ground states with $N = 10$ particles. In general, starting from the unstructured state $L^z = N(N-1)/2$ (top left) with $\nu = 1$, a fraction of atoms starts to arrange itself on a symmetric cluster on the edge leaving a residual "Fermi sea" at the centre. This process progresses up to $L^z = 73$ (2nd center) in such a way that each of the edge atoms, seven in the above case, successively picks up one quantum of angular momentum. For smaller particle numbers, this behavior continues until correlations are homogenously established at $L^z_{1/3} = 3N(N-1)/2$. For larger $N$ though, the energetically favorable configuration changes at some point(2nd right), and a new series with an additional atom on the edge is established. In a dynamical picture this demands a reorganization procedure within the bulk. Speaking of causality, it could be equally stated that a correlation induced change of the bulk configuration of atoms drives a reformation of the edge. In the latter more intuitive picture, a competition between these two bulk phases consecutively occurs. It is manifested by the alternation of edge atoms from $L^z = 73$ to $85$. This competition finally causes a frustration of the system. In this regime of intermediate correlations, fascination states with a density defect at the origin appear. For $N = 10$, they are associated with one of the ground state phases, i.e., seven edge atoms at $L^z = 90$($3^{\text{rd}}$ right) and $95$($4^{\text{th}}$ center), and eight at $L^z = 103$($4^{\text{th}}$ right) and $111$(bottom left), respectively. Since this process is solely observed for a sufficiently large system, true bulk effects are strongly suggested to be its origin. If the rotational frequency is further increased, the system is stabilized at a specific edge configuration, which contributes most of the angular momentum (bottom center), and, finally, the $\nu = 1/3$ dipolar Laughlin state is obtained (bottom right).

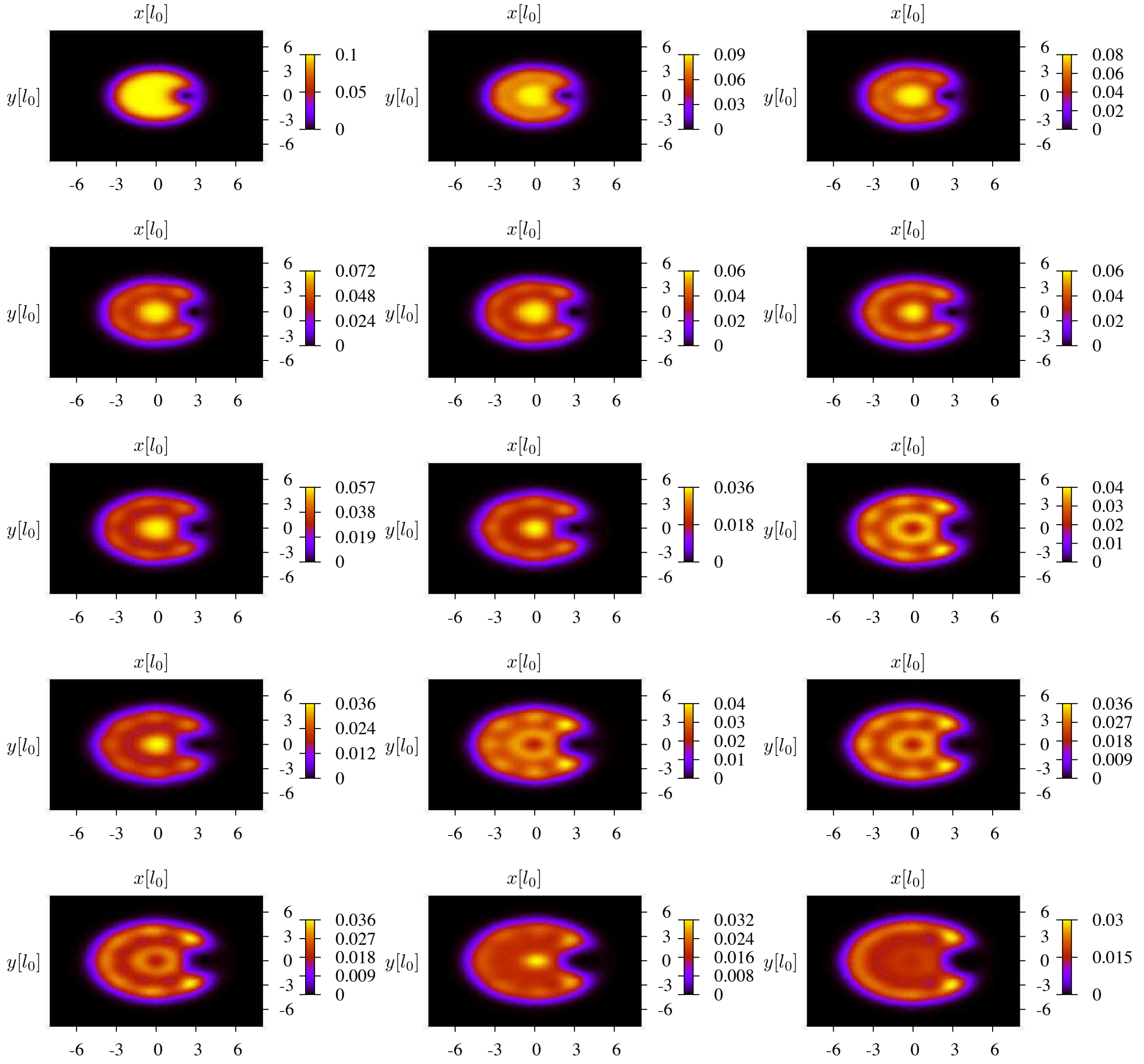

FIGURE 2.18: Ground state density-density correlation functions $\hat{\rho}^{(2)}(\vec{r}, \vec{r}_0)$ for $N = 10$ dipolar fermions. $L^z$ = (top) 45, 52, 59, (2nd) 66, 73, 77, (3rd) 80, 85, 90, (4th) 93, 95, 103, (bottom)111, 117, 135 with $\vec{r}_0$ set to the edge maximum of the density

### 2.3.1 Composite Fermions and Quasi-Holes

In order to solve the puzzle of "magic" $L^z$ numbers, an effective theory has been proposed to model an interacting system of electrons in a quantum dot in terms of non-interacting composite fermions[206, 207, 208]. In this ansatz, each fermion captures an even number of vortices and the wave function reads

$$\Psi_{\mathrm{CF}}(\{z_j\}) = \mathcal{N}\hat{\mathcal{P}}_{\mathrm{LLL}} \left\{ \prod_{j<k}^{N} (z_j - z_k)^{2p} \Psi_{\mathrm{Landau}} \right\} . \tag{2.29}$$

Here, $p$ denotes the number of vortex pairs, $\Psi_{\mathrm{Landau}}$ is an $N$-particle eigenfunction of

$$\mathcal{H}_{\mathrm{Landau}} = \sum_{j=1}^{N} \frac{1}{2M} \left(\vec{p}_j - m\omega_{\perp}\vec{\mathbf{e}}_z \times \vec{r}_j\right)^2 , \tag{2.30}$$

and

$$\hat{\mathcal{P}}_{\mathrm{LLL}} = \prod_{j}^{N} \sum_{k=0}^{\infty} \frac{z_j^k}{\pi k!} \mathrm{e}^{-|z_j|^2/2} \int \mathrm{d}^2 z_j' \, \bar{z}_j'^k \mathrm{e}^{-\left|z_j'\right|^2/2} \tag{2.31}$$

projects the resulting wave function to the lowest Landau level. In the following, only two-flux composite fermions with $p = 1$ will be considered. To identify the proper ground states and establish a link to the interacting system, the angular momentum $L_{\mathrm{cf}}^z$ of $\Psi_{\mathrm{Landau}}$ has to be determined. Even though the quantum numbers $|n,\, m\rangle$ of Landau states are non-negative integers, the explicit $\bar{z}_j$ dependencies of higher order Landau particles actually contribute negative angular momenta, i. e., $m_{\mathrm{cf}}^n = -n,\ -n+1, \ldots$ for the $n$-th Landau level. Thus, the appropriate set of $L_{\mathrm{cf}}^z$ is restricted to the interval $-N(N-1) \leq L_{\mathrm{cf}}^z \leq N(N-1)$, which is mapped to $N(N-1)/2 \leq L^z \leq L_{1/3}^z$ for the dipolar fermions due to the Jastrow prefactor.

As (2.30) is a non-interacting Hamiltonian with equally spaced energy levels $\varepsilon_n = 2\hbar\omega_\perp n + 1$, the spectral analysis is purely combinatoric and straightforward.

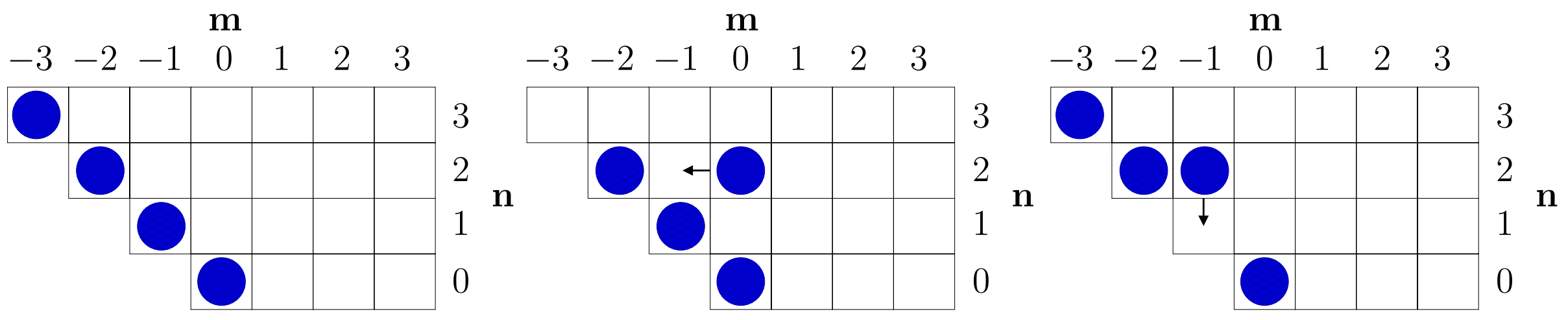

FIGURE 2.19: Ground state configurations of composite fermions. Left: Configuration with each Landau level occupied once corresponds to $\nu = 1$; Center: Inferred configuration from the compact state obtained by $|n=2, m=0\rangle \rightarrow |n=2, m=-1\rangle$; Right: No ground state configuration as particle with $|n=2, m=-1\rangle$ may move to favorable $|n=1, m=-1\rangle$ state

It is clear from $\varepsilon_n$ that the ground state configuration for $L_{\mathrm{cf}}^z = -N(N-1)/2$ is obtained by placing one atom each in the $N$ lowest Landau levels(left picture in Fig. 2.19). This state corresponds to the $\nu = 1$ state at $L^z = N(N-1)/2$ and marks the starting point for the consecutive analysis. To derive the Yrast line within this picture, several conceptual features have to be considered. Given a true ground state configuration with $L_{\mathrm{cf}}^z$, shifting a particle to higher angular momenta preserves the energy if the primary Landau level quantum number remains fixed(central image). This novel configuration can be a ground state in its higher angular momentum manifold, but, to check for lower energies, the analysis can be restricted to compact states. They are defined by the constraint that particles within a Landau level have to occupy the lowest lying $m_{\mathrm{cf}}^n$ states. Further, to be a ground state candidate, the condition has to be met that all occupation numbers $N_j$ of the $j$-th Landau level satisfy $N_j \leq N_{j-1} + 1$. If this is not fulfilled, the $N_j$-th particle may be shifted to the deeper lying level, lowering the kinetic energy (rightmost subfigure). As a consequence, the lowest Landau level has to be occupied at least once, since every other compact state with the above restrictions leaves the $L^z_{\mathrm{cf}}$ range obligatory for the mapping to the dipolar system. The remaining set of configurations has to be calculated and the Yrast line is obtained. Inclusion of the tilting energy yields the composite fermion model for the ground state angular momentum series of ground states as a function of the frequency of rotation.

Motivated by the successful description of microscopic Coulomb interacting electron systems in a quantum dot by this approach, the concept of composite particles has been nicely adapted to bosonic gases with short-ranged interaction[148]. The corresponding composite boson wave function ansatz reads

$$\Psi^{\mathrm{Bose}}_{\mathrm{CF}}(\{z_j\}) = \mathcal{N}\hat{\mathcal{P}}_{\mathrm{LLL}}\left\{\prod_{j<k}^{N}(z_j - z_k)\Psi_{\mathrm{Landau}}\right\} . \tag{2.32}$$

Yrast lines and ground states of the full bosonic Hamiltonian (2.2) were derived in strict analogy. Under consideration of these results, the series of true ground states for dipolar fermions are compared with the composite candidates.

As in both previous approaches, the effective theory provides an excellent prediction in sense of an accurate matching of the set of states where cusps in the Yrast line occur. Furthermore, overlap calculations for electrons and short-range interacting bosons suggest a good approximation of the ground state wave function by (2.29) and (2.32), respectively. Nonetheless, several cusps are omitted by the effective theory for larger particle numbers, and especially, the true ground state series is not appropriately matched in the crossover regime, e. g., for $N = 10$

| | | | | | | | | | | | | | | | | | | | | | |
|---|---|---|---|---|---|---|---|---|---|---|---|---|---|---|---|---|---|---|---|---|---|
| CF | 45 | | 55 | | 63 | | 69 | | 77 | | 83 | | 90 | | | 97 | 103 | 111 | 117 | 125 | 135 |
| true | 45 | 52 | | 60 | | 66 | | 73 | 77 | 80 | | 85 | 90 | 93 | 95 | | 103 | 111 | 117 | | 135 |

This deficiency – though not as extensively manifested – has already been observed and commented for dilute bosonic systems [148]. It suggests that even though basic structural properties of the system are covered in terms of the composite fermion picture, the detailed structure, reorganization phenomena, and bulk correlations are beyond the predictive power of the theory. This is underlined by the fact that deviations are more pronounced in dipolar interacting gases, where stronger correlation effects are expected especially in the crossover regime of bulk frustration. The question remains open, how high overlaps between the true ground state and composite ansatz wave functions may be interpreted. Unfortunately, overlap calculations are so far solely accessible for contact interacting bosons at higher particle numbers. This is due to the unavoidable exponential complexity in the Jastrow prefactor, which is significantly enhanced by the square in (2.29).

Even if the explicit composite fermion wave functions are not available, they can be expected to have high overlaps in the dipolar systems as well. This is supported by the precise matching of ground state candidates. As their bare total angular momentum is dominantly contributed by the edge atoms, the well-established agreement advises that the majority of possible edge configurations is strictly covered by the composite fermions. Then, any sufficiently homogeneous continuation of the density of the state inside the bulk already guarantees a very high overlap. Nonetheless, the appearance and nature of frustrated states as a consequence of inherent correlations is most probably beyond the scope of the simplified non-interacting picture.

Thus, the series of true ground states in Fig. 2.18 is reconsidered, the analysis being focused on the crossover regime. Already at first glance, it seems that the states in the $3^{\text{rd}}$ and $4^{\text{th}}$ row of Fig. 2.18 are somehow related with each other, e. g., $\Psi_{85}$ and $\Psi_{95}$ as illustrated in Fig. 2.20. Instead of interpreting the frustrated state as merely another bulk configuration of two-flux

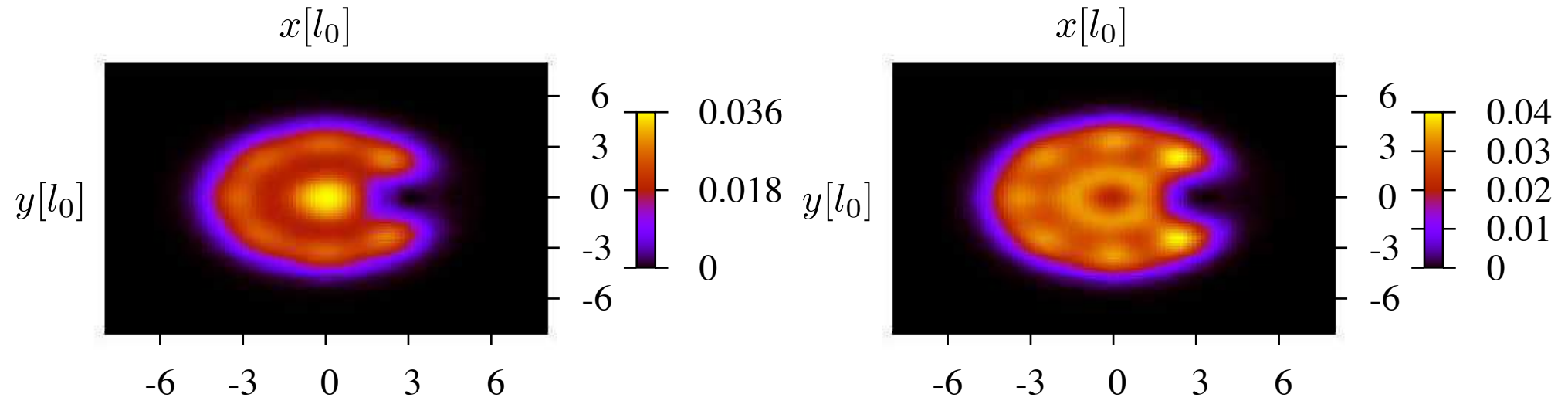

FIGURE 2.20: Ground state density-density correlation functions $\hat{\rho}^{(2)}(\vec{r}, \vec{r}_0)$ for $N = 10$ dipolar fermions at $L^z = 85$ and 95, $\vec{r}_0$ set to the edge maximum of the density

composite fermions, it is appealing to conjecture that the frustrated state is nucleated by a boost of each of the particles. Indeed, all three pairs of states consisting of a frustrated state and one which a ring of edge atoms and a residual "Fermi sea" in the center, are separated by exactly $N$ quanta of angular momentum, i. e., $80 \leftrightarrow 90$, $85 \leftrightarrow 95$, and $93 \leftrightarrow 103$. This corresponds to a boost of every particle by one unit of $L^z$.

Such a process is well-known in the conceptual classification of fractional quantum Hall states. It simply means that a quasi-hole is incipiently nucleated at the origin. In other words, if the primordial state is assumed to be a true composite fermion configuration, the frustrated states are their quasi-hole excitations. This presumption is supported by the occupation of single particle angular momenta and the radial density $\rho(\vec{r}) = \langle \Psi_\mathrm{L} | \hat{\rho}^{(1)}(\vec{r}, \vec{r}_0) | \Psi_\mathrm{L} \rangle$, see Fig. 2.21(l. h. s.).

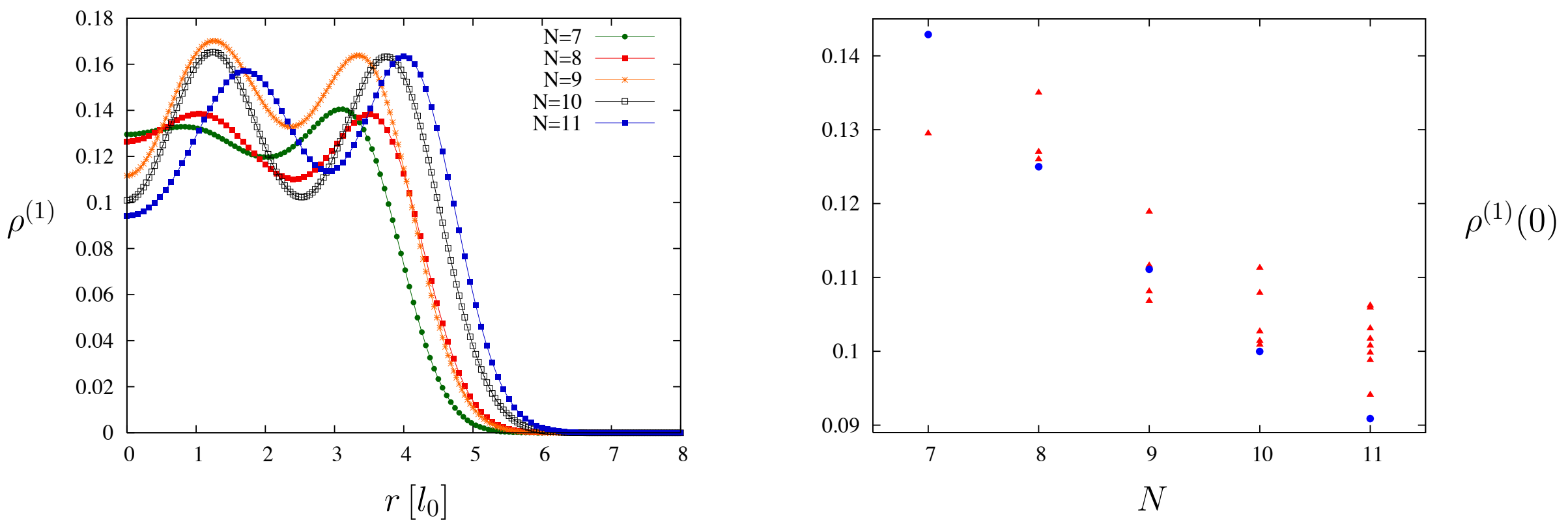

FIGURE 2.21: Radial density $\rho(\vec{r})$ of frustrated density defect states; r. h. s.: $\rho(\vec{0})$ over the number of particles

The quasi-hole excited wave function as originally proposed by Laughlin reads

$$\Psi_\mathrm{qh}(\{z_i\}) = \prod_{j=1}^{N} (z_j - \zeta_0) \Psi_\mathrm{L}(\{z_i\}) \tag{2.33}$$

where $\zeta_0$ denotes the position of the quasi-hole which strongly repulses the particles. If a true excitation of this kind is manifested in the system, the density has to vanish at $\zeta_0$. Certainly, in a microscopic sample, this pure bulk effect is expected to be weakened due to finite size and should behave as $1/N$ on the level of density. A detailed investigation of all $\Psi_\mathrm{L}$ with a cusp in the ground state energy and a density defect at the origin shows a striking agreement as depicted in Fig. 2.21(r. h. s.). Even the only state with these properties for $N = 7$, i. e. $\Psi_{56}$, which possesses but a marginal bulk formation, supports this conjecture. For all of these frustrated configurations, a corresponding primordial state with $\Delta L = N$ exists, which is moreover located at a cusp of the Yrast line. Given this pair of states, they can be related by the prescription of (2.33). Overlap calculations reveal a monotonously increasing agreement between the pairs with respect to the number of particles. The maximum of $\langle \Psi_{qh} | \Psi_\mathrm{L} \rangle$ for all pairs at a given $N$ grows from $0.6$ to roughly $0.7$ despite the significantly increasing basis dimension. The magnitude of these overlaps is even more remarkable as the density defect of $\Psi_{\mathrm{qh}}$ puts a severe constraint on contributing Fock states.

### 2.3.2 Fractional Quantum Hall States and their Excitations

As discussed in previous subsections, specific ground states $\Psi_L$ become stable within a finite range of the rotational frequency $\Omega$ and possess a significant gap with respect to neutral excitations. This guarantees the equilibrated system to have well-defined angular momentum and to be sufficiently stable at low enough temperatures. Especially one state, situated at $L^z = 3N(N-1)/2$, shows a substantially robust gap in the energy spectrum over a wide range of angular momenta. It is the principal dipolar Laughlin state $\Psi^{\mathrm{L}}_{1/3}$ at finite size filling factor $\nu = 1/3$, which has been proposed in the first chapter as a candidate to study the fractional quantum Hall effect in systems of rotating ultracold dipolar Fermi gases. Along the lines of the evolution of ground state structures, $\Psi^{\mathrm{L}}_{1/3}$ becomes stable shortly after the regime of frustrated states is passed. At this level, a concrete configuration of edge particles is energetically favored, and correlations between edge and bulk are homogenously balanced.

The analysis starts from the radial density $\rho^{(1)}(r)$ contrasted in Fig. 2.22 for dipolar and short-ranged interaction. In the latter case, the Laughlin wave function $\Psi_{\mathrm{Laughlin}}$ represents the exact ground state. Both systems are characterized by a ring of edge atoms which is driven outwards

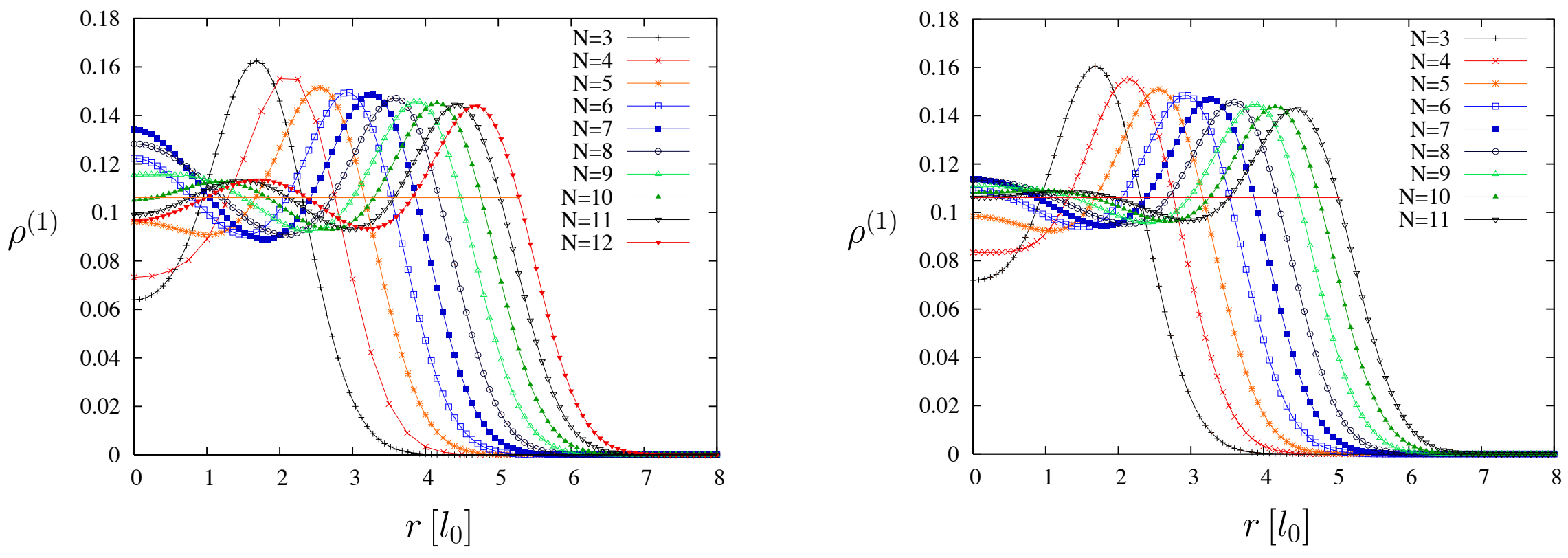

FIGURE 2.22: Radial density $\rho^{(1)}(r)$ of dipolar $\Psi^{\mathrm{L}}_{1/3}$(left) and true Laughlin wave function $\Psi_{\mathrm{Laughlin}}$.

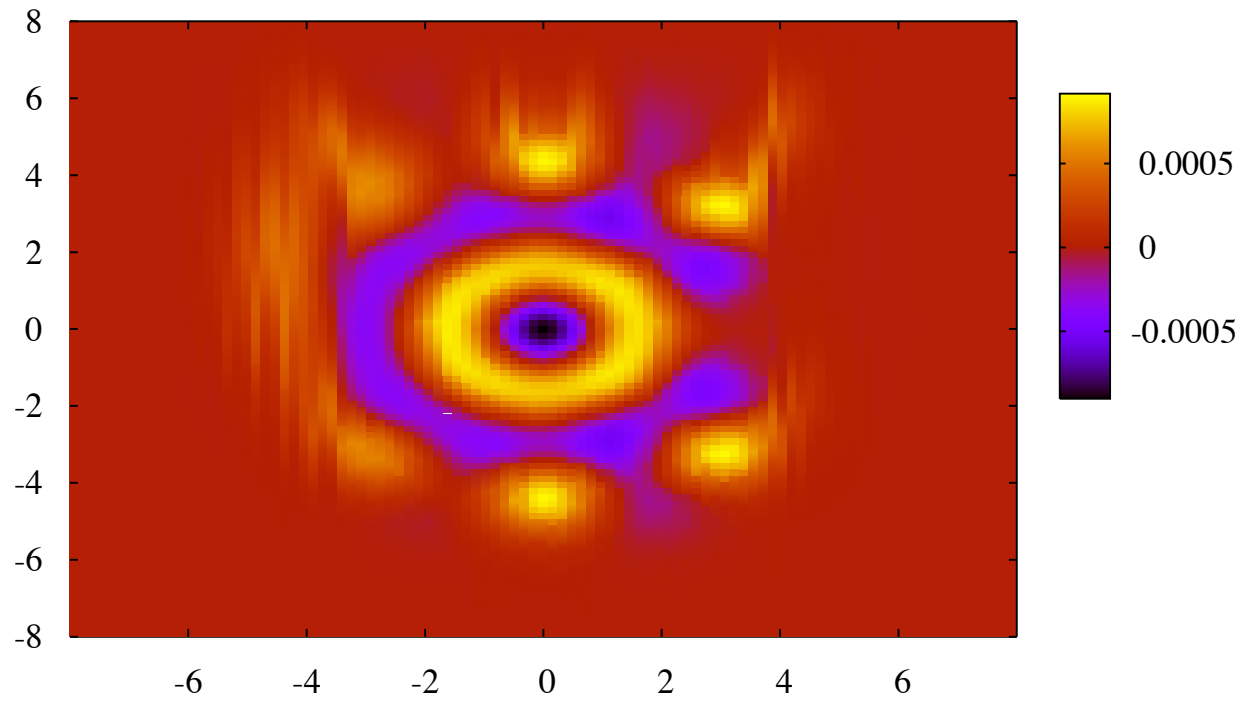

FIGURE 2.23: Difference $\Delta\rho^{(2)}(\vec{r},\, \vec{r}_0)$ between correlation functions of $\Psi^{\mathrm{L}}_{1/3}$ and $\Psi_{\mathrm{Laughlin}}$

for increasing number of particles, while its maximum peak approaches homogenous filling $\rho_\nu = \nu/\pi$. This ring is accompanied by a density valley converging to $\rho_\nu$ from below, which marks the boundary of the inner bulk. Whereas the location of the outer ring almost exactly coincides for both kinds of interaction, the inner structure shows apparent differences. As for the bosonic Laughlin state at half filling, the short-ranged interaction leads to the formation of a homogenous bulk with but a slight oscillation of density and considerably low wave number. Instead, in the dipolar case, these oscillations are significantly pronounced, and alternating domains of density maxima and minima are nucleated. Though the amplitude of oscillation behaves as $1/N$, it is an apparent indicator for enhanced structure and correlated behavior of particles in dipolar systems. Furthermore, it is important to note that even small structures become important in a constant homogenous bulk. This becomes more apparent if the difference between the correlation functions of both states is considered. It is shown in Fig. 2.23. Nonetheless, the validity of the Laughlin wave function ansatz for the calculation of the quasi-hole energy gap and results of Chapter 1 is guaranteed by its overlap with the true dipolar ground state representation $\Psi^{\mathrm{L}}_{1/3}$ illustrated in Fig. 2.24. Despite the effect of bulk deviations, which is also visible on this level for $N > 6$, the agreement is impressive, even more, as the basis dimension for $N = 11$ particles is greater than $10^6$. It must be noted that the above Hilbert space has already been truncated. This is justified due to the existence of a minimum filling factor of the system to be expected. It is imposed by restricting the maximum single particle angular momentum contribution to the $N$-particle Fock states, i. e., $m_{\max} = (n-1)/\nu$. Up to the Laughlin Hilbert sector at $L^z_{1/3}$, this constraint may be realized with $\nu = 1/3$, and the effect on ground state eigenvectors is negligible.

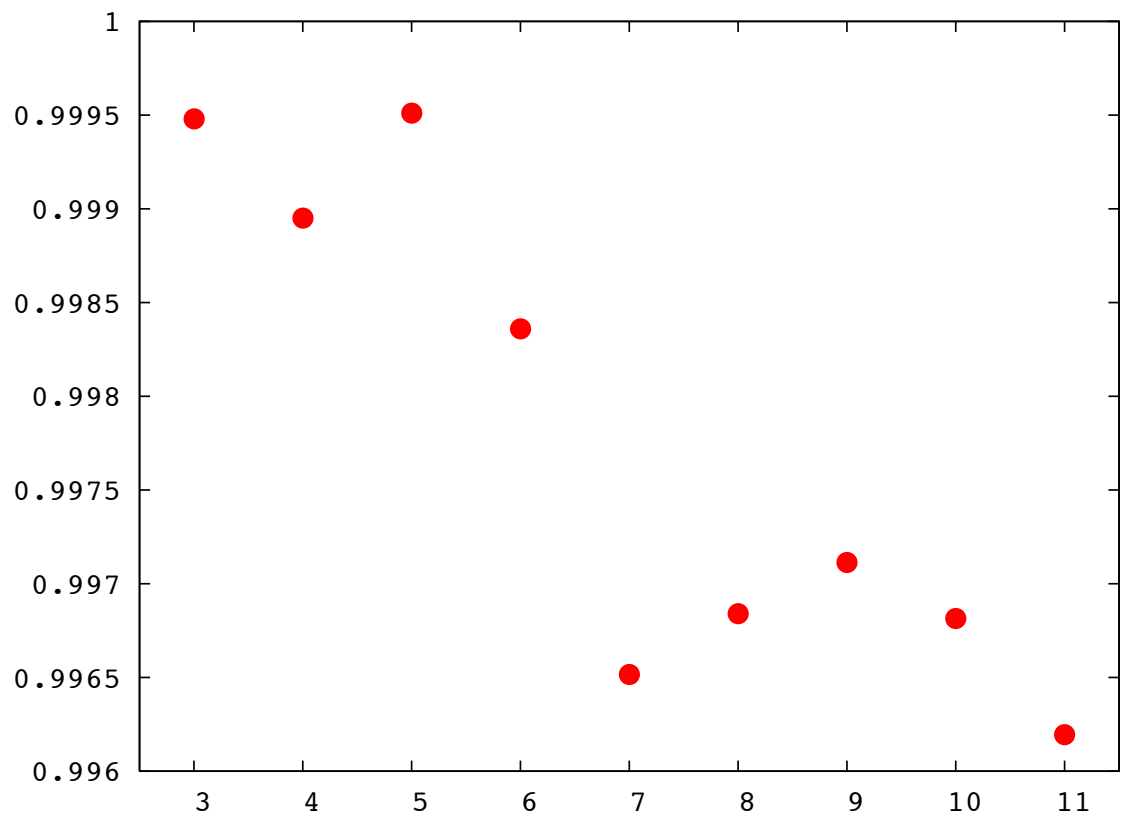

FIGURE 2.24: Overlap $\langle\Psi_{\mathrm{Laughlin}}|\Psi^{\mathrm{L}}_{1/3}\rangle$ between Laughlin wave function and the dipolar ground state.

To get more detailed information on the structure of these dipolar Laughlin states $\Psi^{\mathrm{L}}_{1/3}$, their density-density correlation functions were calculated for up to $N = 12$ particles. As depicted in Fig. 2.25, the structural evolution of $\Psi^{\mathrm{L}}_{1/3}$ while the number of particles increases may be grouped into three phases. The first is almost purely defined by the crystalline arrangement of atoms on the edge as it has been discussed for bosonic gases. The crossover to the second phase occurs between $N = 6$ and $N = 7$ particles where spokes of correlation form ridges from the outer ring to the center and constitute a homogenous bulk. It increases in size up to a point where clear patterns are developed in the third phase. The latter process starts from $N = 10$, which is a strong indicator that significant bulk effects are not expected to be revealed for smaller system sizes.

With this result in mind, the question of quasi-particle excitations can be addressed. For this, the finite size filling factor in the considered disk geometry has to be appropriately defined. This issue has been discussed in detail in Ref. [209]. Accordingly, in the dipolar system, $\nu$ is fixed by imposing the constraint

$$m_\nu \leq \nu^{-1}(N-1) \tag{2.34}$$

on the maximum single particle angular momentum in the $N$-particle Fock basis. It is the same restriction that was previously implemented for the truncation of the $L^z$ Hilbert space. If a quasi-hole or "quasi-electron" is nucleated, the filling factor is lowered and raised by a correction of the order $1/N$, respectively. With respect to (2.34), this yields $m_\nu \rightarrow m_\nu \pm 1$, where $m_\nu + 1$ corresponds to the quasi-hole. Meeting these constraints, the interaction spectrum is calculated as illustrated for $N = 6$ particles in Fig. 2.26. The full spectrum is characterized by the neutral gap and a tail of lower lying states at $L^z > L^{\mathrm{L}}_{1/3}$. This tail is constituted by the set of states with zero energy with respect to the short-ranged pseudo potential which chooses $\Psi_{\mathrm{Laughlin}}$ as the unique ground state at $\Psi^{\mathrm{L}}_{1/3}$. In bosonic gases, these states have been partially identified as edge excitations [163] which carry quanta of angular momentum. This becomes clear when the tilting energy is included in the above spectrum. If the rotational frequency $\Omega$ is tuned to favor $\Psi^{\mathrm{L}}_{1/3}$ as the true ground state, the lowest lying excitations in the quasi-hole subspaces at $L^{\mathrm{L}}_{1/3} < L^z \leq L^{\mathrm{L}}_{1/3} + N$ are gapped by $\Delta\varepsilon \sim 1/N$. If the filling factor is fixed, the branch of edge excitations disappears. Furthermore, all center of mass excitations vanish due to broken translational symmetry. This naturally identifies the dipolar Laughlin state as the unambiguous ground state in a homogenously filled system.

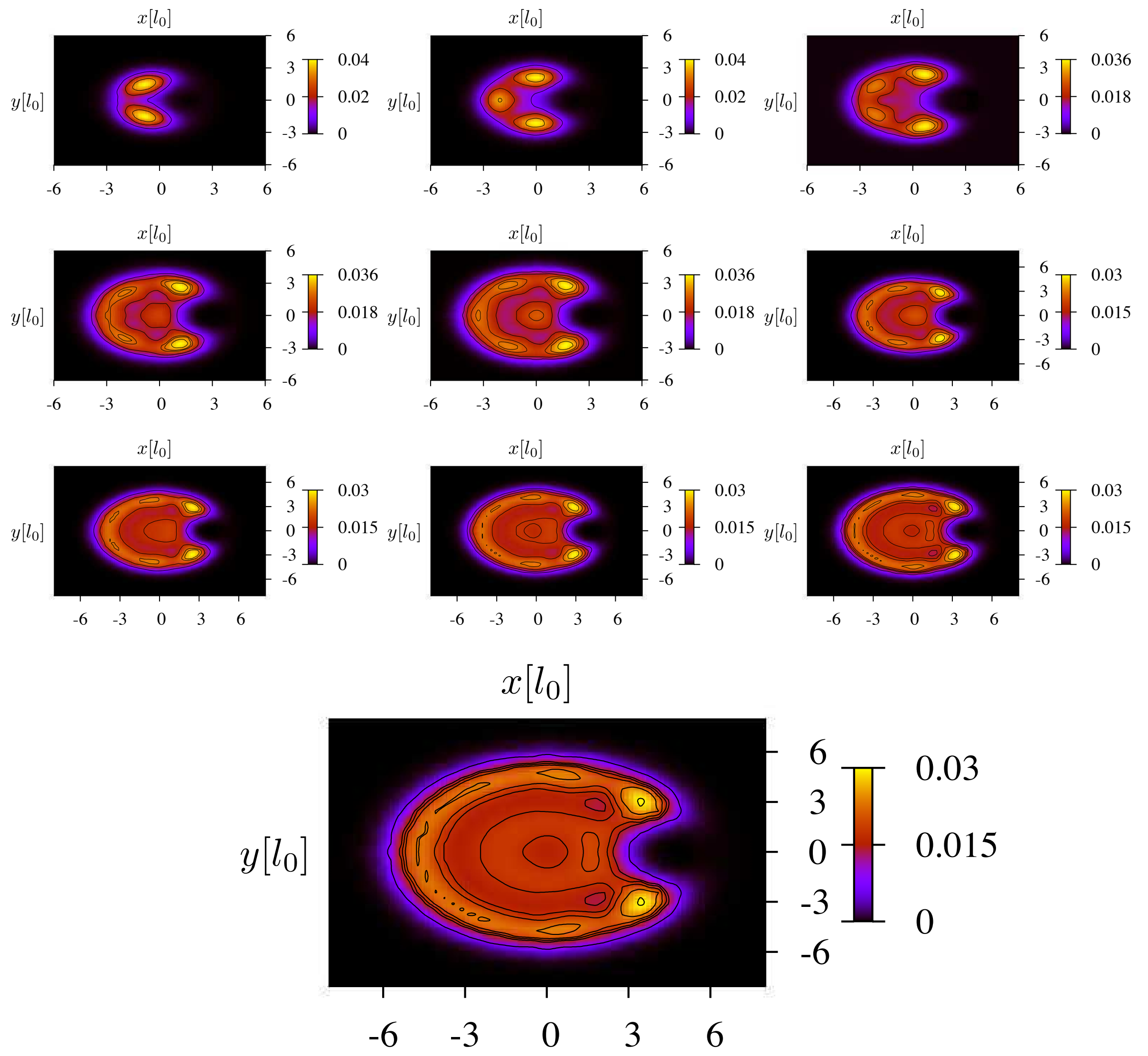

FIGURE 2.25: Ground state density-density correlation functions $\hat{\rho}^{(2)}(\vec{r}, \vec{r}_0)$ of dipolar fermions at $L^z_{1/3} = 3N(N-1)/2$ with $\vec{r}_0$ set to the edge maximum of the density. N=(top)3,4,5, (2nd)6,7,8, (3rd)9,10,11,(large)12.

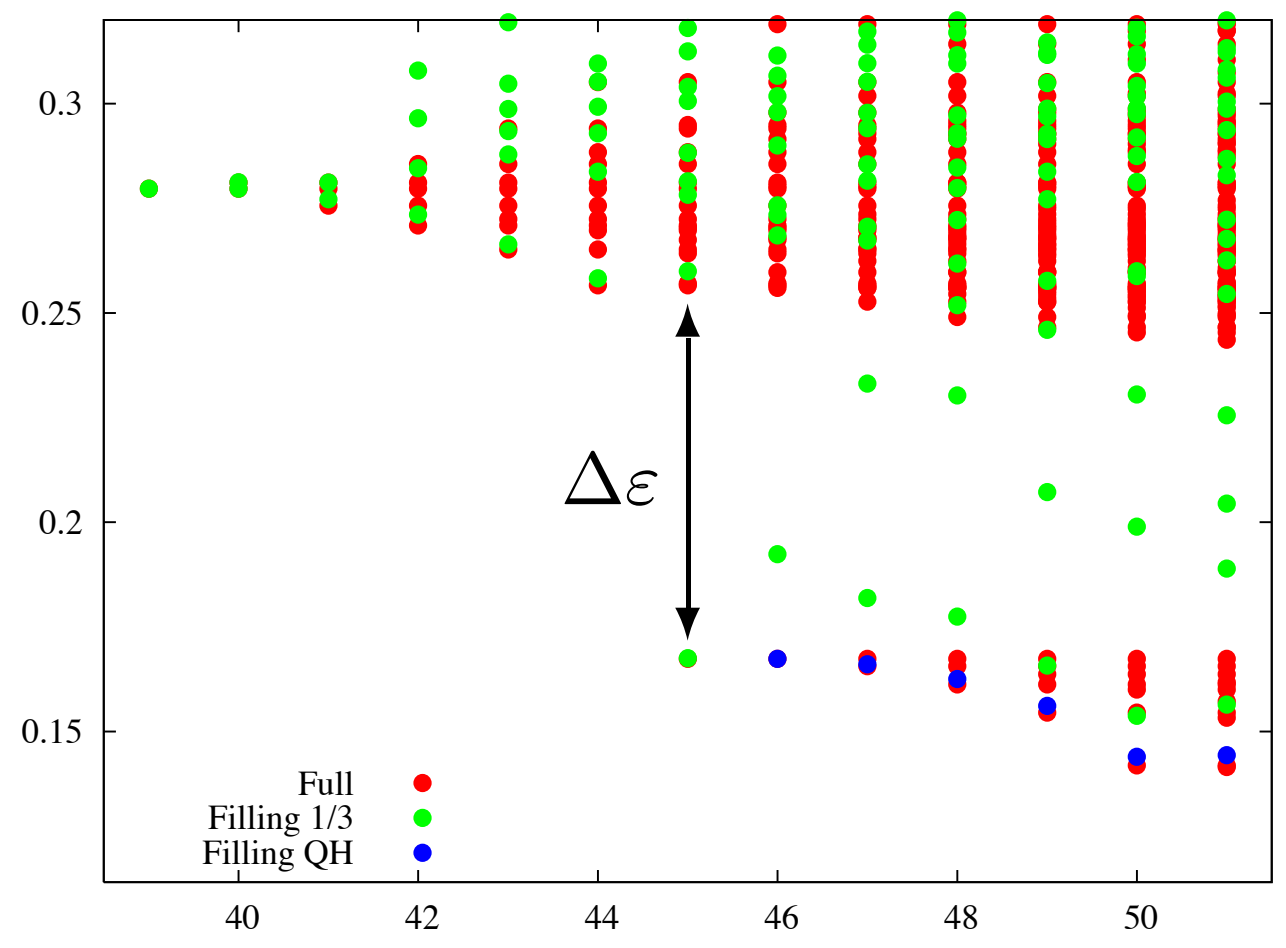

FIGURE 2.26: Interaction spectrum of $N = 6$ particles for the full basis(red) and fixed filling factor $\nu = 1/3$(green), furthermore quasi-hole ground states with $\nu = 1/3 - 1/N$(blue)

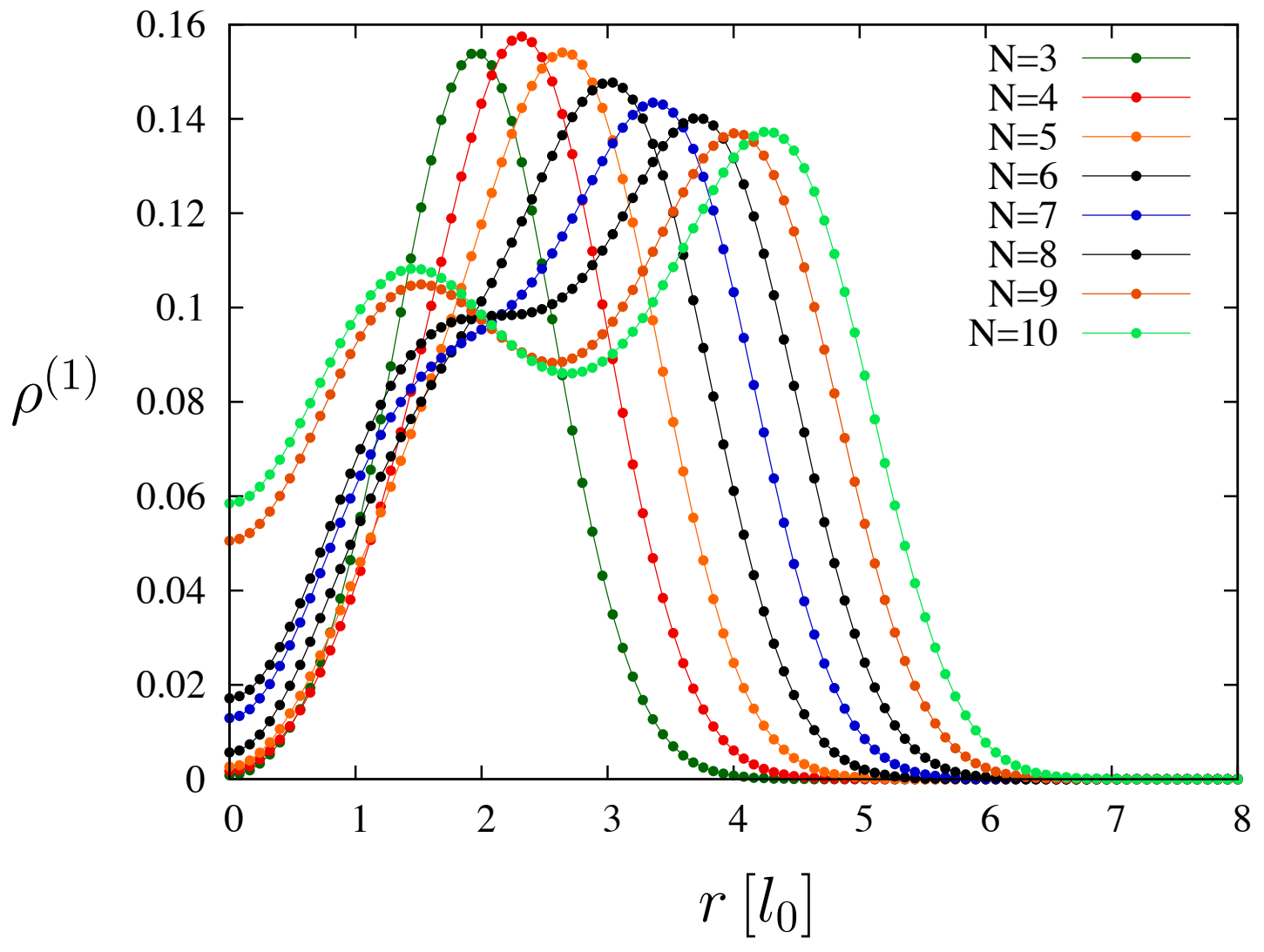

FIGURE 2.27: Radial density $\hat{\rho}^{(1)}(\vec{r})$ at $L^{\mathrm{L}}_{1/3}$ and $L^z_{\mathrm{qh}}$ for $N = 3, \ldots, 10$

To identify quasi-hole excitations located at the origin, the eigenstates at $L^z_{\mathrm{qh}}(N) = L^{\mathrm{L}}_{1/3} + N$ are calculated with adapted $m_\nu + 1$. Apparently, the ground states $\Psi^{\mathrm{qh}}_{\mathrm{L}}$ closely follow the lowest energy of the branch of edge excitations. This behavior holds for all numbers of particles. Thus, $\Psi^{\mathrm{qh}}_{\mathrm{L}}$ represents a finite size edge excitation rather than a quasi-hole. Consequently, the pronounced density defect of $\rho^{(1)}(\vec{r})$ present at the origin for small $N$ disappears as depicted in Fig. 2.27. The analytical shift of $\Psi^{\mathrm{L}}_{1/3}$ via (2.33) does not resolve this issue either. The derived expectation value of energy lies within the edge band and has a large variance. The neutral gap remains as the only reliable and accessible quantity to estimate the quasi-hole energy. It arises from a collective excitation of a quasi-hole-"quasi-electron" pair and is plotted in Fig. 2.28.

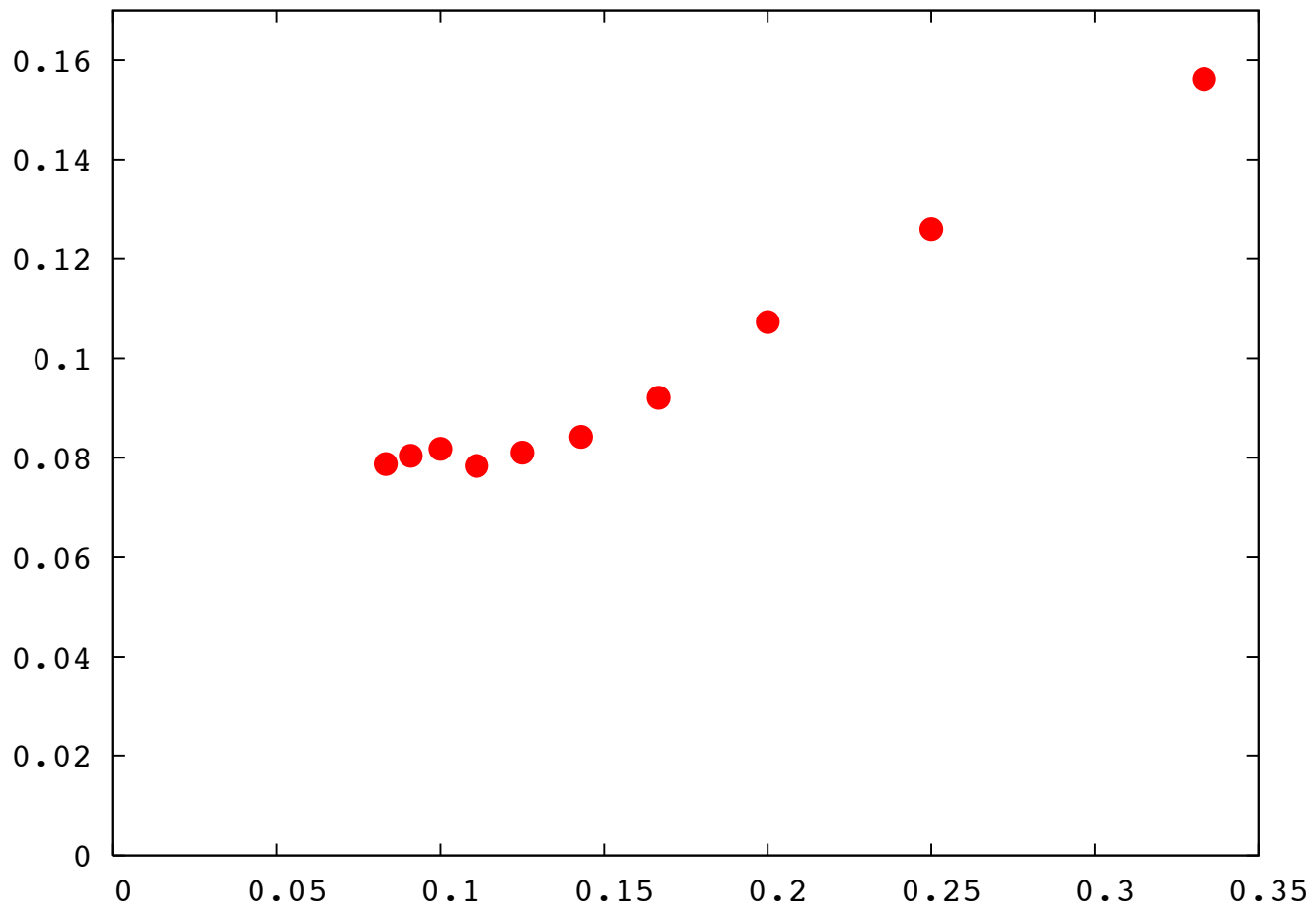

FIGURE 2.28: Neutral energy gap $\Delta\varepsilon$ as a function of $1/N$

Due to the edge, $\Delta\varepsilon$ first decreases, but experiences an upward kink at $N = 10$. This is a direct consequence of the formation of a substantial bulk. Compared to the results of Chapter 1, i.e., for mass $M = 30$amu, dipolar moment $d = 0.5$Debye, and a trap frequency of $2\pi 10^3$Hz, the gap is on the order of $1\%$ of the Landau level spacing $2\hbar\omega_\perp$. This is $2\%$ of the value derived from correlation functions in the thermodynamic limit. It is reasoned by the obvious lack of bulk behavior in microscopic samples, which on top does not allow for reliable extrapolations in the $1/N$ limit.

# CHAPTER 3

# Cold Atoms in Non-Abelian Gauge Potentials

In this chapter, an alternative method is proposed to subject atomic samples to the effects of artificial magnetic vector potentials. If the particles are situated in a two-dimensional optical lattice, it is feasible to experimentally control the phases of tunneling amplitudes in various ways [210, 211, 212]. If these are multiplied around an elementary cell of the lattice, the total phase acquired by the wave function is proportional to the magnetic flux of an effective magnetic field perpendicular to motional plane of the particles. Due to the variety of feasible lattice configurations and additional laser manipulations, which have been discussed in the introductory Sec. (d), the scope of this implementation is enormous. Another highly interesting proposal demonstrates how effective magnetic fields can be employed by electromagnetically induced transparency [213].

Starting from the two-dimensional Hofstadter problem with its famous butterfly spectrum [214], the implementation of this system in a cold gaseous experiment is presented along the lines of the seminal letter by Jaksch and Zoller [210]. The physical features of this system are fascinating and intriguing, but have never been observed in condensed matter experiments. This is due to the available range of lattice spacings in crystalline materials. To access the regime of the butterfly, inaccessibly high magnetic field strengths are necessary. Instead, optical wavelengths allow for suitable lattice spacings in accordance with the artificial magnetic flux. Furthermore, in weakly interacting or weakly disordered systems, interesting spectral modifications occur which allow for a robust detection of the butterfly [210]. If correlations are increased, fractional quantum Hall-like like states are expected in the limit of strong magnetic fields [212].

Nonetheless, the Hofstadter problem is only the basic model for a completely novel class of ultracold atomic systems proposed in the second part of this chapter. Zeeman-level dependent lattice potentials allow to trap atoms with different internal hyperfine states, and state-dependent, laser-assisted tunneling guarantees a coherent transfer between them [215]. By the above method, “isospin”-dependent hopping phases are imprinted on the particles. Thus, they are subject to artificial vector potentials which correspond to non-Abelian unitary groups, e. g. , $U(n)$ and $SU(n)$. The matrix phases of a product around a lattice plaquette are by construction non-trivial. Their mean trace, which corresponds to the Wilson loop in lattice gauge theories (see App. A), is not equal to a simple winding number $n$ [216] and gives rise to the prospect of exciting phases, for instance generalized non-Abelian fractional quantum Hall states.

The analysis continues with finite size realizations and illuminates the differences to the infinite lattice. It turns out that in this approach, additional information on the spectrum is gained and, in contrast to the infinite system, the full range of experimentally feasible configurations of gauge potentials may be studied. In the final part, possibilities to study non-Abelian interferometry are discussed and the question is addressed to what extent the above system may be used as a simulator for lattice of gauge theories.

The proposed system is the first study of matter dynamics driven by designed non-Abelian gauge potentials in the context of atomic, molecular, and optical physics (not much later, an equally promising approach based on electromagnetically induced transparency has been proposed [217]). Although single particles in $SU(n)$ monopole fields have been intensively studied in high energy physics [218, 219], other field configurations did not attract such interest. Specific examples of external gauge fields have also been discussed with respect to NMR, nuclear and molecular physics [269]. However, all of these proposals either do not deal with interacting many-particle systems in such gauge fields, or they do not consider lattice gauge fields of ultra-high strength.

## 3.1 The Hofstadter Problem

The Hofstadter problem considers a single electron in a two-dimensional square lattice of spacing $a$ in the presence of an Abelian magnetic field $B$. When the lattice potential is sufficiently strong, the single particle non-relativistic Hamiltonian in the tight binding approximation is given by the so-called Peierl’s substitution [214] and reads

$$H = -2V_0\Big[\cos\Big(\frac{a}{\hbar}(p_x - \mathrm{i}\frac{\mathrm{e}}{c}A_x)\Big) + \cos\Big(\frac{a}{\hbar}(p_y - \mathrm{i}\frac{\mathrm{e}}{c}A_y)\Big)\Big] \ , \tag{3.1}$$

where $V_0$ is the strength of the optical potential, $\vec{p}$ is the momentum operator, and $\vec{A}$ is the magnetic vector potential. Under the assumption $[p_j, A_j] = 0$, the implicit exponentials of the cosine terms decompose into a potential dependent part and translational operators. The latter can be directly applied, and the single-particle wave equation reads

$$\begin{aligned} e^{-\mathrm{i}\frac{ea}{\hbar c}A_x(y)}\psi(x+a,y) + e^{\mathrm{i}\frac{ea}{\hbar c}A_x(y)}\psi(x-a,y) \\ +e^{-\mathrm{i}\frac{ea}{\hbar c}A_y(x)}\psi(x,y+a) + e^{\mathrm{i}\frac{ea}{\hbar c}A_y(x)}\psi(x,y-a) &= -\frac{E}{V_0}\psi(x,\,y)\,. \end{aligned} \tag{3.2}$$

The choice of $\vec{A}$ determines the magnetic field $\vec{B}$, which in turn determines the behavior of the system. For $\vec{B} = B\vec{e}_z$, one may choose the potential due to gauge freedom as $\vec{A} = (0,\, Bx,\, 0)$. Thus, solely tunnelings in the $y$-direction acquire phases. This reduces the wave equation to an effectively one-dimensional problem which allows for a significantly simplified treatment. After introduction of the natural unit of energy $\varepsilon = -E/V_0$ and discretized lattice coordinates $x = ma$ and $y = na$, the plain wave ansatz $\psi(ma,\, na) = \mathrm{e}^{\mathrm{i}\nu n}g(m)$ transforms (3.2) into Harper's equation [220]

$$g(m+1) - g(m-1) + 2\cos(2\pi m\alpha - \nu)g(m) = \varepsilon g(m)\,, \tag{3.3}$$

where the dimensionless parameter of the magnetic flux $\alpha = \frac{ea^2B}{hc}$ is introduced, about which *"all the fuss is made"* (quoted from [214]). This comment refers to one of the peculiar features of Bloch electrons in magnetic fields, where physics seemingly distinguishes between rational and irrational numbers. The original discussion of the Hofstadter problem intended to resolve this paradox situation as a mathematical artifact without physical manifestation. The regime of the following analysis requires finite values of $\alpha \approx 1$. For a crystal lattice of a few Å, this implies a magnetic field strength $B \approx 10^5$ Tesla since $B \sim 1/a^2$.

Equation (3.3) can be rewritten to recursive form

$$\begin{pmatrix} g(m+1) \\ g(m) \end{pmatrix} = \underbrace{\begin{pmatrix} \varepsilon - 2\cos(2\pi m\alpha - \nu) & -1 \\ 1 & 0 \end{pmatrix}}_{A(m)} \begin{pmatrix} g(m) \\ g(m-1) \end{pmatrix}, \tag{3.4}$$

with the $2\times2$ transfer matrix $A(m)$. A proper eigenfunction of the system has to be bounded for all sites. In an infinite lattice, this puts a constraint on the product of consecutive $A(m)$ matrices.

Let the cosine term be periodic in $m$, which becomes an exact statement if $\alpha \equiv p/q \in \mathbb{Q}$ with $p$ and $q$ being mutually coprime. Then a block transfer matrix can be defined

$$Q(\varepsilon,\, \nu) \equiv A(q) \cdot A(q-1) \cdot \ldots \cdot A(1)\,. \tag{3.5}$$

The physical condition imposed on $Q$ by the demanded boundedness of the wave function is that it possesses a unitary algebraic subspace. This means that an eigenvector with eigenvalue of unit modulus exists. Since

$$\det(Q) = \prod_{j=1}^{q} \det(A(m)) = 1 = \lambda_1 \lambda_2\,, \tag{3.6}$$

where $\lambda_j$ are the eigenvalues of $Q$, this implies that both eigenvalues have to be of unit magnitude. The eigenvalues of the real $2 \times 2$ matrix Q can be expressed as

$$\lambda_{1,2} = \frac{\mathrm{tr}(Q)}{2} \pm \sqrt{\frac{\mathrm{tr}^2(Q)}{4} - 1}\,. \tag{3.7}$$

Thus, the above constraint is equivalent to

$$|\mathrm{tr}\,(Q(\varepsilon,\, \nu))| \leq 2\,. \tag{3.8}$$

It can be proven that (3.8) can be further simplified [221]. As $Q$ is obtained by multiplication of $q$ successive matrices $A(m)$, which equals the full cosine period, the effect of the parameter $\nu$ is simply an oscillation around a mean value of the trace, i.e.,

$$\mathrm{tr}\,(Q(\varepsilon,\, \nu)) = \mathrm{tr}\,(Q(\varepsilon,\, \nu_0)) + 2f(\nu)\,, \tag{3.9}$$

where $\nu_0 = \pi/(2q)$ (there is a misprint in [214] at this point) and $f(\nu)$ is a periodic function of unit amplitude. Thus, the final condition, which defines the energy spectrum of (3.1), reads

$$\left|\mathrm{tr}\left(Q\left(\varepsilon,\, \frac{\pi}{2q}\right)\right)\right| \leq 4\,. \tag{3.10}$$

The result is maybe the most beautiful spectrum in the world of physics, the famous Hofstadter butterfly illustrated in Fig. 3.1. It is highly symmetric and formed by a very subtle topological fractal pattern. In a simple intuitive way, it can be understood by a continued breaking up of the original Bloch band. Since $\mathrm{tr}(Q)$ is always a polynomial of rank $q$, the most obvious behavior is that condition (3.8) is fulfilled around the roots of this polynomial. This simple picture is indeed true and may be extended in terms of continued fractions

$$\alpha = \frac{1}{n_1 + \frac{1}{n_2 + \frac{1}{n_3 + \dots}}}. \tag{3.11}$$

This uniquely defined iteration, which terminates for rational values of the magnetic parameter, motivates that the spectrum for a given $\alpha$ consists of $n_1$ subbands which break up into $n_2$ sub-subbands and so on. Even if this scheme is not perfectly true, it explains the recursive formation of the butterfly quite well. The question remains how physical the butterfly really is, because, even if the spectrum seems sufficiently compact to the slightly unfocused eye, it is highly discontinuous. Though the reader is referred to [214] and the underlying Ph. D. thesis for rigorous proof, it is exactly the above mentioned compact structure which guarantees that the butterfly becomes completely homogeneous for a magnetic field with but the slightest fluctuations.

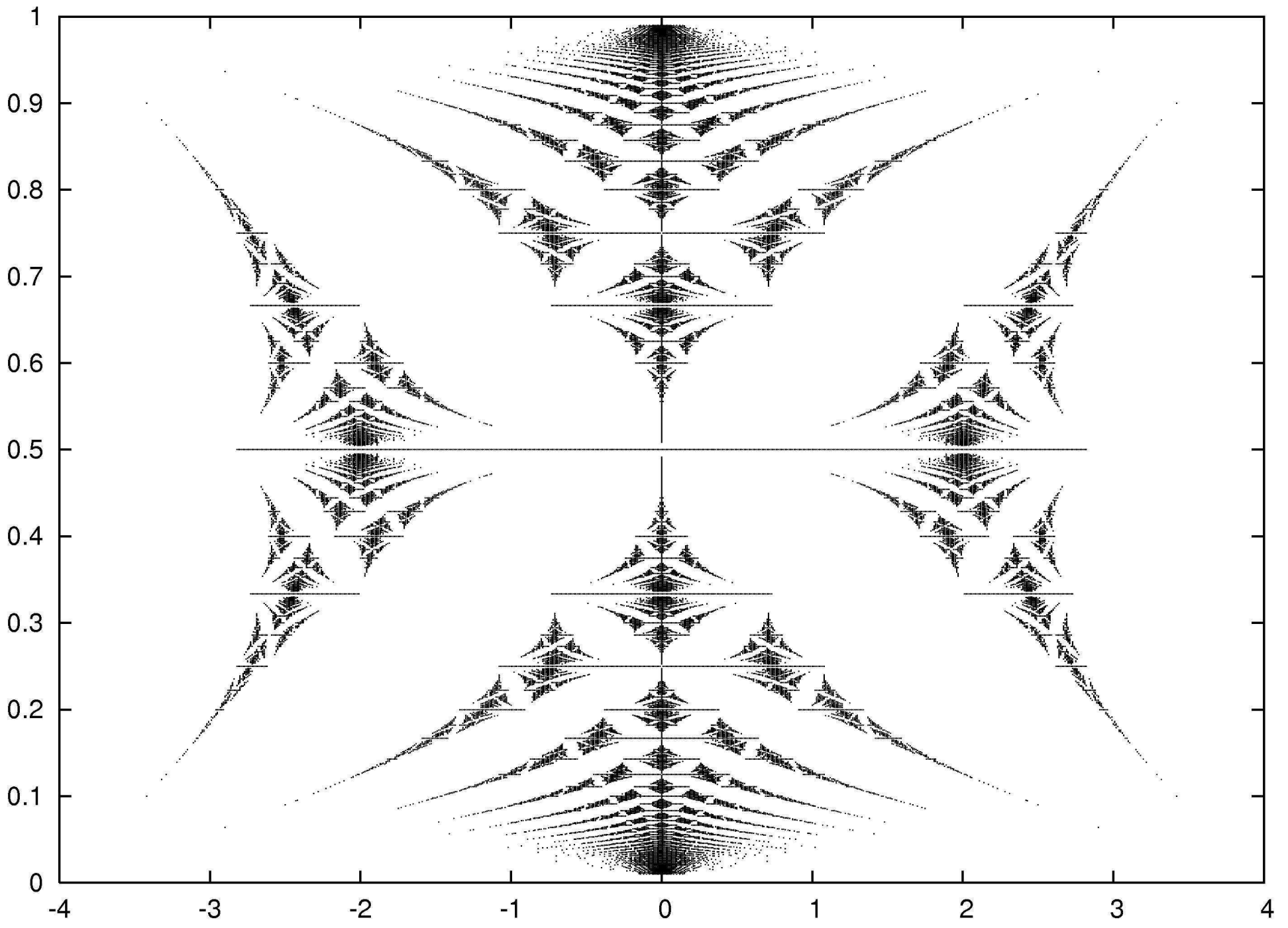

FIGURE 3.1: The Hofstadter butterfly spectrum. Eigenenergies $\varepsilon$ are plotted versus $\alpha = p/q, \in [0,\ 1]$ where $q \leq 150$.

## 3.2 The Generalized Hofstadter Problem

The seminal idea to realize the Hofstadter system in an experiment of cold gases is due to Jaksch and Zoller [210]. They proposed to trap an ultracold atomic gas sample in a three-dimensional optical lattice. Lattice and harmonic confinement in the $z$-direction is assumed to be sufficiently strong to completely suppress tunneling effects in this direction. Thus, effectively, a stack of two-dimensional lattice gases is obtained where only a single copy is considered. The atoms occupy two internal hyperfine states $|g\rangle$, $|e\rangle$, and the optical potential traps them state-dependently in every second column. In discretized lattice coordinates, where $y$ refers to the columns, the state $|g\rangle$ is trapped for $y = \dots, n-1, n+1, \dots$ and $|e\rangle$ for $y = \dots, n, n+2, \dots$, respectively. From now on, both states are no longer distinguished between. Thus, an overall two-dimensional lattice with different spacings $a_x = \lambda/2$ and $a_y = \lambda/4$ is considered. The tunnel rates in the $x$ direction are due to kinetic energy. They are spatially homogeneous and assumed to be equal for both hyperfine states. In the $y$-direction, the lattice is tilted which introduces an energy shift $\Delta$ between neighboring columns. Tilting can be achieved by either an acceleration of the lattice, or by application of a static electric field. For sufficient tilting, standard tunneling rates due to kinetic energy are suppressed. Instead, two pairs of lasers being resonant for Raman transitions between $|g\rangle$ and $|e\rangle$ drive the tunneling in a controlled way between different columns $n \leftrightarrow n \pm 1$. This can be achieved because the offset energy for both transitions is different and equals $\pm\Delta$. The detunings of the lasers have to be adjusted in such a way that the effect of tilting is cancelled in the rotating frame of reference. In this setup, the lasers generate running waves in the $\pm x$-direction. By this, atoms that tunnel between different columns, acquire local phases $\exp(\pm \mathrm{i} q x) = \exp(\pm \mathrm{i} \alpha m)$. These define the magnetic flux parameter $\alpha = p/q$. Due to the local $x$-dependence, the overall phase $\phi_\square$ of an atom which circles around an elementary plaquette of the lattice, does not vanish. The particles "feel" the effect of an artificial magnetic field oriented perpendicular to their $xy$-plane of motion. Its strength is proportional to the magnetic flux associated with the total phase $\phi_\square$. As the latter depends linearly on the row coordinate $x$, the field is homogeneous.

In order to realize artificial non-Abelian fields in a similar scheme, the atoms are assumed to possess degenerate Zeeman sublevels in their hyperfine ground state manifolds labeled by $|g_j\rangle$ and $|e_j\rangle$ with $j = 1, \ldots, n$. These trapped Zeeman states may be thought of as "colors" of the gauge fields. Their degeneracy is experimentally lifted in external magnetic fields. Promising fermionic candidates with the above properties are heavy Alkali atoms, for instance, $^{40}$K atoms in states $F = 9/2, M_F = 9/2, 7/2, \ldots$, and $F = 7/2, M_F = -7/2, -5/2, \ldots$. In particular, they allow for the realization of "spin"-dependent lattice potentials and hopping [215].

Having identified the internal isospin-"color" degrees of freedom, the experimental scheme is slightly modified, as depicted in Fig. 3.2 Laser assisted tunneling rates along the $y$-axis should

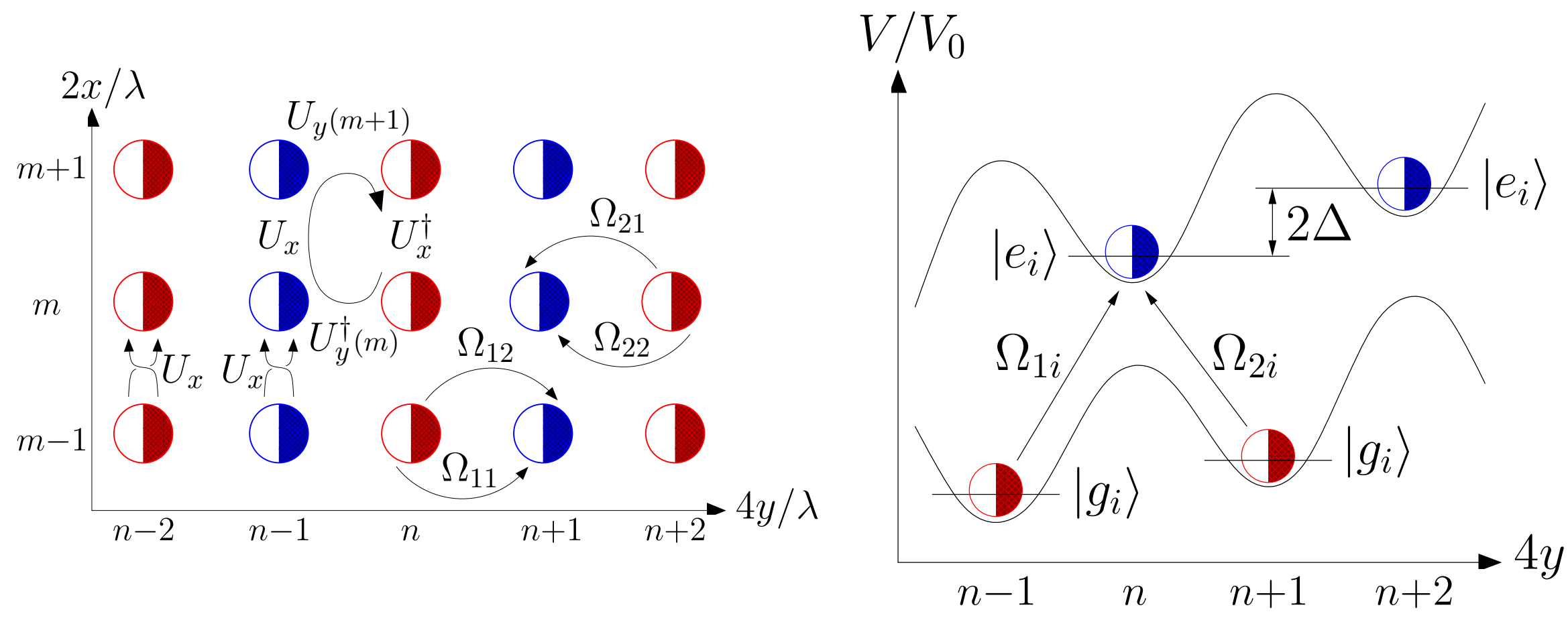

FIGURE 3.2: *Optical lattice setup for U(2) gauge fields:* Red and blue color refer to states $|g_j\rangle$ and $|e_j\rangle$, respectively. Open semi-circles are associated with the first isospin state, i. e., $|g_1\rangle$ and $|e_1\rangle$. Closed semi-circles denote the second one. l. h. s.: Hopping in the $x$-direction is laser assisted and allows for a unitary exchange of colors; it is described by the same unitary hopping matrix $U_x$ for both hyperfine states. Hopping along the $y$-direction is also laser assisted and attains "isospin"-dependent phase factors. r. h. s.: Trapping potential in $y$-direction. Adjacent sites are set off by an energy $\Delta$ due to lattice acceleration or a static inhomogeneous electric field. The lasers $\Omega_{1i}$ are resonant for transitions $|g_{1i}\rangle \leftrightarrow |e_{2i}\rangle$, while $\Omega_{2i}$ are resonant for transitions between $|e_{1i}\rangle \leftrightarrow |g_{2i}\rangle$ due to the offset of the lattice sites. Because of the spatial dependence of $\Omega_{1,2}$ (running waves in $\pm x$ direction) the atoms hopping around the plaquette are unitarily transformed as $\hat{U} = U_y^\dagger(m) U_x U_y(m+1) U_x^\dagger$, where $U_y(m) = \exp(2\pi i m \,\mathrm{diag}[\alpha_1, \alpha_2])$.

depend on the internal state, although not necessarily in the sense of Ref. [215]. For a given link $|g_j\rangle$ to $|e_j\rangle$, tunneling should be realized in such a way that it is represented by a non-trivial unitary matrix $U_y(x)$ which corresponds to the designed "color" group, e. g., $U(n)$ or

$SU(n)$. For unitary groups, the corresponding gauge potential is introduced by the following representation of the tunneling matrix

$$U_y(x) \equiv \exp(\mathrm{i}\tilde{\alpha} A_y(x)) \,. \tag{3.12}$$

Here, $\tilde{\alpha}$ is real, and $A_y(x)$ is a Hermitian matrix from the associated gauge algebra, e.g., $\mathfrak{u}(n)$ or $\mathfrak{su}(n)$. Since transitions from $|g_j\rangle$ to $|e_j\rangle$ correspond to different frequencies for each $j$, they are driven by different running wave lasers, and may attain different phase factors $\exp(\pm \mathrm{i} q_j x) = \exp(\pm \mathrm{i}\alpha_j m)$. Thus, two independent flux parameters are introduced, one for each complex dimension of the group.

In order to create gauge potentials that cannot simply be reduced to two independent Abelian components, tunneling in the $x$-direction has to be also Raman laser assisted and should allow for coherent transfer between internal Zeeman states. For simplicity, it is assumed for now that this is proceeded by the same unitary tunneling matrix $U_x$ for both hyperfine state manifolds, although more general situations are feasible and basically interesting.

A genuine non-Abelian character of the fields is assured by $[U_x, U_y(x)] \neq 0$. It has to be stressed that all elements of the above scheme, as shown in Fig. 3.2, are experimentally accessible. Nevertheless, consistent implementations demand tunneling matrix amplitudes which are controlled with sufficient precision that they strictly belong to the corresponding gauge groups.

The proposed scheme allows for generalizing the Hamiltonian (3.1) to the case of non-Abelian vector potentials. In fact, the components of $\vec{A}$ are simply replaced by the appropriate elements from the group algebra. In particular, the illustrated setup generates "artificial" gauge potentials of the form $\vec{A} = (A_x, A_y(m), 0)$, with

$$\vec{A} = \frac{\hbar c}{\mathrm{e}a} \left( \begin{pmatrix} -\frac{\pi}{2} & \frac{\pi}{2}\mathrm{e}^{\mathrm{i}\phi} \\ \frac{\pi}{2}\mathrm{e}^{-\mathrm{i}\phi} & -\frac{\pi}{2} \end{pmatrix}, \begin{pmatrix} 2\pi m\alpha_1 & 0 \\ 0 & 2\pi m\alpha_2 \end{pmatrix}, 0 \right) . \tag{3.13}$$

More precisely, the scheme associates a unitary tunneling operator with every link, in analogy with standard lattice gauge theory prescriptions [216, 222, 223], as presented in App. A

$$\begin{aligned} U\Big((m-1,\, n) \to (m,\, n)\Big) &\equiv \quad U_x \quad &= \exp\left(+\mathrm{i}eaA_x/c\hbar\right) \\ U\Big((m,\, n) \to (m,\, n+1)\Big) &\equiv \; U_y(m) &= \exp\left(+\mathrm{i}eaA_y(m)/c\hbar\right) . \end{aligned} \tag{3.14}$$

The only difference to be noted is that $\vec{A}$ acquires an overall factor of $\hbar c/ea$. Even though it does not behave well in the continuum limit when $a \to 0$, the "magnetic flux" per plaquette given by $\alpha_{1,2}$ remains finite. Thus, essentially the same limit of ultra-intense fields $B \sim 1/a^2$ is considered as in the original Hofstadter problem of Abelian magnetic fields.

The ansatz $\psi(ma,\, na) = \mathrm{e}^{\mathrm{i}\nu n} g(m)$, leads to a generalized Harper wave equation

$$\begin{pmatrix} g(m+1) \\ g(m) \end{pmatrix} = B(m) \begin{pmatrix} g(m) \\ g(m-1) \end{pmatrix} \tag{3.15}$$

with

$$B(m) = \begin{pmatrix} \begin{pmatrix} 0 & \mathrm{e}^{\mathrm{i}\phi}\varepsilon_{2,\,m}(\nu) \\ \mathrm{e}^{-\mathrm{i}\phi}\varepsilon_{1,\,m}(\nu) & 0 \end{pmatrix} & \begin{pmatrix} -1 & 0 \\ 0 & -1 \end{pmatrix} \\ \begin{pmatrix} 1 & 0 \\ 0 & 1 \end{pmatrix} & \begin{pmatrix} 0 & 0 \\ 0 & 0 \end{pmatrix} \end{pmatrix},$$

where $\varepsilon_{j,\,m}(\nu) = \varepsilon - 2\cos(2\pi m\alpha_j - \nu)$ is the Harper energy term. Both, $\psi(m,\, n)$ and $g(m)$ are now two-component "isospinors". In the particular case of eq. (3.13), when two successive transfer matrices $B(m)$ are multiplied by each other, Eq. (3.15) decomposes into a pair of independent two-dimensional equations

$$B(2k+1)\cdot B(2k) = \left(\begin{array}{c|cc|c} \varepsilon_{1,\,2k}\varepsilon_{2,\,2k+1} - 1 & 0 & 0 & -\mathrm{e}^{\mathrm{i}\phi}\varepsilon_{2,\,2k+1} \\ \hline 0 & \varepsilon_{1,\,2k+1}\varepsilon_{2,\,2k} - 1 & -\mathrm{e}^{-\mathrm{i}\phi}\varepsilon_{1,\,2k+1} & 0 \\ 0 & \mathrm{e}^{\mathrm{i}\phi} f_{2,\,2k} & -1 & 0 \\ \hline \mathrm{e}^{-\mathrm{i}\phi}\varepsilon_{1,\,2k} & 0 & 0 & -1 \end{array}\right).$$

Both of these transfer matrices are closely related with each other. Given one of them, the other is obtained by complex conjugation and additional exchange of $\alpha_1 \leftrightarrow \alpha_2$. The original Hofstadter problem is restored for $\alpha_1 = \alpha_2$ and $\phi = 0$.

Despite their similar structure, both subspaces have to be checked separately to derive the spectrum. The periodicity $\tilde{q}$ of the problem is due to the double-cosine term and equals the least common multiple of the flux denominators $\tilde{q} = \mathrm{lcm}(q_1,\, q_2)$. To make use of the above two-dimensional decomposition, the technical periodicity $q$ has to be even

$$q = \begin{cases} 2\tilde{q} & \text{if } \tilde{q} \text{ odd} \\ \tilde{q} & \text{otherwise} \end{cases}. \tag{3.16}$$

The eigenvalue problem to be solved is again transferred to boundedness of the "isospin"-wave function $\psi(m, n)$. As in the original problem, this is equivalent to the existence of a unitary subspace which boils down to the trace condition (3.8). As a true complication, the reduction of this trace to a one parameter problem, which led to (3.10) for the butterfly, does not work. The basic ingredient of the proof [221], which ensures the simplification, is to find a mean parameter value $\nu_0$ such that an overall oscillating function can be extracted from $\mathrm{tr}\left(Q(\varepsilon, \nu)\right)$. In the original problem with magnetic parameter $\alpha = p/q$, this is possible if and only if there is a natural number $M$ such that

$$pM = 1 \mod q \,. \tag{3.17}$$

Since $p$ and $q$ are mutually coprime, this can always be fulfilled. For the $U(2)$ generalization, there are two magnetic flux parameters $\alpha_j = p_j/q_j$ and their common period equals

$$\tilde{q} = l_1 q_1 = l_2 q_2 \,. \tag{3.18}$$

Then, (3.17) is generalized to a set of two equations

$$\begin{aligned} p_1 M &= l_1 \mod \tilde{q}\,, \\ p_2 M &= l_2 \mod \tilde{q}\,. \end{aligned} \tag{3.19}$$

In general, there is no $M \in \mathbb{N}$ which simultaneously fulfills these conditions. This leads to severe numerical complications as the absolute value of the oscillatory amplitude of $\mathrm{tr}(Q_{1,2})$ with respect to $\nu$ exponentially increases with the period $\tilde{q}$. Here, $Q_j$ is the $j$-th $2\times2$ subblock of the block transfer matrix $Q$, which is derived from the product of $q/2$ successive Harper matrix pairs (3.16). Thus, a dynamic regularization of the trace has to be implemented to calculate the spectrum of $Q_j$ in a properly controlled way.

The full spectrum for $\phi = 0$, in a suitable way depicted in Fig. 3.3 , equals the union of the two sub-spectra and exhibits a very complex formation of holes of finite measure and various sizes, which is named the Hofstadter "moth". This name is not only appropriate from a zoologic point of view with respect to species analogies, but also due to its apparent furryness and the fact that the structure is composed of moth-eaten bands. For fixed $(\alpha_1, \alpha_2)$, it shows a band structure which is intersected by the complex and very fine pattern of the "moth". These bands are bounded by two hyperplanes (depicted in grey in Fig. 3.3). Outside of these hyperplanes, all energies are forbiddden. In fact, it is their structure which is linked with the superfluid-Mott-insulator transition that has been analyzed in [224]. Unfortunately, the analytic form of these hyperplanes is not accessible from the available data. The substantial character of the holes becomes apparent if the spectrum is calculated in microscopic detail by the analysis of very high fractions. Fig. 3.4 displays the isosurfaces which surround the holes in the central region of the "moth". The inner symmetry point $P_0$ is located at $\alpha_1 = \alpha_2 = 1/4$. At this point, all energies in the displayed range belong to the spectrum, as visible in Fig. 3.1. This still holds if only one magnetic flux parameter is locked to $1/4$. Thus, there are two hyperplanes which separate the spectrum in quadrants. Among other very fine structures which are not accessible on this level of resolution, these planes are approached by tubes of spectral holes in this three-dimensional parameter space. As it is clearly visible from inside these tubes, there are no localized eigenenergies, which proves the robustness of the band gap.

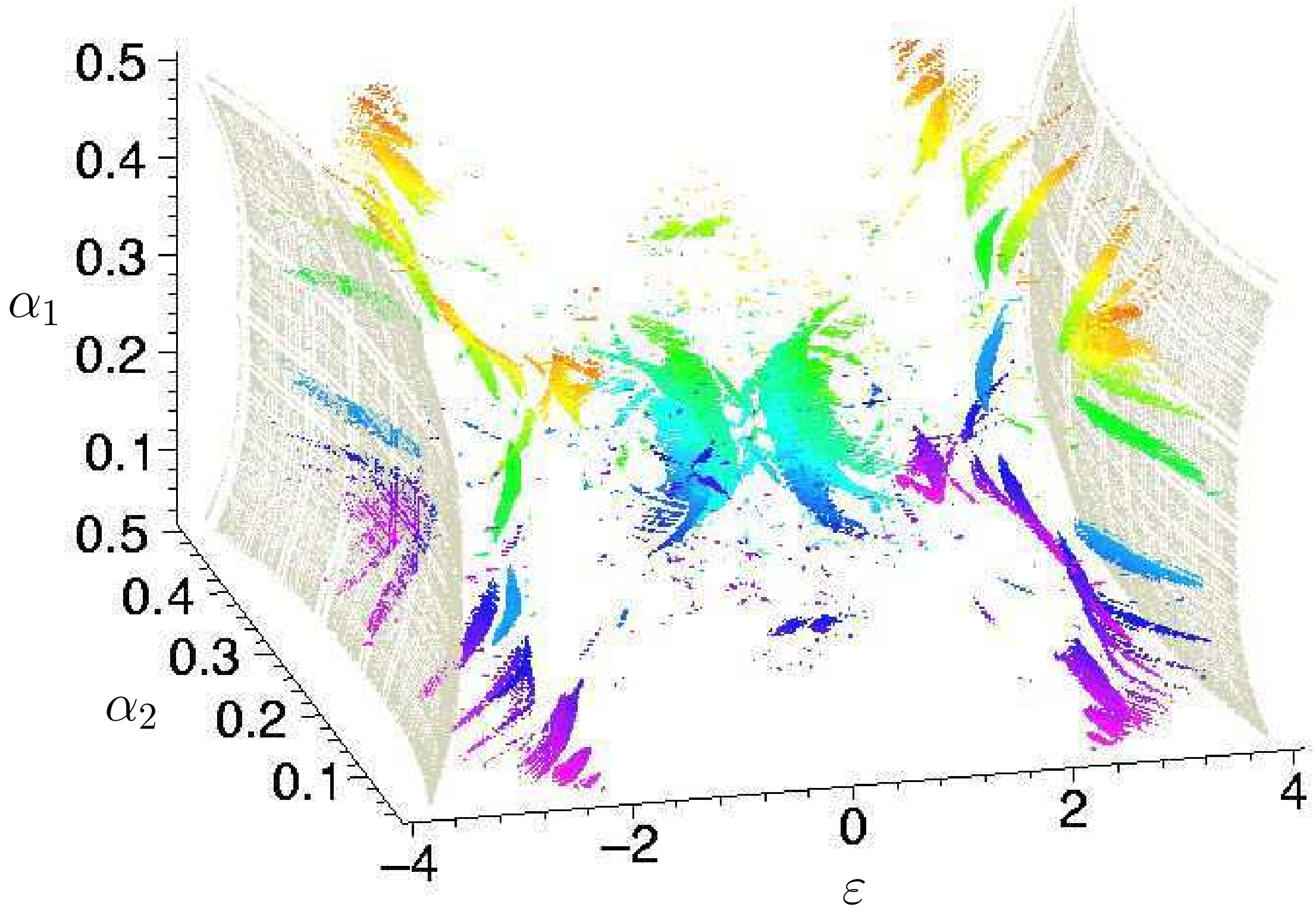

FIGURE 3.3: The Hofstadter "moth" spectrum. Forbidden eigenenergies $\varepsilon$ are plotted versus $\alpha_i = p_i/q_i, \in [0,\, 0.5]$ $(i = 1, 2)$, where $q_i \leq 41$ and $\alpha_1 \neq \alpha_2$. The two hyperplanes depicted in grey form the boundary of the spectrum.

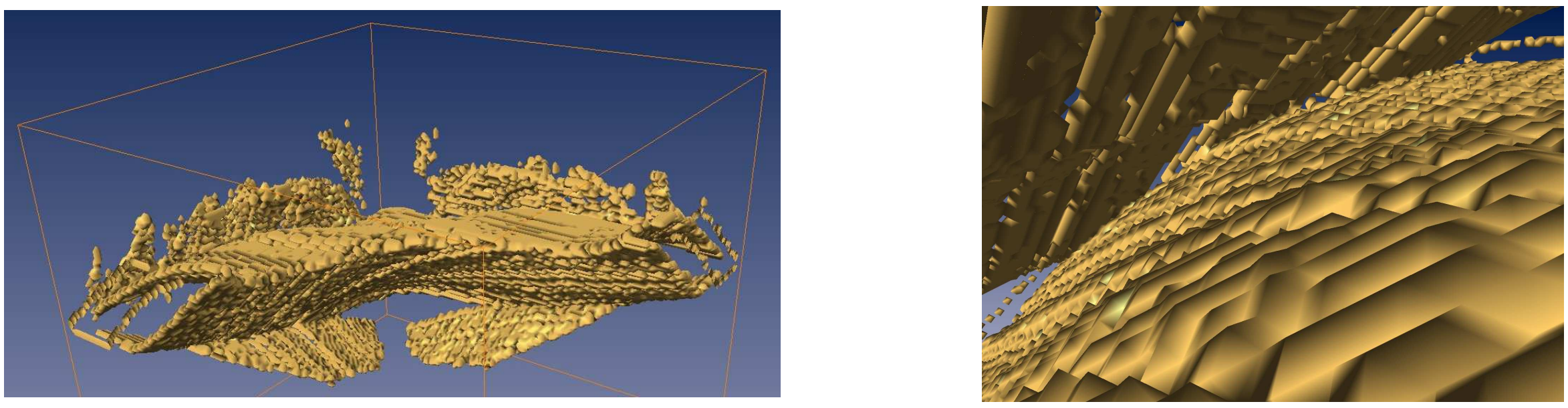

FIGURE 3.4: Isosurface boundaries of the holes for $\alpha_j \in [\frac{1}{4}, \frac{1}{3}]$ and $0 \leq \varepsilon \leq 0.75$. (l. h. s.) view towards the central symmetry point $P_0$ with $\alpha_j = 1/4$; (r. h. s.): view from the vicinity of the symmetry point inside the central hole tube towards the observer of the first figure.

If the "moth" structure is displayed in such detail, very fine and delicate structures appear which may be remnants of the fractal butterfly structure, though a rigorous proof could not be provided so far. Obviously, this fragile structure will be very sensitive to any sort of perturbation on very small scales, for instance due to finite size of the system or external trapping potential. But, since these holes are true three-dimensional objects with finite volume, the spectrum will be more robust on a larger scale to perturbations than in case of the Hofstadter butterfly. This may be revealed by comparing the "moth" with two uncoupled Abelian "butterflies", whose spectrum of energies consists of intersecting lamellas of zero width. Here, the perturbations are required to guarantee allowed regions of finite volume. Deliberately, the spectrum, as it is displayed in Fig. 3.3, does not provide the full information. This has been done for reasons of clarity. First of all, the butterfly is still present on the main diagonal with $\alpha_1 = \alpha_2$. If this gap-dominated spectrum had been displayed, the rest of the structure would have been conceiled by it. But there exist further uncurved hyperplanes which are given by inherent symmetries of the system, e. g., the one displayed in Fig. 3.5 which corresponds to $\alpha_1 = \alpha_2 + 1/4 \mod 1/2$, where *modulo* with respect to $1/2$ is understood as shifted by multiples of $1/2$ unless the result is an element of the semi-open interval $[0, 1/2)$, for example $19/21 \mod 1/2 = 17/42$.

To experimentally detect the "moth"-structures, a dilute Bose condensate could be loaded into the lattice. If the evolution of the particle density is consecutively detected with sufficient precision, the spectral properties could be reconstructed as it it suggested in Ref. [210]. Alternatively, the Fermi energy of an ultra-cold polarized Fermi gas could be measured as a function of the number of particles. To what extent the details of hyperplanes are accessible in a real experiment is hard to guess, as they are objects of zero measure. Thus, magnetic fluctuations will certainly "favor" the surrounding robust moth structure.

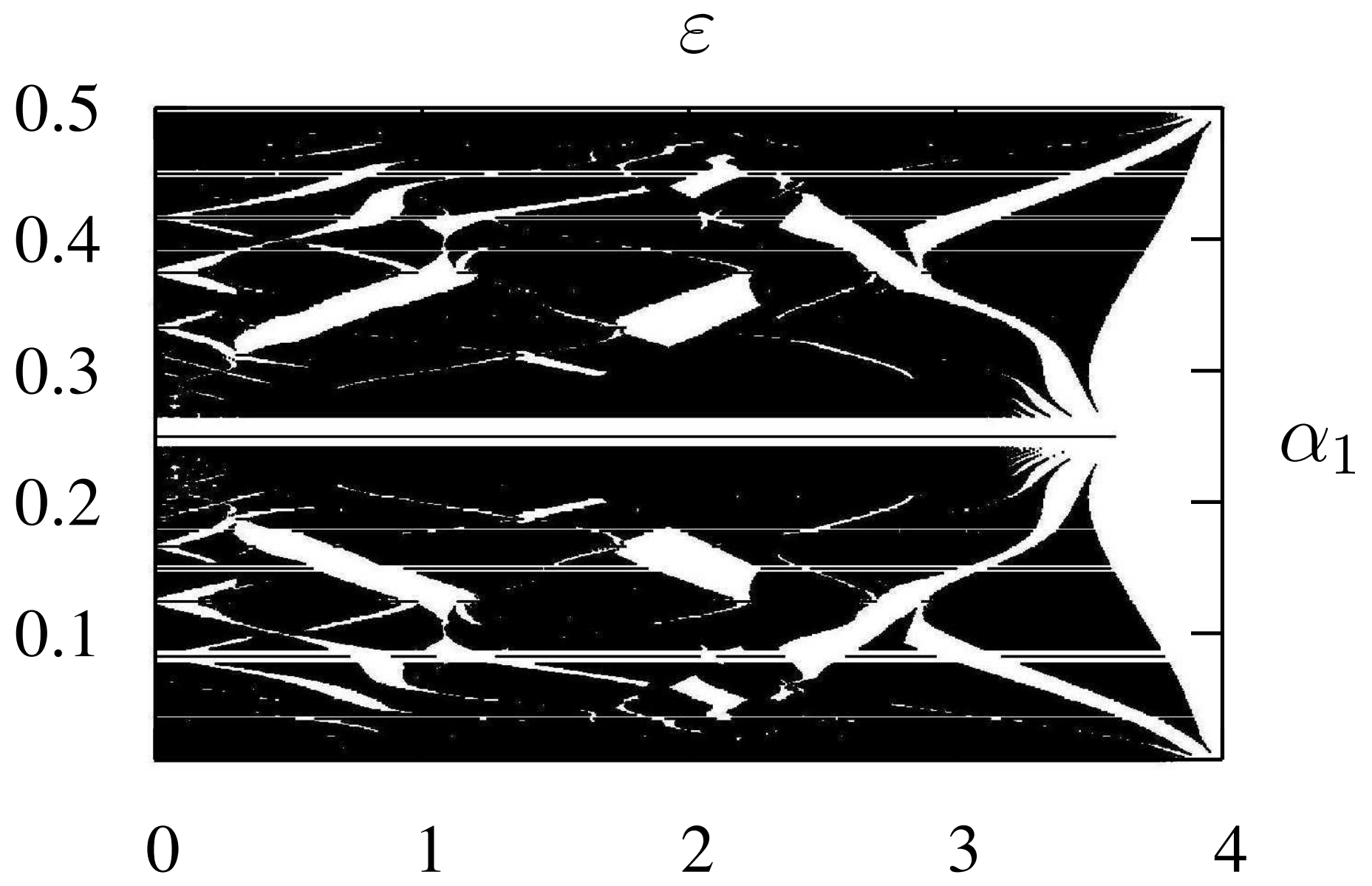

FIGURE 3.5: The Hofstadter "moth" spectrum. Eigenenergies $\varepsilon$ are plotted for $\alpha_1 \in [0,\, 0.5]$, where $\alpha_2 = \alpha_1 + 1/4 \mod 1/2$.

## 3.3 The "Moth" in a Finite System

The above calculations have been performed on an infinite lattice. There, the validity of an eigenenergy of the Hamiltonian (3.1) was constrained by the existence of a unitary subspace of the block transfer matrix $Q$. The necessity of $|\lambda_j| = 1$ for at least one eigenvalue is apparent. If the modulus of an eigenvalue $\lambda_j$ is less than unity, repeated application of $Q$ would identically map the wave function to zero. If on the other hand $|\lambda_j| > 1$, the norm blows up *ad infinitum*. These general conditions drastically change in case of a finite $N_x \times N_y$ optical lattice in a real experiment. There, the original Hamiltonian has to be diagonalized exactly. Still, the plain wave separation ansatz is straightforwardly applicable, which leads to Harper's equation (3.15). Due to the finite extension in the $y$-direction $L_y = N_y a$, the Fourier modes $\exp(\mathrm{i}\nu_k n)$ are now quantized according to the $y$-dimension $N_y$ of the lattice

$$\nu_k = \frac{\pi}{N_y a} k, \qquad k \in \{1, \ldots, N_y\}. \tag{3.20}$$

The remaining part of the "isospin" wave function which corresponds to the $x$-dimension of the lattice becomes a $2N_x$-dimensional vector, and the Hamiltonian $\mathcal{H}^{\text{finite}}_{U(2)}$ is represented as a $2N_x \times 2N_x$ matrix in this basis. It is band diagonal because of the recursive nature of (3.15), e. g., penta-diagonal in the $U(2)$ case, and reads

$$\mathcal{H}^{\text{finite}}_{U(2)} = \begin{pmatrix} C(1) & U_x^\dagger & 0 & \cdots & 0 \\ U_x & C(2) & \ddots & \ddots & \vdots \\ 0 & \ddots & \ddots & \ddots & 0 \\ \vdots & \ddots & \ddots & \ddots & U_x^\dagger \\ 0 & \cdots & 0 & U_x & C(N_x) \end{pmatrix}, \tag{3.21}$$

where

$$C(m) = \begin{pmatrix} 2\cos(2\pi m\alpha_1 - \nu) & 0 \\ 0 & 2\cos(2\pi m\alpha_2 - \nu) \end{pmatrix} \quad \text{and} \quad U_x = \begin{pmatrix} 0 & 1 \\ 1 & 0 \end{pmatrix} \tag{3.22}$$

in accordance with the vector potential (3.13) for $\phi = 0$. This Hamilton operator implicitly depends on the Fourier wave number $\nu_k$. Thus, to derive the total spectrum, (3.21) has to be diagonalized independently for each $\nu_k$. This seeming disadvantage turns out to be beneficial instead. Whatever vector potential has to be analyzed, the numerical difficulty is always the same. The importance of this property is presented from a technical point of view in App. C.

The band structure of the Hamiltonian (3.21) allows for the treatment of really huge systems. More precisely, the calculation can be adapted to any specific experimental system size. With this option, finite size deviations from a pure model are automatically modeled in the appropriate way. Furthermore, a very important quantity is accessible in the finite system approach, i. e., the density of states. It can be quite easily measured in an experiment and allows conclusions about the spectrum if experimental conditions guarantee sufficient precision. This should at least compensate for the fact that a finite size analysis of course lacks the proper insight in some inherent properties, e. g., symmetries revealed in the hyperplanes for fixedly linked magnetic flux parameters.

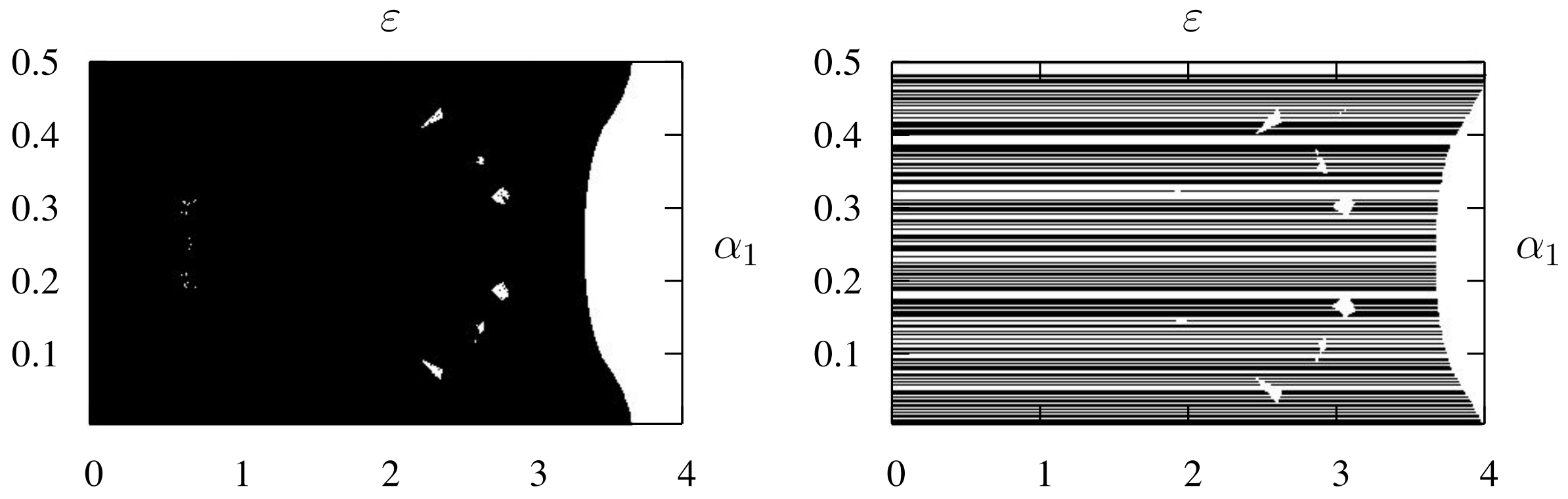

FIGURE 3.6: The finite size Hofstadter “moth” spectrum. l. h. s.: Eigenenergies $\varepsilon$ are plotted for $\alpha_1 \in [0,\ 0.5]$, where $\alpha_2 = 2/5$; r. h. s.: For comparison, the same for an infinite lattice.

The linear density of states $n_1(\varepsilon)$ is obtained from a simple averaging process. In the first step, the full spectral width $\Delta\varepsilon = 4$ is divided into $N$ sufficiently small intervals which may correspond to experimental resolution. Then, (3.21) is diagonalized for all $\nu_k$. The complete set of eigenenergies $\varepsilon_{k,j}$ where $j = 1, \ldots, 2N_x$ is now distributed over these intervals. The result is illustrated in Fig. 3.6 for the particular case of $N_x = 1000$, $N_y = 10$, and $\alpha_2 = 2/5$ with the gauge potentials chosen as before.

The “moth” holes are apparently reproduced with several small artifacts inside. These are merely finite size effects and distributed more or less randomly. Further structures are revealed around the symmetry axis $\alpha_1 = 1/4$, which share similarities with a cut through the profile of the “moth”. They can be interpreted as regions of low density of states, which become forbidden regions if $\alpha_2$ varies. Thus, the density of states provides the necessary ingredients to understand the formation of the “moth”. Holes in the spectrum are linked by a web of fine-structured bands. Bearing this in mind, a rigorous constraint to detect the underlying structure is to impose a condition of a minimal $n_1^{\min}(\varepsilon)$. A good choice is to demand

$$n_1(\varepsilon) > n_0 = N_x N_y / N\,, \tag{3.23}$$

where $n_0$ is half the homogeneous density This removes artifacts caused by finite size in a controlled way, as random homogeneous pair state distributions are not considered. The result is an impressively detailed spectrum which exhibits recursive, seemingly fractal structures. These are separated by a complex net of regions with high densities of states. The whole spectrum is so rich in structure that it stimulates associations of various kind. The arguably most beautiful one is the impression of six-fingered "hands" reaching out for an Aztech-like "man" wearing a helmet, whose mouth, nose, and shadowed eyes are clearly visible.

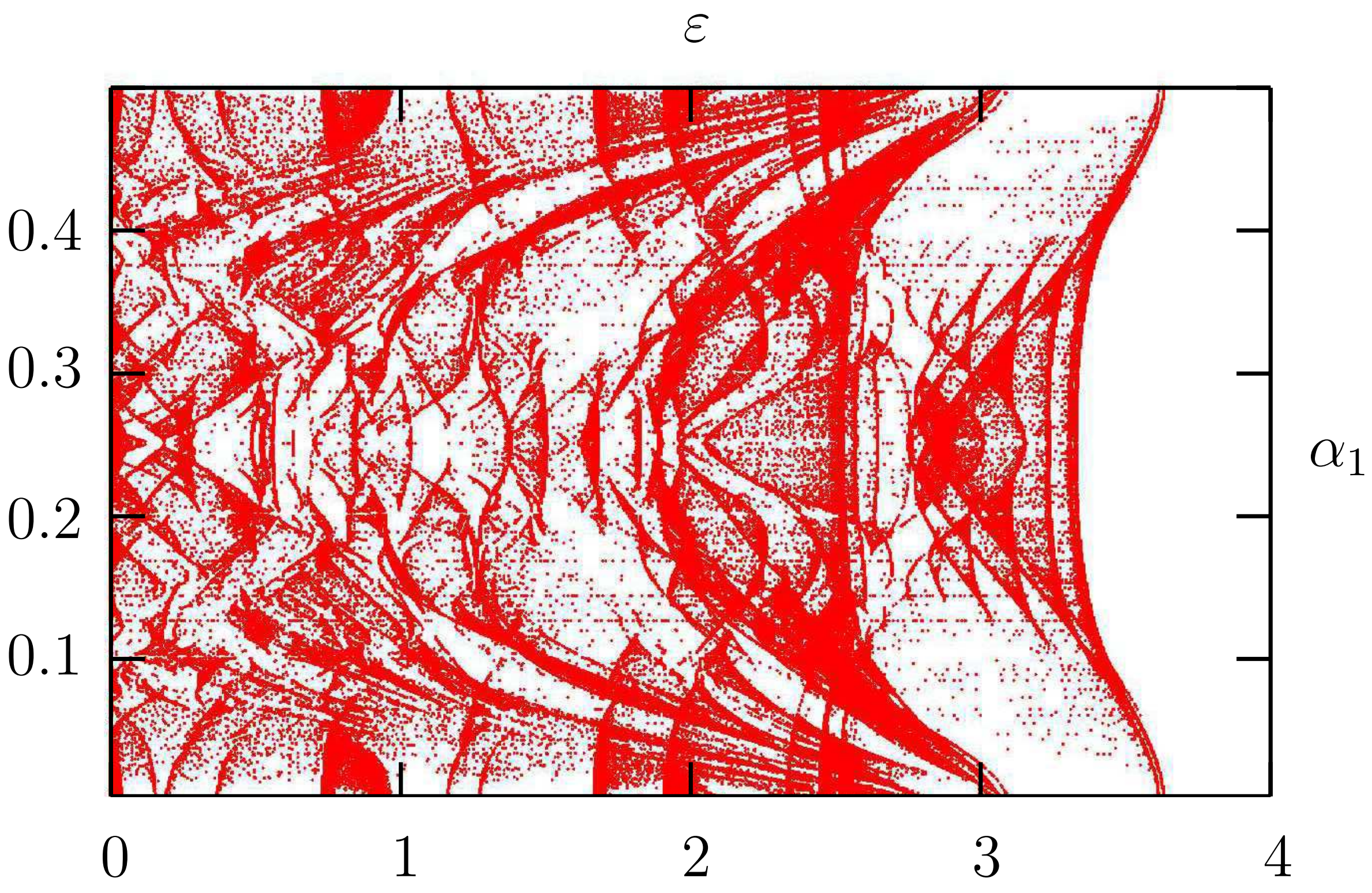

FIGURE 3.7: The finite size Hofstadter "moth" spectrum. Eigenenergies $\varepsilon$ are plotted for $\alpha_1 \in [0,\, 0.5]$, where $\alpha_2 = 2/5$.

Already on the level of this single plane of the full spectrum, the apparent complexity gives rise to further detailed analyses of three-dimensional spatial dynamics of these structures. Certainly, the "moth" is only the tip of the iceberg. One of the open and maybe even more exciting questions is how the spectrum changes due to continuous variations of the gauge potentials, especially those who may be associated with ordered phases of the similarly structured gauge theories.

## 3.4 Non-Abelian Interferometry

It is interesting to consider yet another effect that becomes particularly spectacular in the limit of ultra-intense gauge fields, and that can be measured in the proposed system: a non-Abelian

Aharonov-Bohm effect which is considered as an example of non-Abelian interference. In order to realize it, a weakly interacting Bose condensate should be prepared in a definite internal state $|\psi_0\rangle$ around a location $P_1$. Then, the condensate or parts of it should be split by Raman scattering, and consecutively be dragged piecewise to a meeting point $P_2$ on two distinct paths $\mathcal{P}_{1,2}$. This can be performed by appropriate laser tweezers for instance. The paths $\mathcal{P}_j$ correspond to some unitary transporters or Wilson lines $U_1$ and $U_2^\dagger$, respectively. A measurement of the density of atoms at $P_2$ will reveal a non-Abelian interferometric signal, i. e., it will detect the interference term

$$n_{int} \propto \langle \Psi_0 | U_2 U_1 | \psi_0 \rangle \,. \tag{3.24}$$

If, for simplicity, a rectangular loop is chosen which consists of $L_{x_1}$ steps in the $-x$-direction, $L$ steps in the direction of $y$, $L_{x_1} + L_{x_2}$ steps in the $x$-direction, $L$ steps in $-y$ directions, and finally $L_{x_2}$ steps in $-x$-direction, the interference signal in case of the gauge potentials (3.13) reads

$$n_{int} \propto \begin{cases} 1 & \text{if } L_{x_1} + L_{x_2} \text{ even} \\ \langle \Psi_0 | \exp\left[\pm 2\pi i \hat{\alpha} L (L_{x_1} + L_{x_2})\right] | \psi_0 \rangle & \text{if } L_{x_1} + L_{x_2} \text{ odd} \end{cases} \tag{3.25}$$

where $\hat{\alpha} = \mathrm{diag}(\alpha_1, \alpha_2)$. The signal is thus extremely sensitive to $L_{x_1}$ and $L_{x_2}$. If in another attempt, phase shifts were caused by obstacles placed on the $y$-arms of this interferometer, the measurement would strongly depend on the $x$-coordinate of the obstacles, and the location of $P_1$, $P_2$. This would be a direct impact of the non-Abelian external gauge potential.

Obviously, the properties of the considered system in the limit of ultra-intense fields are quite complex. Though, to get a better intuition within this scheme, it is useful to consider the "continuum" limit $a \to 0$ with $V_0 \to a^2/m$. Then, the Hamiltonian reduces to Landau form $\mathcal{H} = (\vec{p} - e/c\vec{A})^2/2m$, though with non-Abelian $\vec{A}$. A natural question to ask is what kinds of gauge fields with "normal", i.e., $a$-independent strength, can be realized. In other words, which set of artificial gauge potentials is experimentally feasible? In the most general scheme, phase factors caused by running wave vectors can be introduced for all tunneling matrices:

$$\vec{A}(\vec{r}) = \frac{c}{el}(M_1 + [M_2(\frac{x}{l}) + M_3(\frac{y}{l})], N_1 + [N_2(\frac{x}{l}) + N_3(\frac{y}{l})], 0) \,. \tag{3.26}$$

Here, $l$ is the characteristic length on which $\vec{A}$ is assumed to vary. $M_i, N_i$ are arbitrary and in general non-commuting, dimensionless matrices, which are independent from $a$, and belong to the gauge algebra $\mathfrak{u}(2)$. On top, local disorder may be introduced in a controlled way that allows for small fluctuations of the matrices $M_i$. In particular, disorder can be made annealed. Thus, it changes on a time scale comparable with relevant time scales of the system and mimics thermal fluctuations in a very nice fashion. It can be of significant amplitude, provided that it does not drive the assisting lasers system out of resonance. For instance, wave vectors of running waves can strongly fluctuate and force the magnetic flux to fluctuate correspondingly. Even more complicated spatial dependencies, e.g., piecewise linear components of $\vec{A}$ are feasible: These potential configurations may be achieved by an additional static electric fields or the implementation of laser-induced potentials. If there is still room on the optical table, further lasers may introduce local and in general time-dependent unitary operators. Such transformations would generate arbitrary local temporal components of the gauge potential $A_0(x, y)$. Although this component may be gauged out for 2+1 dimensional Yang Mills fields, if it is fixed to Weyl or strict temporal gauge [225], the corresponding gauge transformations may introduce more complex spatial and temporal forms of the remaining two components of $\vec{A}$.

## 3.5 Ultracold Gases and Lattice Gauge Theories

Another fascinating question concerns the possibility to what extent the proposed scheme is capable to study lattice gauge theories in 2+1 dimensions motivated by apparent similar characteristics, which have been addressed in the previous sections. By contrast, this part of the chapter investigates important conceptual and physical differences between both approaches. Possible resolutions to these obstacles are proposed and concrete realizations of specific field theories are discussed in this context.

As presented in App. A, a lattice gauge theory is obtained by projection of a continuum gauge theory on a Euclidian space-time lattice. Diagrammatic amplitudes and order parameters are hereby calculated in analogy to methods of statistical physics, such as Monte Carlo sampling techniques, which sum over all configurations of gauge and matter fields. Apparently, gauge fields are essential and inherently dynamical variables, whereas they are obviously not in the proposed scheme. Moreover, the latter is naturally realized in real rather than in imaginary time. On the other hand, this deficiency is softened by the advantage that, given a concrete gauge potential configuration in an optical lattice, the dynamics of matter fields in real time are directly accessible without any sampling process. Nonetheless, is has to be stressed that any simulation is restricted to specific limits of the associated gauge theory as some effects are definitely out of the scope of any atom experiment, for instance pair-creation processes.

But, if various configurations of gauge fields are created by repeated experiments or dynamical variation of the phase-driving lasers, a Monte Carlo sampling could be "mimicked"to a certain extent. Averaging over both, annealed disorder and quantum fluctuations should at least somehow approximate the statistical average in lattice gauge theories. However, this inevitably requires that generated configurations represent characteristic or statistically relevant phases.

Although the gauge fields accessible in the proposed scheme are limited, at least some of them share characteristics of the above phases, e. g., an area law fulfilled by Wilson loops in the confinement sector. Indeed, the $U(2)$ gauge potential of (3.13) with fluctuating, but anticorrelated fluxes $\alpha_1 = -\alpha_2$ yields an area law for loops in the $xy$-plane, provided the probability distribution of the fluxes is Lorentzian. However, it would be desirable to create configurations that exhibit other characteristics of the confinement phase such as appropriate distributions of center vortices, Abelian magnetic monopoles, instantons, merons, calorons etc. [226, 227, 228].

The situation is much more promising when Abelian gauge theories are considered. These, at least in some concrete and interesting limits, may be reduced to models with ring exchange interactions which involve products of operators over an elementary plaquette of the underlying lattice. The first proposal for such interactions was formulated in [229]. There, a spin model for a multi-component Bose or Fermi gas is derived in the Mott insulator limit with one particle per site. In this scenario, $3$-spin interactions are obtained in the third order of $t/U$ expansion, because of the two different tunneling paths from one site to another are possible in a triangular lattice, either directly or via the third site. The crucial disadvantage of this proposal is that it involves very small energy scales and long time scales, since in the Mott limit $(t/U)^3 \ll 1$.

In another approach, Büchler et al. [230] consider a standard single component Bose-Hubbard model in a square (cubic) lattice coupled to diatomic "molecules"which are trapped in the centre of the lattice plaquette. In that way, the latter form a lattice of their own. The coupling to the molecular state, which is assumed to be $d$-symmetric, takes the form

$$H_m = \nu \sum_{\square} m^\dagger_\square m_\square + g \sum_{\square} \left[ m^\dagger_\square (b_1 b_3 - b_2 b_4) + h.\ c. \right] , \tag{3.27}$$

where $\sum_\square$ implies the sum over all plaquettes. Two atoms may perform a Raman transition to a molecular state of coupling $g$, and $\nu$ denotes the detuning of this transition. Perturbative elimination of the molecules leads to the effective ring exchange Hamiltonian for bosons:

$$H_{RE} = K \sum_{\square} \left( b^\dagger_1 b_2 b^\dagger_3 b_4 + b^\dagger_4 b_3 b^\dagger_2 b_1 - n_1 n_2 - n_3 n_4 \right) . \tag{3.28}$$

More precisely speaking, the above system can be realized when atoms with two internal states are used. One of them is trapped in the square (cubic) lattice and described by the standard Bose-Hubbard model, and the other is confined in the centers of plaquettes in a site potential that is not symmetric, but rather shares a point symmetry of the lattice. “Molecules”are formed by two atoms in this second internal state, and have the necessary $d$–symmetry. This model of a Bose-Hubbard Hamiltonian extended by $H_m$ is a promising candidate for a deconfined quantum critical point in two dimensions [231]. This conjecture is supported by the fact that the latter theory is supposed to undergo a quantum phase transition from the bosonic superfluid phase to a molecular “density wave”, a so-called stripe phase. Superfluidity occurs for $\nu \gg g$ and $J \gg g^2/\nu$ and breaks the $U(1)$-symmetry. By contrast, the stripe phase breaks translational symmetry and is observed in the opposite limit $\nu < 0$ and $|\nu| \gg J, g$.

Both, in two and three dimension, the above model describes a $U(1)$ gauge theory, and is likely to exhibit a variety of quantum phases, for instance a deconfined insulator [232, 233].

Very recently, a dipolar Bose gas was proposed to simulate a compact $U(1)$ lattice gauge theory [234]. The underlying model is properly described by an extended Bose-Hubbard Hamiltonian, as discussed in a preceding paper [235]. To assure isotropy of dipolar interactions between nearest neighbors, the lattice geometry should be either a two-dimensional kagomé or a three-dimensional pyrochlore lattice. Several possibilities to create a kagomé-shaped optical potential are discussed in [80]. In the limit where the strength of on-site interactions $U$ is comparable to the nearest-neighbor interactions $V$, and both are sufficiently strong, the physics is dominated by configurations, in which fluctuations of the number of atoms are zero in all elementary plaquettes. To the lowest relevant order of perturbation theory ($(t/U)^3$), a Hamiltonian with “ring exchange”terms is obtained on a dual hexagonal lattice. This model directly reduces to a $U(1)$ lattice gauge theory, and allows for the realization of various fractionalized topological phases. Moreover, concrete methods to detect signatures of the emergent $U(1)$ Coulomb phase are proposed in the context of [234], which is called the “emergence of artificial light in an optical lattice”. As most of these approaches concern lower space-time dimensions, it should be stressed that Yang-Mills theories in 2+1 dimensions are in the center of interest in high energy physics, as they describe the high temperature behavior of four-dimensional models [236]. Recently, progress has been made in the understanding of these theories [225, 237].

# CHAPTER 4

# Conclusion

This work has been motivated by the impressive experimental progress in quantum optics and atomic and molecular physics over the recent years. The joint efforts of these physical disciplines nowadays allow for quantum engineering on an unprecedented level. In this context, the results of this thesis are meant to be understood as proposals for future experiments.

Experiments on rapidly rotating short-range interacting Bose gases will most probably conquer the lowest Landau level regime in the next few years. From the present point of view, the most promising approach employs multiple copies of spinning microsystems. It has been demonstrated that in these small samples, ordered structures manifest themselves in a modified way. Finite size effects significantly restrict the parameter space, in which structures of broken symmetry are observable. Incipient vortex patterns are only nucleated in narrow frequency windows. A deeper successive analysis of these issues is provided in [256]. If the spinning frequency is increased, the bosonic Laughlin wave function at half-filling becomes the exact ground state. For very small system sizes, its structure is dominated by an ordered ring pattern of atoms “localized” at the edge of the system. This ”crystallization” of atoms, which makes an experimental distinction between the Laughlin state and a finite-sized Wigner crystal inherently difficult, disappears, when the bulk structure becomes sufficiently important. This happens for larger particle numbers, where the degree of correlation is more pronounced, and the ground state is expected to behave as a true quantum liquid. A major definite obstacle to the observation and identification of fractional quantum Hall features in case of contact interactions is the very small quasi-particle excitation gap. Though its magnitude may be enhanced by Feshbach resonances, this gain is at the expense of particle losses which have to be suitably controlled.

Dipolar interacting Fermi gases have been proposed as natural candidates to circumvent these difficulties. For these systems, the excitation gap is proven to be of substantial size provided the dipole moment of the particles is strong enough. However, the chances that fractionally charged quasi-particles might be detected in dipolar chromium are rather limited. This does not mean that the incompressibility gap is not observable at all, but very low temperatures of the order of a few nK are required. Yet, exciting perspectives are opened by strong efforts towards stable trapping of polar molecules in their vibronic ground state. There, dipole moments of several Debye shift the gap temperature to the $\mu$K scale.

Investigations comparing between dipolar Laughlin states and their famous electronic counterparts reveal small oscillatory density variations which survive on the level of second order correlation functions. Though of small amplitude, these are signatures of the long-ranged character of interaction which mediates correlations in the homogeneous bulk liquid. Indications of elementary quasi-hole excitations prove to be difficult to observe. A detailed investigation of density defects in low-lying states, at appropriately chosen angular momenta for a quasi-hole at the origin, shows that these defects decay as $1/N$. Thus, the corresponding states have to be associated with edge excitations of the system. Apart from dipolar Laughlin states, the evolution of ground state structures towards the strongly correlated regime was investigated. In particular in the crossover regime, fascinating structures are observed. For sufficiently large systems, frustration of competing series of states with distinct bulk order leads to stable configurations with a quasi-hole in the center. These may be linked with proposals of an effective theory of composite fermions which should be addressed in future research. To prove that the above phenomena are not due to finite size effects, diagonalizations should be repeated with periodic boundary conditions. Still, efforts to discover quasi-particles in the disk geometry are certainly worthwhile. Preliminary results in this direction provide clues of their existence.

Another fascinating problem is the competition between Laughlin states and the Wigner crystal. Though it is rather daring to speak of a crystal in the case of a dozen particles, the study of dipolar ground states at $\nu = 1/5,\ 1/7,\ 1/9,\ \ldots$ reveals significant bulk deviations from the Laughlin state. At lower fillings, the Laughlin wave function is obtained by exact diagonalization methods with properly defined Haldane pseudopotentials. An analytic decomposition of the Jastrow factors into Fock-Darwin states already fails for moderate particle numbers $N > 7$ due to the exponential growth of partitions in terms of monomials. Numerous other directions of research are conceivable. Recently, there is a growing interest in the mysterious states at even denomina-tor fillings. The most prominent state at $\nu = 5/2$ is assumed to exhibit non-Abelian excitations and is apparently well-described by rational conformal field theories [257, 258]. Because of this feature, these states have been subsequently discussed as suitable candidates for quantum computational methods [259]. However, not long ago, a controversial proposal was made [260], and exact numerical methods arguably prove that excitations are well-behaved and of Abelian nature [261]. For bosons, a promising non-Abelian candidate is the $\nu = 3/2$ state which belongs to the Read-Rezayi sequence of incompressible correlated liquids. There are promising indications that this state becomes stable in rapidly rotating Bose gases interacting via contact forces with a moderate amount of dipolar interactions [262]. There is hope that these fractional quantum Hall states will provide experimentally feasible examples of non-Abelian anyons.

A further unresolved puzzle concerns the "strange" fillings, for instance $\nu = 4/11$ [89]. These states exceed the scope of the widely accepted and prominent model of non-interacting composite fermions. Soon after the first indications of their existence had been observed, their strangeness was supported by a novel hierarchical conformal field theory approach [263]. A little later, final experimental proof for an incompressible state at $4/11$ was obtained [264]. It is proposed that in these states, interacting composite fermions form correlated pairs [265].

Exciting research concerning atomic "isospin" dynamics in non-Abelian vector potentials has just started. The scheme to implement the generalized Hofstadter problem as it was presented in Chapter 3 was the first proposal in this direction. It was soon accompanied by a complementary ansatz which employs electromagnetically induced tranparency [217]. The single particle spectrum of the generalized Hofstadter problem has been investigated for the case of $U(2)$ vector potentials in the present work. It shows a very complex formation of holes of finite measure and various sizes. The term "moth" has been chosen for this spectrum to emphasize the analogy to the Hofstadter "butterfly". The structural depth and complexity of the "moth" holes is astonishing. Calculations on finite lattice systems are shown to be efficient tools to analyze the "dynamics" of these spectral patterns with respect to variations in the gauge potentials. Furthermore, the density of states is accessible in this approach. It gives insight into impressively detailed spectral substructures which exhibit recursive, maybe fractal properties. By contrast, infinite lattice calculations provide important information on inherent symmetries. Unfortunately, they are solely tractable for very specific potential configurations. The classification of the set of all potentials is a challenging task for future research. It may shed light on the puzzling question to what extent this novel class of atomic systems may be suitable to study lattice gauge theories in 2+1 dimensions.

Apart from this fundamental scope of investigations, atoms subject to non-Abelian vector potentials offer numerous other possibilities. As discussed in Chapter 3, generalized Aharonov-Bohm interferometry suggests promising applications. Further research in this direction in a simplified experimental setup has been proposed [266]. Additionally, generalized integer quantized Hall states are expected to appear [267]. The development of schemes to detect non-Abelian variants of fractional quantum Hall states is in progress [268]. This direction of research is particularly interesting with respect to fractionally "charged" non-Abelian excitations. At present, their existence and properties are intensively discussed. If a quasi-particle of this kind is moved around the origin, it induces a non-Abelian Berry "phase" described by non-trivial matrix operators [269]. Particularly interesting in this context are generalizations of methods used by [270] in the classical Abelian scenario. Studies of such systems, which are related to bilayer Laughlin quantum liquids, have recently been initiated [271, 272].

The full experimental setup, required to simulate the proposed $U(2)$ lattice scheme, is rather intricate. However, a simpler method to simulate the particle dynamics of the generalized Hofstadter problem can be employed. It consists of a one-dimensional optical lattice with standard kinetic tunneling and a superimposed spin-dependent super-lattice of a spatial period which equals the inverse of the magnetic flux parameters. As a a step in this direction, a scheme towards the realization of a non-Abelian Pfaffian quantum Hall state has been recently proposed for a one-dimensional Bose gas in the Tonks-Girardeau regime [273].

It is apparent that the "moth" has inspired investigations in various directions. Besides the examples discussed above, the scope even includes exotic proposals towards a possible connection between ultracold atoms and string theory [274]. No matter, how miraculous and ambitious these directions of research may seem, they clearly indicate that ultracold atomic gases subject to artificial gauge potentials in optical lattices are a promising field of modern quantum optics. Despite the fact that certainly not all expectations will be fulfilled, it is definitely true that forthcoming studies will lead to many fascinating results.

# APPENDIX A

# Gauge Theories and Fields

In the approach of Chapter 3, a novel species of atomic systems subject to non-Abelian vector potentials was introduced. The question has been addressed how such a system can be seized as a quantum simulator for fundamental gauge theories. To put this discussion in a proper context, the basic concepts of gauge fields and theories are introduced. The analysis starts from classical electrodynamics, before the fundamental idea of local symmetry is introduced. Finally, the mechanism to properly represent gauge invariant theories on a lattice is motivated. For further details, the reader is referred to the following excellent books [222, 238, 239, 240, 241]. A very nice pedagogical review for the explicit (2+1)D Yang-Mills case is given by [225].

## A.1 Classical Electrodynamics

The fundamental equations for the dynamics of the electric and magnetic field and their interaction with charged particles were proposed by James Clerk Maxwell in 1865

$$\begin{aligned} \nabla \cdot \vec{E} &= 4\pi\rho & \nabla \times \vec{E} &= -\frac{1}{\mathrm{c}}\partial_t \vec{B} \\ \nabla \cdot \vec{B} &= 0 & \nabla \times \vec{B} &= \frac{1}{\mathrm{c}}\partial_t \vec{E} + \frac{4\pi}{\mathrm{c}}\vec{j}\,, \end{aligned} \tag{A.1}$$

where $\vec{E}$ and $\vec{B}$ denote the electric and magnetic field respectively, $\rho$ is the charge density and $\vec{j}$ is the current density [11]. Maxwell himself conceived the origin of the velocity constant $c$ which appears in his equations:

> *"This velocity is so nearly that of light, that it seems we have strong reason to conclude that light itself (including radiant heat, and other radiations if any) is an electromagnetic disturbance in the form of waves propagated through the electromagnetic field according to electromagnetic laws."*

This understanding successively stimulated research to perceive the deep link between fields and their underlying symmetry, which finally led to the discovery of relativistic invariance [12].

The apparent symmetric nature of (A.1) is understood in a more modern approach based on the principle of least action. On this level, the classical fields originate from vector potentials which constitute the relativistic Lagrangian function. Integrated over time, the latter yields the action functional

$$S[\vec{v},\, \vec{A},\, \phi] = \int_{t_1}^{t_2} \left( -mc^2\sqrt{1-\vec{v}^2/\mathrm{c}^2} + \frac{\mathrm{e}}{\mathrm{c}}\vec{A}\cdot\vec{v} - \mathrm{e}\phi \right) \mathrm{d}t\,, \tag{A.2}$$

where $\phi$ and $\vec{A}$ are the scalar potential and vector potential, respectively. They are related to the observable fields via

$$\vec{B} \,=\, \nabla\times\vec{A} \qquad \vec{E} \,=\, -\frac{1}{c}\partial_t\vec{A} - \nabla\phi\,. \tag{A.3}$$

Obviously, both potentials do not appear in (A.1), neither in any classical observable of the system. As the fields are fixed by (A.3), there is an equivalence class of potentials associated with the same observables. Under the simultaneous transformation

$$\vec{A} \,\rightarrow\, \vec{A} + \nabla\lambda \qquad \phi \,\rightarrow\, \phi - \frac{1}{\mathrm{c}}\partial_t\lambda\,, \tag{A.4}$$

where $\lambda(\vec{r},\, t)$ is an arbitrary scalar function, the physical properties of a given system remain unaffected. The specific choice of $\lambda$ fixes the system to the corresponding gauge class.

In terms of the potentials $\vec{A}$ and $\phi$, the Maxwell equations can be written in Lorentz symmetric four-vector formalism, which is needed in the following sections. For this reason, the field tensor is defined as

$$F^{\mu\nu} = \partial^\mu A^\nu - \partial^\nu A^\mu\,, \tag{A.5}$$

where $A^\mu \equiv (\phi,\, \vec{A})$ and $\partial^\mu = g^{\mu\nu}\partial_\nu = (\partial_{\mathrm{ct}}, -\nabla)$. The metric $g^{\mu\nu}$ is defined as $(1,\, -1,\, -1,\, -1)$ on the diagonal and $0$ otherwise. In this form, (A.1) reads

$$\partial_\mu F_{\nu\rho} + \partial_\nu F_{\rho\mu} + \partial_\rho F_{\mu\nu} \,=\, 0 \qquad \partial_\mu F^{\mu\nu} \,=\, \frac{4\pi}{\mathrm{c}} j^\nu\,. \tag{A.6}$$

## A.2 Non-Abelian Gauge Theories

Historically, the relativistic quantum treatment of single particle wave equations of Klein-Gordon or Dirac type led to unphysical inconsistencies as negative energy states. To overcome these difficulties, the field theoretical viewpoint is appealing. Under consideration of Einstein's equivalence principle of mass and energy, any process taking place at sufficiently high energies allows for the creation of particle-antiparticle pairs. Thus, a multi-particle treatment is obligatory. It took the conceptual ingenuity of Feynman who grasped this idea in his diagrammatic representation of perturbative field theory. To ensure a proper handling and control of the inherent divergencies, a detailed understanding of renormalization methods was necessary. Among others, Schwinger, Tomonaga, and Feynman put the early ideas in a strict conceptual context [241]. Two decades later, the underlying fundamental methods were formulated by Wilson in the much broader context of renormalization group [26].

Pushed forward by these pioneering works, a vast progress in the understanding of fundamental interactions between elementary particles was made. Yet, it was restricted to scalar field theories as Yukawa theory and Quantum Electrodynamics. Nonetheless, it turned out that a huge set of processes detected in particle accelerator experiments could not be explained by imaginable theories of this kind. The breakthrough occured in 1954, when Yang and Mills proposed a seminal theory based on the very principle of local gauge equivalence [242]. From today's point of view, the latter certainly is one of the most deeply rooted axioms in the modern physical understanding of nature.

The motivation to search for reasonable extensions of known scalar theories by use of gauge equivalence arose from the study of Quantum Electrodynamics. There, the problem of negative-norm states was resolved by the Ward identity [28], which indeed follows from the local invariance of the Lagrangian function. Bearing this concept in mind, the derivation starts from the postulate that the desired theory should be invariant under local U(1) transformation. For the complex valued fermionic Dirac fields, this reads

$$\psi(x) \to \mathrm{e}^{\mathrm{i}\alpha(x)}\psi(x)\,. \tag{A.7}$$

Obviously, the mass term $m\bar{\psi}\psi(x)$ is invariant under this transformation. Here, $\bar{\psi} \equiv \psi^\dagger\gamma^0$, where $\gamma^0$ belongs to the Dirac $\gamma$-matrices satisfying $\{\gamma^\mu, \gamma^\nu\} = \gamma^\mu\gamma^\nu + \gamma^\nu\gamma^\mu = 2g^{\mu\nu}$. With respect to kinetic terms, the transformation (A.7) demands a sensible geometric redefinition of derivatives. As the latter depends on the field variation in a given direction, a suitable Lorentz invariant scalar operator $U(x, y)$ has to be defined that takes care of the local phase transformation at distinct points. Its transformation law has to satisfy

$$U(x,\, y) \rightarrow \mathrm{e}^{\mathrm{i}\alpha(x)} U(x, y) \mathrm{e}^{\mathrm{i}\alpha(y)}\,, \tag{A.8}$$

for which it is sufficient to define the comparator $U(x,\, y)$ as a pure phase. With this operator, the so-called covariant derivative in the direction of $\hat{u}$ can be defined in a geometrically reasonable way

$$\hat{u}^\mu \mathrm{D}_\mu \psi = \lim_{\epsilon \to \infty} \frac{1}{\epsilon} \big[ \psi(x + \epsilon \hat{u}) - U(x + \epsilon \hat{u},\, x) \psi(x) \big]\,. \tag{A.9}$$

In the infinitesimal limit, $U(x,\, y)$ is expanded to linear order, and the covariant derivative is obtained

$$\mathrm{D}_\mu \psi(x) = \partial_\mu \psi(x) + \mathrm{i}e A_\mu \psi(x)\,. \tag{A.10}$$

In the above derivation, the vector field $A_\mu(x)$ is defined by the coefficients of the linear displacement $\hat{u}$. The nomenclature of this so-called connection and its coefficient of proportionality $\mathrm{e}$ is chosen for obvious reasons. It is the photonic vector field of the Lagrangian function of Quantum Electrodynamics.

The consequences of this simple result are enormous. The very existence of the electromagnetic vector potential $A_\mu$ and even its transformation law are direct consequences of the postulate of local gauge equivalence. Thus, its internal structure is strictly determined by the inherent properties of the group of local symmetry transformations.

The above results and corresponding transformation laws

$$\mathrm{D}_\mu \psi(x) \quad \rightarrow \quad \mathrm{e}^{\mathrm{i}\alpha(x)} \mathrm{D}_\mu \psi(x) \tag{A.11a}$$

$$A_\mu(x) \quad \rightarrow \quad A_\mu(x) - \frac{1}{e} \partial_\mu \alpha(x) \tag{A.11b}$$

$$[\mathrm{D}_\mu,\, \mathrm{D}_\nu] = \mathrm{i}e F_{\mu\nu} \quad \rightarrow \quad \mathrm{i}e F_{\mu\nu} \tag{A.11c}$$

allow for a general procedure to construct invariant Lagrangians. They have to be constituted of quadratic terms in the fields $\psi$, their covariant derivatives $\mathrm{D}_\mu \psi$, the electronic field tensor $F_{\mu\nu}$ and its derivatives. Restrictions appear due to renormalizability which only tolerates operators up to a maximum mass dimension and by the necessity not to violate the discrete symmetries of charge, parity, and time. The result is remarkable, as these postulates and deeply rooted constraints uniquely lead to the Lagrangian of Quantum Electrodynamics

$$\mathcal{L}_{\mathrm{Maxwell-Dirac}} = \bar{\psi}(\mathrm{i}\gamma^\mu \mathrm{D}_\mu)\psi - \frac{1}{4}F_{\mu\nu}F^{\mu\nu} - m\bar{\psi}\psi\,. \tag{A.12}$$

The success of this approach motivated Yang and Mills to construct more general theories based on local gauge invariance with respect to higher dimensional groups. The conceptual changes are illustrated for the case of $SU(2)$ and can be analogously generalized to more complex groups. As the complex dimension of $SU(2)$ equals two, its elements act on an abstract doublet of Dirac spinors which transform as

$$\begin{pmatrix}\psi_1(x)\\ \psi_2(x)\end{pmatrix} \quad\rightarrow\quad \exp\left(\mathrm{i}\alpha^j(x)\frac{\sigma^j}{2}\right)\begin{pmatrix}\psi_1(x)\\ \psi_2(x)\end{pmatrix}. \tag{A.13}$$

Here, $\sigma^j$ are the Pauli sigma matrices which generate the group. To define the covariant derivative, the comparator $\tilde{U}(x,\, y)$ is introduced in strict analogy

$$\tilde{U}(x,\, y) \quad\rightarrow\quad \exp\left(\mathrm{i}\alpha^j(x)\frac{\sigma^j}{2}\right)\tilde{U}(x, y)\left[\exp\left(\mathrm{i}\alpha^j(y)\frac{\sigma^j}{2}\right)\right]^\dagger. \tag{A.14}$$

This time, $\tilde{U}(x,\, y)$ can be represented as a unitary matrix. As $\tilde{U}(x, x) = 1$, it can be expanded in terms of the generators of the group for infinitesimal separation of the points $x$ and $y$

$$\tilde{U}(x + \epsilon\hat{u},\, x) = 1 + \mathrm{i}\epsilon g\hat{u}^\mu A^j_\mu\frac{\sigma^j}{2} + \mathcal{O}(\epsilon^2)\,. \tag{A.15}$$

Thus, three vector fields are required to represent the geometric properties of the group, one for each Hermitian generator. This is an important consequence, as any candidate which is to be properly described, e. g., the theory of weak interactions, has to match the number of vector bosons which appear in the theory, which equals three in the above case.

The covariant derivative directly follows from (A.15)

$$\mathrm{D}_\mu = \partial_\mu - \mathrm{i}gA^j_\mu\frac{\sigma^j}{2}\,. \tag{A.16}$$

This time, the transformation rules cannot be as easily calculated as in the $U(1)$ case. This is due to the non-commutative space-dependent operators of (A.14). For infinitesimal transformations though, the vector potentials behave as

$$A^j_\mu(x)\frac{\sigma^j}{2} \;\to\; A^j_\mu(x)\frac{\sigma^j}{2} + \frac{1}{g}(\partial_\mu \alpha^j(x))\frac{\sigma^j}{2} + \mathrm{i}\left[\alpha^j(x)\frac{\sigma^j}{2},\, A^k_\mu(x)\frac{\sigma^k}{2}\right]. \tag{A.17}$$

Finally, the non-Abelian field strength tensor is obtained from the commutator of (A.16) and the group algebraic relations. Apart from being a pure four-dimensional curl, the geometric structure of $SU(2)$ leads to a further term quadratic in the components of the potential

$$F^j_{\mu\nu} = \partial_\mu A^j_\nu - \partial_\nu A^j_\mu + g\epsilon^{jkl}A^k_\mu A^l_\nu\,. \tag{A.18}$$

As a consequence, $F^j_{\mu\nu}$ is no longer gauge-invariant, since different directions of rotation in the isospin space of the spinor doublet possess independent field strengths. Nonetheless, the trace over all three directions leads to a gauge invariant kinetic term of the potentials

$$\mathcal{L}_{\mathrm{SU(2)-Maxwell}} = -\frac{1}{2}\mathrm{tr}\left(F^j_{\mu\nu}\frac{\sigma^j}{2}\right)^2 = -\frac{1}{4}\left(F^j_{\mu\nu}\right)^2. \tag{A.19}$$

In this way, the Lagrangian function of $SU(2)$ Yang-Mills theory is obtained

$$\mathcal{L}_{\mathrm{Yang-Mills}} = \bar{\psi}(\mathrm{i}\gamma^\mu \mathrm{D}_\mu)\psi - \frac{1}{4}\left(F^j_{\mu\nu}\right)^2 - m\bar{\psi}\psi\,. \tag{A.20}$$

Due to the quadratic terms in (A.18), the interaction vertices of Yang mills theory contain cubic and even quartic terms in the gauge potentials. In this manner, (A.20) describes a nontrivially interacting field theory.

The quantization process of the above Lagrangian has to be performed with great care in order to properly define functional integrals of the gauge fields. Within this procedure, several novel phenomena appear which are accompanied by intricate divergencies. These are related to the unphysical degrees of freedom of the Lagrangian fields and have to be mapped out by cancelation processes similar to the Ward identity in case of Quantum Electrodynamics. It took several years before a consistant renormalization could be formulated by t'Hooft [243, 244].

The above Lagrangian is but the simplest version of a non-Abelian gauge theory, and it was a rough way before the Standard Model of Elementary Particles could be formulated, based on a Yang-Mills theory with symmetry group $SU(3) \times SU(2) \times U(1)$.

## A.3 Lattice Gauge Theories

To tie up to the concepts of Chapter 3, it is more suitable to start from a different but equivalent definition of gauge theories. There, the interaction of an elementary particle with a gauge field is represented by path-dependent phase factors acquired by the particles [245, 246]. This effect is expressed in terms of the comparator $U(x,\, y)$ which was introduced in the previous section. In contrast to infinitesimal separation between $x$ and $y$, which has been considered for the derivation of the covariant derivative, a finite transformation is properly expressed via

$$U_{\mathcal{C}}(x_2,\, x_1) = \exp\left(-\mathrm{i}e\int_{\mathcal{C}} \mathrm{d}x^\mu A_\mu(x)\right)\,, \tag{A.21}$$

where the path $\mathcal{C}$ goes from $x_1$ to $x_2$. In non-Abelian theories, these phases become unitary operators acting on the isospin degrees of freedom. A straightforward generalization of (A.21) is nontrivial, as the potentials are matrix-valued fields that do not necessarily commute. This problem is resolved by a proper path-ordering prescription, similar to the well-known time-ordering in a perturbative treatment or radial ordering in conformal field theories. Let $\mathcal{C}(s)$ denote the parametrized path with $\mathcal{C}(0) = x_1$ and $\mathcal{C}(1) = x_2$, the comparator is formally written as

$$U_{\mathcal{C}}(x_2,\, x_1) = \mathcal{P}\left\{\exp\left(\mathrm{i}g\int_0^1 \mathrm{d}s\frac{\mathrm{d}x^\mu}{\mathrm{d}s}A^j_\mu(x)t^j\right)\right\}\,. \tag{A.22}$$

Here, $t^j$ are the Hermitian generators of the symmetry group $\mathcal{G}$ of dimension $d(\mathcal{G})$, and the exponential is to be read as its Taylor series, where products of group elements $G_\mu(s) \equiv A^j_\mu(s)t^j$ are ordered as

$$\mathcal{P}\Big\{G_\mu(s_1)G_\mu(s_2)\Big\} = \begin{cases} G_\mu(s_1)G_\mu(s_2) & s_1 > s_2 \\ G_\mu(s_2)G_\mu(s_1) & \text{otherwise}\,. \end{cases} \tag{A.23}$$

Under local gauge transformations, the so-called Wilson line (A.22) is only sensitive to the gauge elements at the extremal points $x_j$ of the path, e. g., in the case of $SU(2)$

$$U_{\mathcal{C}}(x_2,\, x_1) \;\;\rightarrow\;\; \exp\left(\mathrm{i}\alpha^j(x_2)\frac{\sigma^j}{2}\right)U_{\mathcal{C}}(x_2,\, x_1)\left[\exp\left(\mathrm{i}\alpha^j(x_1)\frac{\sigma^j}{2}\right)\right]^\dagger\,. \tag{A.24}$$

An immediate consequence of (A.24) is the gauge invariance of the Wilson loop for closed paths, reading

$$\mathcal{W}(\mathcal{C}) \equiv \mathrm{tr}\Big(U_{\mathcal{C}}(x_1,\, x_1)\Big) \tag{A.25}$$

which is independent from the starting point $x_1$. It is the underlying geometrical picture of $\mathcal{W}(\mathcal{C})$ by which the gauge invariance of (A.19) can be understood.

With the above tools, an elegant formulation of gauge fields on a Euclidian space-time lattice is accessible which has been developed by Wilson in the context of Quantum Chromodynamics [247]. It is based on the concept of gauge fields as path-dependent phase operators in a way which exactly preserves local symmetry. In this approach, each nearest-neighbor pair of lattice sites $(j,k)$ is associated with an element $U_{jk}$ of the gauge group. The inverse path is defined by $U_{kj} = (U_{jk})^\dagger = (U_{jk})^{-1}$. These unitary operators represent elementary Wilson lines on the lattice and are assumed to be path-ordered. A vector potential is consistently introduced by

$$U_{jk} = \exp(\mathrm{i}gA_\mu a)\,, \tag{A.26}$$

where $a$ denotes the lattice spacing and $\mu$ coincides with the orientation of the link $(j,\ k)$. The space-time coordinate of $A_\mu$ may be defined half-way on the bond. This convention becomes irrelevant in the continuum limit. The goal is to formulate a Lagrangian that reduces to the gauge part of (A.20). As the field strength tensor is a four-dimensional curl, Wilson proposed the following action

$$S = \sum_{\Box} S_\Box\,, \tag{A.27}$$

where $S_\Box$ denotes the action of each elementary lattice plaquette, given by the trace of the counter-clockwise Wilson loop around it. By virtue of (A.26), it can be proven that

$$S_\Box = \frac{2d(\mathcal{G})}{g^2}\left[1 - \frac{1}{d(\mathcal{G})}\Re\big\{\mathrm{tr}(U_{jk}U_{kl}U_{lm}U_{mj})\big\}\right] \tag{A.28}$$

reduces to the Yang-Mills action for the gauge potentials.

The definition of fermionic spinors on a lattice still lacks complete understanding and is accompanied with difficulties that are beyond the scope of this thesis. It is sufficient to note that discretization demands for counterterms which preserve physical features but destroy inherent symmetries. Apart from this, the process is quite similar to that of gauge fields, though the fermionic fields are situated on the lattice sites instead of the bonds. Kinetic elements of the Lagrangian become hopping terms, and the covariant derivative couples the isospin degrees of freedom with the gauge fields. In this manner, the full interacting action of gauge fields and fermionic spinors is obtained

$$S = \sum_{\Box} S_{\Box} + \frac{1}{2}\mathrm{i}a^3 \sum_{\langle j,k\rangle} \bar{\psi} j \big(1 + \gamma_\nu \hat{u}_\mu\big) U_{jk}\psi_k + (a^4 m + 4a^3) \sum_j \bar{\psi}_j \psi_j \,. \tag{A.29}$$

Here, $\langle j,\ k\rangle$ denotes the sum over pairs of nearest neighbors and $\hat{u}_\mu$ is the unit vector in the $\mu$-th direction.

Starting from (A.29), the partition function is derived from the path integral in which the integration is taken over all field configurations and the compact group measure. Distinct quantum phases correspond to discontinuities in the correlation functions of order parameters. As discussed in Chapter 3, one prominent order parameter for the pure gauge sector is the Wilson loop. It distinguishes between weakly interacting and confined phases. The correlation functions of such an order parameter rely on the full integration of all dynamical variables, which include fermionic spinors and the potentials themselves. The question remains open to what extent it is feasible to sample a representative statistical average of various important Wilson line configurations in an experiment of ultracold atomic gases.

# Appendix B

# Numerical Methods

Numerical methods are efficient tools for calculations on quantum many-body systems [248]. Unfortunately, the limits of computational ressources are soon within reach and shortly later exceeded. This imposes severe constraints on accessible system sizes and demands for a sophisticated treatment of the problem in focus. This especially applies to exact approaches which have been extensively used to derive the results of Chapter 2. As the name implies, these methods implement a complete quantum dynamical treatment of the system. To get an impression of the meaning of this statement, consider a one-dimensional half-integer spin-chain with but four different electronic configurations per lattice site: $|0\rangle, |\uparrow\rangle, |\downarrow\rangle, |\uparrow\downarrow\rangle$. Thus, the Hilbert space of such a chain with $M$ sites has the dimension

$$d_M = 4^M \,.$$

The assignment of all Hamilton matrix elements with double precision on a 32-Bit machine requires $8(d_M)^2$ Bytes of memory, i. e., slightly more than 2GB for $M = 7$. Obviously, such a straightforward approach is somewhat hopeless.

A way to reduce this problem of dimensionality is the identification of symmetries. The corresponding similarity transformations lead to a block-diagonal Hamiltonian which may be diagonalized piecewise. Since the complexity of a full diagonalization scales cubically in the basis dimension, the linear block-by-block scheme saves a lot of computational runtime. It should be kept in mind that, though the complexity of the problem is reduced, the program code gets more and more restricted to the set of physical problems which share all implemented symmetries. This loss of generality should be balanced with the gain in speed and resources.

Even if just a single specific problem has to be solved, the number of conserved quantities is finite. Otherwise, the problem is integrable and the above numerical approach *a priori* inappro-priate. Thus, even if the rule of Moore's law is given for granted, and the computational power is consecutively doubled in constant intervals of time, the programmer still has to be very patient to treat one more lattice site. The simple but rigorous conclusion is that the dream of a complete numerical solution has to be abandoned.

Fortunately, there are numerical algorithms, which, despite this, allow for exact calculations in considerably larger systems, though at the expense of being restricted to the lowest lying energy levels [249, 250]. In many problems of interest, the temperatures considered are very low compared to the energy scales of excitations in the system. Then, the statistical weight of the accessible states becomes sufficiently large to allow for an excellent perturbative analysis. These so-called block-diagonalization routines boil down to the "effective" implementation of a matrix-vector multiplication. As it can be guessed, the interpretation of "effective" is a very specific issue. Generally, the best choice is a good balance between memory and processor usage.

The easiest way to save memory is achieved when matrix elements do not have to be stored but can be calculated "on the fly". Very efficient algorithms have been developed for this purpose which cover a large class of systems [251]. A very crucial feature of these routines is the efficient representation and identification of basis elements. This is demonstrated for the concrete example of the fermionic dipoles in Chapter 2. There, all matrix elements are calculated from the elementary terms

$$\langle m_1, \ldots, m_N | \, a_i^\dagger a_j a_k^\dagger a_l \, | n_1, \ldots, n_N \rangle \,. \tag{B.1}$$

The basis elements are $N$-particle Fock states, $m_j$, $n_j$ denote the single-particle angular momenta, and $a_j^\dagger$, $a_j$ create and annihilate a particle with angular momentum $j$, respectively. To derive (B.1), the following steps have to be implemented. First, it has to be checked, if $a_l$ is contained in the ket-state. If $k = j$, this check equally applies to $a_j$. Second, both angular momenta have to be replaced by $a_i$ and $a_k$. Then, the Pauli principle is consulted to check if the result is nonzero. These manipulations are repeated billions of times, so optimization is worthwhile. A very efficient basis representation is obtained by mapping the Fock state to an integer

$$|m_1, \ldots, m_N\rangle \rightarrow N_m = \sum_{j=1}^{N} 2^{m_j} \,. \tag{B.2}$$

It allows for bitwise check and shift operations between distinct state vectors, which are the fastest manipulations available. Of course, $N_m$ exceeds the range of large numerical integers at some point. Thus, effectively, each basis element is mapped to an array of integers. As a consequence, the transformation (B.2) is only efficient for sufficiently small $m_j$. The next vastly time-consuming step is the identification of the matching bra-vector. If the Hilbert space is very large, a suitable index set helps to overcome this problem though at the expense of memory. The latter step may of course be applied to every kind of basis.

Even if the above effort is made to optimize the derivation of matrix elements, the problems of Chapter 2 do not allow for "on-the-fly" implementations, as the calculational time for a single Hamilton element exceeds the I/O speed of available hard disks. As a consequence, the complete set of matrix elements has to be stored somehow. This drawback is not as bad as it seems. The true advantage of block-diagonalization methods is that they solely rely on an exact matrix-vector multiplication routine. Thus, only non-vanishing matrix elements have to be stored whereas the complete diagonalization needs the full $d_M^2$-sized array in memory. As a result, the size of a numerically tractable problem is determined by the sparseness of the Hamilton alone. The significance of this innocent issue is enormous in the considered two-dimensional problems. For the largest solved system of $N = 12$ dipolar fermions, it reduces the storage size of the Hamilton matrix from $525000$ GB to $35$ GB.

## B.1 Sparse Matrices

An efficient implementation of sparse matrices is a crucial ingredient for block-diagonalization codes, if matrix elements cannot be proceeded "on the fly". In this context, efficiency means the optimization of memory usage and the availability of a fast matrix-vector multiplication routine. The plainest thinkable method is to store each element $A_{jk}$ of a $D \times D$ matrix as a coordinate triple

$$A_{jk} \rightarrow \{j,\, k,\, A_{jk}\}\,. \tag{B.3}$$

In this way, three arrays are stored and the memory usage equals $3N_A\langle\text{typesize}\rangle$, where $N_A$ denotes the number of matrix elements. One array $R$ is associated with the row, the second $C$ with the column, and the third $V$ yields the value. For a random matrix $A$, this method is suitable, if $3N_A \leq D^2$, otherwise full storage is favored. Simultaneously, (B.3) is the optimal choice for a random matrix with $N_A \leq D$.

If symmetries are present, the balance between storage optimization and speed of multiplication becomes important. The most prominent symmetry, which is in fact the only one used in the coded methods of Chapter 2, is hermiticity. Then, only one triangle of $A$ has to be stored. Still, the storage scheme (B.3) remains optimal for a random upper/lower diagonal with $N_A^{\mathrm{ud/ld}} \leq D$. For a larger number of elements, the storage size can be further decreased. This is achieved by so-called column or row compression. In this representation, one of the arrays $R$ and $C$ is compressed. Without loss of generality, the upper diagonal of a Hermitian matrix $A$ and the row column $R$ are considered. Then, $R_j$ is simply defined as the index of the first upper diagonal element of the $j$-th row. To address this row, the start index $R_j$ and the end index $R_{j+1} - 1$ have to be evaluated. Then, for each of these indices, $C_{R_j}$ denotes the column and $V_{R_j}$ the value of the corresponding element. Apparently, to work properly for the $D$-th row, $R_{D+1}$ has to be defined as the index of the last upper diagonal nonzero element of $A$ added by one. This equals the total number of nonzero entries if the index range is chosen to start from zero. This sparse matrix format requires a memory usage of $(2N_A^{ud/ld} + D + 1)\langle\text{typesize}\rangle$ and is far more efficient than (B.3) for typical matrices. For clarity, the following explicit, though non-sparse, example can be consulted. All indices are assumed to start from zero.

$$\begin{pmatrix} 1 & 2 & 3 & 4 \\ 2 & 5 & 6 & 7 \\ 3 & 6 & 8 & 9 \\ 4 & 7 & 9 & 10 \end{pmatrix} \longrightarrow \begin{array}{rcl} R & = & \{0,\, 4,\, 7,\, 9,\, 10\} \\ C & = & \{0,\, 1,\, 2,\, 3,\, 1,\, 2,\, 3,\, 2,\, 3,\, 3\} \\ V & = & \{1,\, 2,\, 3,\, 4,\, 5,\, 6,\, 7,\, 8,\, 9,\, 10\} \end{array} \tag{B.4}$$

Apparently, the addressment of a specific matrix element is non-trivial. If $A_{jk}$ is needed, the only way to obtain its value is to scan the set $\{C_l\}$ with $R_j \leq l < R_{j+1}$. If one of these $C_l$ equals $k$, the value of the matrix entry is $V_l$ and $0$ otherwise. Fortunately, the multiplication

$$\vec{y} = A \cdot \vec{x} \tag{B.5}$$

of a vector $\vec{x}$ by such a row-compressed and upper diagonally stored matrix $A$ does not rely on this and may be easily implemented in an efficient way. For this, it only has to be understood that each non-diagonal element of the matrix contributes two elementary summations to the resulting vector $\vec{y}$ due to hermiticity, whereas each diagonal element only yields one. Thus, all matrix elements are directly processed row by row. Given an element of the $j$-th row with index $R_j \leq l < R_{j+1}$, $C_l$ denotes its column coordinate. The contribution to (B.5) reads

$$\begin{array}{rccccl} & V_l x_{C_l} & \stackrel{+}{\longrightarrow} & y_l & & \\ \text{and} & V_l x_l & \stackrel{+}{\longrightarrow} & y_{C_l} & \text{if } l \neq C_l & . \end{array} \tag{B.6}$$

In this way, the multiplication routine works straightforward and is very fast.

## B.2 The Davidson Algorithm

This section presents a type of "exact" algorithm which is suitable to address very large matrix eigenvalue problems that can not be diagonalized completely. "Exact" means that it is based on the knowledge of the full Hamilton operator. Thus, matrix elements either have to be calculated whenever needed or have to be stored in a suitable sparse format, as discussed previously. The only manipulation needed is matrix multiplication. As stated above, this may be processed independently from the storage format in a term by term scheme (B.5). The presentation will be without proof and is only meant to provide a conceptual approach. The Hamiltonian is considered to be symmetric and real, which applies to the operators treated in Chapter 2.

Given the above requirements, all of these so-called block-diagonalization routines are based on the same idea. They calculate an intermediate set of vectors $\widehat{u}_j$, whose linear combinations are meant to posses a good overlap with the true eigenvectors $\vec{\lambda}_j$ of the problem. If these basis vectors are iteratively derived in a clever way, the approximate expansion of $\vec{\lambda}_j$ in terms of $\widehat{u}_j$ will converge to the true eigenvectors. The astonishing result is that the above scheme even works if the dimension $D$ of the Hilbert subblock spanned by the $\widehat{u}_j$ is much smaller than the total Hilbert space dimension $N$.

The method starts with the definition of the reduced eigenvalue problem in the $D$-dimensional $\langle \widehat{u}_j \rangle$ subspace, matrix operators are denoted by bold letters for reasons of clarity,

$$(\mathbf{U}^T \mathbf{H} \mathbf{U}) \vec{a}_k = \mu_k^D \mathbf{U}^T \mathbf{U} \vec{a}_k \,, \quad k = 1, \ldots, D \,. \tag{B.7}$$

Here, the $j$-th column of $\mathbf{U}$ equals the coefficients of $\widehat{u}_j$, and $\vec{a}_k$ is the $k$-th eigenvector of the problem with eigenvalue $\mu_k^D$. The index $D$ is meant to remind of the dimensionality of the subspace. The dimensionality $D$ is hereby assumed to be sufficiently small, so (B.7) can be completely diagonalized by standard routines.

The approximate expansion of the true eigenvectors in terms of these $\vec{a}_k$ is given by

$$\vec{x}_k = \sum_{j=1}^{D} (\vec{a}_k)_j \widehat{u}_j \,. \tag{B.8}$$

The eigenvalues $\mu_k^D$ of the problem (B.7) are always assumed to be ordered in $k$. Due to the variational principle of Hylleraas-Undheim-MacDonald [252, 253],

$$\lambda_j \leq \mu_j^D \,. \tag{B.9}$$

Furthermore, if the Hilbert subspace $\langle \widehat{u}_j \rangle$ is truncated by a single eigenvector $\widehat{u}_j$, it follows

$$\mu_j^D \leq \mu_j^{D-1} \,. \tag{B.10}$$

In this way, if the subspace dimension $D$ is successively increased, a monotonously decreasing series of eigenvalues is obtained, which apparently converges to $\mu_j^D = \lambda_j$ for $D = N$.

The question remains, which vectors $\widehat{u}_j$ are a good initial choice and which vectors should be iteratively added to the set. A general answer to this problem is not available, as it strongly depends on the specific problem to be solved. Of course, if the initial set of vectors is chosen orthogonal to the lowest lying eigenvectors, the algorithm will not work or converge to some local minimum of the energy. To avoid this, several initial guesses are probed independently. Their (hopefully coinciding) results and the speed of convergence generally provides useful information for other quantum numbers. This, of course, only applies, if the Hamiltonian could be split to independent diagonal blocks due to a symmetry of the problem, e. g., blocks of different $L^z$ from those Chapter 2. A good first guess are always the canonical unit vectors in direction of the lowest diagonal entries of the Hamiltonian matrix.

More subtle on the mathematical level is the proper choice for the extension of the $\widehat{u}_j$ subspace after each step of iteration. So far, there is, to the author's knowledge, no rigorous proof of optimal, or rather say "stable" convergence, at least on the level of a sufficiently small subspace dimension $D$. However, the specific block-diagonalization algorithm invented by Davidson, which is extensively discussed in [254], relies on a very natural ansatz. Given the approximate expansion (B.8), this will somehow deviate from the true eigenvectors

$$\vec{\lambda}_k = \vec{x}_k + \vec{\delta}_k \,. \tag{B.11}$$

These vectors of deviation $\vec{\delta}_k$ point into the abstract "direction" of the wanted eigenvectors and are thus good candidates. They are of course unknown, because otherwise, the $\vec{\lambda}_k$ would be known. Yet, they can be approximated by the parameters obtained from the last iteration. Starting from the exact residual vectors

$$\vec{r}_k \;=\; (\mathbf{H} - \lambda_k \mathbf{1})\vec{x}_k \;== -(\mathbf{H} - \lambda_k \mathbf{1})\vec{\delta}_k \,, \tag{B.12}$$

the eigenvalues are replaced by the Raleigh quotients

$$\widetilde{\lambda} = \frac{\vec{x}_k^T \mathbf{H} \vec{x}_k}{\vec{x}_k^2} \,. \tag{B.13}$$

Then, $\mathbf{H}$ is considered to first order, i. e., replaced by its diagonal $\mathbf{D}$

$$(\mathbf{H} - \widetilde{\lambda}_k \mathbf{1}) \approx (\mathbf{D} - \widetilde{\lambda}_k \mathbf{1}) \,. \tag{B.14}$$

With the above ingredients, the $\vec{\delta}_k$ are approximated by

$$\vec{\delta}_k \approx -(\mathbf{D} - \widetilde{\lambda}_k)^{-1} \vec{r}_k \,. \tag{B.15}$$

The next puzzle is to choose an optimal number of vectors by which the subspace is extended. In the original proposal of Davidson, solely one vector was added per iteration. If sufficient memory is available, the efficiency can be increased [255], when one vector is added for each of the $N_0$ desired eigenvectors to the block basis $\widehat{u}_j$ in each step. In real problems, memory is often the limiting quantity. Thus, a maximum subspace dimension $D_{\max}$ has to be defined. If the latter is exceeded, only the eigenvectors $\widehat{u}_k$, corresponding to the lowest $N_0$ eigenvalues, are kept. Then, the process is repeated until a self-defined criterion of convergence is met.

A schematic proposal for a time-efficient implementation of this Davidson-Liu algorithm is presented in [254].

# APPENDIX C

# The Necessity of a Finite Size Approach to the "Moth"

Several advantages and disadvantages of the transfer matrix and finite size approach to the generalized Hofstadter problem were discussed in Chapter 3.

Of the addressed criteria, the equivalence of distinct $\vec{A}$ with respect to numerical difficulty is the most important argument in favor of the finite size treatment. In an infinite system, only a few suitable gauge potentials allow for the reduction of the full $4 \times 4$ transfer matrix problem to lower dimensions. Even if this property seems to be but of marginal importance, it rules out the efficient analysis of a wide class of gauge potentials.

Given a $4 \times 4$ matrix eigenvalue problem, it seems quite harmless indeed. The characteristic polynomial is of rank $4$, and basic algebra promises its full separability. Unfortunately, this does not help at all. This algebraic aproach can only be seized if the elements of the block transfer matrix $Q$ are analytically known. As the latter results from the product of $q$ successive matrices $A(m)$, the number of summands grows exponentially with the period $q$. Thus, manipulations depending on the coefficients of the characteristic polynomial are not tractable. Furthermore, as it can be guessed from the fact that $4$ is the highest rank for a definitely separable polynomial, the dynamics of eigenvalues is already highly complicated on this level, especially as $Q$ is not Hermitian. Since $\det(Q) = 1$, the product of all four complex eigenvalues equals

$$\prod_{j=1}^{4} \lambda_j = 1\,. \tag{C.1}$$

It was mentioned previously that the absolute value of the oscillatory amplitude of $\mathrm{tr}(Q)$, as a function of the Fourier mode $\nu$, grows exponentially with the period $q$. Since

$$\mathrm{tr}(Q) = \sum_{j=1}^{4} \lambda_j\,, \tag{C.2}$$

at least one of the eigenvalues has to enormously large on a wide parameter range in order to satisfy (C.1).

The severe consequences of this behavior are illustrated for the specific case of the $U(2)$ "moth" problem. There, the eigenvalues move pairwise in absolute value due to the separation of $Q$ into two more or less equivalent $2 \times 2$ subproblems. Already for small periods $q \approx 50$, which correspond to very basic magnetic flux parameters, e. g., $\alpha_1 = 1/5$ and $\alpha_2 = 1/9$, the absolute value of the trace oscillates between $10^{20}$ and $10^{-20}$ for $\nu \in [0, \pi]$. This demands for a numerical precision of at least $40$ digits.

Even worse, the two pairs of eigenvalues approach unit modulus on an interval of negligible size for most of the "moth" subbands, cycle diametrally around the complex unit sphere and drift off towards $10^{20}$. At this point, even arbitrary precision libraries are no more dynamically controllable on a sufficient level. Calculations were performed for periods up to $q = 100$, which allows for only a handful of fractions, which may be studied. There, typical $\nu$ intervals, which have to be detected to guarantee the existence of an eigenenergy, are of the order of $10^{-80}$, whereas $\max_{j,\nu} |tr(Q)| = 10^{100}$. The slowdown due to a number precision of at least $300$ digits to compensate for numerical errors in matrix multiplication and other routines is enormous. If these arguments have not yet ruled out the above approach, this is finally accomplished by the necessity to analyze periods $q \approx 10^3 - 10^4$ to allow for a sufficiently fine-grained set of data points. This is obligatory in order to draw general conclusions which do not distinguish between the rational and irrational nature of the magnetic flux parameters $\alpha_j$.

As the above arguments apply for general vector potential configurations, the finite size analysis is not only more closely related to experimental conditions but also the solely tractable approach.

# Acknowledgments

First of all, I would like to express my gratitude to Maciej Lewenstein, my “Doktorvater”, who gave this word a meaning. I feel honored that he gave me the opportunity to work in his group and to share the spirit of AGL. To a vast extent it is his physical intuition, which inspired the essence of this work. I am very much indebted to Walter Apel, Mikhail Baranov, Nuria Barbeán, Luis Santos, and Eric Jeckelmann for countless fruitful discussions and for sharing their experience with me. I am very grateful to Holger Frahm who very kindly accepted being the “godfather” of the project and welcomed me to his group.

A lot of ideas of this work were inspired by scientific exchange with colleagues from all over the world. This would not have been possible without the enthusiasm of researchers who take a lot of effort to establish network projects and keep them vividly alive. For this reason, I would like to thank Wolfgang Ertmer, Olaf Lechtenfeld and Jan Arlt.

If there is a “father” of a group, there has to be a “mother”. It was Christel Franko who took care of all of us and never hesitated to spend far more time than necessary to find a way out of every thinkable problem. I thank you very much.

If almost the whole institute meets on the corridors, a very likely reason is that the file server is down. Thus, at this point I want to thank all administrators for keeping a continuously changing system running. They have been doing indeed a great job during the last few years. The same applies to the administrative staff.

I am furthermore very grateful to Hans-Ulrich Everts, Eric Jeckelmann, and Anna-Katharina Kraemer for reviewing and proof-reading this work. I thank Christian Sämann and Benjamin Knispel for the very comfortable and joyful atmosphere in the office, Alex Cojuhovschi for establishing a very efficient coffee-sharing system, Hans-Ulrich Everts for at least half a ton of apples and joyful discussions on life, the universe and everything, Paolo Pedri for countless competitions in quantitative communication theory, Henning Fehrmann, Jarek Korbicz, and Alem Mebrahtu for the great time in Berlin, Oleksiy Kolezhuk and Laurent Sanchez-Palencia for sharing a good glass of wine with me after a long day, Carsten Luckmann for his ambitious fight against the “wrong operating system”, Helge Kreutzmann for encrypting everything, Kai Eckert for a lot of fun in Frankfurt, Philipp Hyllus for his “TGRBT” button, Florian Hulpke for destructive criticism on movies, Guillaume Palacios for philosophical debates on soccer, Ujjwal and Aditi Sen(De) for delicious meals and really hot peppers, Anna Sanpera for the hospitality in Barcelona, the “stringons” for contributing exciting topics to the PSO(12) seminar and their hilarious spirit, Martin Paech for figure artwork, Carsten von Zobeltitz and Jörn Bröer for cookies and cappuccino, and the incredible soccer team who fought for the trophy.

Finally, I would like to thank my parents and my brother, Gabriele and Matthias Kraemer, and, most of all, Anna-Katharina for their unconditional encouragement and support.

# Selbständigkeitserklärung

Hiermit versichere ich, die vorliegende Dissertation selbständig und unter ausschließlicher Verwendung der angegebenen Hilfsmittel angefertigt zu haben.

I hereby declare having conducted this thesis entirely on my own using solely the quoted resources.

Hannover, den 2. November 2006

# Curriculum Vitae

## Personal Data

| | |
|---|---|
| Name | Klaus Osterloh |
| Date of Birth | April 15, 1977 in Herford |

## University Education

| | |
|---|---|
| 10/1997 — 10/99 | Undergraduate studies of Mathematics and Physics at the University of Hannover |
| 09/1999 | *Vordiplom* in Physics, |
| 10/1999 — 11/02 | Graduate Studies of Physics at the University of Hannover |
| 10/2001 — 11/02 | Diploma thesis project “*Conformal Field Theory Bulk Wavefunctions with Application to the Fractional Quantum Hall Effect*” in the group of Prof. Dr. O. Lechtenfeld at the Institute for Theoretical Physics in Hannover |
| 11/2002 | *Diplom* in Physics |
| 2003 — 2006 | PhD thesis project “*Ultracold Atomic Gases in Artificial Magnetic Fields*” in the group of Prof. Dr. M. Lewenstein at the Institute for Theoretical Physics in Hannover |

## School Education

| | |
|---|---|
| 1983 — 1987 | Primary School Babbenhausen/Oberbecksen |
| 1987 — 1996 | Immanuel-Kant-Gymnasium Bad Oeynhausen |
| 06/1996 | High school graduation, *Abitur*, Major subjects: Physics and Mathematics |

## Military Service

| | |
|---|---|
| 09/1996 — 07/97 | Artillery battalion in Dülmen |